\documentclass[a4paper]{article}
\usepackage{fullpage}
\usepackage[utf8]{inputenc}
\usepackage{newpxtext}
\usepackage{authblk}
\usepackage{graphicx}
\usepackage{subcaption}
\usepackage[font=small]{caption}
\usepackage{url}
\usepackage{textgreek}

\title{Reading from External Memory}
\author{Ruslan Savchenko \\ {\small\texttt{ruslan.savchenko@gmail.com}}}
\affil{Yandex School of Data Science}
\date{}

\begin{document}
\maketitle

\begin{abstract}
Modern external memory is represented by several device classes.
At present, HDD, SATA SSD and NVMe SSD are widely used.
Recently ultra-low latency SSD such as Intel Optane became available on the market.
Each of these types exhibits it's own pattern for throughput, latency and parallelism.
To achieve the highest performance one has to pick an appropriate I/O interface provided by the operating system.
In this work we present a detailed overview and evaluation of modern storage reading performance with regard to available Linux synchronous and asynchronous interfaces.
While throughout this work we aim for the highest throughput we also measure latency and CPU usage.
We provide this report in hope the detailed results could be interesting to both researchers and practitioners.
\end{abstract}

\section{Introduction}
\label{sec:introduction}

During the last 10 years external memory market has experienced an unprecedented revolution.
Well known hard disk drives has been present since 1970s and their performance has constantly being improved.
To grasp the pace of this progress, bandwidth increased by a factor of four since the beginning of the century.
Today typical speed is about from 100 to 200 megabytes per second.
Solid state drives appeared in 2009 and raised external memory speed up to 300 megabytes per second.
Later NVMe SSD increased the speed further up to 3 gigabytes per second.
For manufactures it makes more sense to sell NVMe SSD so vendors are switching to this interface.
Industry data-centers are switching to NVMe SSD as well.

Typical NVMe SSD throughput for reading is about 3 gigabytes per second.
This is quite a large value for modern system.
For example it is close to the bandwidth of 25 gigabit network interface.
Thus it is enough to saturate modern network interface with a single NVMe SSD.

It is not only bandwidth which solid state drives improve but also latency.
Hard disk drive requires about 12 milliseconds to read a random block.
Solid state drive needs only a few hundreds of microseconds to get the job done.
Recently ultra-low latency SSD pushed this even further to astonishing single-digit number of microseconds.
As a result, external memory latency dropped by a factor of thousand in just a decade.

The random read workload throughput depends on latency.
If one reads small random blocks from HDD then all the time would be wasted on the arm positioning and instead of 200 MB/s the resulting data transfer speed would be less than a megabyte per second.
It is a common technique to increase block size and use cache when fetching data from HDD.
In contrast, one can saturate the whole NVMe SSD bandwidth with random reads even if the block size is only 4 kilobytes.

This revolution in hardware capabilities requires the software interfaces to change as well.
The well-known \texttt{read} interface form 1960s is not useful for anything serious anymore.
It was replaced by \texttt{pread} which allows one to specify the offset within the same system call.
Next was \texttt{preadv} which makes possible to fetch data into multiple buffers (yet all the data should be continuous on the drive).
It followed by \texttt{preadv2} which introduced some flags to ask kernel for enhancements (we are interested in \texttt{RWF\_HIPRI} to make reading done via polling and save CPU time form unnecessary interrupts).

Although all above system calls are synchronous input/output operations are asynchronous by their nature.
There were several attempts to express this in programming interface.
POSIX AIO made it possible to make several requests and execute them in separate threads in parallel.
It was going to be replaced by Linux aio which used a kernel queue for requests.
Linux aio requests are gathered in kernel to be executed asynchronously and even out of order.
Finally Linux 5.4 introduced \texttt{io\_uring} interface which addresses some drawbacks of Linux aio and aims to be the major asynchronous userspace data transfer interface.
Both Linux aio and \texttt{io\_uring} introduced new system calls.
On the opposite POSIX AIO is implemented with just a few threads defined in GNU libc.
In fact, many applications implement their own I/O thread pool which works like POSIX AIO.

For data managing software all this means that old fashioned way to read large blocks and store them in a cache has to be reconsidered.
The amount of data is constantly increasing and sometimes it is impossible to keep even the most relevant things the main memory.
The popularity of distributed key-value databases nudges service developers to request random keys during online processing.
As a result we have a huge amount of random external memory fetches.

In this work we present our experiments with different types of external memory devices.
Our main goal is to choose the best way to read if reading is always performed from a storage.
So we disable the Linux filesystem cache by specifying \texttt{O\_DIRECT} flag.
The cache is wonderful when hit, however it introduces additional latency when data is fetched from a device.
To stress this argument we also show a few measurements taken without \texttt{O\_DIRECT} flag.

The paper is organized as follows.
Sections~\ref{sec:overview},~\ref{sec:configuration} give a brief overview of hardware and specify models used in our experiments.
Section~\ref{sec:single} studies single read latency.
Section~\ref{sec:sequential} studies synchronous reading from a single thread.
Section~\ref{sec:multithread} explores extension to multiple threads.
This concludes synchronous interfaces. Further we study study asynchronous I/O.
Section~\ref{sec:aio} evaluates Linux aio interface.
Section~\ref{sec:aio54} compares Linux aio performance in 4.19 and 5.4 kernels.
Section~\ref{sec:uring} gives benchmarks for the \texttt{io\_uring} interface.
Section~\ref{sec:uringtune} investigates \texttt{io\_uring} extended features.
Section~\ref{sec:summary} summarizes all our experiments.
Section~\ref{sec:relatedwork} covers related work.

\section{Hardware Overview}
\label{sec:overview}

In this section we briefly describe modern external memory hardware.
We have HDD, SATA SSD, NVMe SSD, and Intel Optane installed in our testing machine.
While it is impossible to cover all the details in a single section we try to give a grasp of what they are and how they differ from each other.
For more information refer to operating systems textbook~\cite{ArpaciDusseau18-Book} or materials covered in section~\ref{sec:relatedwork}.

\subsection{Hard Disk Drive}

Probably the most well-known external memory device is the hard disk drive.
Internal structure of HDD resembles that of gramophone record: the information is stored on tracks on aluminum platter which rotate at high speed (however HDD tracks are concentric circles while gramophone record has a single spiral track).
To read or write data a magnetic head is used.
The magnetic head is attached to a mechanical arm which is used to position the head over the right track.
To access data the drive first has to move the arm over the right track and then wait for the platter rotation to bring the data right under the head.

It is easy to calculate how long it takes to wait for the right angle.
The platter rotation speed is usually included in a drive title as RPM (rotations per minute) keyword.
For a typical drive which rotates at speed 7200 RPM a single rotation takes 8 milliseconds.
This means that we wait 4 milliseconds on average for our data to appear under the head.
As we will see in the next sections it takes 12 milliseconds on average to access random block on HDD.
Given these two numbers we get that arm positioning (which is usually called track seeking) takes 8 milliseconds.

For small blocks the data transfer time is much less than seeking time.
Only if block is large enough it becomes noticeable.
For our HDD this happens when the block size is about 256 kilobytes.
It means that for a random access there is no difference if the block size is 256 kilobytes or less.
Note that this is quite large compared to both filesystem block size (4 kilobytes) and traditional database block size (about 8 kilobytes).

The latency of 12 milliseconds imposes a huge restriction on the number of operations per second.
It is easy to see that if every access is expected to be random then the drive can perform only 80 operations per second.
In practice this can be a bit higher because the OS driver or the drive internal controller itself can reorder requests to reduce the total distance of arm movements.
Nevertheless the number of requests per seconds is far too lower than what is expected from modern information systems.

\subsection{Solid State Drive}

Flash memory has been developed since 1980s.
It avoids some problems of the HDD but introduces some others.
There are no mechanical parts in the SSD therefore a random read doesn't have to wait for slow arm motion or platter rotation.
It just need to access proper memory cell electrically and data transfer can be performed immediately.
Thus SSD suits random reads better. 

Nevertheless SSD is exposed to another problem.
It happens that one cannot change flash memory cell value easily.
To be able to write data the SSD first needs to clear a large area of memory called block in SSD terminology.
Usually block size is in range form 128 to 512 kilobytes.
After this SSD block is cleared small portions of it called pages can be used to write new data.
Typical page size is 4 kilobytes which exactly matches the filesystem block.
After data is written into a page this page becomes read-only.
The only way to write new data to the page is to first clear the whole block that contains it.

It is not surprising that SSD internally remaps pages.
When a filesystem logically overwrites a page an SSD allocates a new page and writes data into it instead of rewriting the whole block.
After a while the SSD has to do defragmentation: used pages are gathered together and old blocks are cleared.
So unused pages become available for new writes.

The background defragmentation process affects SSD write performance.
Sometimes write latency is so high that it looks more like SSD is frozen for a while.
Moreover the latency pattern is unpredicted and can depend on factors such as free space.
We won't consider write in this report but we believe reader should be aware of this problem.

\subsection{Non-Volatile Memory Express}
\label{subsec:nvme}

Serial ATA interface has limited capacity: 3 gigabits per second for SATA II and 6 gigabits per second for SATA III.
This allows SATA to transfer 384 or 768 megabytes per second. 
To bypass this restriction vendors started attaching SSD to PCI-E bus directly.
Early devices implemented their own protocol and required proprietary OS drivers.
Later this approach evolved into industry standard which is now called non-volatile memory express (NVMe).

It is not only physical interface that has been changed but the protocol itself.
The NCQ extension for SATA allowed a host to issue up to 32 commands to a device to be executed in parallel (or to be reordered according to device geometry in case of HDD).
The NVMe interface allows the queue length to be up to 65536 commands.
Moreover a single device can support many queues, up to 65536 queues are allowed by the standard.
Real-world devices have much fewer queues.
The ones we consider in this report have 32 and 128 queues.

Each queue consists of two ring buffers: one for commands submitted to the device and another one to notify the host about command completion.
These buffers act asynchronously: each participant adds entries to the head and another reads requests form the tail.
This looks similar to network interface devices with their Tx and Rx queues.

Modern NVMe SSDs are more advanced than old SATA SSDs and are able to sustain stable write speed for a long period of time.
If one wishes to evaluate an SSD write performance we strongly suggest to consider NVMe SSD.
Again, we don't do writes in our experiments.

\subsection{Intel Optane}

Recently Intel\footnote{All product and company names are trademarks or registered trademarks of their respective holders. Use of them does not imply any affiliation with or endorsement by them.} announced a new persistent memory type.
This memory has two advantages over prevalent SATA and NVMe SSDs.
First, pages can be overwritten directly and it is no longer required to clear an entire block first.
Second, latency is lower by a factor of magnitude.
We were able to read 4 kilobyte block from it in just 12 microseconds.

Latency as low as 4 microseconds has been reported when SPDK is used.
SPDK is a library which allows one to bypass the kernel and access a drive directly from the userspace.
However when raw device is exposed to the userspace the filesystem should also be implemented in the userspace.
There is one for SPDK called BlobFS however it doesn't look as mature as kernel filesystems.
For example there is no journal in BlobFS which means it cannot be used when durability is of concern.

There are several products all marked as Optane memory.
To avoid confusion we list here all the items available on Intel website at the time this report was written.

\textit{Intel Optane SSD DC D4800X Series.}
This device is made completely from Optane memory and has NVMe interface.
The size differs from 375 gigabytes up to 1.5 terabytes.
Intel suggests it for datacenters. 

\textit{Intel Optane Memory (M10) Series.}
An Optane memory device of size from 16 to 64 gigabytes attached via M.2 interface.
The main purpose is to be used as a filesystem cache in desktops and laptops.

\textit{Intel Optane Memory H10 with Solid State Storage.}
A combined Optane and QLC NAND flash memory device.
The Optane part is 16 or 32 gigabytes.
The NAND flash part can be from 256 gigabytes to 1 terabyte.
It  attaches via M.2 interface and an be used in desktops and laptops as a complete external memory device with a fast internal Optane cache.

\textit{Intel Optane DC Persistent Memory.}
An Optane-backed DIMM module of size 128, 256 or 512 gigabytes.
Recall that DIMM is interface for main (volatile) memory.
As the first one, this device is intended for datacenters.

The first one looks like a more advanced NVMe SSD.
In fact there is a new class of such devices called ultra-low latency SSD in general.
This is exactly the device we examine in our report.
The second and the third devices look like a solution for laptops.
Probably one could reduce OS boot time if it is put onto Optane memory cache.
The last one looks really interesting however it is the least available.
It has been made it into the market only in 2019.
Unfortunately at the time we performed our experiments we didn't have access to this kind of device.
Nevertheless it has been extensively studied recently and well-written reports are available in the literature~\cite{Basic-Optane-PMM},~\cite{OSTI-Optane-PMM}.

\section{Testing Configuration}
\label{sec:configuration}

We run all our tests on a single machine which has all four storage device types installed.
The machine has 512 gigabytes of RAM, two Intel Xeon CPU E5-2660 v4 processors running at clock speed 2.00GHz with HyperThreading enabled.
Most tests run on Linux kernel 4.19.
To test \texttt{io\_uring} we use Linux kerenl 5.4.
Most of the experiments were executed in november and december of 2019 therefore the kernel version choice.
Even at the time of writing many production servers still use earlier Linux kernel versions.
So we hope our evaluations are still relevant.

The testing machine has the following storage hardware installed:

\begin{itemize}
\item\textit{HDD:} Western Digital/HGST Ultrastar He12 7200 rpm 12 Tb.

\item\textit{SATA SSD:} Micron 5200 PRO 1.92 Tb.

\item\textit{NVMe SSD:} Micron SSD 3.2TB U.2 MLC NVME 12V 3.2 Tb.

\item\textit{NVMe Optane SSD:} Intel Optane SSD DC P4800X Series 750 Gb.
\end{itemize}

The main goal of this work is to evaluate the hardware and not the filesystem.
On each device we create a file of size approximately 90\% that of a device size.
We fill it with specific data to be sure that the space is allocated and that the reading takes place.
To assert the latter we verify that received bytes match the ones we expect.

\begin{figure}
    \begin{minipage}[c]{.47\linewidth}
        \includegraphics[width=\linewidth,trim=25 5 70 15, clip]{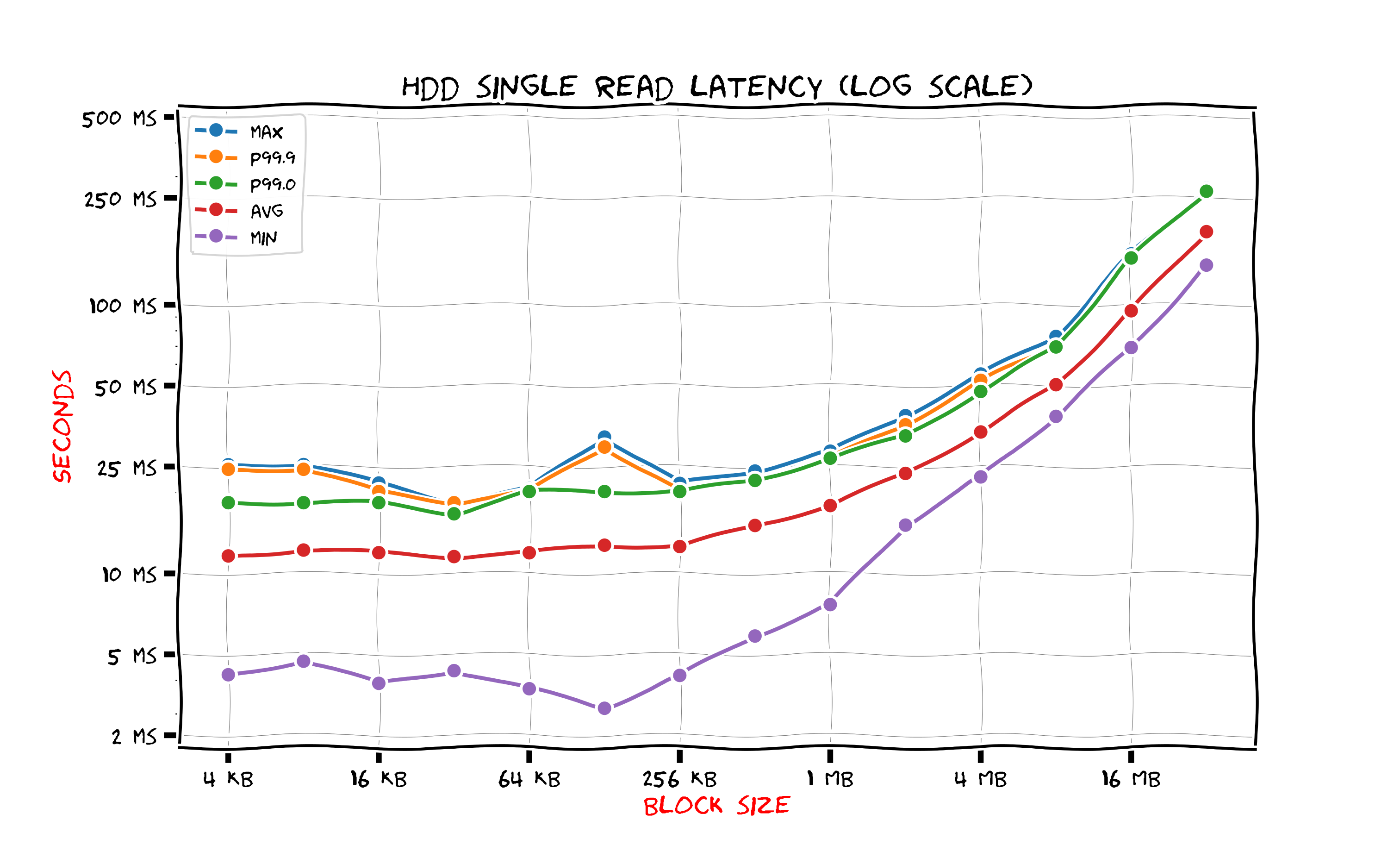}
        \caption{HDD single read latency.}
        \label{fig:single:hdd}
    \end{minipage}
    \hfill
    \begin{minipage}[c]{.47\linewidth}
        \includegraphics[width=\linewidth,trim=25 5 70 15, clip]{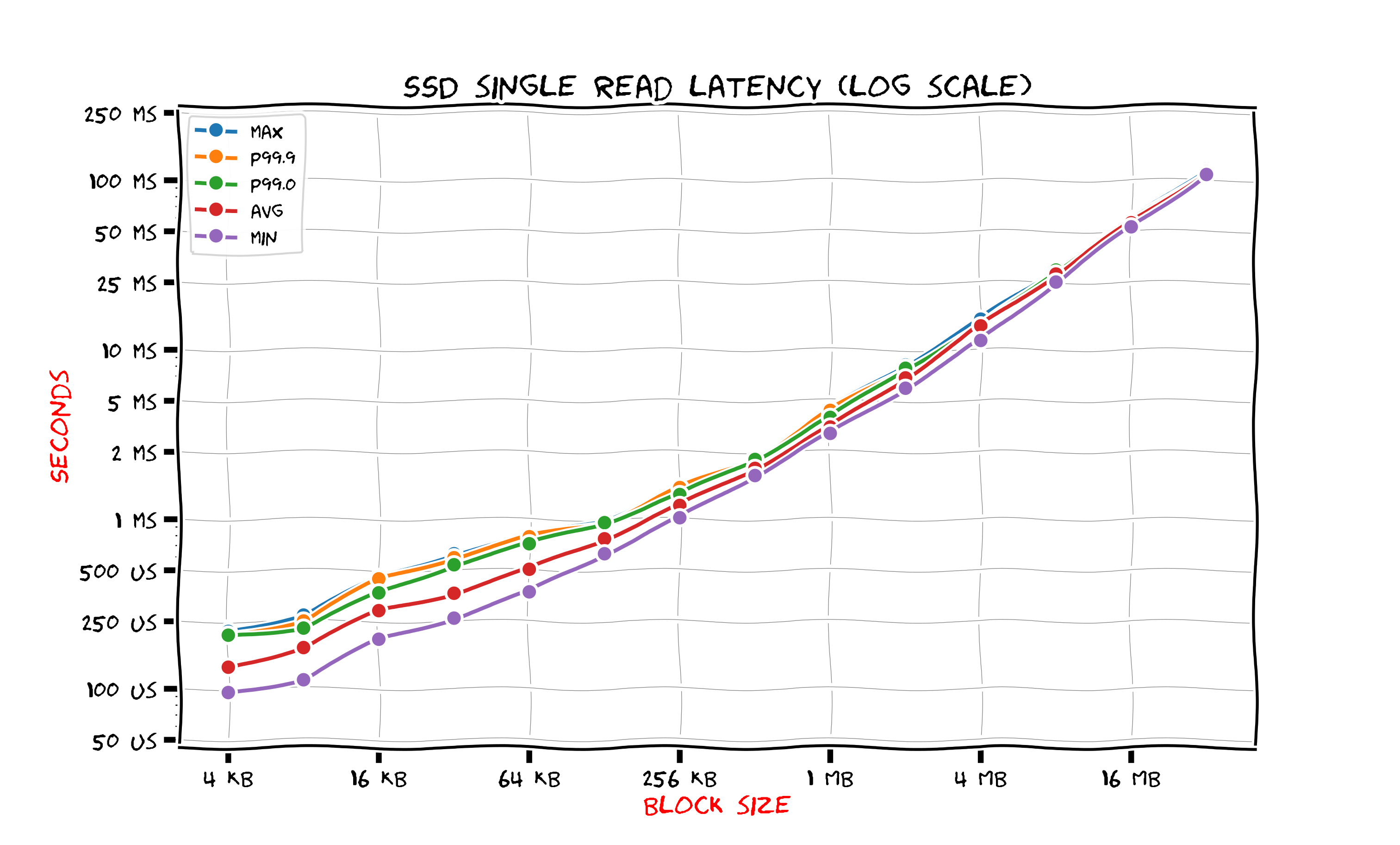}
        \caption{SATA SSD single read latency.}
        \label{fig:single:ssd}
    \end{minipage}
\end{figure}

\section{Single Random Reading}
\label{sec:single}

We start with reading a single block.
Our goal is to pick the best block size for a random read.
An application (or filesystem) can pick any block size and access data with respect to this block size.
We vary block size from 4 kilobytes up to 32 megabytes.
For each block size we make some random reads.
Among these reads we calculate average, minimum and maximum latency as well as 99,0 and 99,9 percentiles.
We use system call \texttt{pread} in this experiment.
We believe that \texttt{lseek} followed by \texttt{read} should have the same performance since the observed storage access time is far longer than a system call.
Later we will read from multiple threads and \texttt{pread} will be the only choice.
We use the same interface to make our experiments comparable to each other.

Hereinafter we use ``SSD'' and ``NVMe'' notions for SATA SSD and NVME SSD.
It looks like this jargon appears in some programmers communities and doesn't lead to confusion at present.
However one should keep in mind that NVMe is just an interface and more storage devices could use it in future.
Thus said, Intel Optane SSD uses NVMe interface.
We use ``Optane'' notion when we talk about Intel Optane SSD.
These notions are used in all figures throughout our report.

\subsection{HDD}

Figure~\ref{fig:single:hdd} shows results for HDD.
The latency is almost the same for all block sizes smaller than 256 kilobytes.
This happens because seek time is much larger than the data transfer time.
The seek time includes arm positioning to find the right track and awaiting for platter rotation to bring data under the head.
A simple consequence is that for a HDD random read one should use blocks of size at least 256 kilobytes.
Even if an application use smaller blocks the drive access time would be the same.
However one could still decide to use smaller blocks for better cache utilization: if the amount of data per request is small and is expected to fit in cache then storing a large block along with the requested data would actually make cache capacity smaller in terms of useful data.

The 256 kilobyte block read takes 12 milliseconds on the average.
We experienced variations from 4 milliseconds up to 25 milliseconds. 
This is really a huge amount of time for a computer.
For example the typical process scheduling quantum is just a few milliseconds.
An operating system can (and in fact does) execute other processes while our process waits for the data to arrive from the hard drive.

\begin{figure}
    \begin{minipage}[c]{.47\linewidth}
        \includegraphics[width=\linewidth,trim=25 5 70 15, clip]{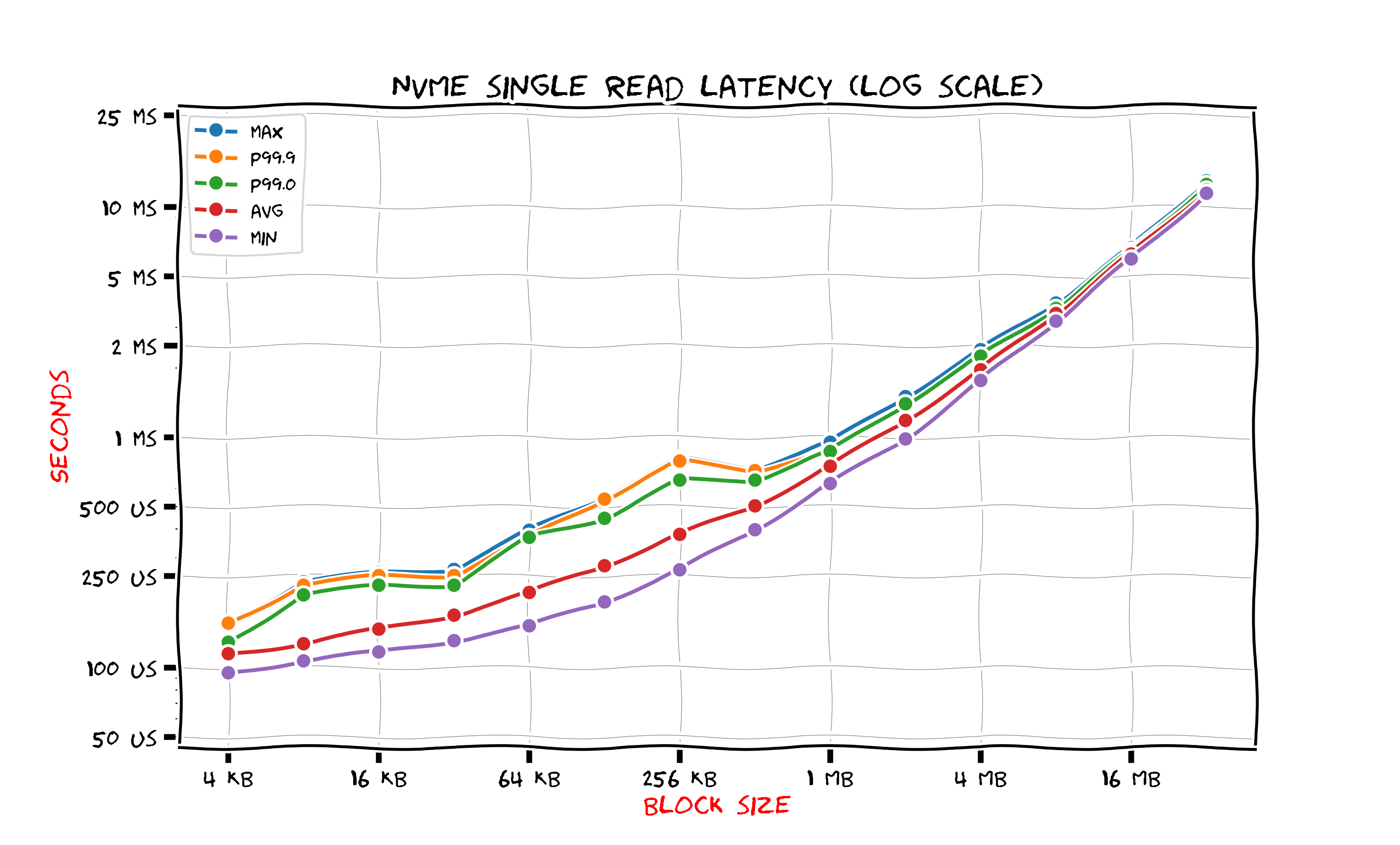}
        \caption{NVMe SSD single read latency.}
        \label{fig:single:nvme}
    \end{minipage}
    \hfill
    \begin{minipage}[c]{.47\linewidth}
        \includegraphics[width=\linewidth,trim=25 5 70 15, clip]{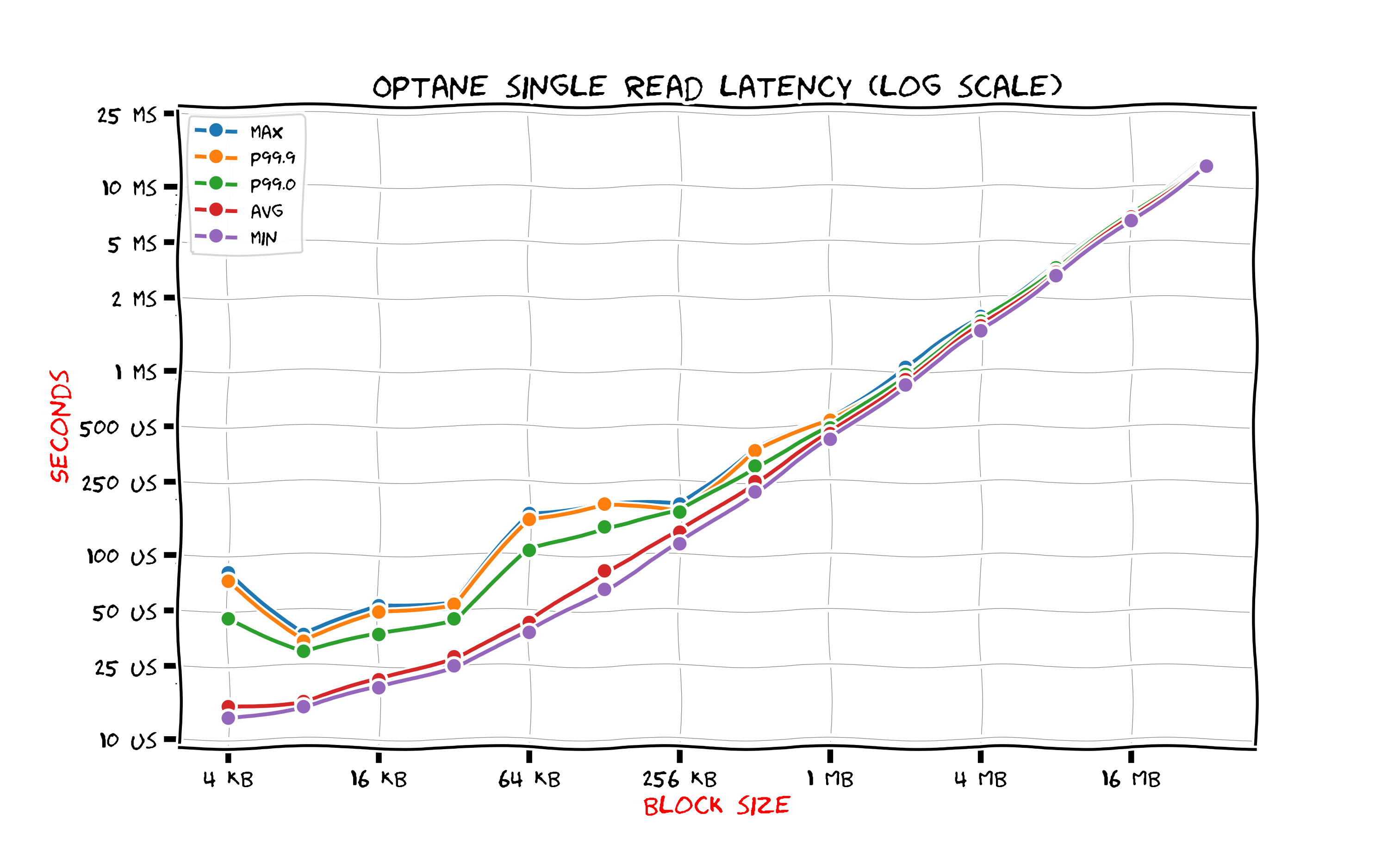}
        \caption{Intel Optane SSD single read latency.}
        \label{fig:single:optane}
    \end{minipage}
\end{figure}

\begin{figure}
\begin{center}
    \begin{minipage}[c]{.47\linewidth}
        \includegraphics[width=\textwidth,trim=25 5 70 15, clip]{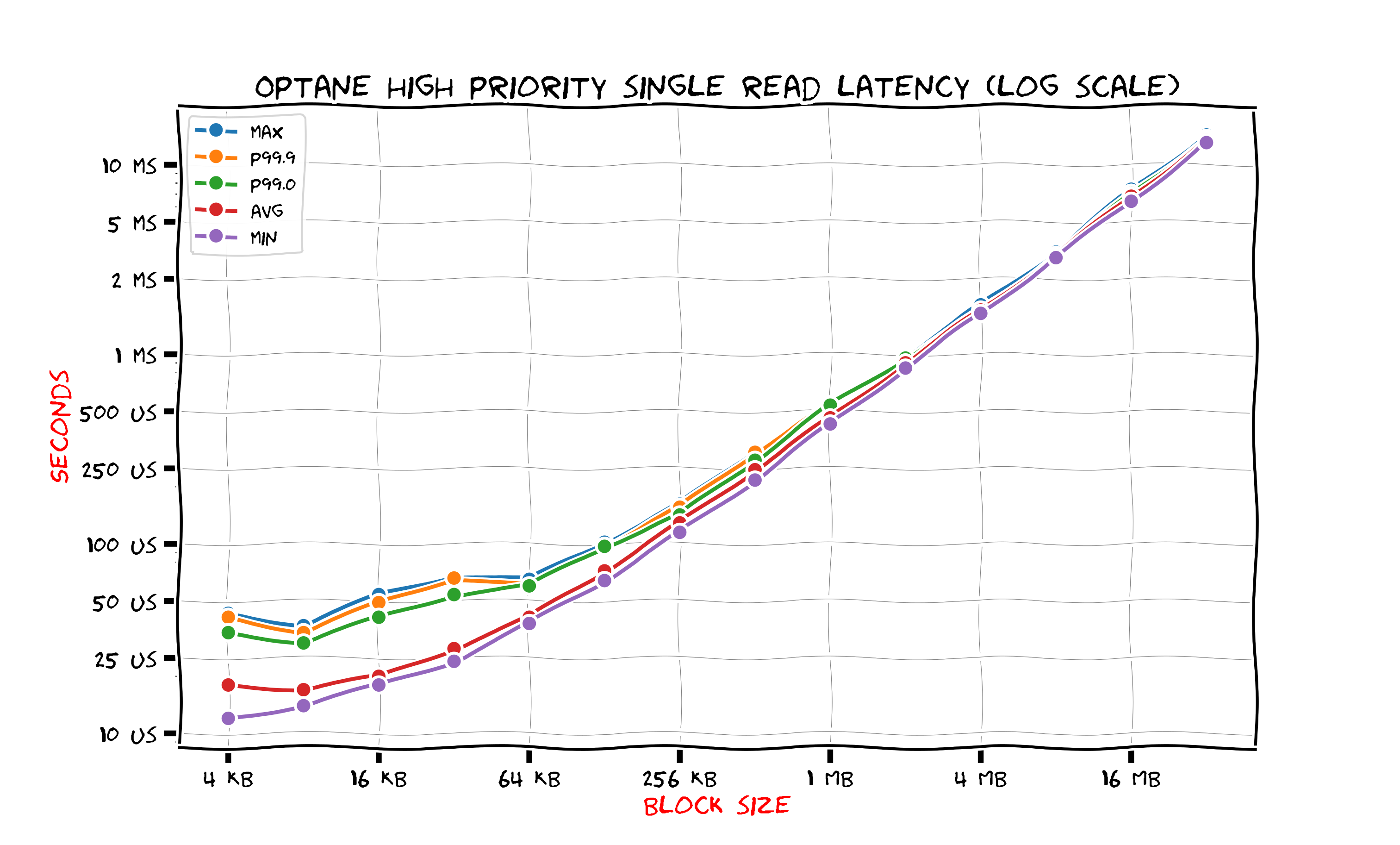}
        \caption{Intel Optane SSD single read latency in poll mode.}
        \label{fig:single:optaneprio}
    \end{minipage}
\end{center}
\end{figure}

\subsection{SSD}

Figure~\ref{fig:single:ssd} shows SATA SSD read latencies.
Note that the time at the lower part of the figure is in microseconds (we use standard shortenings ms for milliseconds and us for microseconds).
Reading block of size 4 kilobytes takes 140 microseconds on the average and the time growth is linear when the block size increase.
Compared to HDD reading a 4 kilobyte block from SSD is 80 times faster.
For a 256 kilobyte block SSD is ten times faster than HDD.
When block size is large enough (starting from 4 megabytes) SSD is only two times faster than HDD.

\subsection{NVMe}

Figure~\ref{fig:single:nvme} shows results for NVMe SSD.
The latency is better than those for SATA SSD.
For a 4 kilobytes block size the average time improved only a little, but the 99 percentile is two times lower.
It takes less than millisecond to read a megabyte block from NVMe SSD.
For SATA SSD it took 3 milliseconds.
As we see, upgrade from SATA SSD to NVMe SSD is not as dramatic as upgrade from HDD to SATA SSD.
This is not surprising since both SATA and NVMe SSD are based on the same thechnology.
Only interfaces differ.

\subsection{Optane}

Figure~\ref{fig:single:optane} shows results for Intel Optane SSD.
Minimal latency is 12 microseconds whih is 10 times lower than those of NVMe SSD.
Average latency is 1000 lower than those of HDD.
There is quite large variation for small block read latency: even though the average time is quite low and close to minimal latency the maximum latency and even 99 percentile are significantly worse.
If somebody looks at these results and wishes to create an Optane-bases service with 12 microsecond latency for reads they would have to install larger number of Optane drives or consider providing more realistic timings.

When latency is so small overheads of context switching and interrupt handling become noticeable.
One can use polling mode to gain some improvement.
In this mode the Linux kernel monitors the completion queue instead of switching to some other job and relying on hardware interrupt with interrupt handler to notify about completion.
Clearly, it is considerable to use the polling mode only when hardware response is expected to arrive fast enough.

The polling mode is used when an application calls \texttt{preadv2} system call with \texttt{RWF\_HIGHPRI} flag.
The result of using this call for Intel Optane is shown on figure~\ref{fig:single:optaneprio}.
Compared to usual \texttt{pread} the polling mode lowers the maximum latency by a factor of two for block sizes up to 256 kilobytes.

\subsection{Summary}

\begin{figure}
    \centering
    \includegraphics[width=\textwidth]{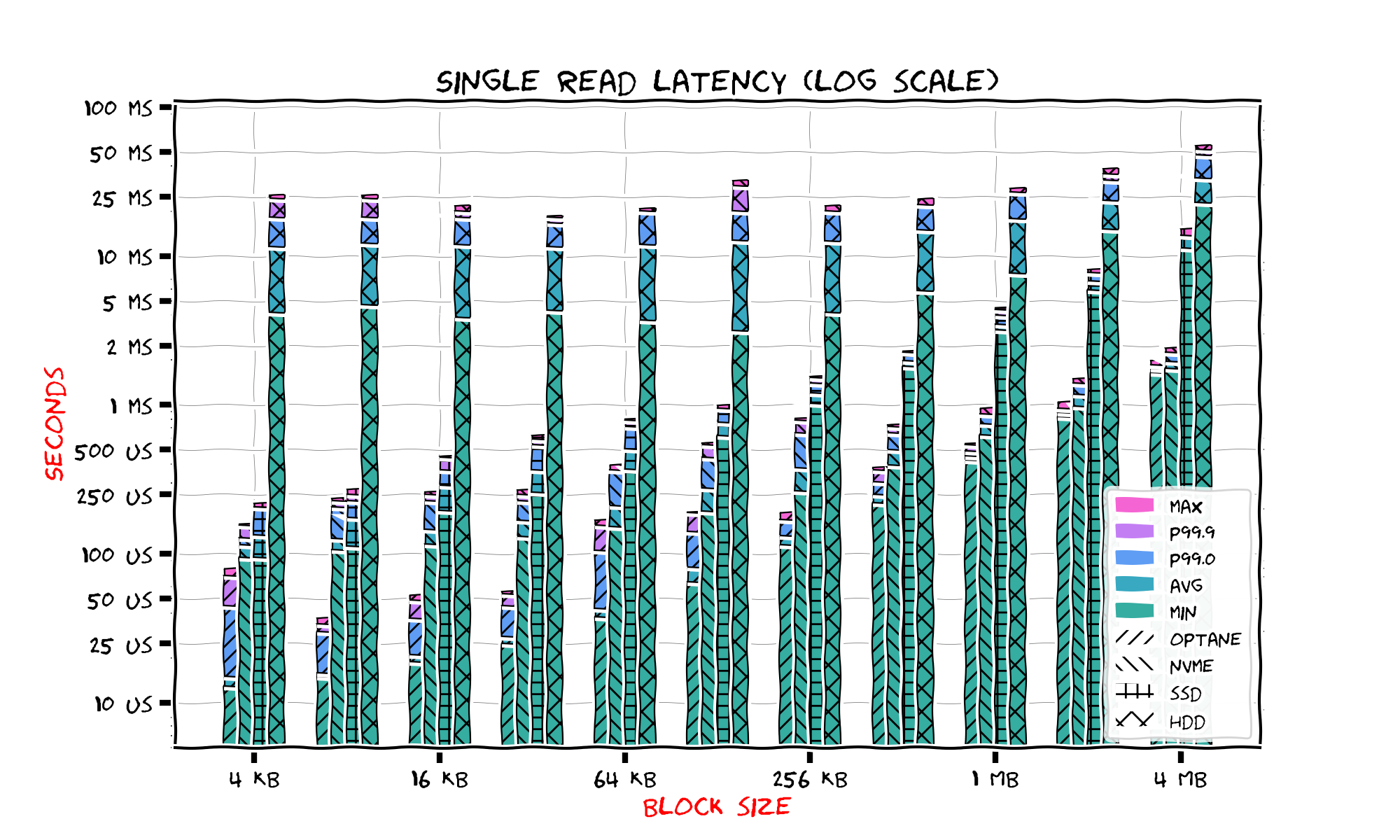}
    \caption{Single read latency for HDD, SATA SSD, NVMe SSD and Optane.}
    \label{fig:single:all}
\end{figure}

Figure~\ref{fig:single:all} shows single read latencies for all four storage types on a single chart.
Starting from~4 megabytes the latency is easily predicted by linear extrapolation so we don't show larger blocks here.
To show everything on a single figure we are forced to use quite an overloaded legend.
We use vertical level to show the latency and we iterate the block size horizontally.
For each block size we show four bars, from left to right: for Intel Optane, NVMe SSD, SATA SSD, and HDD.
Storage type is represented by hatch and the latency by color.

We see that solid state device latencies are far better than HDD.
For a single read the leader is Intel Optane, however as we shall see later it has it's own drawback compared to NVMe SSD.
NVMe SSD and SATA SSD look quite close to each other when the block size is small.
Our observations show that the best block size for random read is 256 kilobytes for HDD, 4 kilobytes for NVMe and SATA SSD and 8 kilobytes for Intel Optane.

\begin{figure}
    \begin{minipage}[c]{.47\linewidth}
        \includegraphics[width=\linewidth,trim=25 5 70 15, clip]{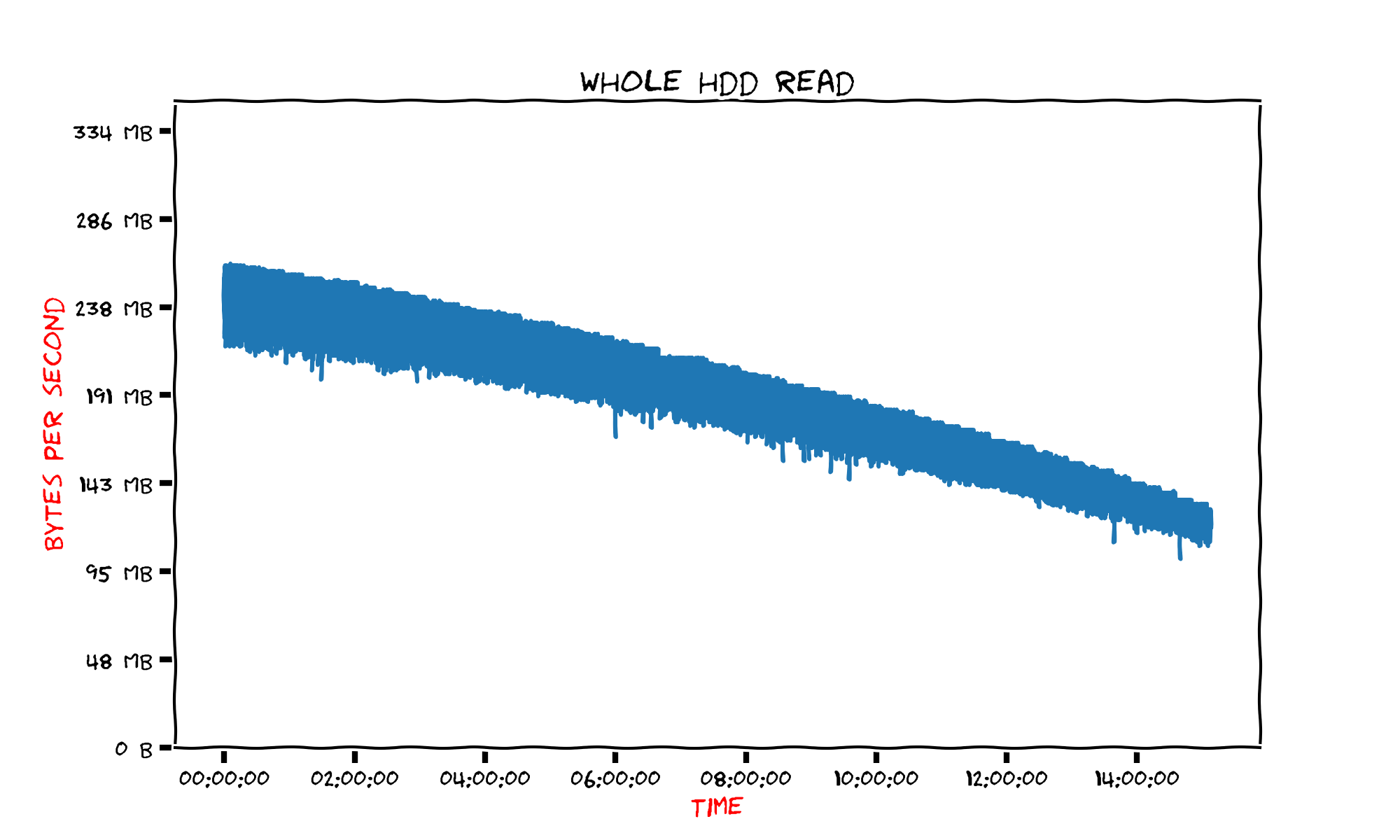}
        \caption{Whole HDD read.}
        \label{fig:whole:hdd}
    \end{minipage}
    \hfill
    \begin{minipage}[c]{.47\linewidth}
        \includegraphics[width=\linewidth,trim=25 5 70 15, clip]{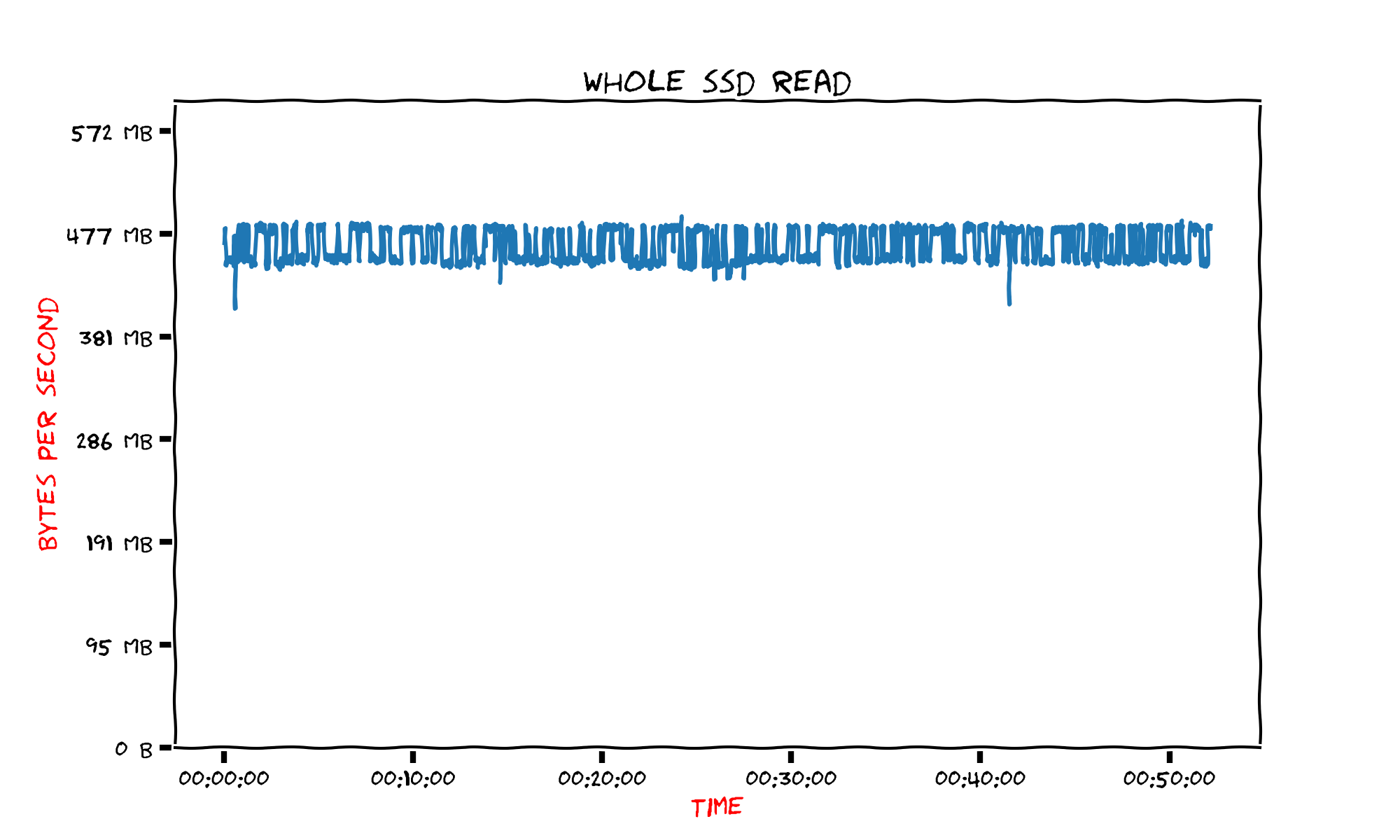}
        \caption{Whole SATA SSD read.}
        \label{fig:whole:ssd}
    \end{minipage}
\end{figure}
\begin{figure}
    \begin{minipage}[c]{.47\linewidth}
        \includegraphics[width=\linewidth,trim=25 5 70 15, clip]{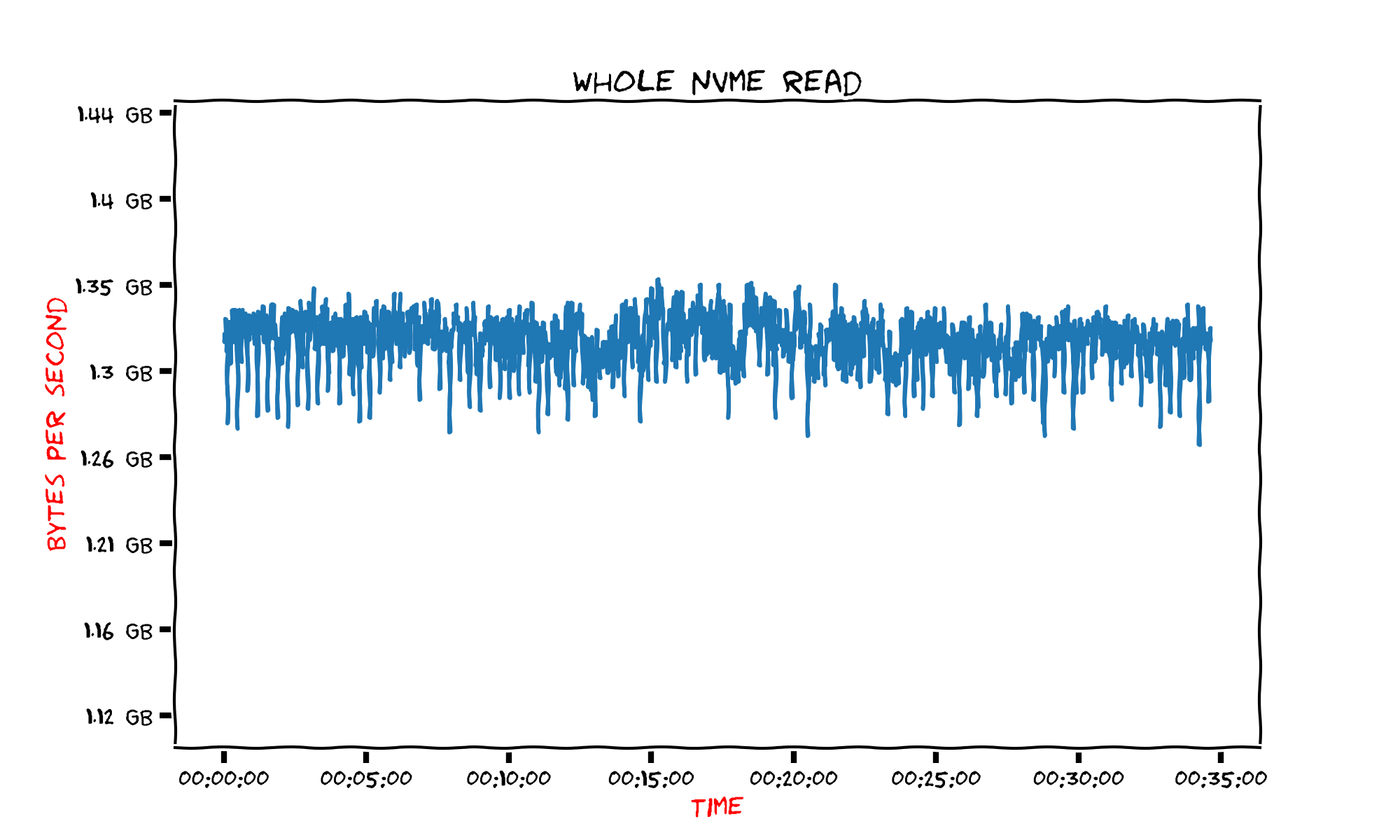}
        \caption{Whole NVMe SSD read.}
        \label{fig:whole:nvme}
    \end{minipage}
    \hfill
    \begin{minipage}[c]{.47\linewidth}
        \includegraphics[width=\linewidth,trim=25 5 70 15, clip]{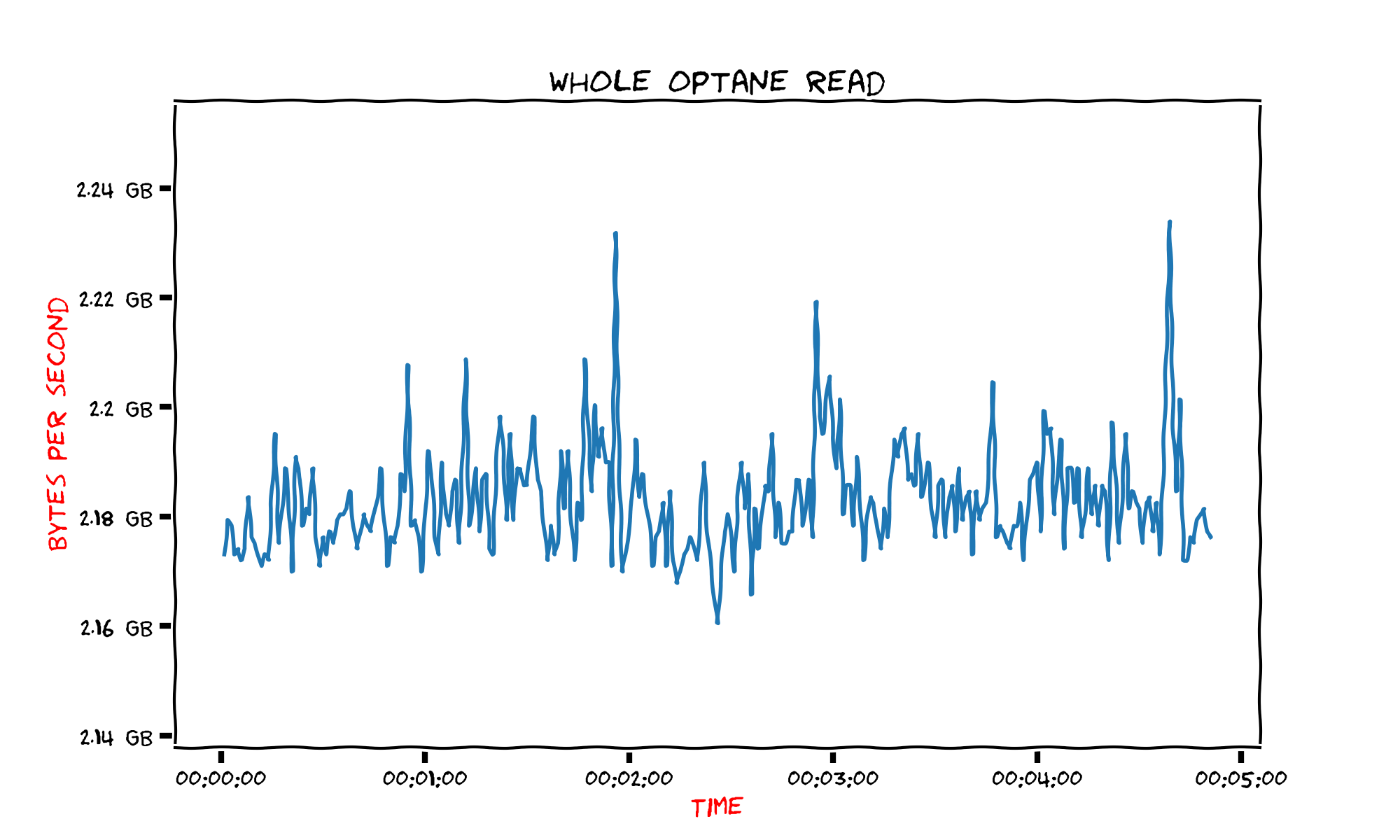}
        \caption{Whole Intel Optane SSD read.}
        \label{fig:whole:optane}
    \end{minipage}
\end{figure}

\section{Single-threaded Reading}
\label{sec:sequential}

In previous section we studied reading just a single block from a storage device.
In real life disk I/O operations are happening all the time.
When a large amount of data is read we usually talk about a stream of data.
Under the hood any stream is just a set of blocks if we look at it from point of view of a storage device.

A stream is usually measured by throughput which is an amount of data transferred during a particular amount of time.
Typically we talk about megabytes per second.
Since any stream is just a set of blocks we can look at blocks separately and measure latency.
These latencies can be aggregated just like in the experiments from the previous section.

In this section we study reading a single stream of data (from a single CPU thread).
We consider both sequential and random streams.
We measure throughput and aggregated latency for each stream.

\subsection{Reading the whole device}

Hard disk drive bandwidth depends on physical data location.
Imagine an HDD platter.
It seems natural that the outer track contains more data than the inner track.
However rotation speed is constant.
Thus the further the head is from the center the more is the bandwidth.
Filesystems exploit this phenomena and put important metadata close to the outer side of the disk to improve performance.

To demonstrate the difference in bandwidth we tried to read the whole drive in one pass.
To alleviate the effect of disk rotation we used block size of 1 megabyte in our calls.
Also we disabled \texttt{O\_DIRECT} and hoped that OS disk scheduler would execute requests more carefully.
It took 15 hours to read 10 terabytes of data from HDD.
Figure~\ref{fig:whole:hdd} shows the result.
It is easy to note that the bandwidth goes down throughout this experiment.
At the beginning the data transfer speed was about 250 megabytes per second.
While the head moved to the end of disk the speed felled down to 150 megabytes per second.

For solid state devices the bandwidth should not depend on data position.
Nevertheless when we did our experiments we noticed some fluctuations.
Unfortunately the origin is unknown.
It could be from internal buffering inside the device but it is only a speculation.

It turned out that with \texttt{O\_DIRECT} the bandwidth is increased
for NVMe SSD by a factor of two.
As for SATA SSD and Optane it is slightly worse to use \texttt{O\_DIRECT} but we enable it anyway just to make our experiments consistent with each other.

Figure~\ref{fig:whole:ssd} shows results for SATA SSD.
The bandwidth is about 450 megabytes per second which is 2-3 times higher than those of HDD.
The pattern is the same as the reading goes from the beginning of the device to the end.
This is expected since SSD doesn't have anything which makes speed to depend on data location.
We believe that the observed fluctuations are due to caching and predictions made by internal device controller.

NVMe SSD exhibits bandwidth three times larger than those of SATA SSD in this experiment.
Figure~\ref{fig:whole:nvme} shows results for NVMe SSD.
The observed bandwidth is about 1.3 gigabytes per second and also oscillates.
The pattern looks different from SATA SSD though.

Figure~\ref{fig:whole:optane} shows the result of this experiment for Intel Optane.
This device demonstrated the highest bandwidth of 2.2 gigabytes per second.
Since the speed is so high and the device size is the smallest (recall our model capacity is only 750 GB) it took only 5 minutes to read the whole storage.
Compared to 15 hours of reading from HDD this is indeed a dramatic difference.

\begin{figure}
    \begin{minipage}[c]{.47\linewidth}
        \includegraphics[width=\linewidth,trim=25 5 70 15, clip]{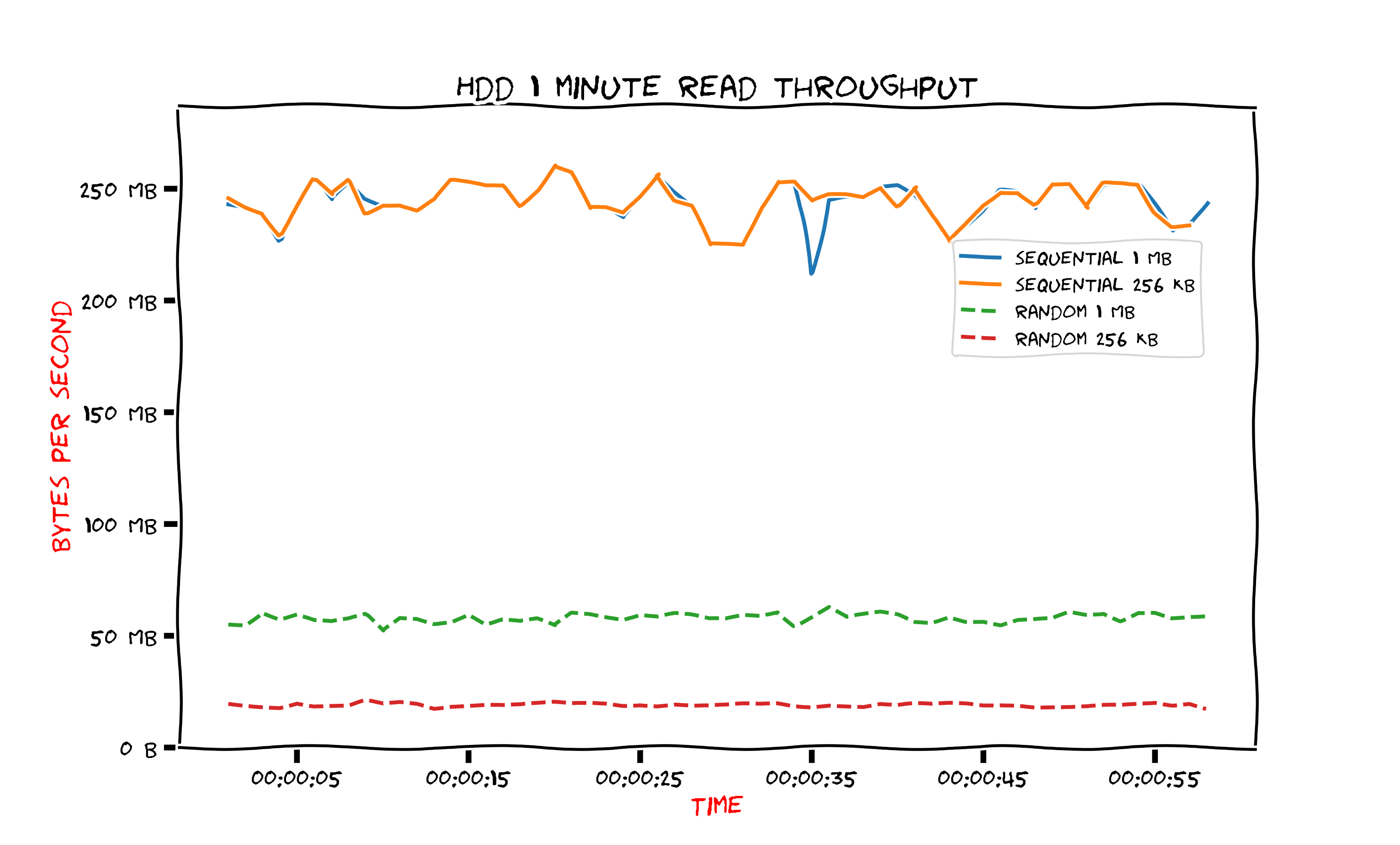}
        \caption{Comparison of sequential and random reading throughput for HDD.}
        \label{fig:seqrand:hdd}
    \end{minipage}
    \hfill
    \begin{minipage}[c]{.47\linewidth}
        \includegraphics[width=\linewidth,trim=25 5 70 15, clip]{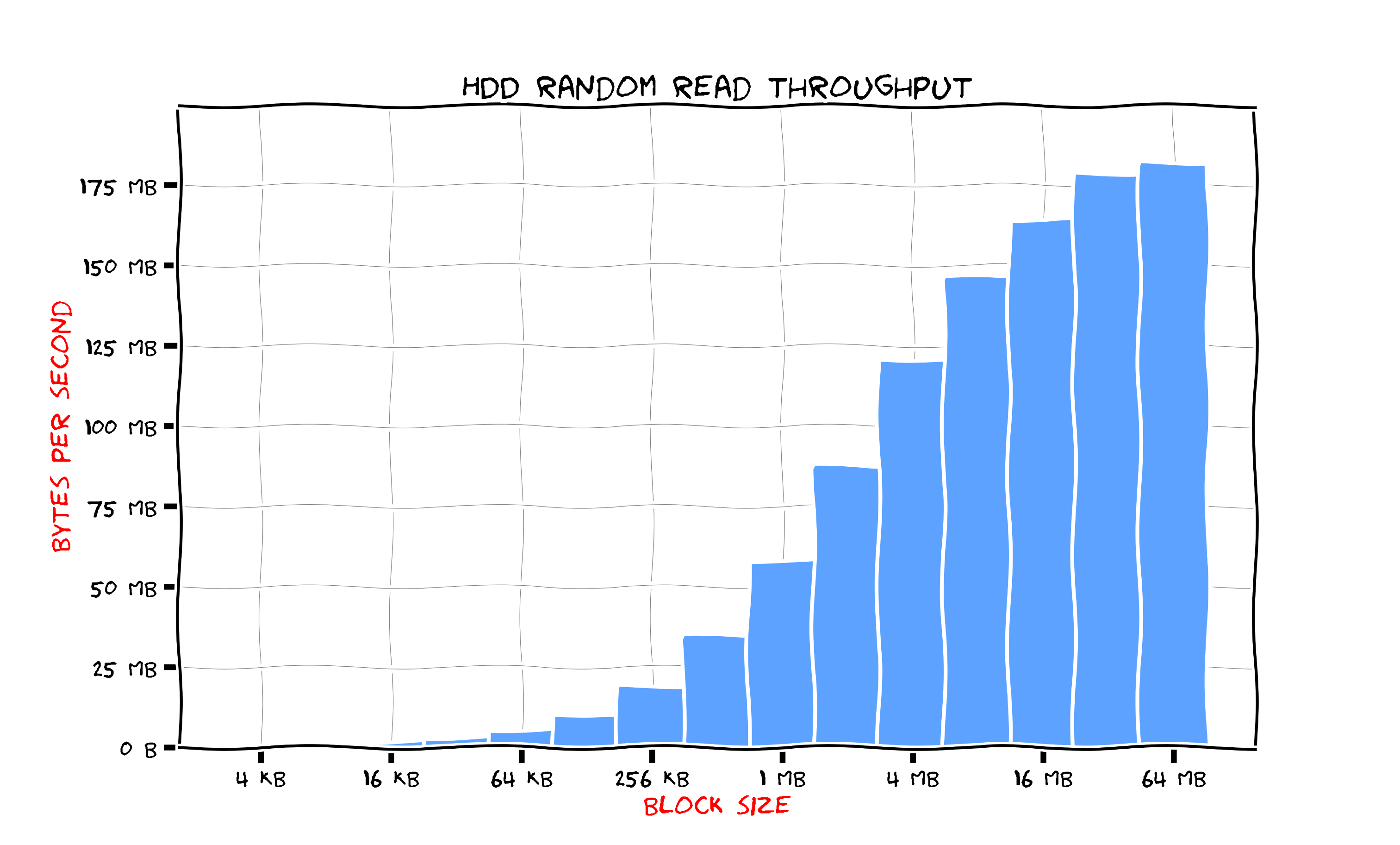}
        \caption{Random HDD reading throughput.}
        \label{fig:random:hdd}
    \end{minipage}
\end{figure}

\subsection{HDD}

Previously we observed how reading a whole drive at once performs.
This experiment is actually quite far from real world workload.
We used large blocks and read them sequentially.
Tyical application breaks one of these conditions (or both).

Next we study reading from a single thread in more details.
We pick different block sizes and execute requests for a minute in a loop.
We choose next block both in sequential and random manner.
This is more likely to how a real application behaves: sometimes large amount of data should be read seqentially
and sometimes blocks of interest are scattered across the device.
We are not aiming at bandwidth saturation in this section, rather we wish to highlight some interesting properties of single stream reading.

There is a tremendous difference between sequential and random reading from HDD.
If reading is sequential the drive doesn't have to move the arm to fetch next block since it will be right under the head when the current block ends.
If blocks are picked at random, then an application would have to wait for the head to move to the next block before the drive could read it.
As observed earlier this takes 12 milliseconds on average.

An easy consequence is that random reading leads to performance degradation if the block size is small.
If reading is sequential then the maximum bandwidth is achieved for a 16 kilobyte blocks.
On the other hand random reading comes close to the maximum bandwidth only when the block size is tens of megabytes.

Figure~\ref{fig:seqrand:hdd} shows bandwidth for sequential and random readings with block sizes 256 kilobytes and 1 megabyte.
Lines for sequential readings are solid and look almost identical which means that the bandwidth is saturated.
Lines for random readings are dashed and it is easy to see that they are way below the maximum.

Nevertheless one can saturate the bandwidth even when blocks are picked at random if sufficiently large block size is used.
Figure~\ref{fig:random:hdd} shows how bandwidth depends on the block size for random readings.
It is almost unnoticeable for blocks smaller than 64 kilobytes and reaches slightly above 175 megabytes per second when the block size goes up to 64 megabytes.
Recall that peak bandwidth depends on the position on the platter and varies from 250 to 140 megabytes per second.
Reaching 175 megabytes per second when blocks are scattered across the whole drive seems like a fair result.

In the introduction we discussed the \texttt{O\_DIRECT} flag.
Recall that this flag controls whether kernel page cache is used when file is accessed.
Cache brings significant benefit for sequential reading.
When page cache is used (e.g. file has been opened without \texttt{O\_DIRECT} flag) kernel fetches subsequent data in advance.
One can specifically notify the kernel that reading pattern is sequential by calling \texttt{posix\_fadvise()} with \texttt{POSIX\_FADV\_SEQUENTIAL} flag.
Reading data that will be requested in the future allows kernel to saturate drive bandwidth even if user calls \texttt{read} with small block size.

\begin{figure}
    \begin{minipage}[c]{.47\linewidth}
        \includegraphics[width=\linewidth,trim=25 5 70 15, clip]{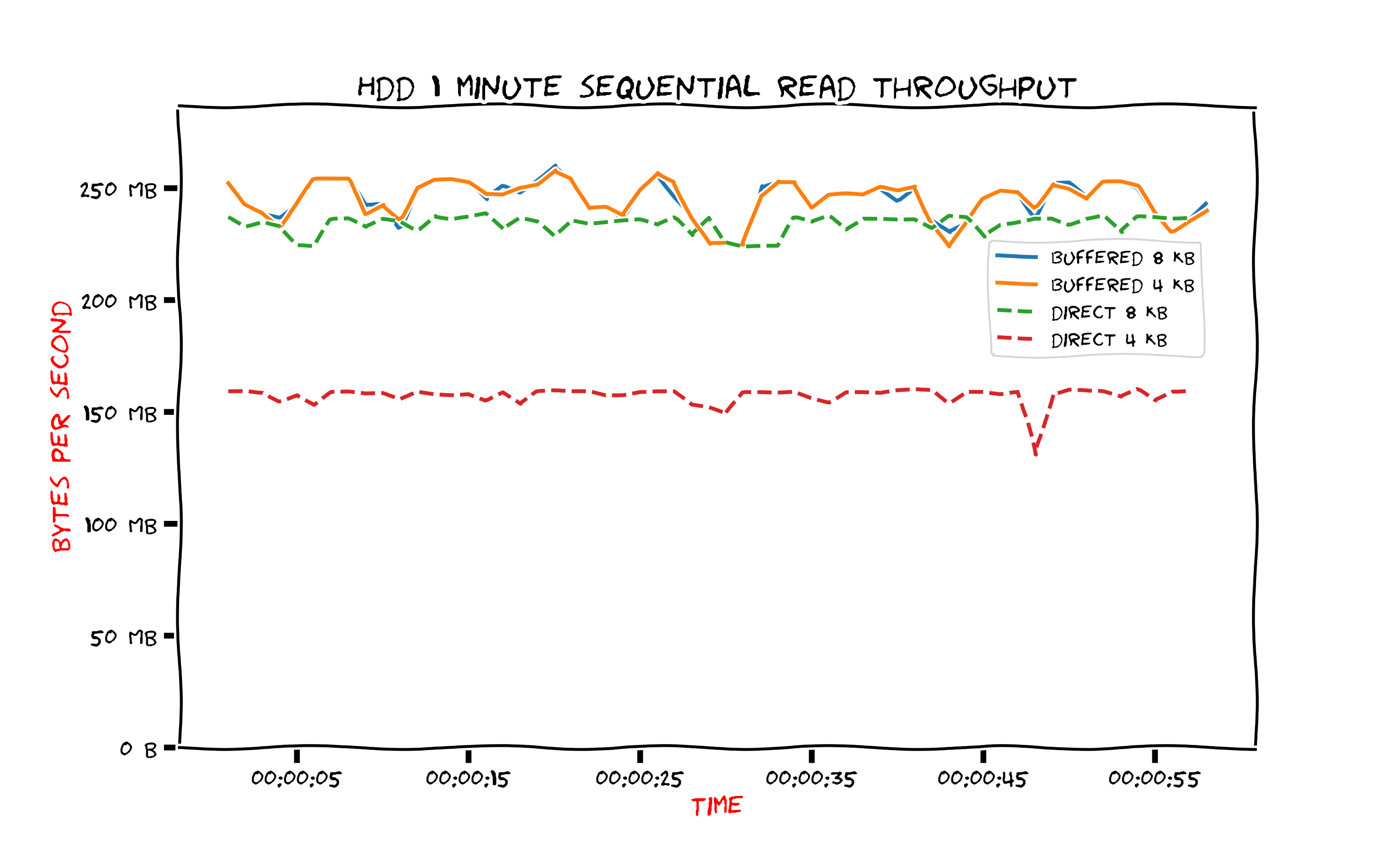}
        \caption{Comparison of buffered and direct reading throughput for HDD.}
        \label{fig:buffdirect:hdd}
    \end{minipage}
    \hfill
    \begin{minipage}[c]{.47\linewidth}
        \includegraphics[width=\linewidth,trim=25 5 70 15, clip]{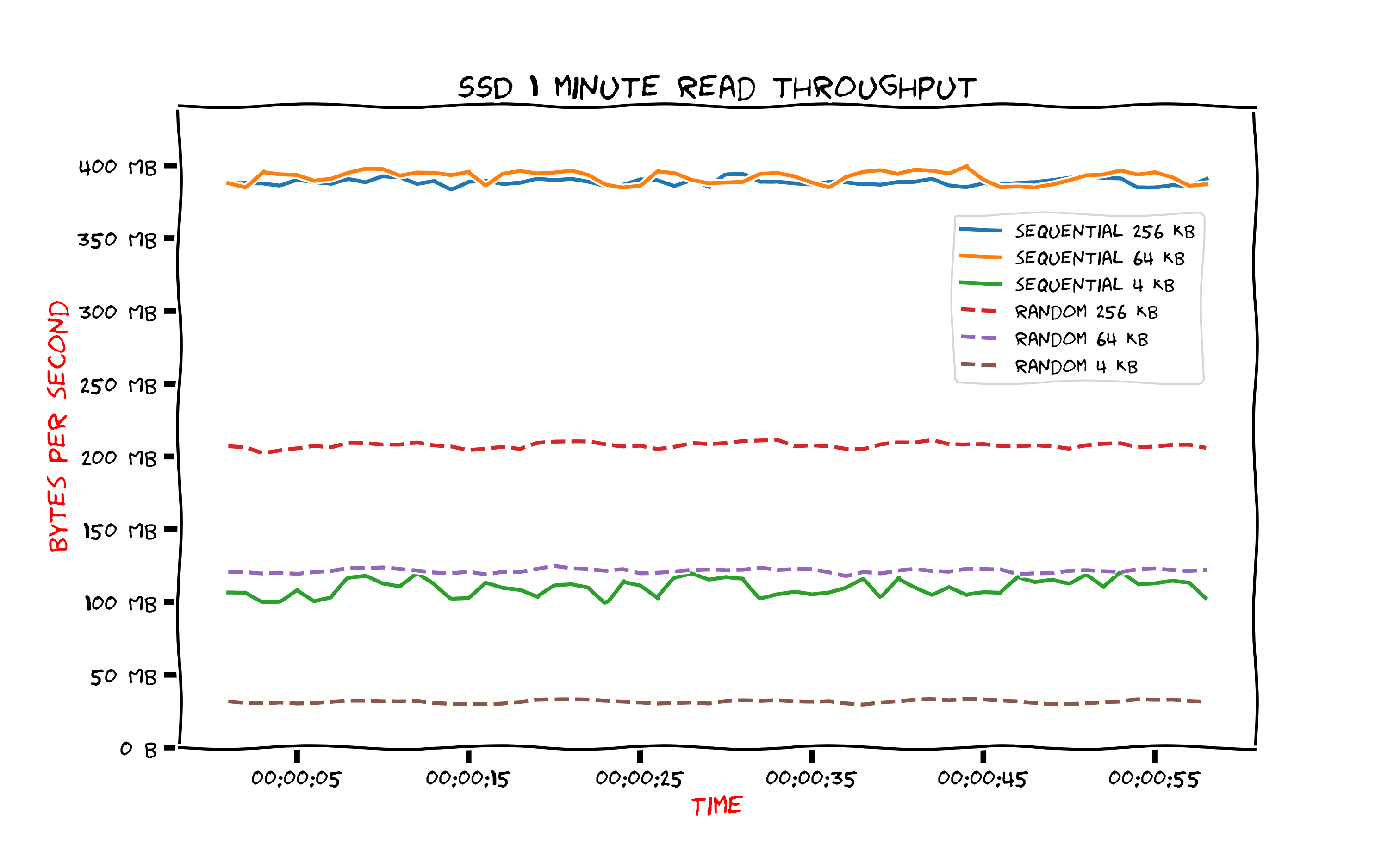}
        \caption{Comparison of sequential and random reading throughput for SATA SSD.}
        \label{fig:seqrand:ssd}
    \end{minipage}
\end{figure}

Figure~\ref{fig:buffdirect:hdd} shows both buffered and direct sequential readings form the beginning of the file.
Block sizes of 4 and 8 kilobytes are considered.
When block size is 4 kilobytes reading in direct mode reaches bandwidth of 150 megabytes per second while the maximum bandwidth for this part of the disk is 250 megabytes per second.
On the other hand when block size is 8 kilobytes bandwidth is close to the maximum even for unbuffered reading.
If we had picked block size equal 16 kilobytes then the line would coincides with those of buffered reading so it is omitted to keep the chart readable.

\subsection{Flash devices}

When reading from SSD the behavior is quite steady.
Figure~\ref{fig:seqrand:ssd} shows some sequential and random reads for SATA SSD.
The chart clearly could be replaced with just bars for each block size.
So we won't study SATA, NVME SSD and Optane in details here and move to combined results for all four storage types.
We present charts for both latency and throughput.

\begin{figure}
    \includegraphics[width=\linewidth]{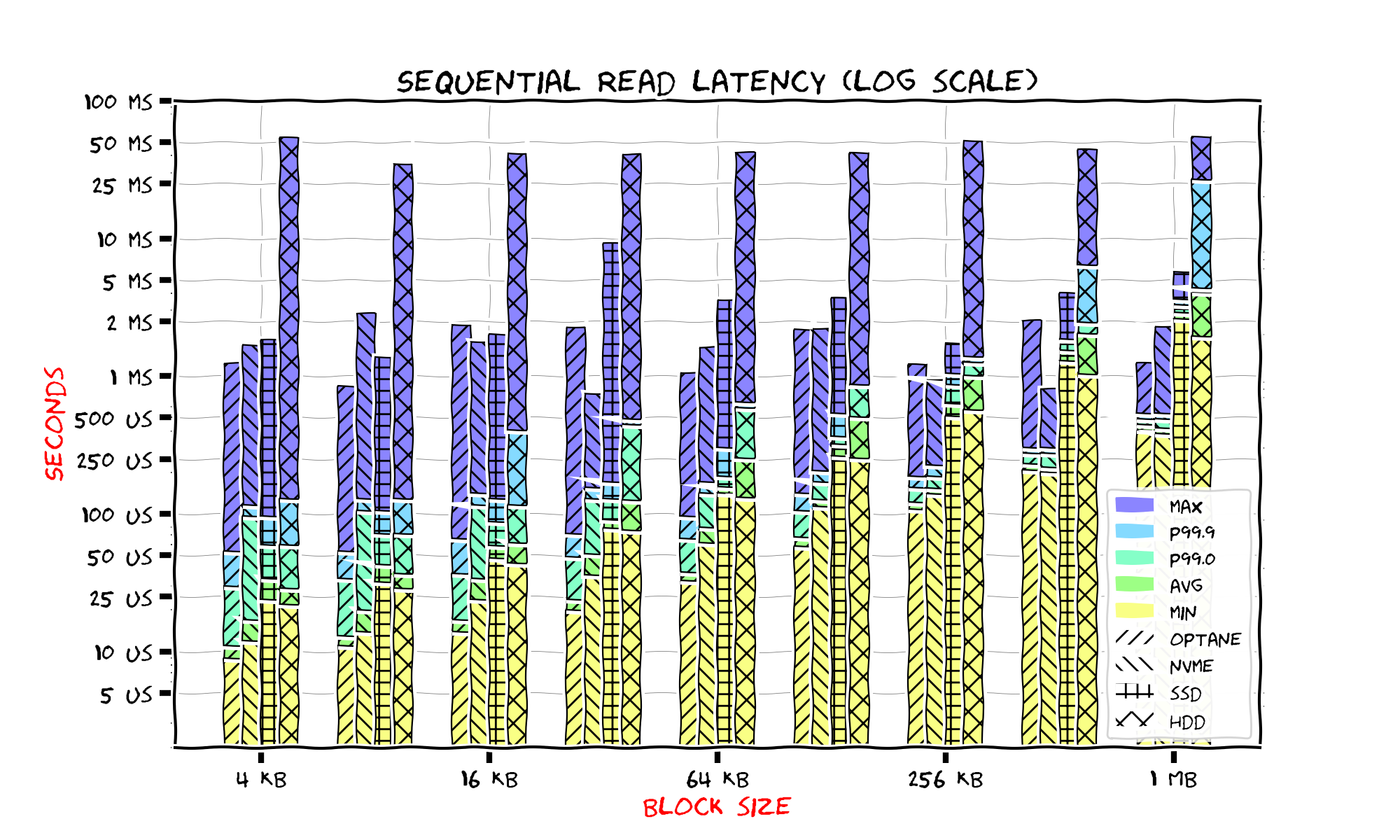}
    \caption{Latency for all storage types sequential blocks reading.}
    \label{fig:latencyseq:all}
\end{figure}

\begin{figure}
    \includegraphics[width=\linewidth]{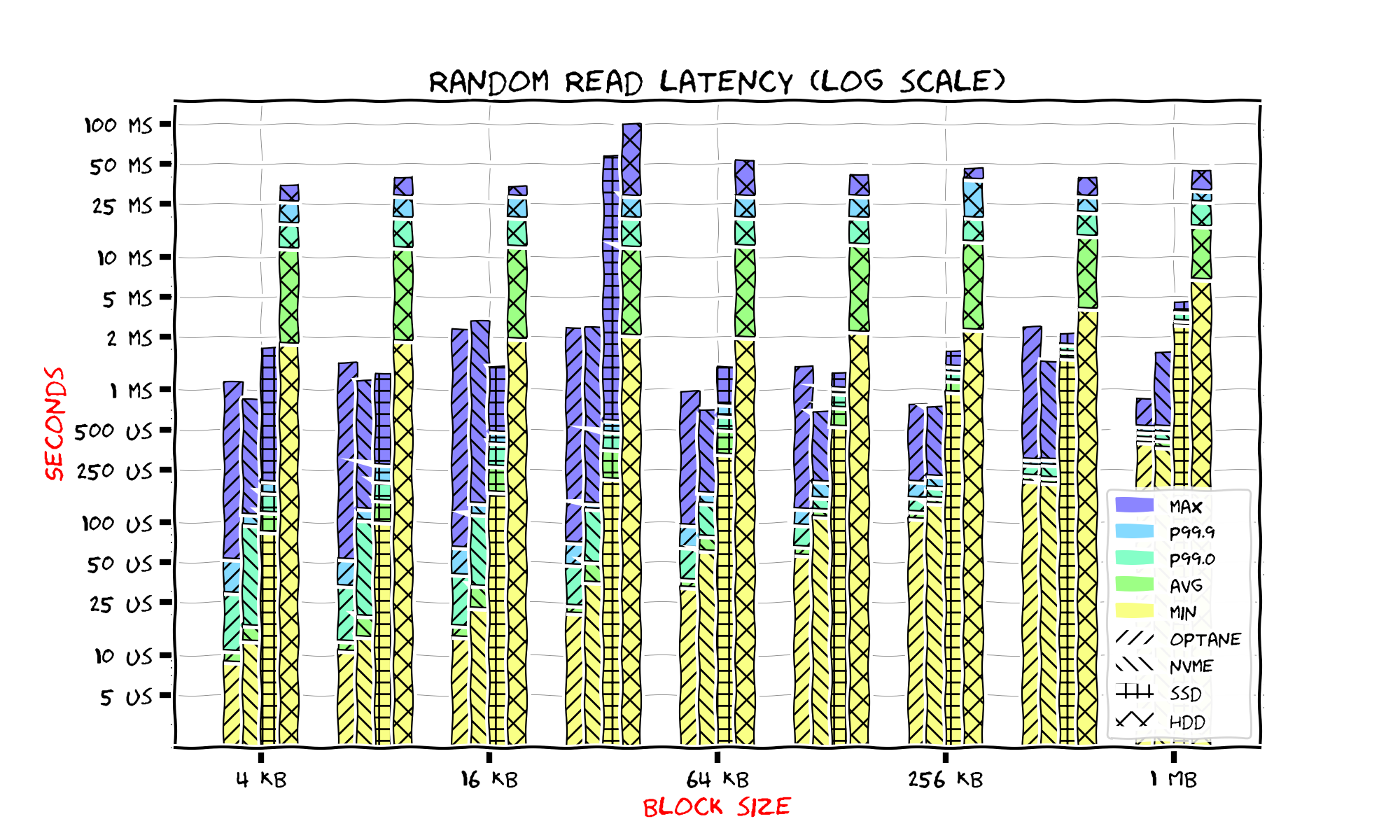}
    \caption{Latency for all storage types random blocks reading.}
    \label{fig:latencyrand:all}
\end{figure}

Figures~\ref{fig:latencyseq:all} and~\ref{fig:latencyrand:all} shows latencies observed in our experiments for sequential and random block sequences.
Each chart is similar to the latency chart from section~\ref{sec:single}.
The difference is that now we evaluate statistics from replies observed during one minute of continuous execution.
It is noticeable that for sequential reading there is a huge difference between maximum value and 99.9 percentile for all storage devices.
Most likely this happens due to sequential reading recognition and internal buffering in the device.

For sequential reading the minimum latency for all four devices is rather close to each other.
Even HDD behaves nearly as good as SATA SSD in terms of minimum and average latency when reading is sequential.
This matches our hypothesis about internal buffering: once data is read form the platter it is stored in an internal device memory.
Subsequent requests ask device for the next blocks.
Since they already reside in the internal device memory only SATA interface latency contributes to the total latency of the transfer.
Hence HDD and SATA SSD behave similarly.

NVMe SSD and Optane perform close to each other as well though Optane is faster for small blocks on average.
It is interesting that for both sequential and random reading the maximum time of all three solid memory devices is much higher and reaches 1-2 milliseconds which is way higher than the 99.9 percentile.
A possible explanation could be the context switch.
Since our test runs as usual Linux process it should be suspended from time to time to give some other processes and the kernel to execute on the same CPU.
The timing matches what a gap should be on a rather idle CPU and of cause there is a possibility for such event to occur.
Anyway we test not only a device but \texttt{pread} interface as a whole.
This is an observed behavior and one better be prepared for it.

When block sequence is random HDD performs as bad as it should be and the latency is similar to what we observed in section~\ref{sec:single}.
It is interesting to note that the be behavior of NVMe SSD and Intel Optane looks quite similar with a little difference when the block size is small.
Intel Optane is faster but the difference in the performance under load is not as dramatic as for a single request. 
On average Optane is two times faster than NVMe SSD while the minimum latency is almost the same.
On the other hand SATA SSD is worse than NVMe SSD by a noticeable factor of 2-4 times.
Compared to solid state devices HDD is worse by two orders of magnitude.
These days one should not use HDD if laency is at stake.

Figure~\ref{fig:latencyrand:all} also shows an annoying property of SATA SSD which is discussed quite often.
For block size 32 kilobytes maximum latency reaches almost 50 milliseconds.
This just is as bad as HDD.
As we discussed in~\ref{sec:overview} SSD sometimes need to rearrange internal data layout and this requires both resources and time.
To cope with this problem one could issue trim requests to the device when it is unused but it sounds rather involved to schedule such maintainance at a cluster scale.
Another way would be to accept that SSD can respond unbearably slow from time to time.
If system is disk-based and tries to achieve low latency the request should be sent to multiple replicas.
The good news is that we don't observe such spikes for NVMe SSD and Optane.

Next we discuss throughput for both sequential and random reading.
Note that for NVMe SSD and Intel Optane there is almost no difference from sequential and random reading.
As for HDD and SATA SSD random reading performs significantly worse.
When the block size is small the throughput is so small that the bars are invisible.

It is the first time we see that NVMe SSD beats Intel Optane: throughput is higher for the former when block size is 16 megabytes or higher.
There is a limit of 2.3 gigabytes per second for Intel Optane.
For NVMe SSD it looks like the limit is yet unreached.
The bar for block size of 64 megabytes is almost 2.9 gigabytes per second
For small blocks Intel Optane throughput is better than those of NVMe SSD by a factor of two.
Recall that the average latency is also lower by a factor of two so these results match each other.

Sequential reading througput form both SATA SSD and HDD doesn't depend on block size and reaches 250 megabytes per second.
This experiment is not quite representative for HDD: the reading starts from the beginning of the disk rather than from a random point where throughput is lower.
For a small block size random reading throughput is so low for both SATA SSD and HDD that the bars are invisible.
However for sufficiently large blocks SSD achieve almost 300 megabytes per second while HDD reaches 175 megabytes per second.

\subsection{Summary}

In this section we observed how reading from a single thread behaves for both sequential and random block sequenceses on all four storage types.
To saturate the bandwidth one should pick up an appropriate block size.
If data is on HDD block size should be 16 megabytes or higher for random reading.
For SATA SSD it suffices to use 1 megabyte blocks.
For Intel Optane block size should be 4 megabytes to saturate the 2.3 gigabyte per second bandwidth.
If one uses NVMe SSD it is hard to reach the bandwidth with a single thread.
Even block size of 64 megabytes is not sufficient.
In the next section we will study reading form multiple threads.
As it will turn out three threads are enough to saturate NVMe SSD in terms of throughput.


\begin{figure}
    \includegraphics[width=\linewidth]{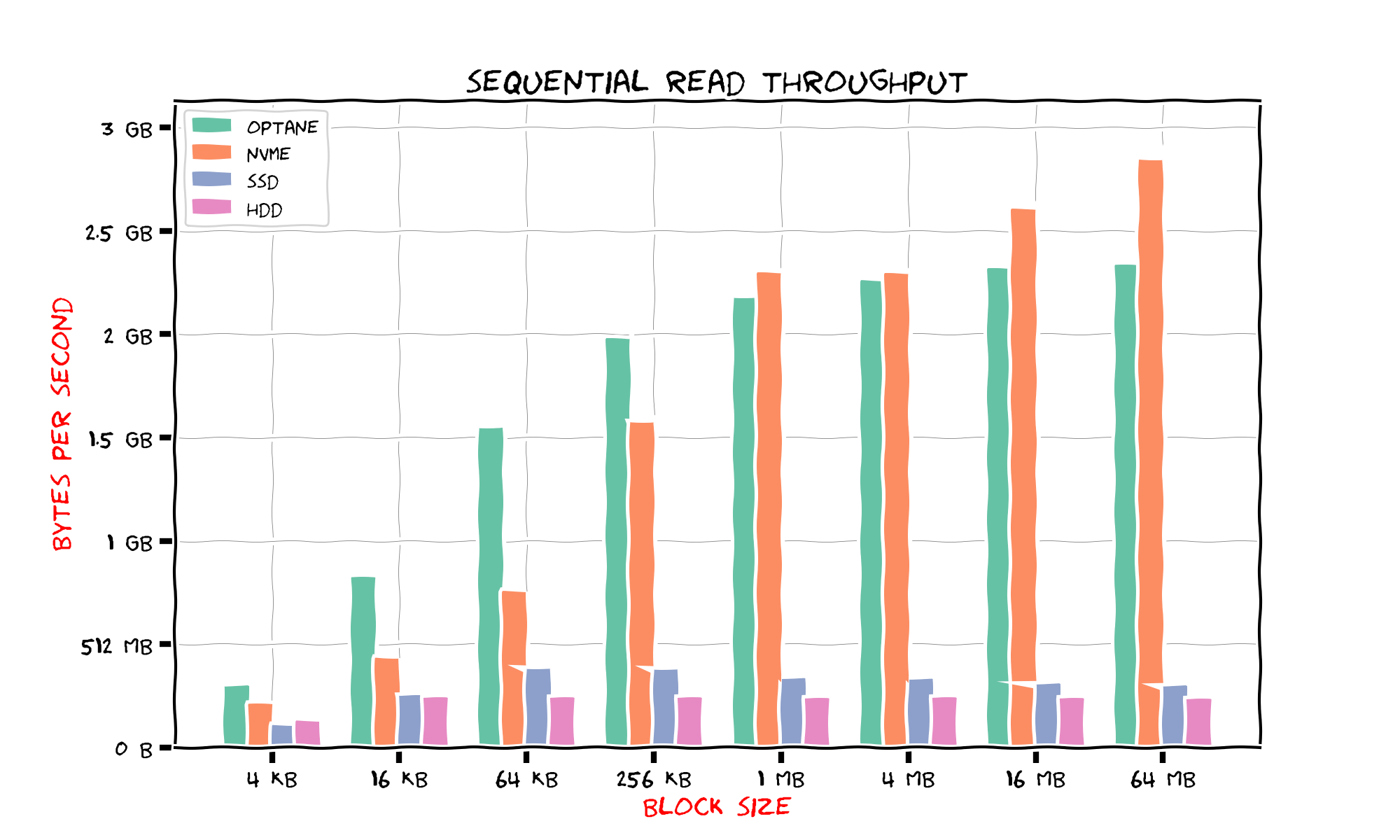}
    \caption{Throughput for all storage types sequential blocks reading.}
    \label{fig:throughputseq:all}
\end{figure}
\begin{figure}
    \includegraphics[width=\linewidth]{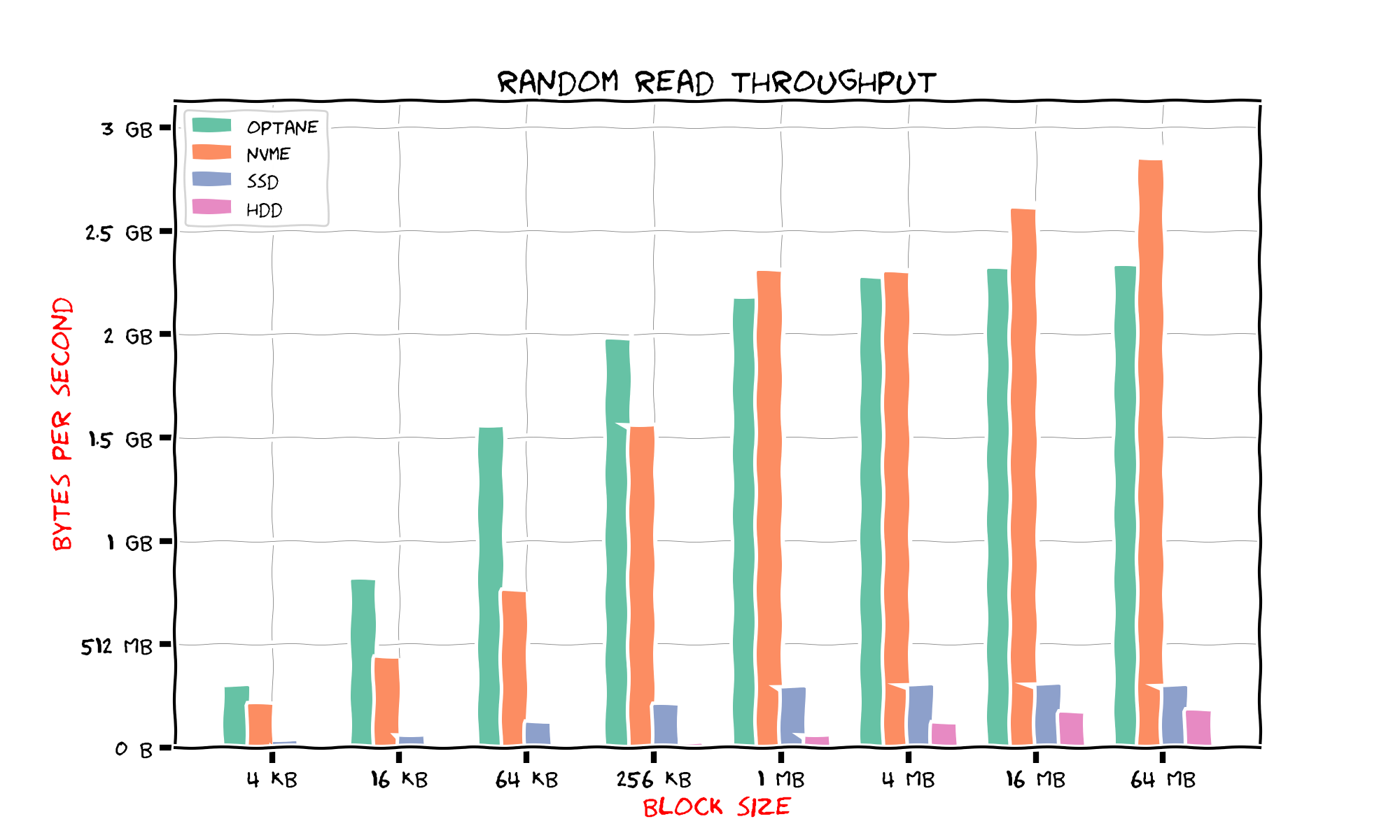}
    \caption{Throughput for all storage types random blocks reading.}
    \label{fig:throughputrand:all}
\end{figure}

\section{Multi-threaded Reading}
\label{sec:multithread}

Finally we start talking about how real production systems work with storage.
As shown in section~\ref{sec:sequential} reading from a single thread cannot saturate a modern device to its full potential.
Even for HDD it may be better to execute several requests simultaneously to allow them to be reordered in kernel thus reducing head movements between tracks.

In this section we consider random reading from multiple threads.
This corresponds to what many applications do.
Typically a dedicated thread pool is created and I/O requests are served by threads from this pool.
In fact, this is how POSIX AIO works in Linux by using thread pool created inside GNU Libc.
It is worth to note that we don't study here the cost of transferring the task into the thread pool.
We measure only the work inside the thread pool.
The results can be compared with those of asynchronous I/O in the next sections.

\begin{figure}
    \includegraphics[width=\textwidth]{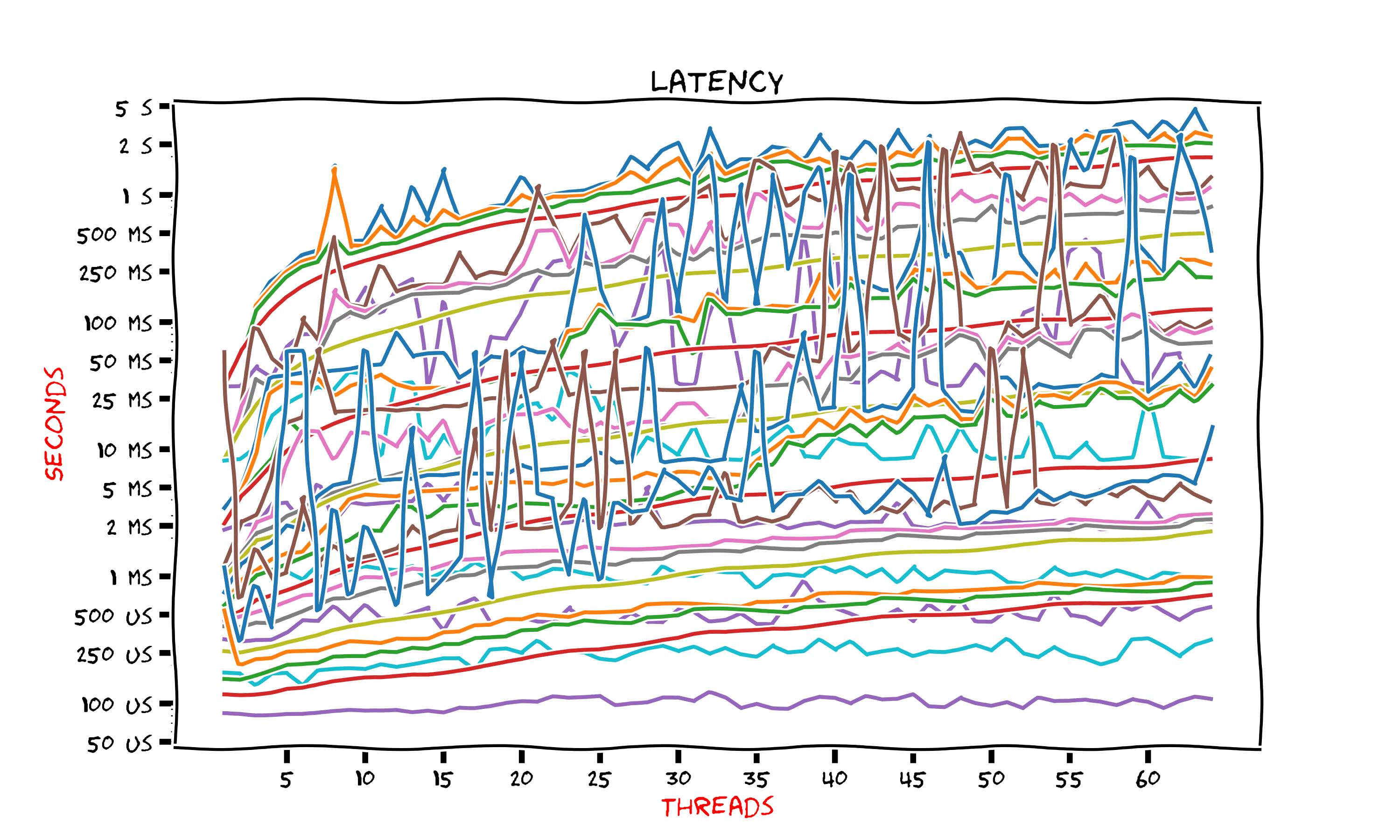}
    \caption{A failed attempt to show multiple results on a single chart.}
    \label{fig:allmultithread}
\end{figure}

There will be lots of figures in this section. Showing more information on a single chart looks like figure~\ref{fig:allmultithread}.

\subsection{HDD}

It's been long since hard disks and OS drivers learned a simple optimization: requests can be grouped together and reordered to save time spent on disk's head movements between tracks.
If it is necessary to visit several places on a platter it makes sense to reorder them by a track number and visit in a single sweep.
More intelligent schedulers consider not only time required to move head to the next point's track but the time which will be necessary to wait until required sector appears below the head due to disk rotation.
Thus the total seek plus rotation time between two points is taken into the account at the time of head route planning.
Obviously the firmware must receive several requests in advance to make the calculations.
NCQ extension for SATA protocol allow up to 32 commands to be sent to a device.

Before being sent to firmware requests are kept in a queue inside the kernel.
The kernel scheduler can reorder requests and merge them if blocks are consequent.
Therefore even for such sequential device as HDD making requests from several threads can result in better overall bandwidth compared to single-threaded execution.

\begin{figure}
    \begin{minipage}[c]{.47\linewidth}
        \includegraphics[width=\linewidth,trim=25 5 70 15, clip]{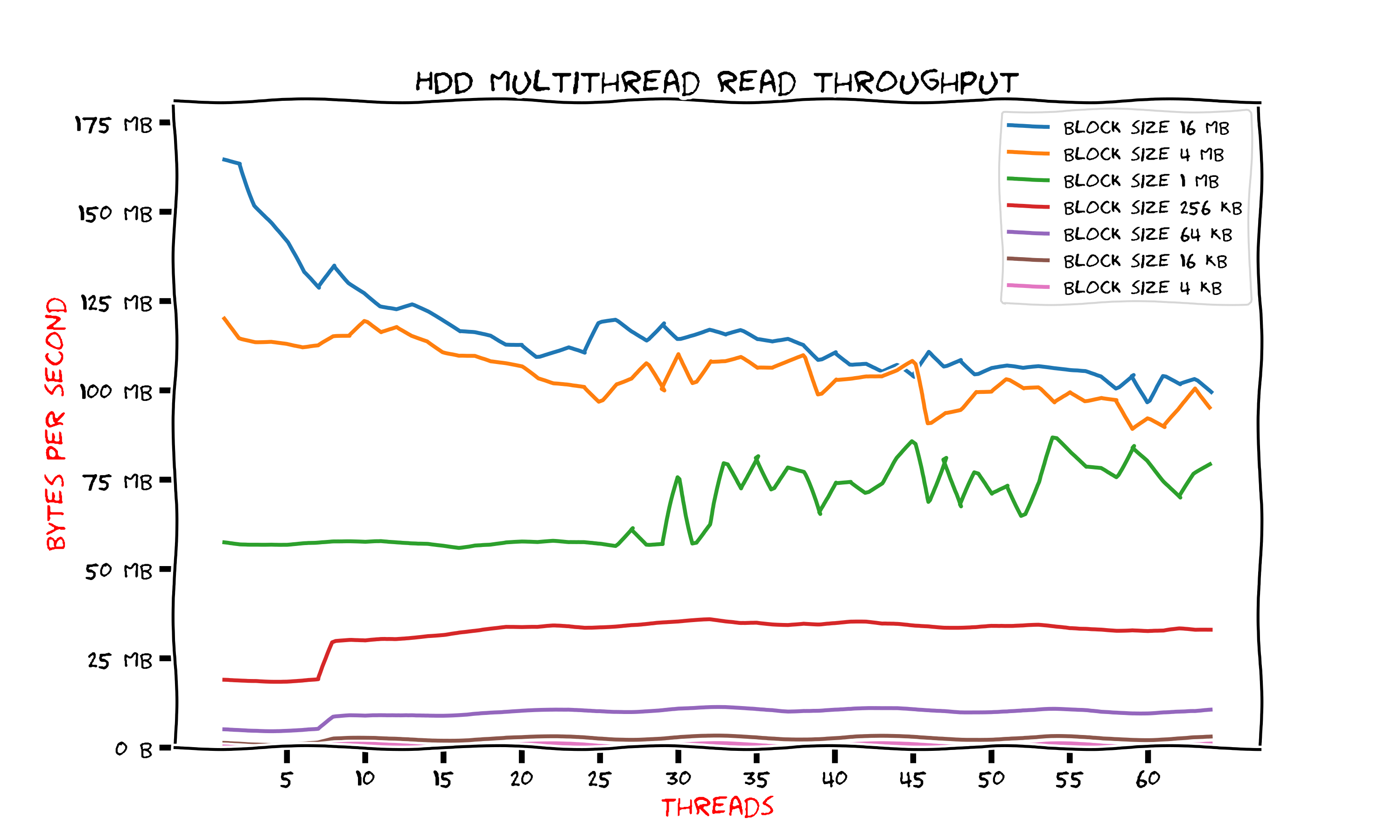}
        \caption{HDD reading throughput depending on number of threads.}
        \label{fig:multbw:hdd}
    \end{minipage}
    \hfill
    \begin{minipage}[c]{.47\linewidth}
        \includegraphics[width=\linewidth,trim=25 5 70 15, clip]{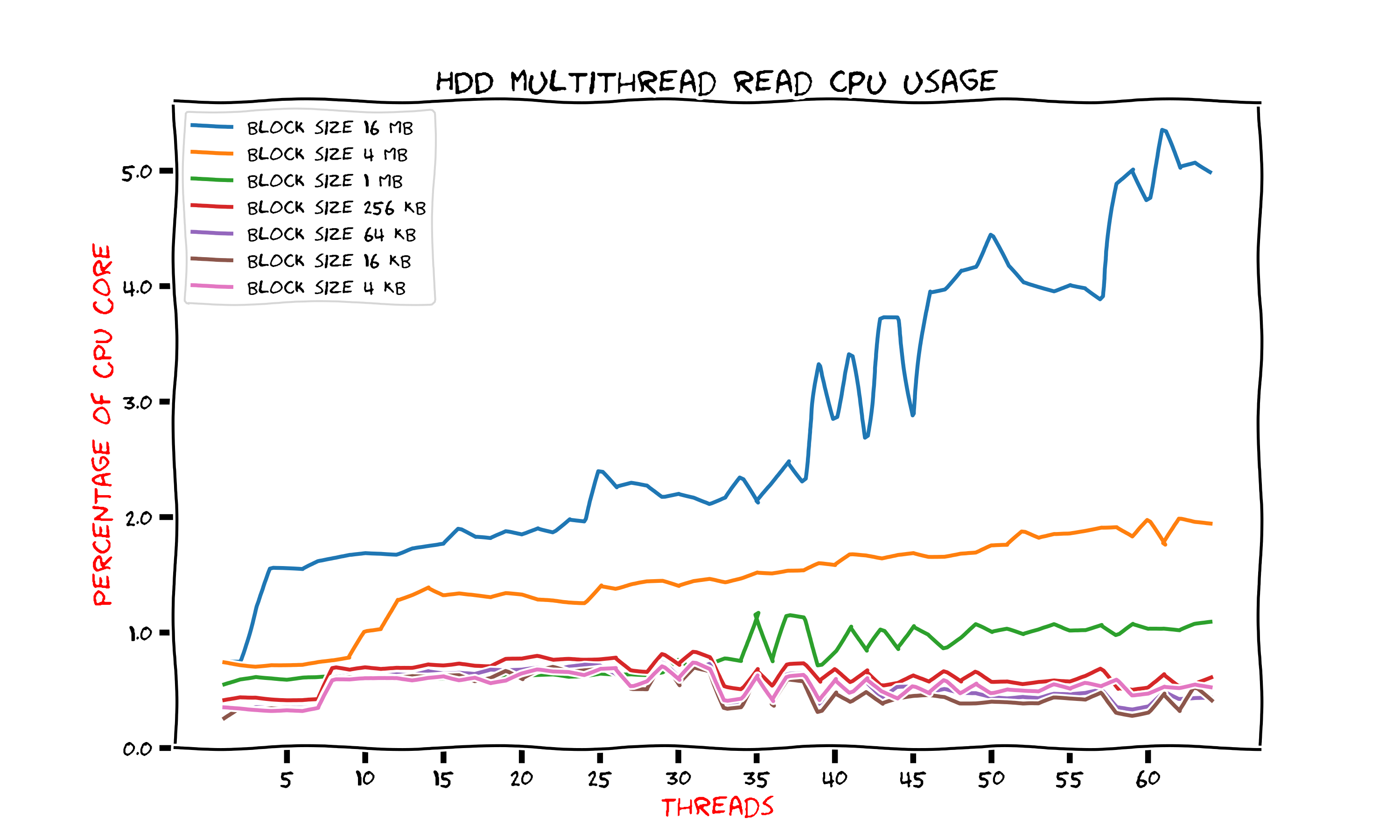}
        \caption{HDD reading CPU usage depending on number of threads.}
        \label{fig:multcpu:hdd}
    \end{minipage}
\end{figure}

Figure~\ref{fig:multbw:hdd} shows how throughput changes when the number of threads increases.
Different lines correspond to different block sizes.
First of all we should notice that for large blocks throughput decreases when the number of threads increases.
It looks a bit strange: although latency decrease is absolutely expected in this case it should be possible to sustain throughput.
Second, there is a really sharp step for 64 and 256 kilobytes blocks when the number of threads switches form 7 to 8.
It seems like heuristics in the scheduler depend on constants and are triggered in one case but not in the other.
In the next subsection we will look a bit into this.

Next we look at CPU usage.
This is also a resource that needs to be accounted even when we talk about reading from a storage.
It turns out CPU usage can be annoyingly noticeable.
In the next sections CPU usage will be one of the parameter to compare between synchronous and asyncrhonous interfaces.

Figure~\ref{fig:multcpu:hdd} shows CPU usage for readings with different block sizes from a different number of threads.
As on the previous figure lines correspond to different block sizes and the number of threads increases from left to right.
CPU usage is shown vertically and the value is the percentage of a single core usage.
This should be familiar to what usually performance utilities such as \texttt{top} show.
Reading form HDD takes no more than 5\% of a single core and the reader could ask why do we even care about such a small value.
However when we look at NVMe SSD a little further we will see saturation of several cores.
So an architector of resource-intensive application should take this value into account.
For completeness of the presentation we show CPU usage for HDD too.

\begin{figure}
    \begin{minipage}[c]{.47\linewidth}
        \includegraphics[width=\linewidth,trim=25 5 70 15, clip]{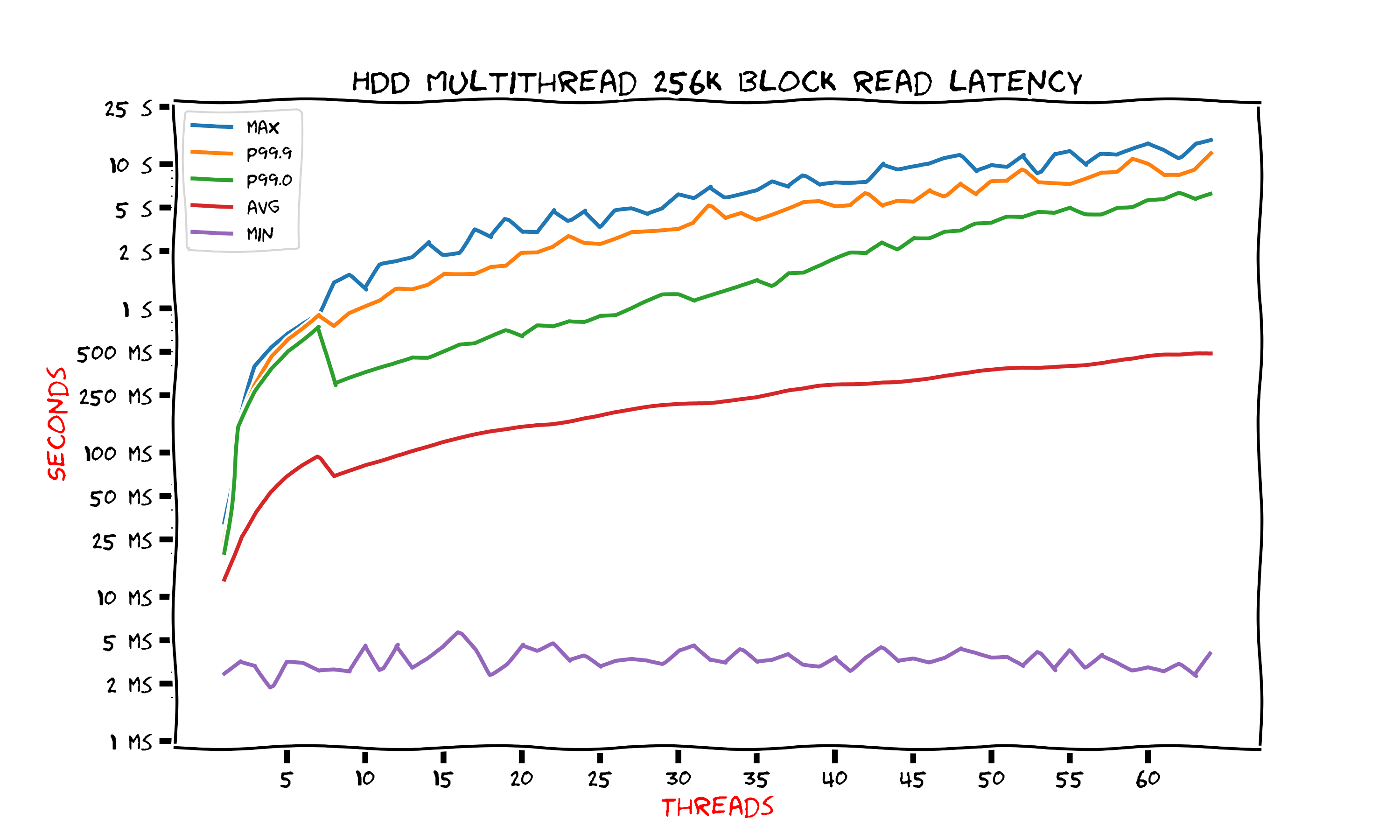}
        \caption{HDD multithread reading latency for 256 kilobytes block.}
        \label{fig:multlat256k:hdd}
    \end{minipage}
    \hfill
    \begin{minipage}[c]{.47\linewidth}
        \includegraphics[width=\linewidth,trim=25 5 70 15, clip]{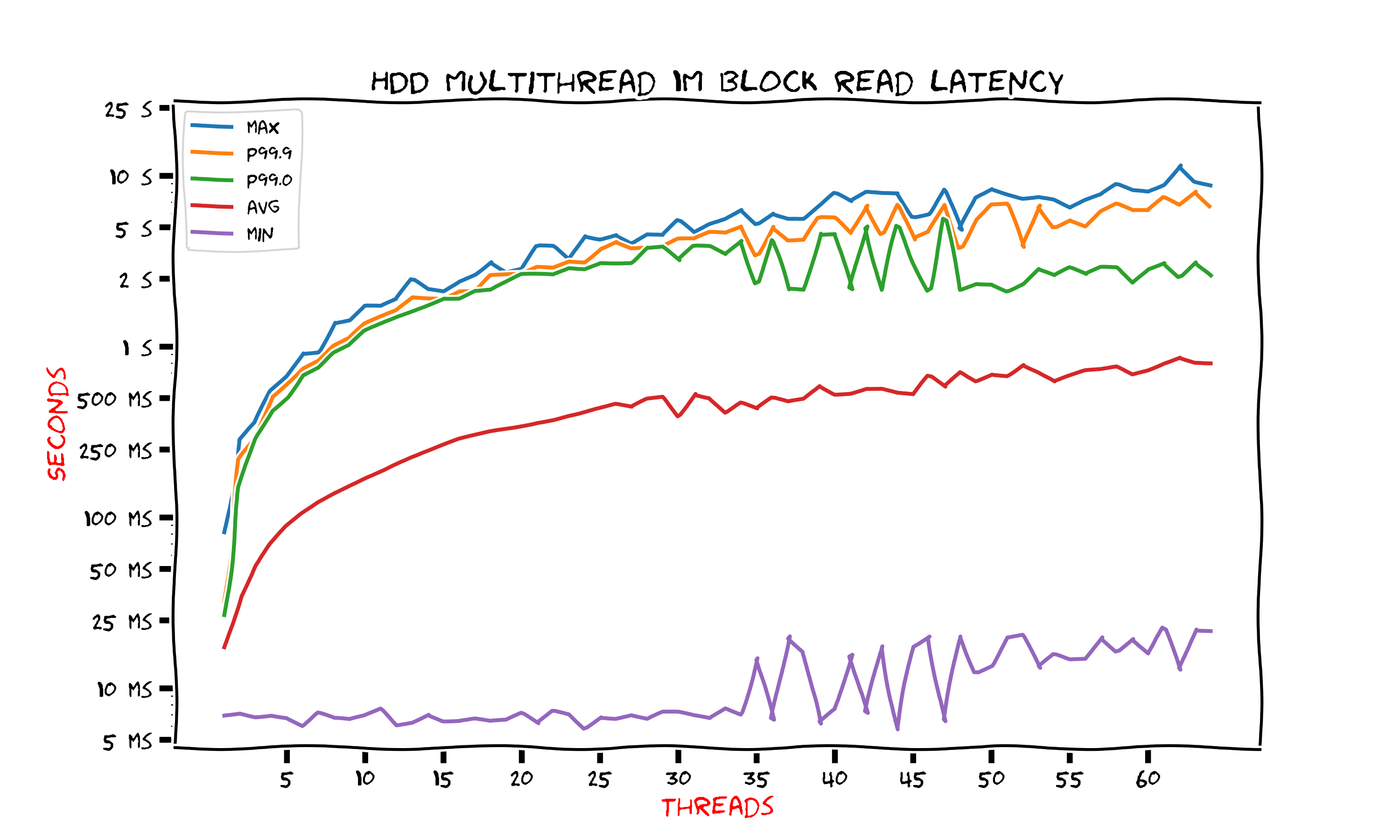}
        \caption{HDD multithread reading latency for 1 megabyte block.}
        \label{fig:multlat1m:hdd}
    \end{minipage}
\end{figure}

\begin{figure}
    \begin{minipage}[c]{.47\linewidth}
        \includegraphics[width=\linewidth,trim=25 5 70 15, clip]{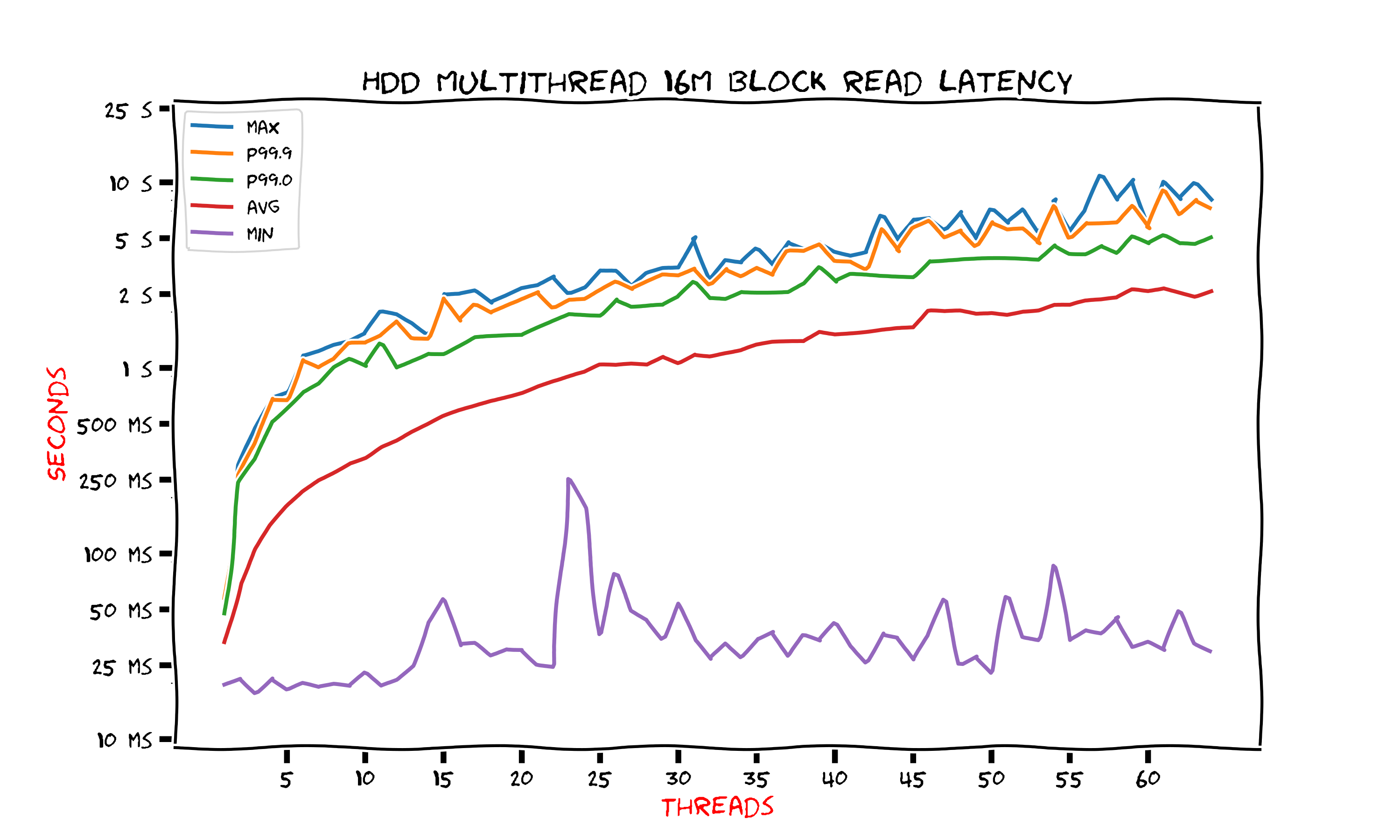}
        \caption{HDD multithread reading latency for 16 megabytes block.}
        \label{fig:multlat16m:hdd}
    \end{minipage}
    \hfill
    \begin{minipage}[c]{.47\linewidth}
        \includegraphics[width=\linewidth,trim=25 5 70 15, clip]{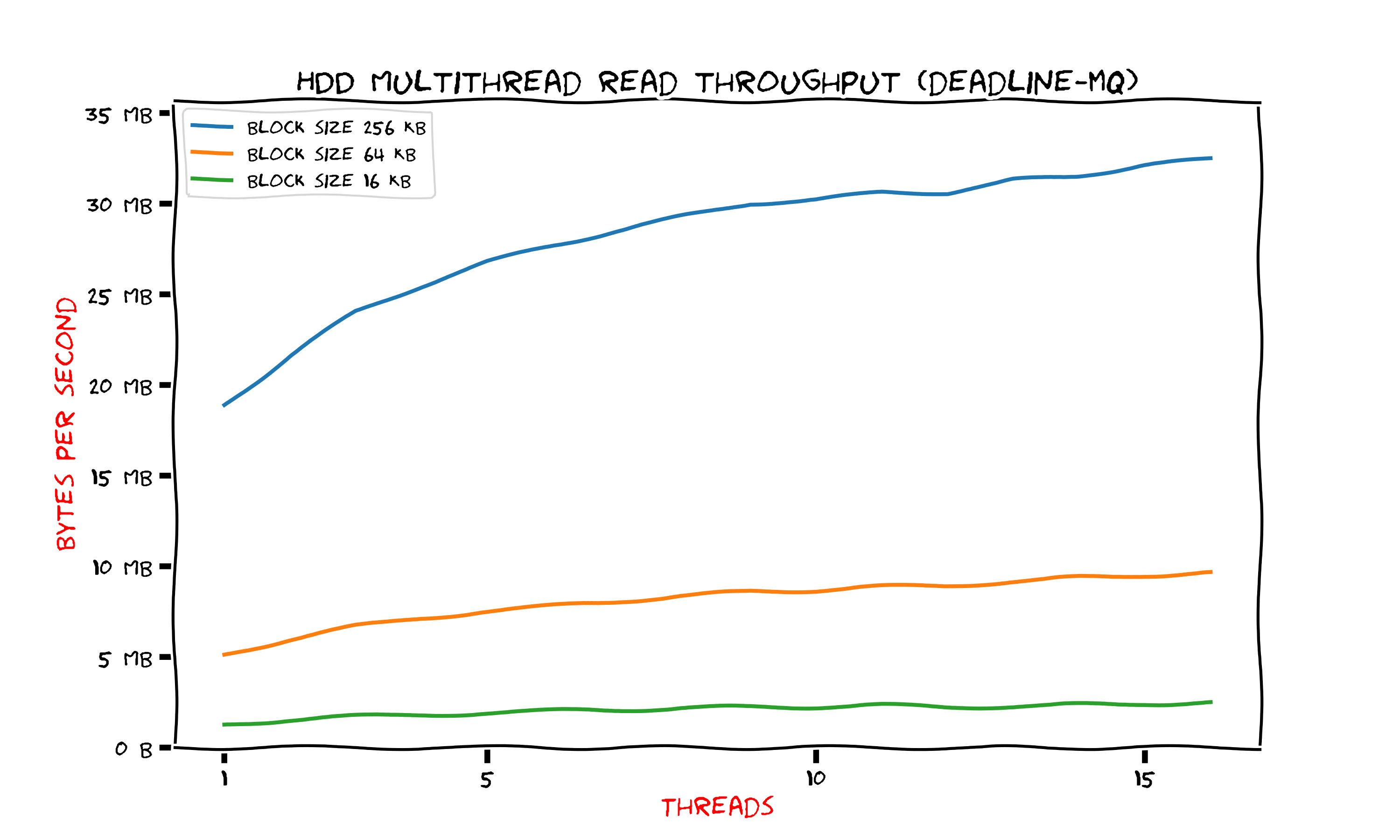}
        \caption{HDD multithread reading throughput when mq-deadline scheduler is enabled.}
        \label{fig:multmqbw:hdd}
    \end{minipage}
\end{figure}

Finally, we look at third interesting measurement~--- latency.
Latency also changes when the number of threads increase.
What makes latency different from throughput and CPU usage is that we need not only one value but several of them: minimum, maximum, average and percentiles.
We show them on the same chart dedicated for a certain block size.
Figure~\ref{fig:multlat256k:hdd} shows latency for 256 kilobytes block readings, figure~\ref{fig:multlat1m:hdd} for 1 megabyte block and figure~\ref{fig:multlat16m:hdd} for 16 megabyte block.
As we discussed earlier smaller blocks are not very meaningful for HDD random reading.

Surprisingly, when the number of threads changes from 7 to 8 in case of 256 kilobytes block size not only throughput increases but also latency decreases.
The value drops by a factor of two for the 99 percentile.
However the benefit is not so effective for the 99.9 percentile.

For a 1 megabyte block size latency behaves as expected when the number of threads increases from 1 up to 32.
It just raises smoothly.
Recall that throughput is constant for such block size.
This meets our expectations: there is a limited capacity of a hard drive to read random blocks.
If we increase the number of readers then they all just wait longer.
However if we pick a smaller block size or inadequate number of threads there will be artifacts on the chart.
Nevertheless when the number of threads is 55 and higher the lines become more smooth.

For a block size of 16 megabytes everything is pretty straightforward.
When the number of threads increases we see that latency raises and throughput decreases.
An obvious conclusion is that when a block size is large it is better to just use a single thread.

It is easy to see that if we want to get the maximum possible throughput we need to read large blocks from a single thread.
For 16 megabyte block size throughput as high as 170 megabytes per second is achievable even for a random blocks.
Each read requests takes 10 milliseconds in 99.9 percentile and about 30 milliseconds on the average.
Taking into account the hard disk geometry this result is pretty close to optimum.

If the product logic dictates as small block size as 256 kilobytes and it is necessary to maximize throughput 8-9 threads is the best choice.
The throughput is 30 megabytes per second which is nearly 6 times worse than the maximum average throughput.
The latency is even worse: 50 milliseconds on the average and more than half a second in 99.9 percentile.
It may be better to lower the expectations and agree to 20 megabytes per second when reading from a single thread
but on the other hand achieve more reasonable latency of 25 milliseconds in 99.9 percentile.

\subsection{I/O Scheduler}

\begin{figure}
    \begin{minipage}[c]{.47\linewidth}
        \includegraphics[width=\linewidth,trim=25 5 70 15, clip]{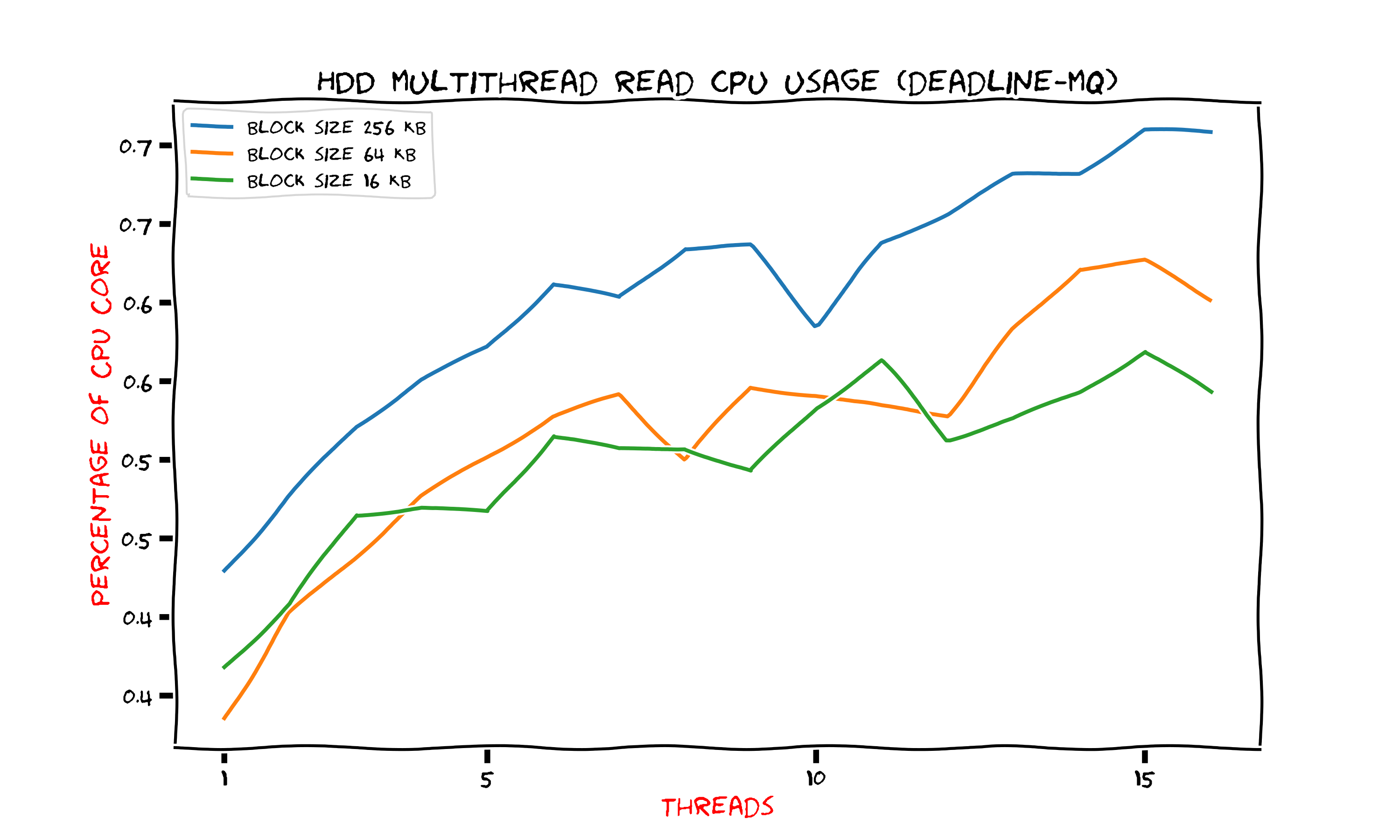}
        \caption{HDD multithread reading CPU usage when mq-deadline scheduler is enabled.}
        \label{fig:multmqcpu:hdd}
    \end{minipage}
    \hfill
    \begin{minipage}[c]{.47\linewidth}
        \includegraphics[width=\linewidth,trim=25 5 70 15, clip]{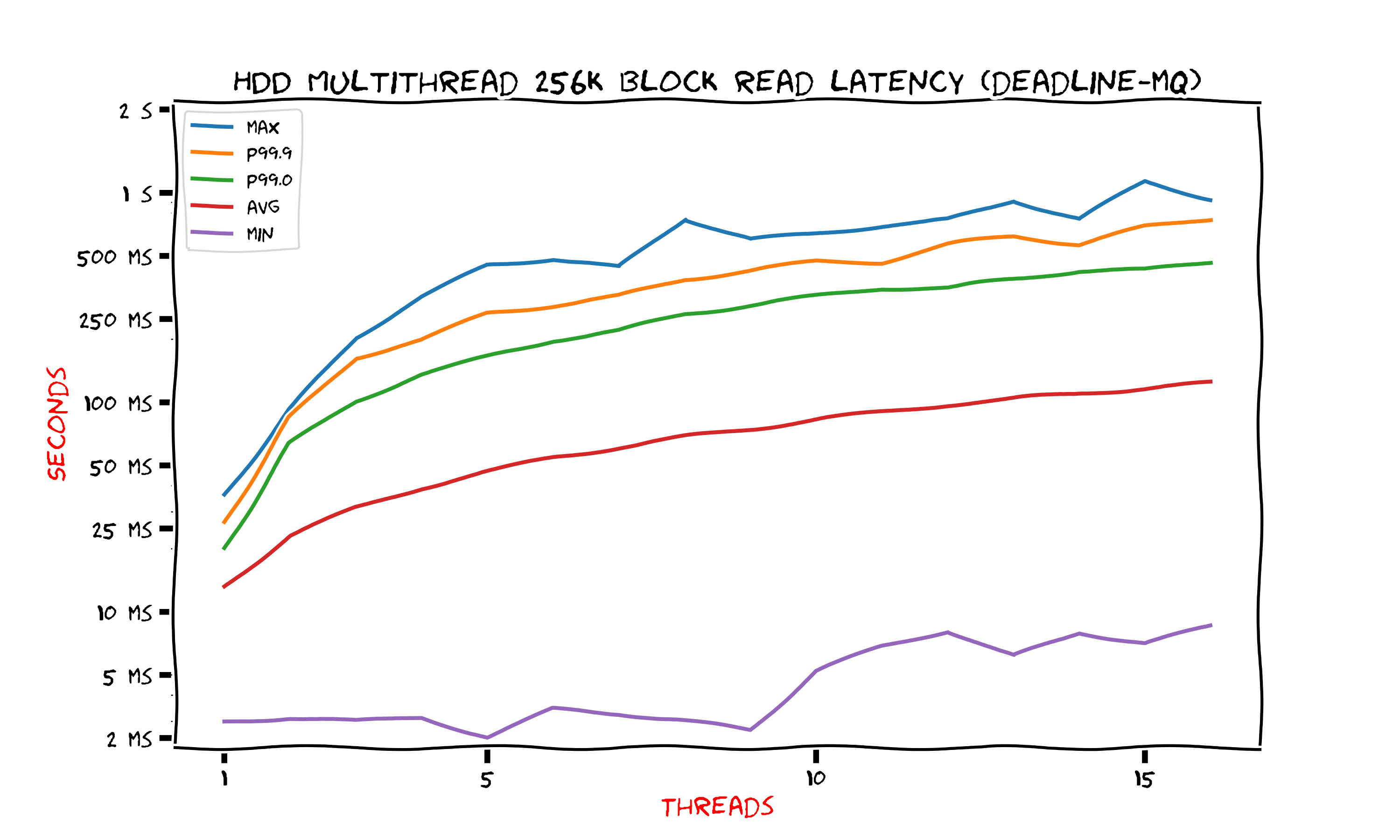}
        \caption{HDD multithread 256k blocks reading latency when mq-deadline scheduler is enabled.}
        \label{fig:multmqlat:hdd}
    \end{minipage}
\end{figure}

Previously we observed a sharp step in throughput when the number of threads increased from 7 to 8.
This happens because of kernel I/O scheduler.
Before application request is sent to a device it is kept inside the kernel queue.
Requests in that queue could be rearranged to achieve better performance because of locality.
The behavior is defined by particular I/O scheduler.
There are different schedulers for different device types.
In our case the default scheduler is Budget Fair Queuing~\cite{BFQ2010}.
To find out which scheduler is currently used one can execute the following command:

\begin{verbatim}
    $ cat /sys/block/sdb/queue/scheduler
    mq-deadline kyber [bfq] none
\end{verbatim}

BFQ scheduler is rather complex and depends on heuristic based on the number of blocks.
Another scheduler could be used, for example the one which implements deadline policy.
Such a scheduler tries to make sure that requests don't stay in the queue for too long.
To change current scheduler we execute the following command:

\begin{verbatim}
    # echo mq-deadline > /sys/block/sdb/queue/scheduler
\end{verbatim}

When \texttt{mq-deadline} scheduler is enabled results are more smooth.
Figures~\ref{fig:multlat16m:hdd},~\ref{fig:multmqcpu:hdd}, and~\ref{fig:multmqlat:hdd} show throughput, CPU usage and latency for the interval in question.
Number of threads is increased from 1 up to 16.
For the sake of presentation size latency is shown only for 256 kilobyte block size.
Latency for 64 kilobyte block size looks almost the same.
Also, as we saw earlier, making block size smaller than 256 kilobytes is not particularly helpful when HDD latency is at stake.

It is easy to notice that throughput now raises smoothly while the number of threads increases.
For a 256 kilobyte block size the line reaches 33 megabytes per second though it seems unlikely that it will cross the 35 megabytes per second even if we add more threads.
For 12 threads throughput of 30 megabytes per second is reached while latency stays below half a second in 99.9 percentile.
This is strictly better than the results for BFQ scheduler.
Nevertheless the resulting latency is quite high.

\subsection{SATA SSD}

In this section we look at similar charts for SATA SSD.
Reading from SSD can be done in much smaller blocks even when access pattern is random.
Actually it is in our interest to reduce the block size since.
Consider for example an application that requires access to random small chunks of information.
Naturally its authors would like to fetch data in 4 kilobyte blocks.
Therefore we study latency for block size of 4, 16 and 256 kilobytes.

\begin{figure}
    \begin{minipage}[c]{.47\linewidth}
        \includegraphics[width=\linewidth,trim=25 5 70 15, clip]{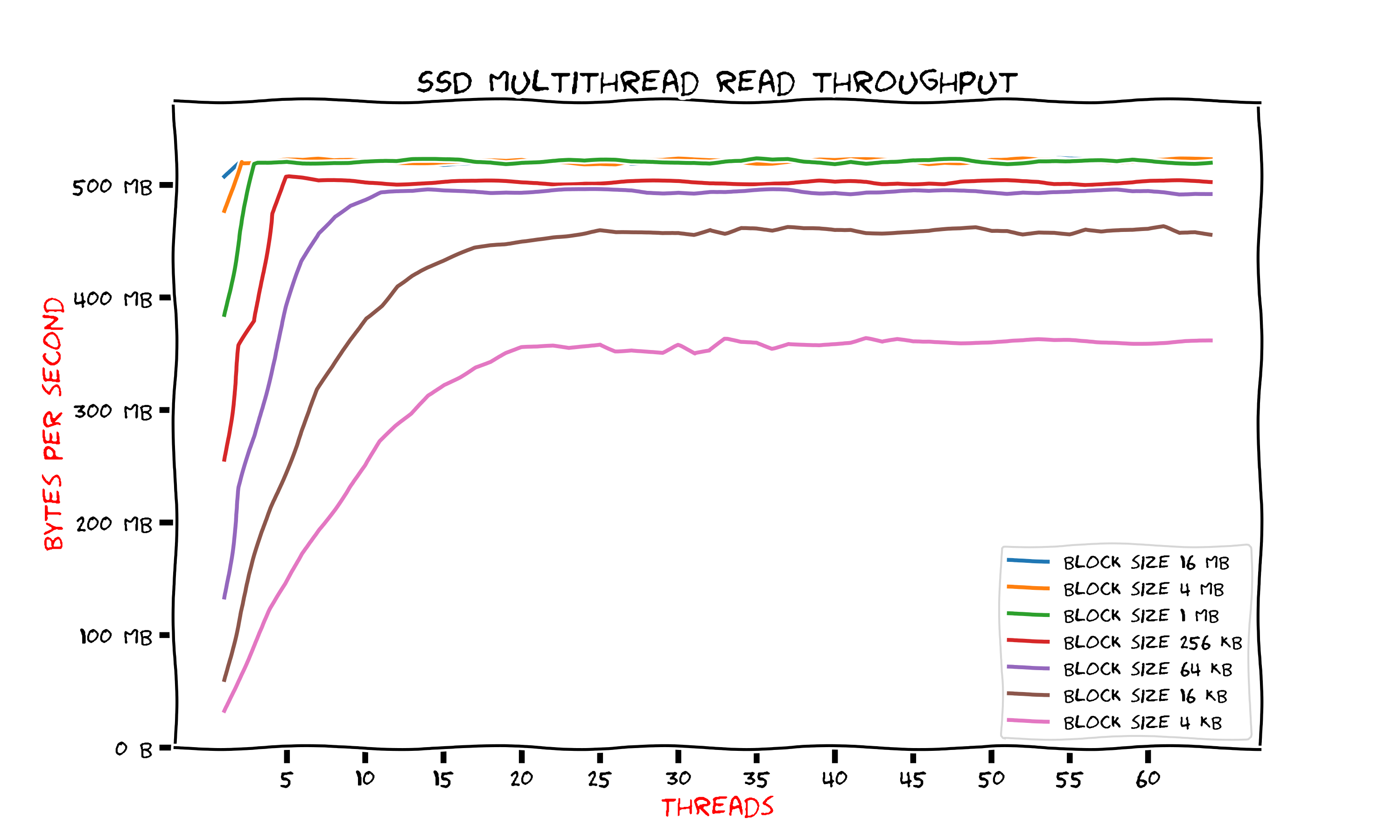}
        \caption{SSD multithread reading throughput.}
        \label{fig:multbw:ssd}
    \end{minipage}
    \hfill
    \begin{minipage}[c]{.47\linewidth}
        \includegraphics[width=\linewidth,trim=25 5 70 15, clip]{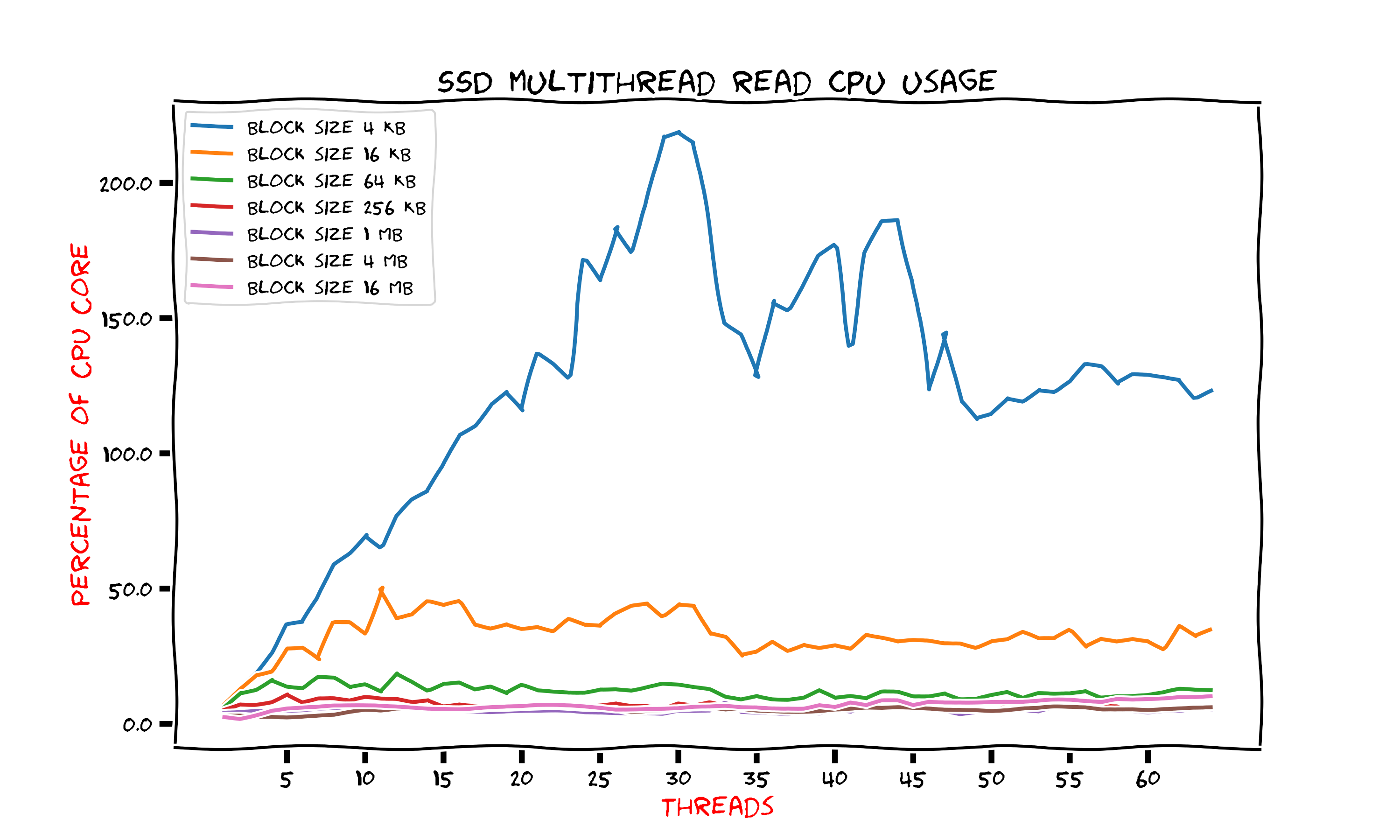}
        \caption{SSD multithread reading CPU usage.}
        \label{fig:multcpu:ssd}
    \end{minipage}
\end{figure}

At first we look at throughput.
Figure~\ref{fig:multbw:ssd} shows throughput for these three block sizes.
All lines behave similarly: at first they raise really fast and quite soon reach a plateau after which additional threads give absolutely no effect.
The overall maximum for all block sizes is 500 megabytes per second.
However reaching this throughput requires reading in rather large blocks: 256 kilobytes and higher.
If the block size is small the throughput is significantly smaller and could not be improved even with more threads.

Next we study CPU usage shown on figure~\ref{fig:multcpu:ssd}.
It turns out that when we read in blocks of size 16 kilobytes a significant amount of CPU time is spent on SSD read requests processing.
We talk about several tenths of a fraction of a single core.
If the block size is 4 kilobytes we need to allocate almost two cores only for SSD reading requests processing.
Needless to say this is a noticeable amount if our application needs to carry a CPU-intensive workload and CPU is carefully accounted.

Figure~\ref{fig:multbw:ssd} shows that if we want to achieve the maximum possible throughput for a block size of 4 kilobytes we need 21 threads.
For a block size of 64 kilobytes we need 13 threads for that and for a blocks of size 256 kilobytes we need only 3 threads.
According to figure~\ref{fig:multcpu:ssd} the CPU usage for these block sizes is respectfully 120\%, 10\% and 5\% of a single core.

\begin{figure}
    \begin{minipage}[c]{.47\linewidth}
        \includegraphics[width=\linewidth,trim=25 5 70 15, clip]{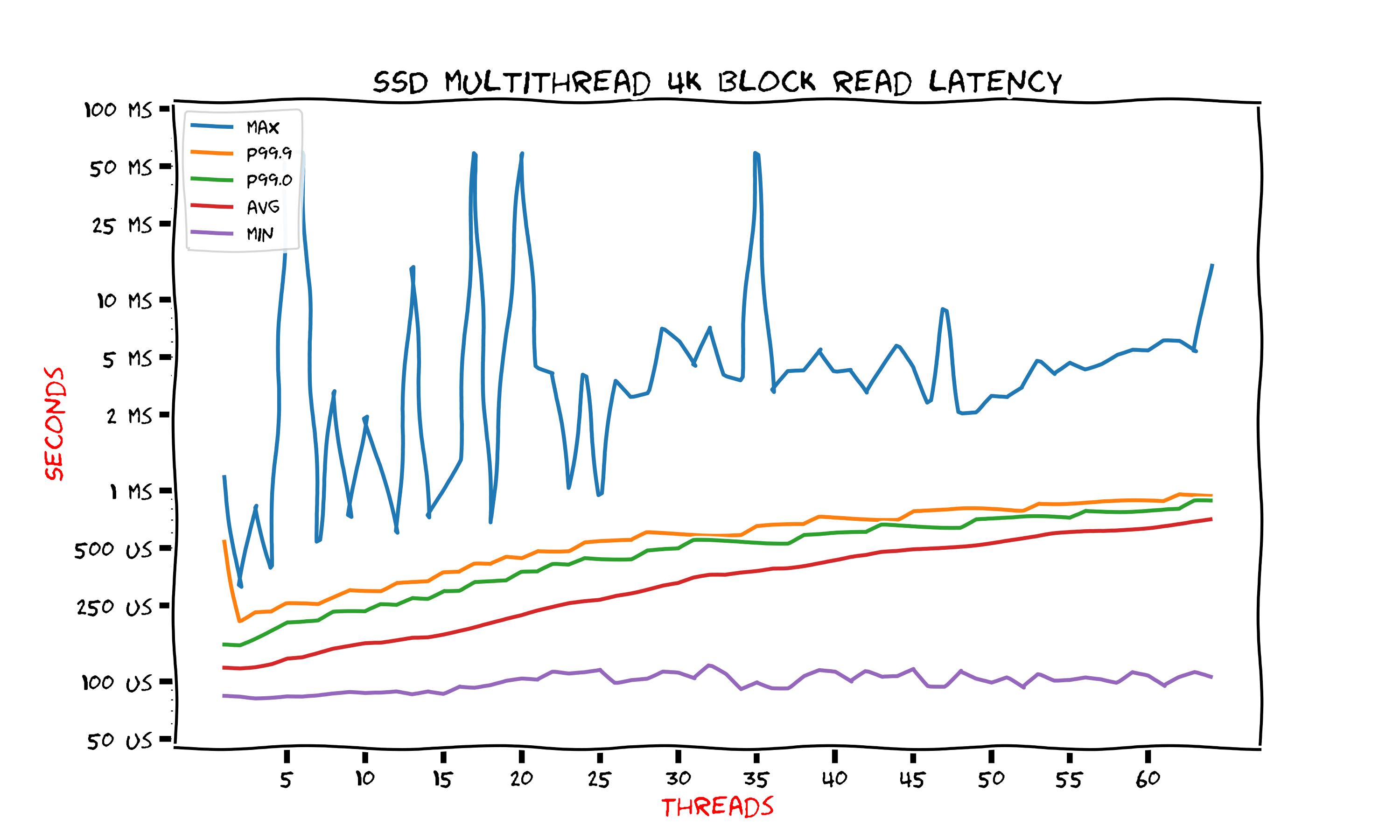}
        \caption{SSD multithread reading latency for 4 kilobytes blocks.}
        \label{fig:multlat4k:ssd}
    \end{minipage}
    \hfill
    \begin{minipage}[c]{.47\linewidth}
        \includegraphics[width=\linewidth,trim=25 5 70 15, clip]{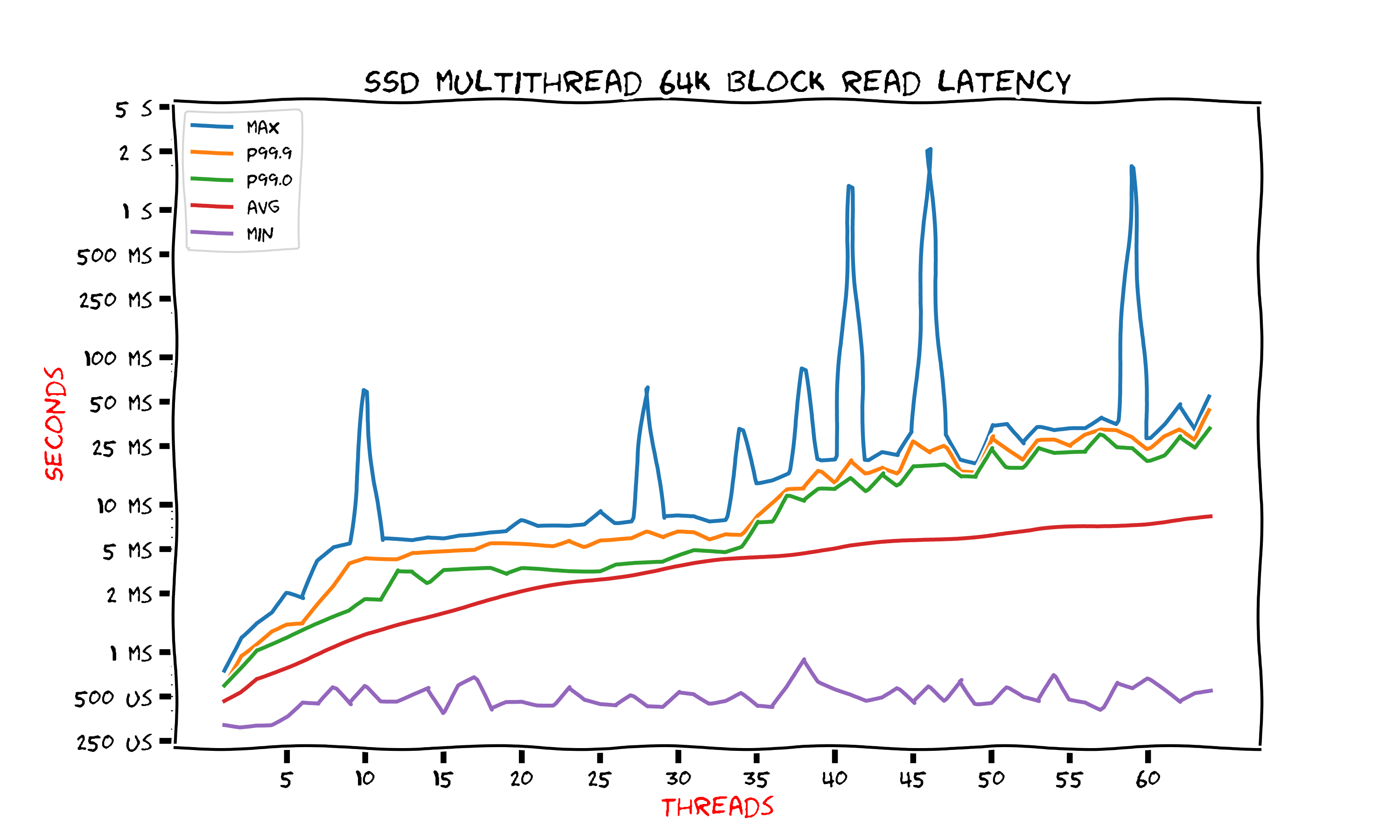}
        \caption{SSD multithread reading latency for 64 kilobytes blocks.}
        \label{fig:multlat64k:ssd}
    \end{minipage}
\end{figure}
\begin{figure}
    \begin{center}
    \begin{minipage}[c]{.47\linewidth}
        \includegraphics[width=\linewidth,trim=25 5 70 15, clip]{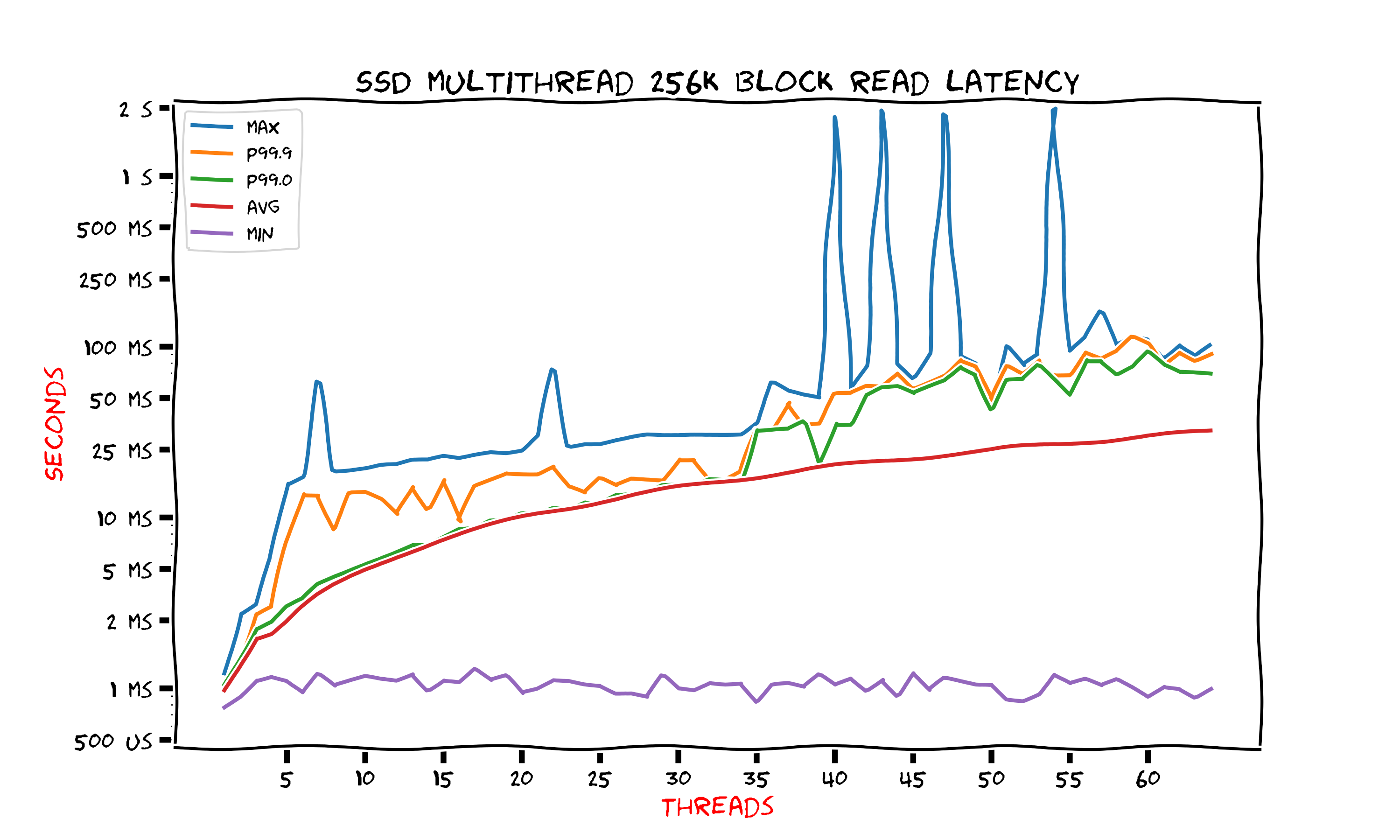}
        \caption{SSD multithread reading latency for 256 kilobytes blocks.}
        \label{fig:multlat256k:ssd}
    \end{minipage}
    \end{center}
\end{figure}

Now lets look at the latency and find out which latencies we get for the above number of threads (picked to optimize throughput).
On figures~\ref{fig:multlat4k:ssd},~\ref{fig:multlat64k:ssd}, and~\ref{fig:multlat256k:ssd} we see that it will take on average 200 microseconds for 4 kilobytes blocks, and between 1 and 2 seconds for blocks of size 64 and 256 kilobytes.

So, for 4 kilobytes blocks we need 21 threads.
Figure~\ref{fig:multlat4k:ssd} shows nearly 500 microseconds latency in 99.9 percentile.
This is almost twice as bad as latency for a single thread.
There is a peak at 1 thread for a 99.9 percentile latency, however it may not be the real behavior but an artifact of cold start.
For each test there is a warming period when requests are being made but no measurements are taken and it may be the case that in this test there were not enough requests at a warming period.

To get maximum throughput for 64 kilobytes blocks we need 13 threads.
As we see on figure~\ref{fig:multlat64k:ssd} the latency is horrifying 5 milliseconds in 99.9 percentile.
This is ten times worse than the latency for a single thread.
However reading from a single thread would give us four times less throughput.

Generally speaking if we agree with 5 milliseconds latency we can try reading larger blocks from a single thread.
Truly the single thread reading latency is only 2 milliseconds even for a megabyte-sized blocks.
However throughput will be smaller by nearly 100 megabytes per second.
Moreover we need to keep in mind that smaller granularity has its own advantages and reading a few randomly distributed smaller blocks could not be compared to reading one large block in terms of data placement and access.

If we select block size of 256 kilobytes then the maximum throughput is achievable by only three threads.
The value is almost 500 megabytes per second.
CPU usage is about 5\% of a single core which is really a small amount.
The latency happens to be 5 milliseconds in 99.9 percentile.

Recall that when we tried to read from HDD in 256 kilobyte blocks we achieved only 20 or 30 megabytes per second.
Thus our results for SSD show about twenty times better throughput.
Latency for HDD would be 25 milliseconds even for a single thread which is five times worse.
And that would be 20 megabytes per second for HDD.
As for 30 megabytes per second which we see in multi-threaded setting the latency is tremendously bad.
By reaching almost half a second it is hundred times worse than one for SSD.

\subsection{NVMe SSD}

\begin{figure}
    \begin{minipage}[c]{.47\linewidth}
        \includegraphics[width=\linewidth,trim=25 5 70 15, clip]{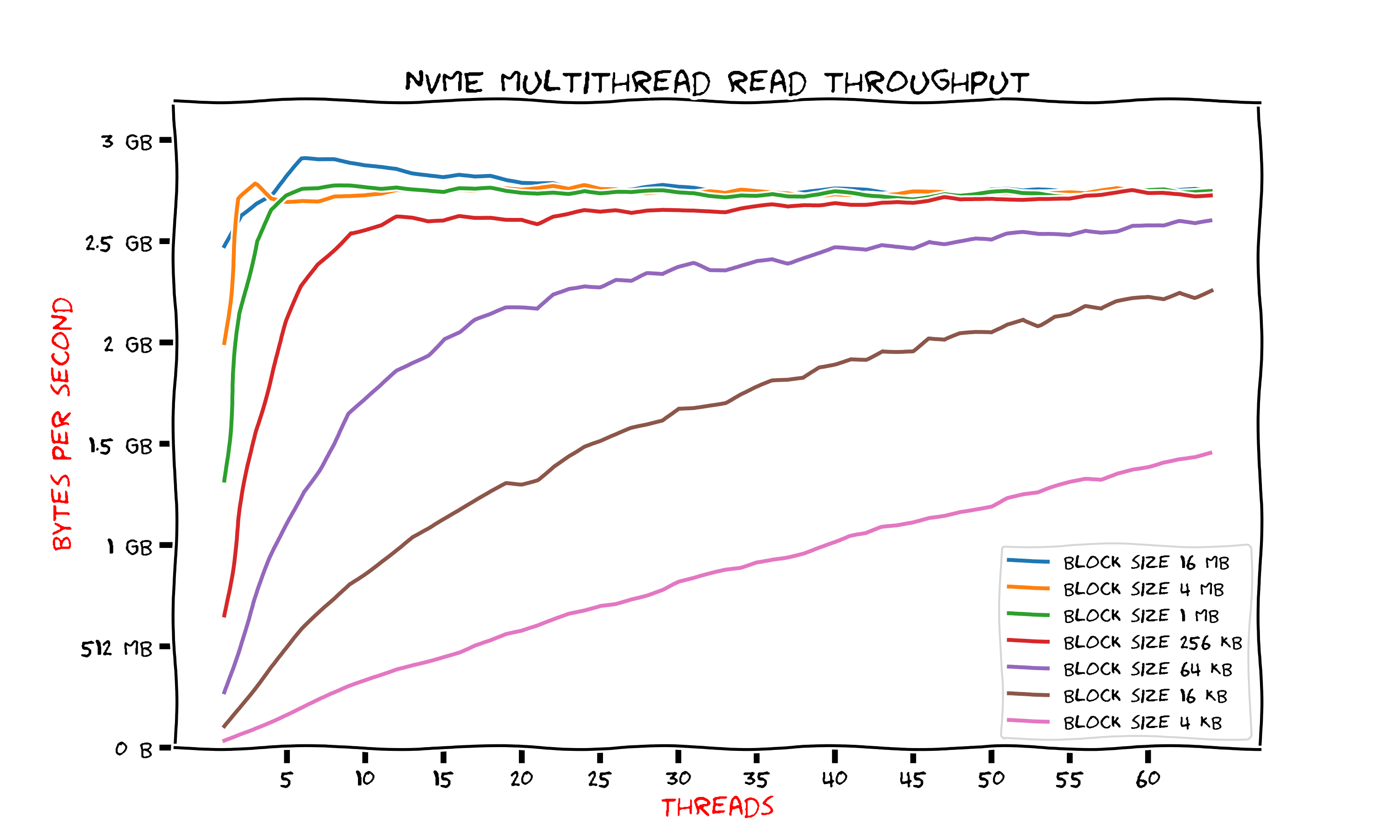}
        \caption{NVMe SSD multithread reading throughput.}
        \label{fig:multbw:nvme}
    \end{minipage}
    \hfill
    \begin{minipage}[c]{.47\linewidth}
        \includegraphics[width=\linewidth,trim=25 5 70 15, clip]{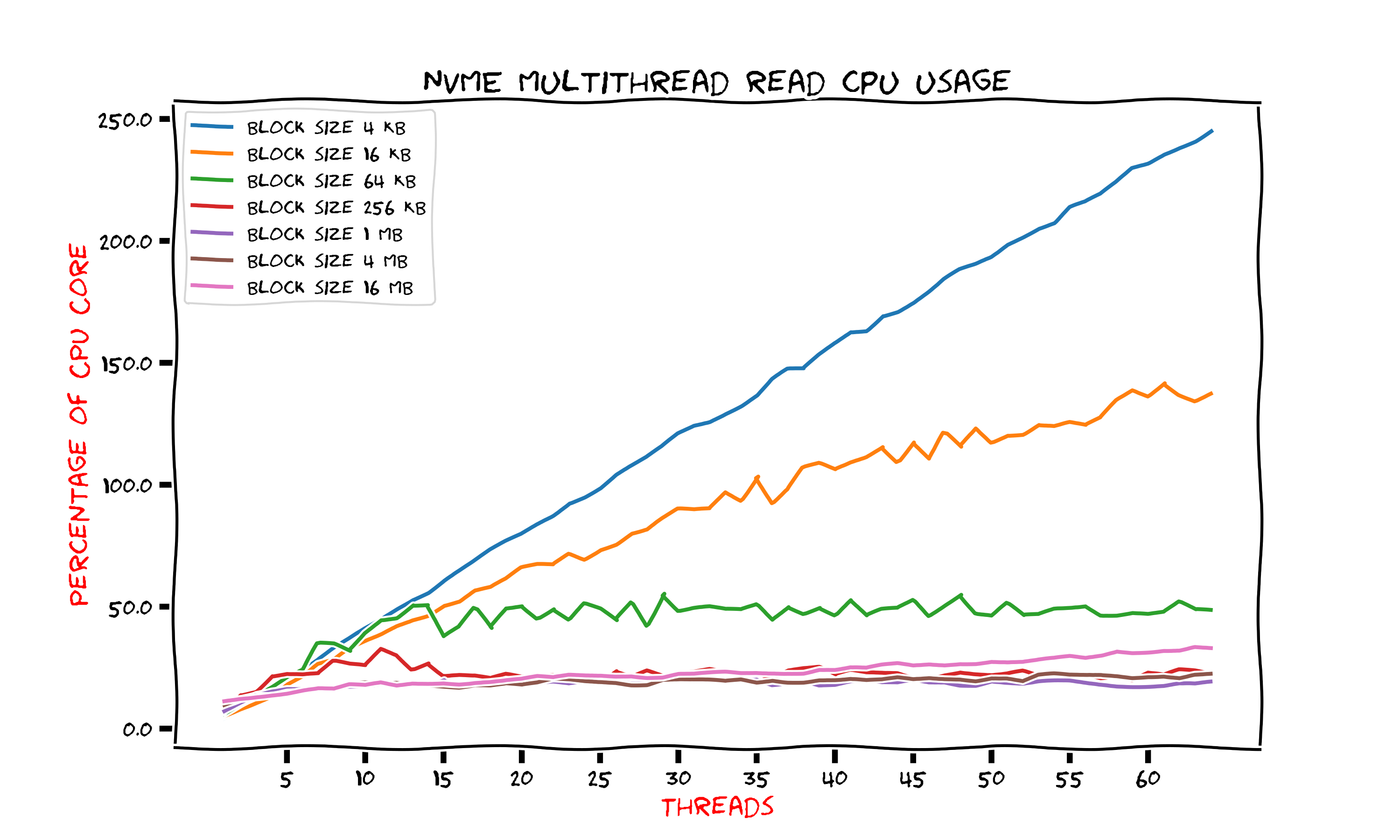}
        \caption{NVMe SSD multithread reading CPU usage.}
        \label{fig:multcpu:nvme}
    \end{minipage}
\end{figure}

Now we move to NVMe SSD and look at throughput on figure~\ref{fig:multbw:nvme}.
Throughput lines appear to behave similarly to SSD: many lines reach plateau really fast and don't change after that despite of how many new threads were added.
For NVMe SSD the plateau is several times higher and is about 2.7 gigabytes per second.
However all of this is not true for small blocks.
For example the line for 4 kilobytes blocks is only half of it's way to the maximum.
We will take a closer look at small blocks in the next subsection.
As for now let's look at large blocks.

For block size 1 megabyte and higher it is sufficient to have five threads to saturate the bandwidth.
For seven threads and 16 megabyte blocks we observe the absolute maximum of nearly 2.9 gigabytes per second.
Surprisingly when the number of threads is increased the throughput gradually reduces to the same 2.7 gigabytes per second.

CPU usage is shown on figure~\ref{fig:multcpu:nvme}.
For large blocks it is about 20\% of a single core.
For small blocks there is a linear growth which continues even after the right border of the chart.
It is expected since the throughput also grows.
But we should note that even this unsaturated CPU usage is quite high.
It's already higher than 100\% of a single core for 16 kilobyte blocks and more than two full cores for 4 kilobytes blocks.

\begin{figure}
    \begin{minipage}[c]{.47\linewidth}
        \includegraphics[width=\linewidth,trim=25 5 70 15, clip]{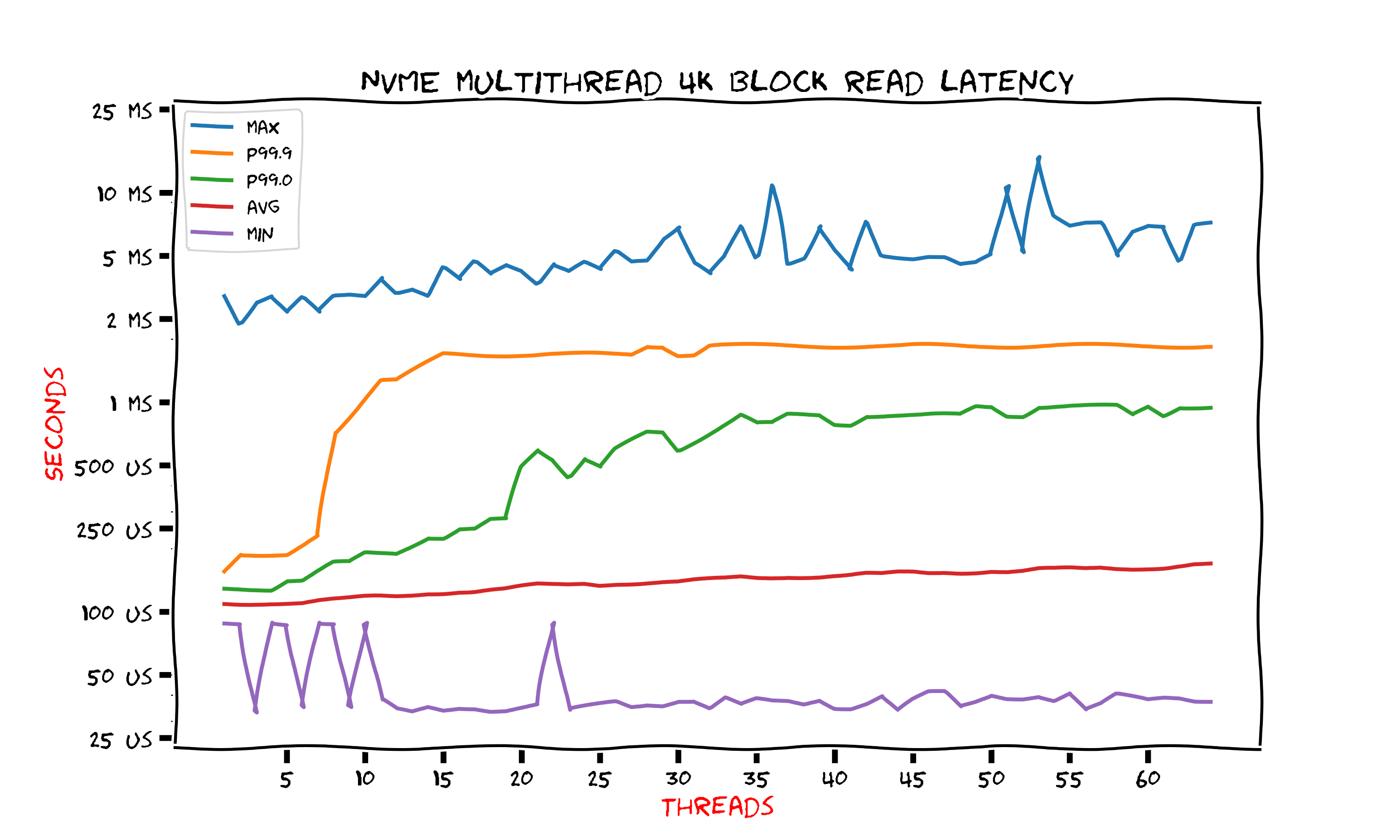}
        \caption{NVMe SSD multithread reading latency for 4 kilobytes blocks.}
        \label{fig:multlat4k:nvme}
    \end{minipage}
    \hfill
    \begin{minipage}[c]{.47\linewidth}
        \includegraphics[width=\linewidth,trim=25 5 70 15, clip]{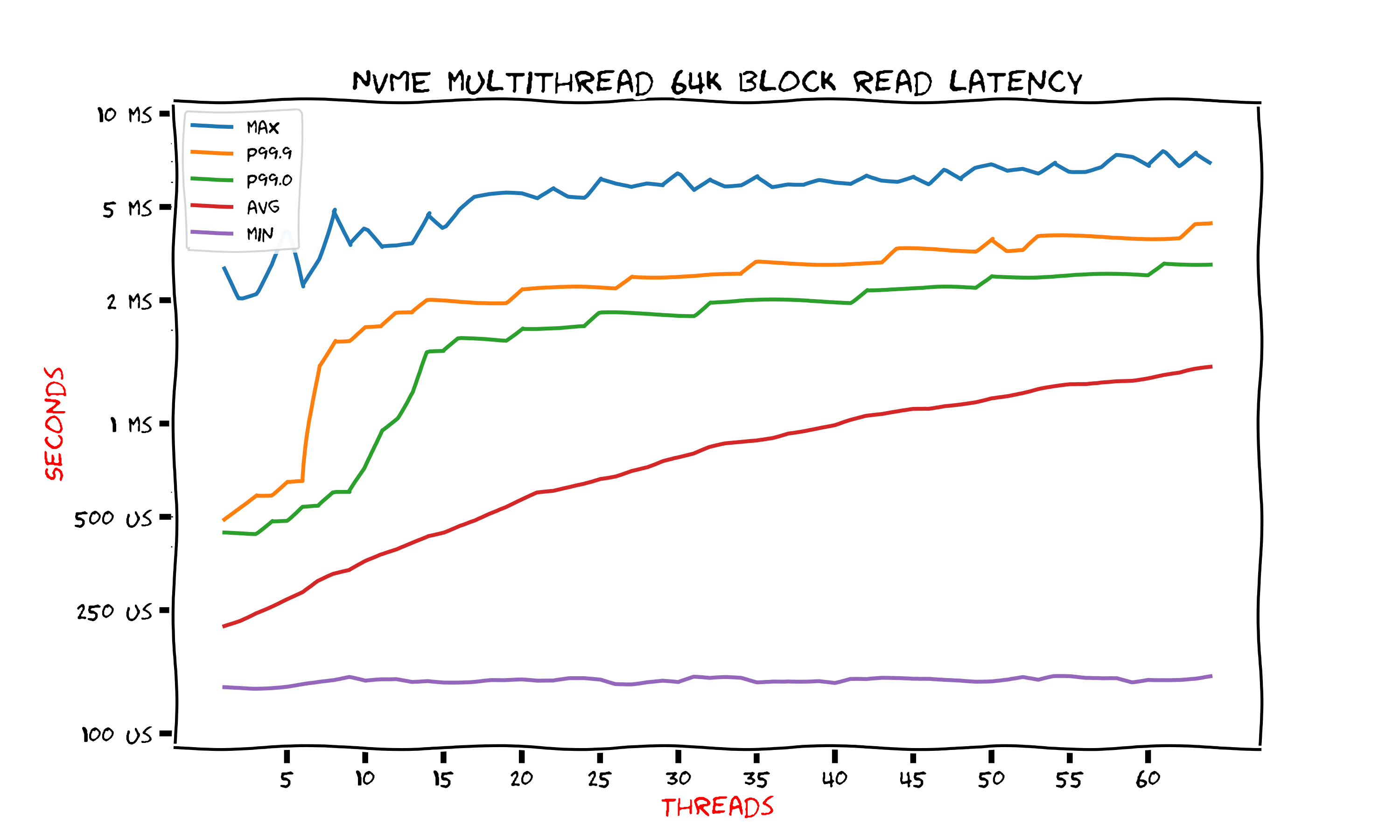}
        \caption{NVMe SSD multithread reading latency for 64 kilobytes blocks.}
        \label{fig:multlat64k:nvme}
    \end{minipage}
\end{figure}

Figure~\ref{fig:multlat4k:nvme} shows latency for 4 kilobytes block size.
As we see, higher throughput comes with the cost by latency.
The 99.9 percentile happens to be be between 1 and 2 milliseconds.
It is quite interesting that it raises to a certain value and doesn't grow more.
Surprisingly the point when this value is reached is when the throughput matches the SATA SSD maximum throughput of 500 megabytes per second.
SSD latency for the same block size is half a millisecond while the maximum throughput of roughly 300 megabytes per second.
Remember that it is impossible to saturate SATA SSD bandwidth with 4 kilobyte blocks therefore this is not quite fair competition.

Another unexpected fact is that the minimum latency significantly reduces when the number of threads is increased.
With the value of nearly 30 microseconds it starts to resemble Intel Optane.
Unfortunately we don't have the explanation for this phenomena.
This probably has to do something with caching.
However it is hard to imagine a cache with such a hit rate for the whole device.
One possibility is that this is an observation of a Flash Translation Layer cache.
Flash Translation Layer is a part of SSD firmware that is responsible to translate LBA sector number (which a device receives in a request from operating system) into internal pages on the flash memory itself.
Due to physical limitations of NAND memory this is quite a complicated and carefully optimized software.

\begin{figure}
    \begin{minipage}[c]{.47\linewidth}
        \includegraphics[width=\linewidth,trim=25 5 70 15, clip]{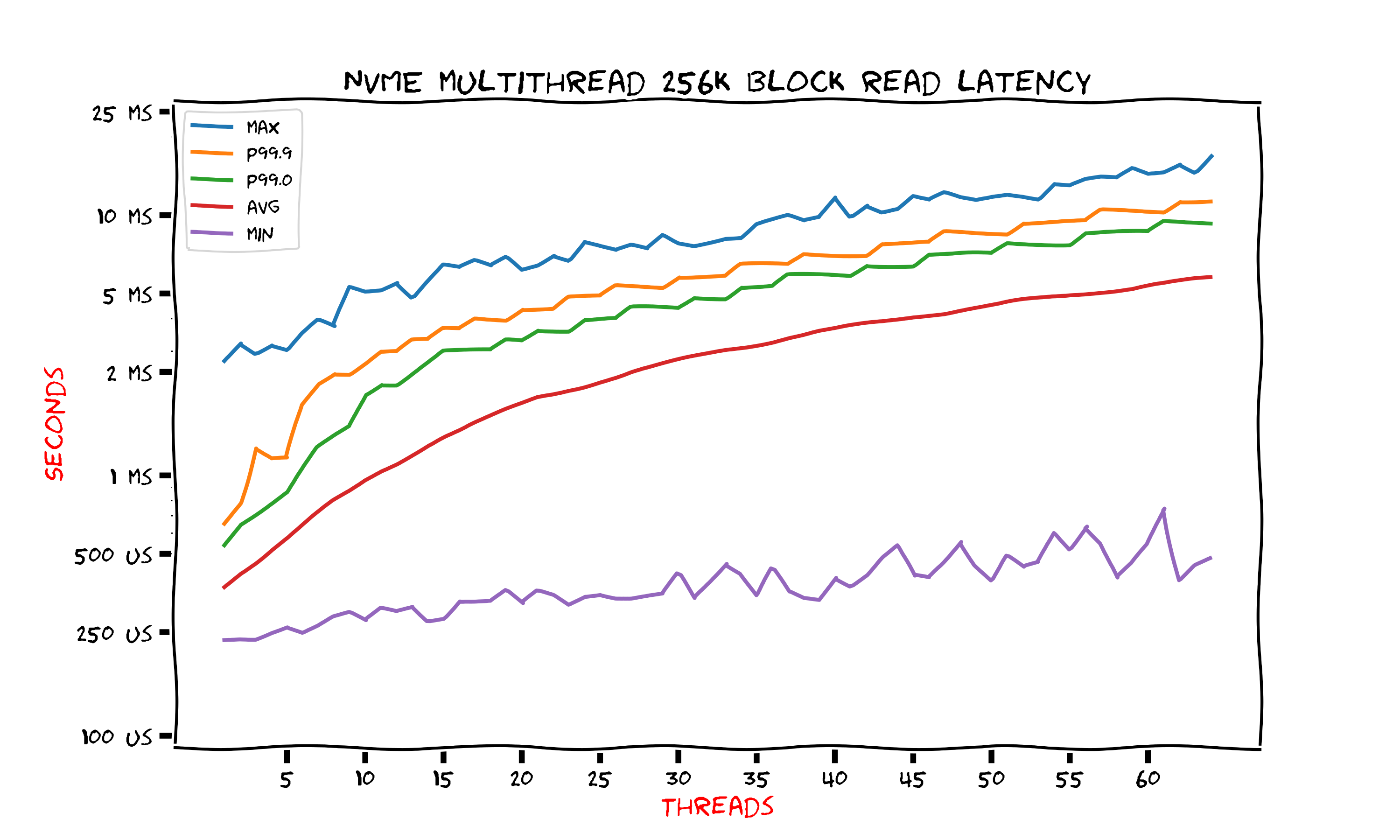}
        \caption{NVMe SSD multithread reading latency for 256 kilobytes blocks.}
        \label{fig:multlat256k:nvme}
    \end{minipage}
    \hfill
    \begin{minipage}[c]{.47\linewidth}
        \includegraphics[width=\linewidth,trim=25 5 70 15, clip]{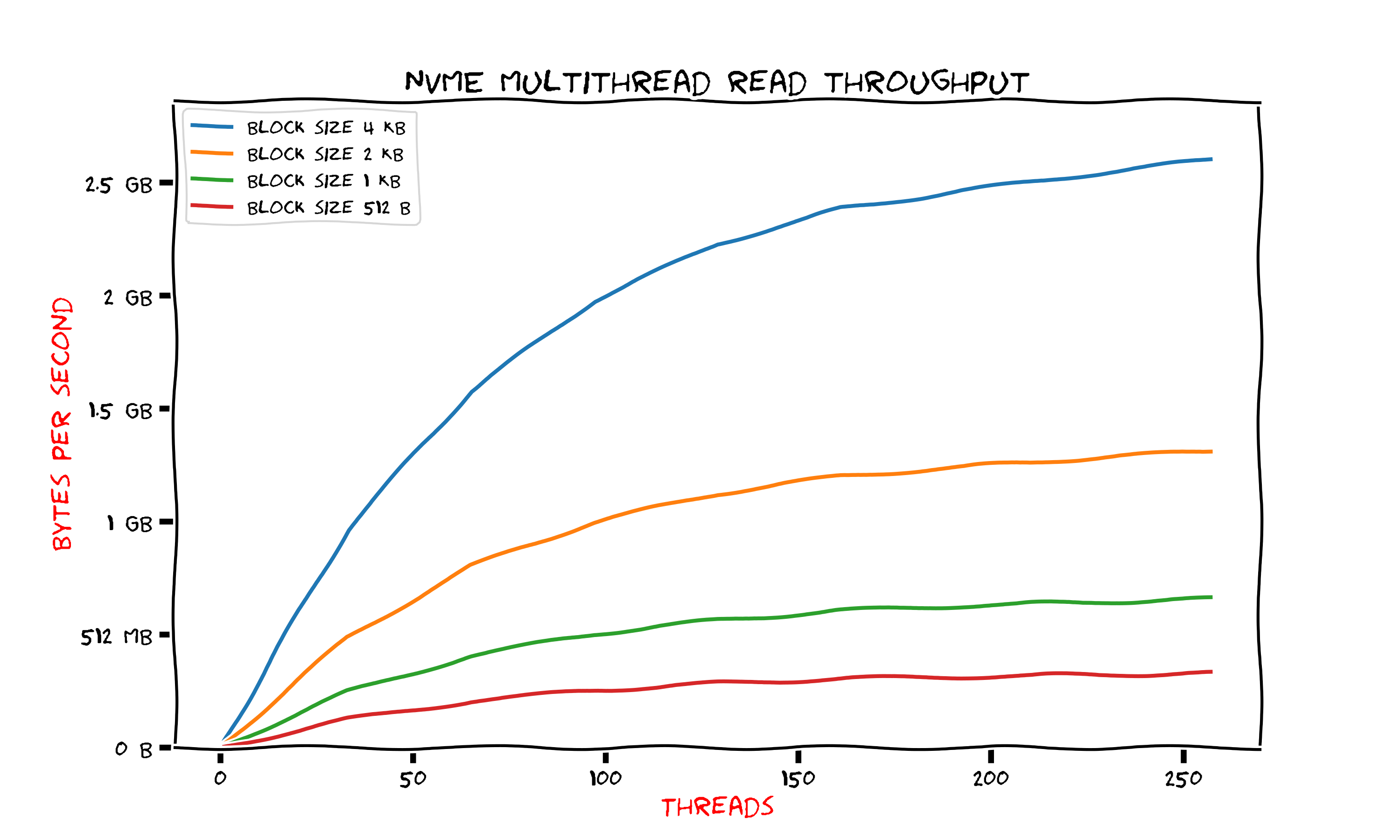}
        \caption{NVMe SSD multithread reading throughput for 4 kilobytes blocks and up to 256 threads.}
        \label{fig:256tbw:nvme}
    \end{minipage}
\end{figure}

When reading with 64 kilobytes blocks the latency also grows along with the number of threads increase.
The growth itself is not so high: it raises smoothly around 2 milliseconds when the number of threads is increased from 9 to 64.
The corresponding throughput growth is from 1.5 gigabytes per second to 2.5 gigabytes per second.
It is worth to note that the CPU usage is nearly a core and a half which is already noticeable and should be taken into account.

Reading with blocks of 256 kilobytes behaves as expected.
The latency grows as a logarithm on a logarithmic scale chart and in total is a few times larger.
Generally speaking we can say that the reading is saturated when the number of threads reach 12.
The throughput growth continues even after but is already quite slow.
We can say that when throughput is 2.5 gigabytes per second the latency is almost 2 milliseconds in 99.9 percentile.
With SATA SSD the latency for the same block size is about 1.5 milliseconds but the transmitted amount of data was five times smaller.

\subsection{NVMe SSD and small blocks}

As we have seen in previuous subsection it is worth to increase the number of threads when we read 4 kilobyte blocks.
Just in case here we also look at smaller blocks.
It appears that even if we specify smaller block size the device itself behaves as if whole 4 kilobyte page has been read.

\begin{figure}
    \begin{minipage}[c]{.47\linewidth}
        \includegraphics[width=\linewidth,trim=25 5 70 15, clip]{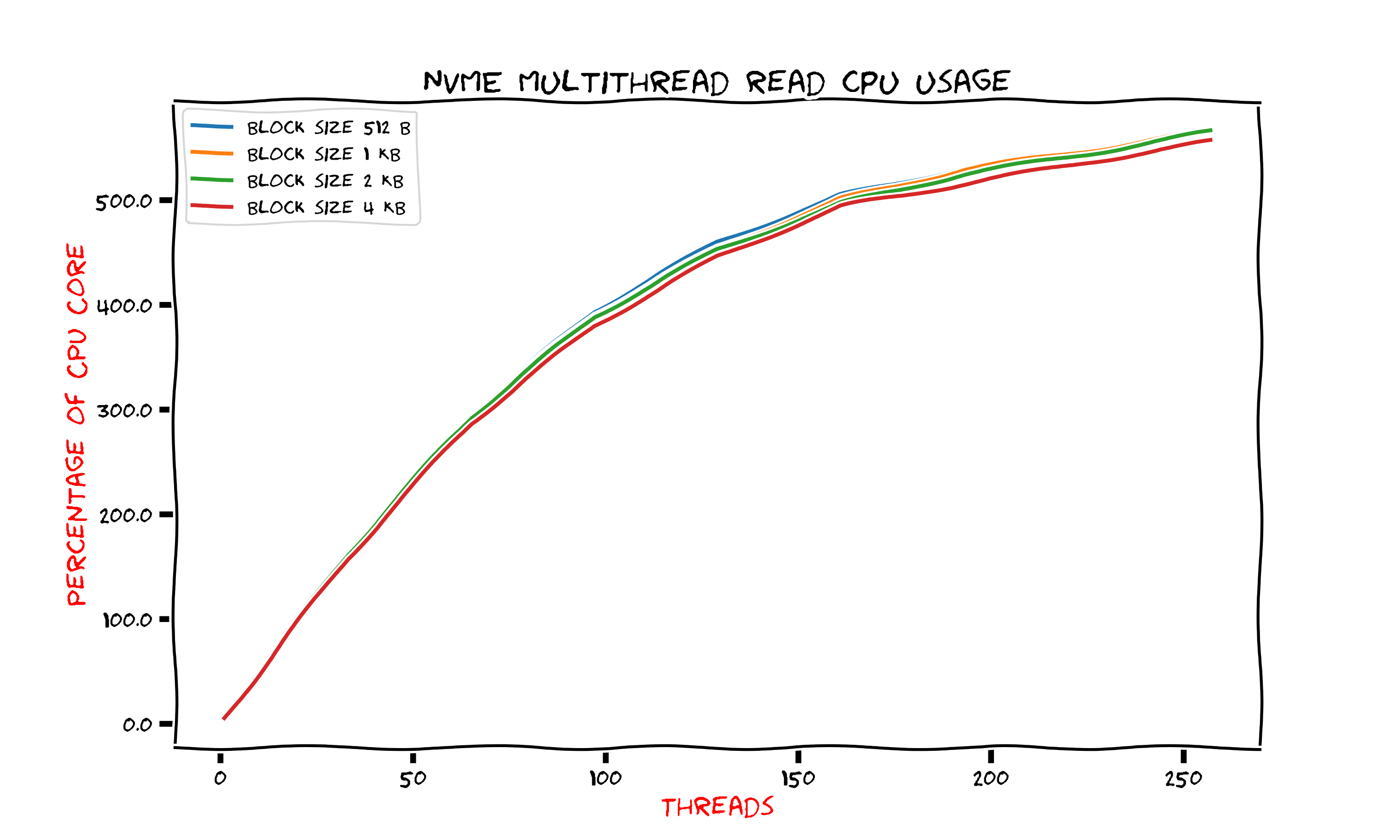}
        \caption{NVMe SSD multithread reading CPU usage.}
        \label{fig:256tcpu:nvme}
    \end{minipage}
    \hfill
    \begin{minipage}[c]{.47\linewidth}
        \includegraphics[width=\linewidth,trim=25 5 70 15, clip]{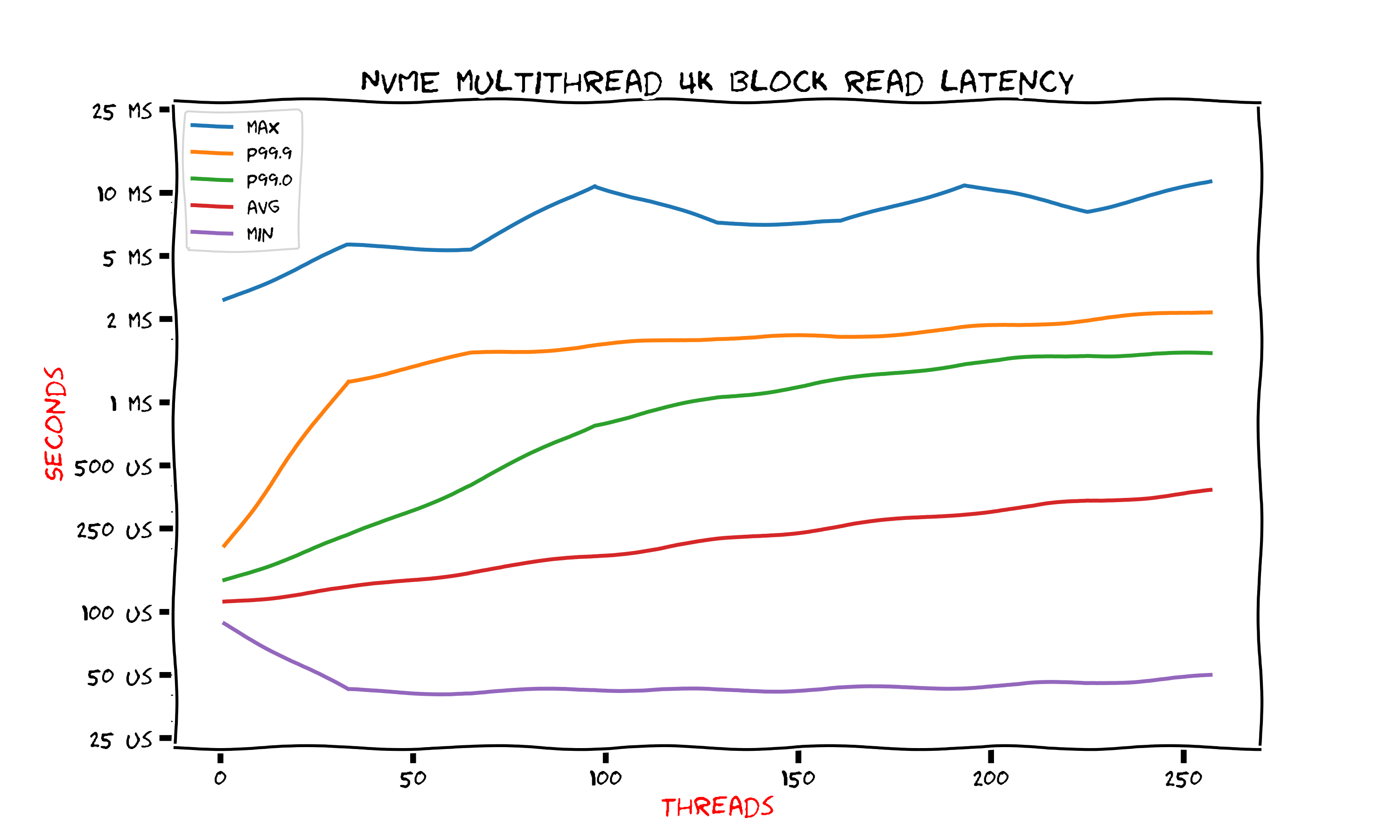}
        \caption{NVMe SSD multithread reading latency for 4 kilobytes blocks.}
        \label{fig:256tlat4k:nvme}
    \end{minipage}
\end{figure}

Figure~\ref{fig:256tbw:nvme} shows the resulting throughput when the number of threads raises up to 256.
Now we see that it is possible to transfer 2.5 gigabytes per second even in 4 kilobytes blocks.
This is obviously the result we couldn't achieve in previous experiments.
However we have to use 256 threads which is five times higher than the number of (hyperthreaded) cores on our testing machine.
Unfortunately this result is good only as a joke since no real world application would use so many threads to read from a single NVMe SSD.

Despite the enormous number of threads the CPU usage is quite reasonable. 
It turns out that in order to read with maximum throughput with 4 kilobytes blocks we need to pay CPU time in equivalent of six full cores.

And the final bad thing here is latency.
Figure~\ref{fig:256tlat4k:nvme} shows that even 99.0 percentile is larger than 1 millisecond.
This is ten times worse than a single read from NVMe SSD.
Probably this is not what we expected when we first heard about single request latency and peak NVMe SSD bandwidth.

\subsection{Optane}

\begin{figure}
    \begin{minipage}[c]{.47\linewidth}
        \includegraphics[width=\linewidth,trim=25 5 70 15, clip]{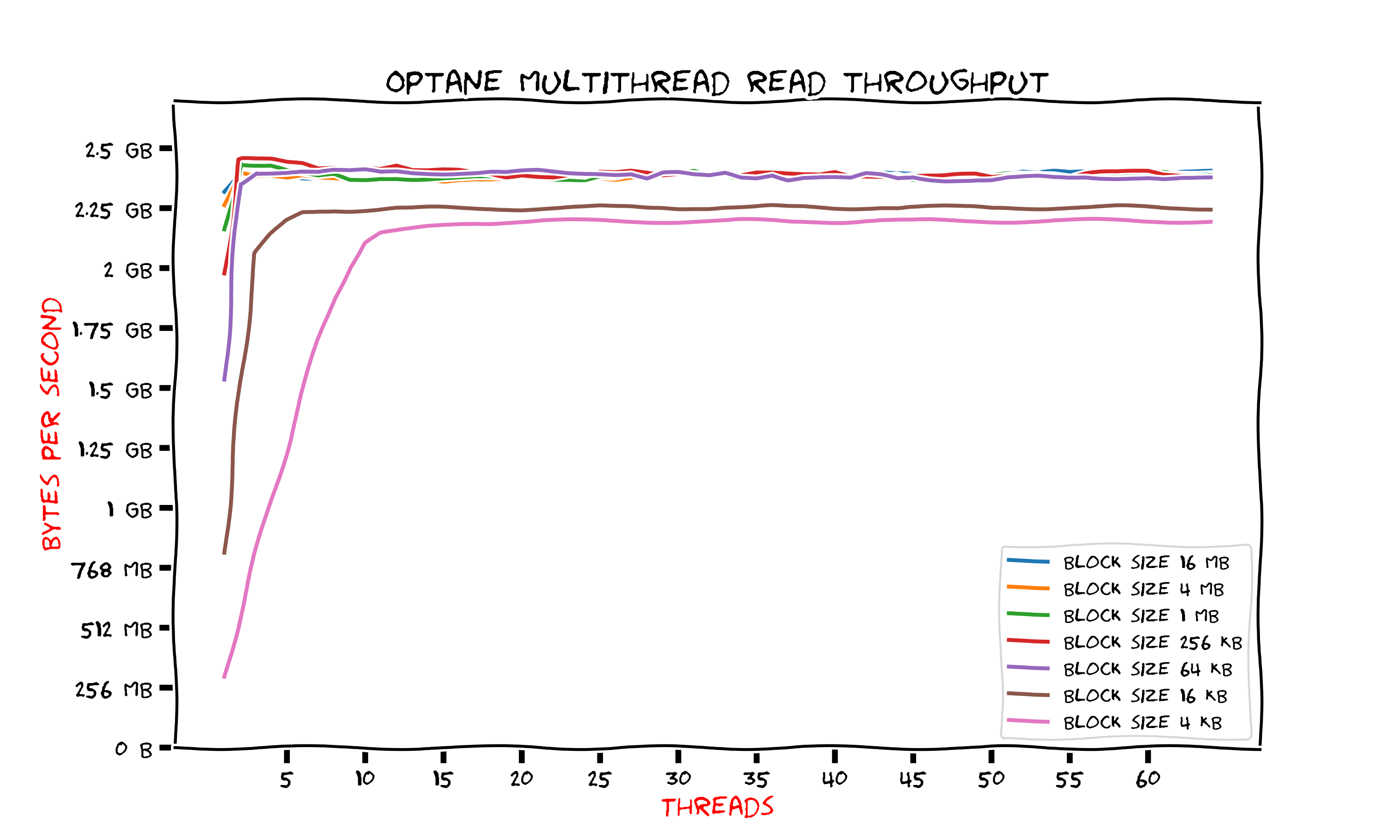}
        \caption{Optane multithread reading throughput.}
        \label{fig:multbw:optane}
    \end{minipage}
    \hfill
    \begin{minipage}[c]{.47\linewidth}
        \includegraphics[width=\linewidth,trim=25 5 70 15, clip]{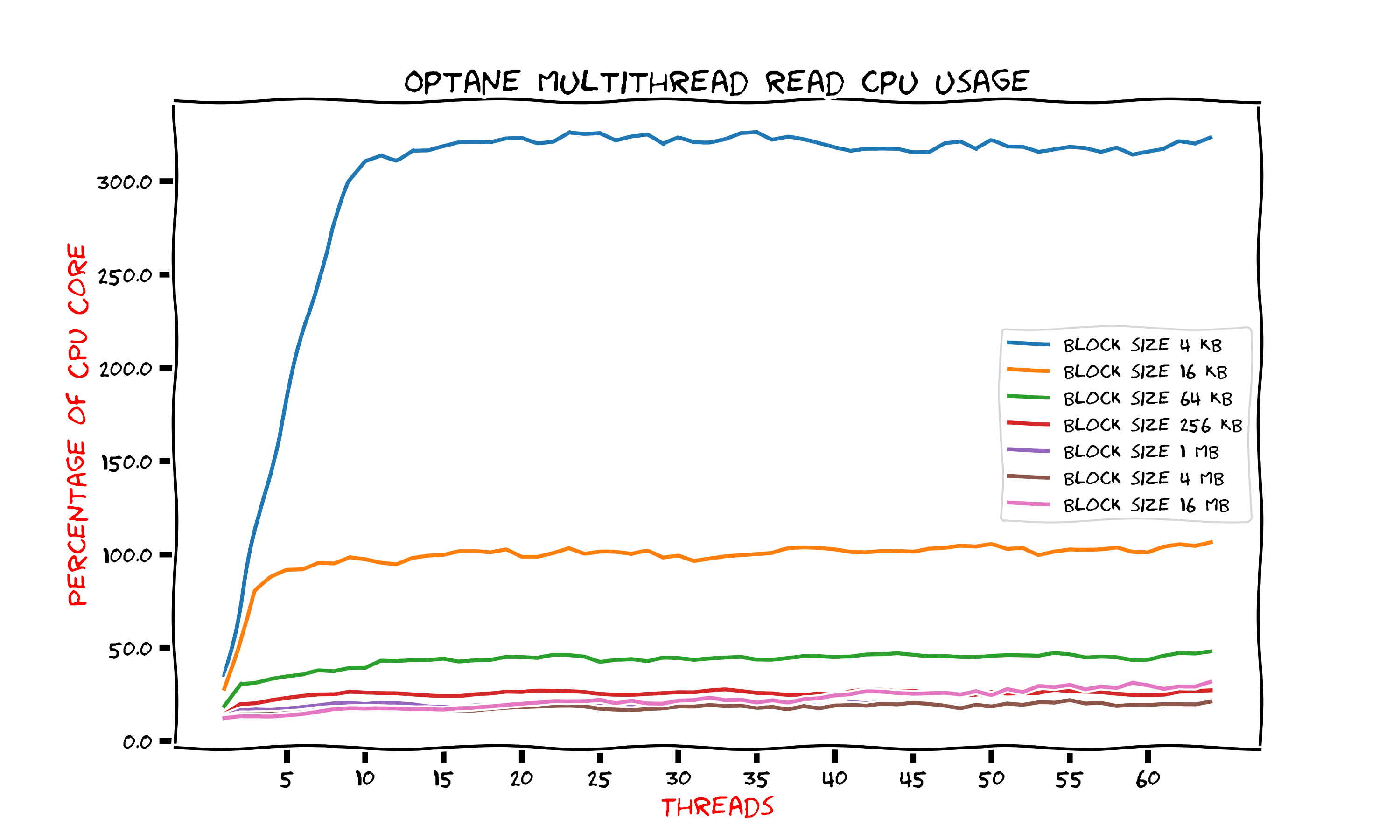}
        \caption{Optane multithread reading CPU usage.}
        \label{fig:multcpu:optane}
    \end{minipage}
\end{figure}

Optane, in contrast to NVMe SSD, turned out out be quite good to read from it from multiple threads.
Starting from block size of 64 kilobytes the maximum throughput is achieved by only two threads.
The resulting throughput is about 2.4 gigabytes per second as we can see on figure~\ref{fig:multbw:optane}.
For blocks of 16 kilobytes and less the maximum throughput is about 2.2 gigabytes per second.
To reach it one needs seven threads for 16 kilobytes blocks and eleven threads for 4 kilobyte blocks.
Sounds quite reasonable and realistic to use in an application.
At least compared to 256 threads for NVMe SSD.

CPU usage is shown on figure~\ref{fig:multcpu:optane} and seems comparable to NVMe SSD.
To read with 4 kilobytes block size we need a little bit more than three fully utilized cores.
For 16 kilobytes blocks it is sufficient to allocate CPU time equivalent to a single core and for larger blocks only 10 to 20 percents of a core are used.

\begin{figure}
    \begin{minipage}[c]{.47\linewidth}
        \includegraphics[width=\linewidth,trim=25 5 70 15, clip]{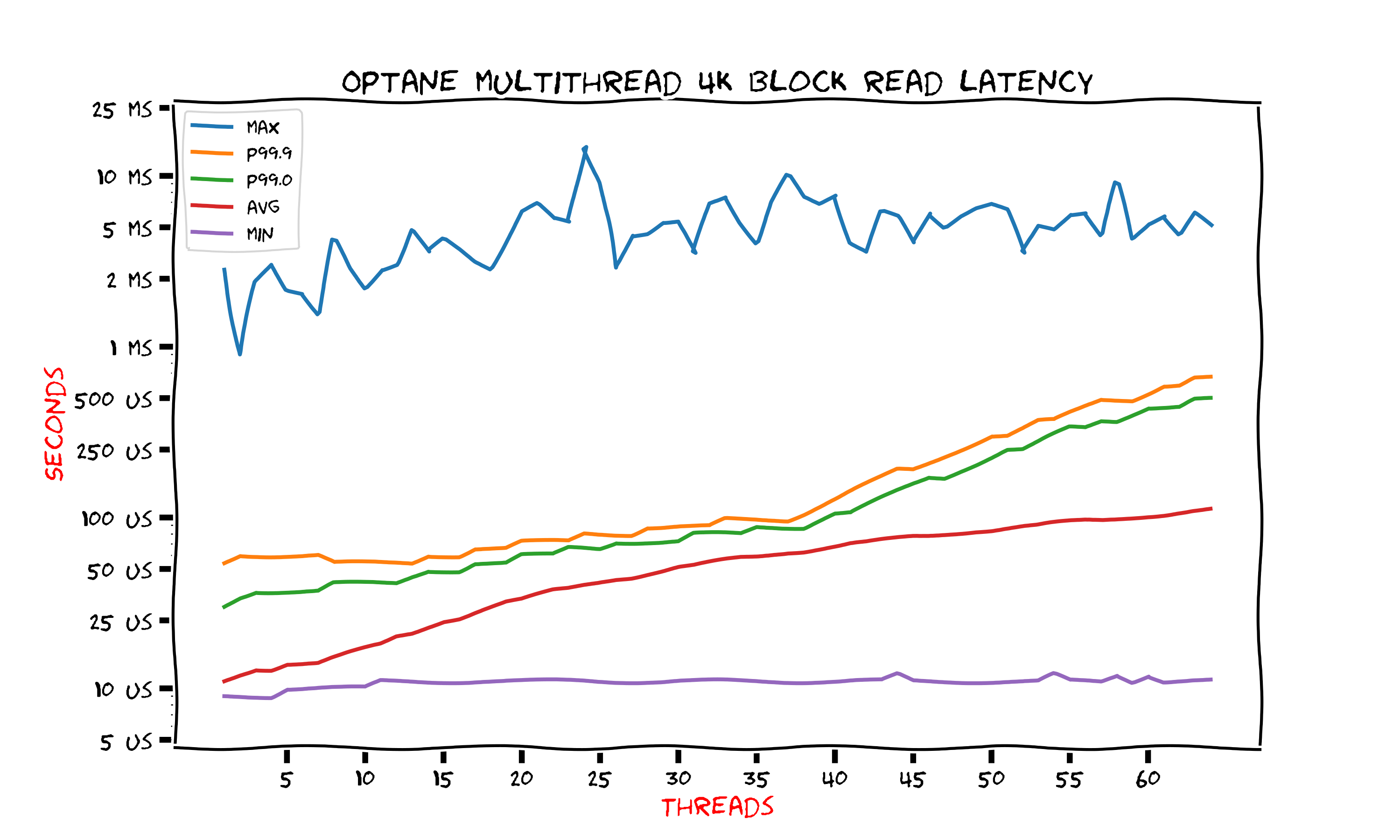}
        \caption{Optane multithread reading latency for 4 kilobytes blocks.}
        \label{fig:multlat4k:optane}
    \end{minipage}
    \hfill
    \begin{minipage}[c]{.47\linewidth}
        \includegraphics[width=\linewidth,trim=25 5 70 15, clip]{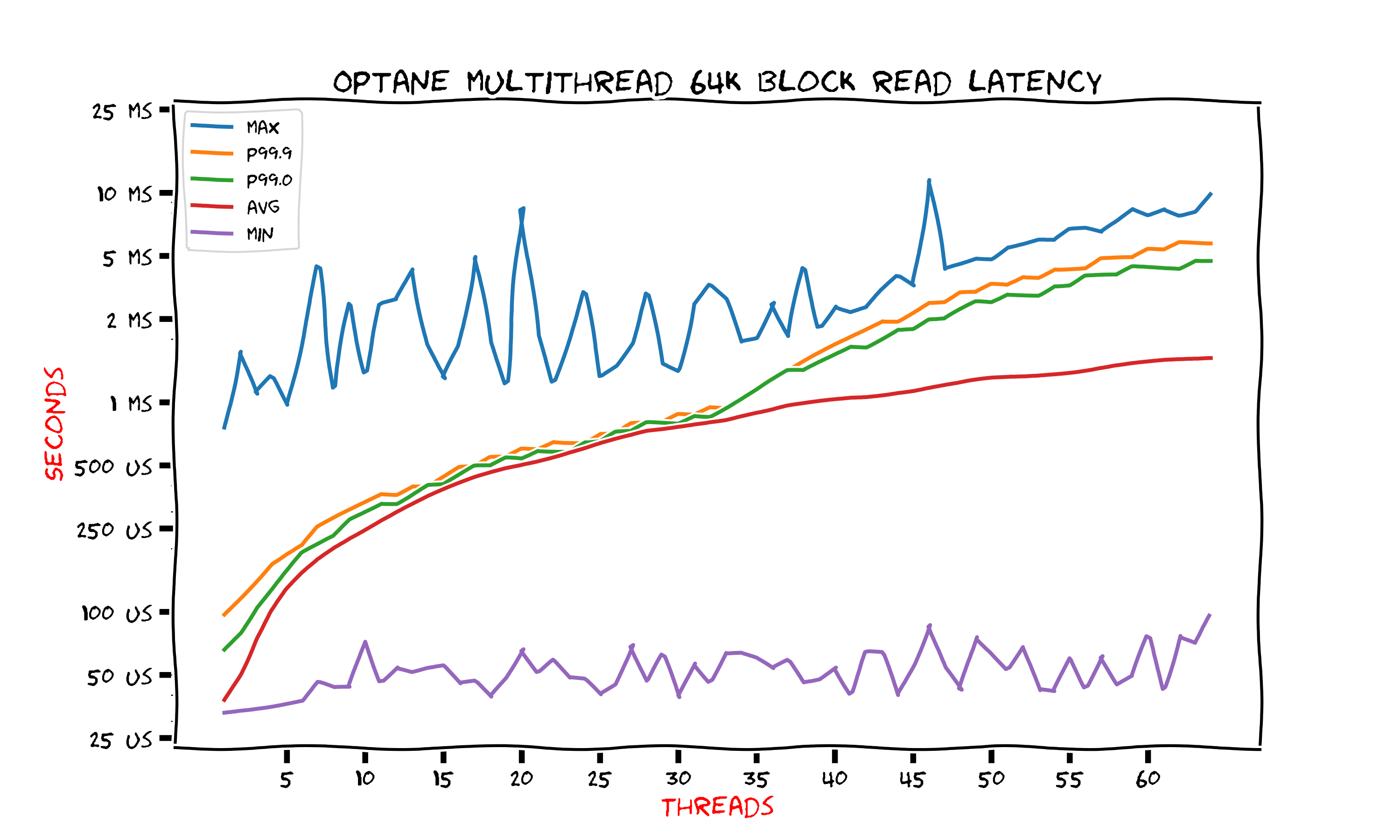}
        \caption{Optane multithread reading latency for 64 kilobytes blocks.}
        \label{fig:multlat64k:optane}
    \end{minipage}
\end{figure}

For 4 kilobytes block size we observe a very good latency.
Figure~\ref{fig:multlat4k:optane} shows a bit higher than 50 microseconds in 99.9 percentile.
Compared to 2 milliseconds of NVMe SSD this is a tremendous speedup.
If we leave out the maximum latency and consider only 99.9 percentile we get that current latency is similar to those of a single read.
Recall that bandwidth is saturated when reading with only eleven threads so we don't need to take into account a sharp growth when the number of threads reach 37.
This is a really good result.
As we will see further it is really hard to reproduce such a low latency with asynchronous interfaces.
In fact we will often fail to achieve it.

\begin{figure}
    \begin{center}
    \begin{minipage}[c]{.47\linewidth}
        \includegraphics[width=\linewidth,trim=25 5 70 15, clip]{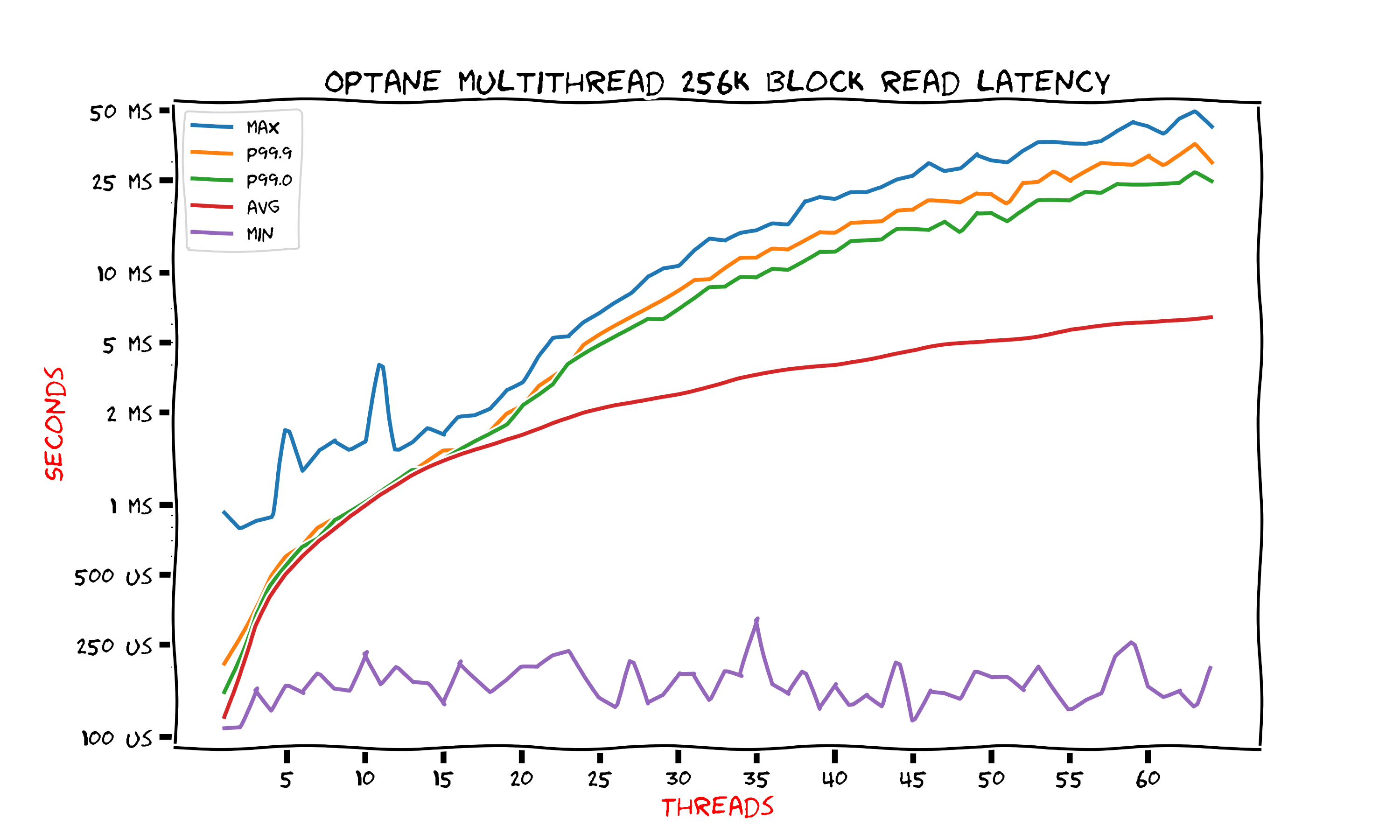}
        \caption{Optane multithread reading latency for 256 kilobytes blocks.}
        \label{fig:multlat256k:optane}
    \end{minipage}
    \end{center}
\end{figure}

For larger blocks latency is also pretty good.
For 64 kilobytes block size we see about 200 microseconds on figure~\ref{fig:multlat64k:optane}.
This is worse than a single read and even a single-threaded read.
On the other hand it not really much worse.
Anyway it is ten times faster than reading with the same block size form NVMe SSD.

Finally it is easy to remember 256 microseconds in 99.9 percentile for blocks of 256 kilobytes.
With this block size we observe bandwidth saturation with only two threads so the whole figure~\ref{fig:multlat256k:optane} may be interesting only from a theoretical point of view.
There is no reason at all to use so many threads in practice.

In general Intel Optane shows pretty good results when being read from multiple threads.
If we could use it everywhere we probably won't even get a desire to look into asynchronous interfaces.

\subsection{Summary}

We evaluated synchronous reading from multiple threads for all our four types of storage devices.
In spite of single-threaded appearance of HDD physics we observed that in some cases reading from multiple threads allowed us to increase throughput.
However this was paid by tremendous worsening of latency.
Also we observed that to extract the maximum throughput from either HDD or SATA SSD we need to read with block size large enough.

As for NVMe SSD and Intel Optane we succeeded in reaching nearly maximum bandwidth with 4 kilobytes blocks.
To read from Optane with the highest throughput we need only eleven threads.
For NVMe SSD we required an enormous number of 256 threads for that.
Obviously nobody sane would use this amount of threads.
At least while the number of CPU cores on a machine is several times smaller.
If the number of active threads exceeds the number of cores then unexpected latency spikes here and there are unavoidable due to a thread being on a standby while all cores execute other threads.
Moreover an application has absolutely no control for these spikes since their origin is inside the kernel scheduler.

Nevertheless it is possible to read from NVMe SSD in 4 kilobytes blocks and fully utilize the bandwidth while keeping the number of threads reasonable.
However we require asynchronous input-output interface for that and this is exactly the topic of the next sections.

\section{Asynchronous Input-Output}
\label{sec:aio}

Linux kernel has asynchronous I/O interface for a long time already~\cite{Bhattacharya2003AsynchronousIS}.
It works as follows: a process creates a queue inside the kernel and then inserts I/O requests into that queue.
Requests specify type of operation, file descriptor, offset and data buffer.
Conceptually queue insertion is non-blocking.
However the Linux aio implementation could block sometimes (this is one of critiques of the interface).
After request is inserted the process continues its execution and the request will be performed asynchronously by the kernel at some point in the future.
Further process can ask kernel for a list of completed requests.
During this call the process can additionally specify the minimum number of requests that should be completed.
If the number of already completed requests is smaller the process will block until enough requests are finished.

Namely there are four system calls: \texttt{io\_setup}, \texttt{io\_destroy}, \texttt{io\_submit}, and \texttt{io\_getevents}.
First two manage the existence of the in-kernel asynchronous requests queue itself.
The third adds requests to the queue and the last one is used to get the results.
Refer to the man pages for the details how to use these system calls.
Rather extensive description based on user experience can be found in~\cite{Cloudfare-AIO}.

\subsection{Reading with asynchronous interface}

\begin{figure}
    \includegraphics[width=\textwidth]{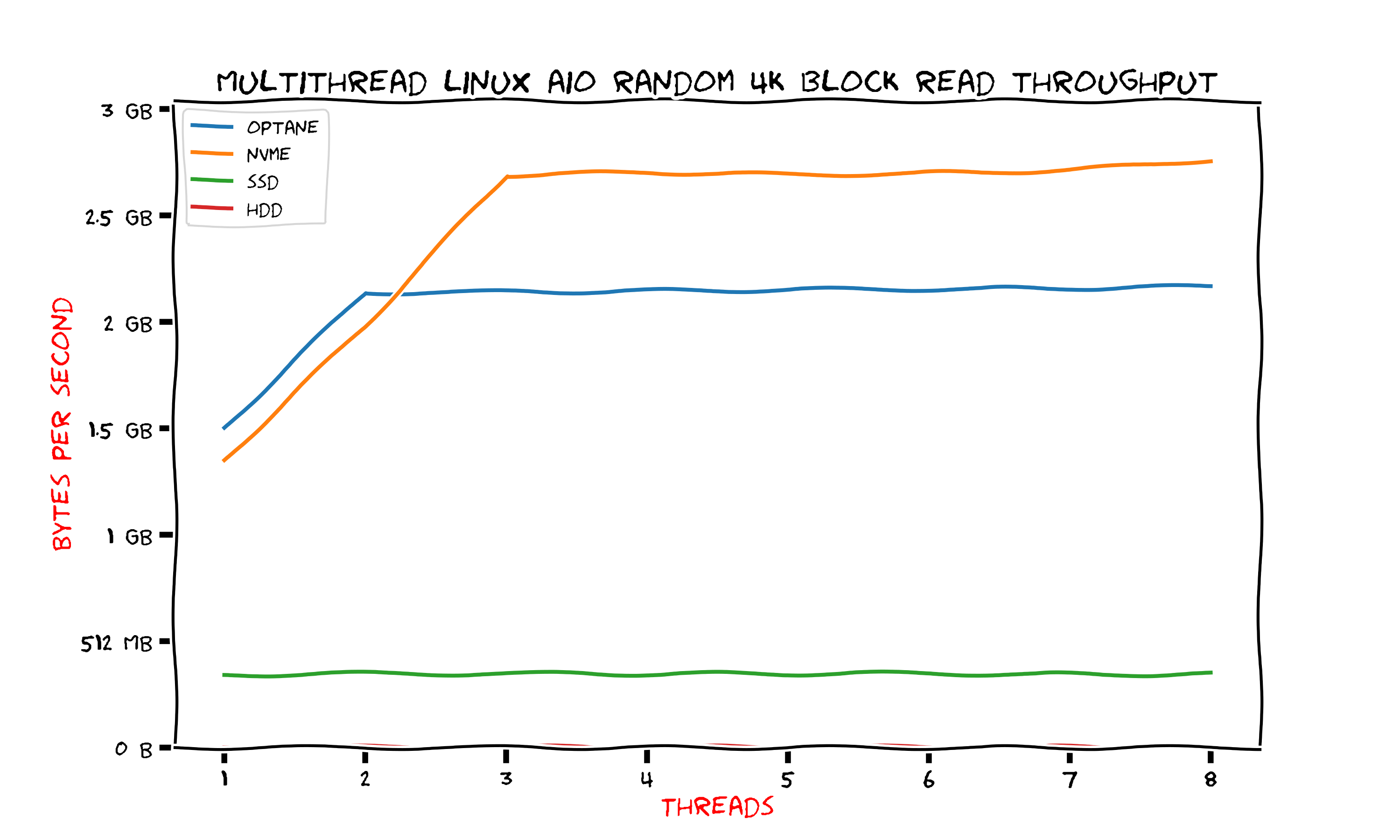}
    \caption{Asynchronous reading throughput.}
    \label{fig:allaio}
\end{figure}

Figure~\ref{fig:allaio} shows that even with asynchronous interface single thread is not enough to saturate NVMe SSD bandwidth.
However we can achieve the goal with only three threads even when 4 kilobytes block size is used.
Recall that for synchronous interface we required an unpleasant number of 256 threads.

Here we create a queue for each thread.
We should mention however that Linux aio interface allows different threads access a single queue.
Maybe there is a good way to exploit it but we won't try to investigate it here.
Our goal is to push throughput to the maximum and alongside it latency to the minimum.
All multi-thread programming experience says that accessing a single object from several threads would be an obstacle for such a goal.

Next we show latency, throughput, and CPU usage with respect to block size.
Asynchronous interface exhibits much higher throughput than synchronous one.
However for a small block size even asynchronous interface is not enough to saturate NVMe SSD bandwidth form a single thread.
We need to use more threads.
Here we demonstrate results for one, two, and three of them.

\begin{figure}
    \includegraphics[width=\linewidth]{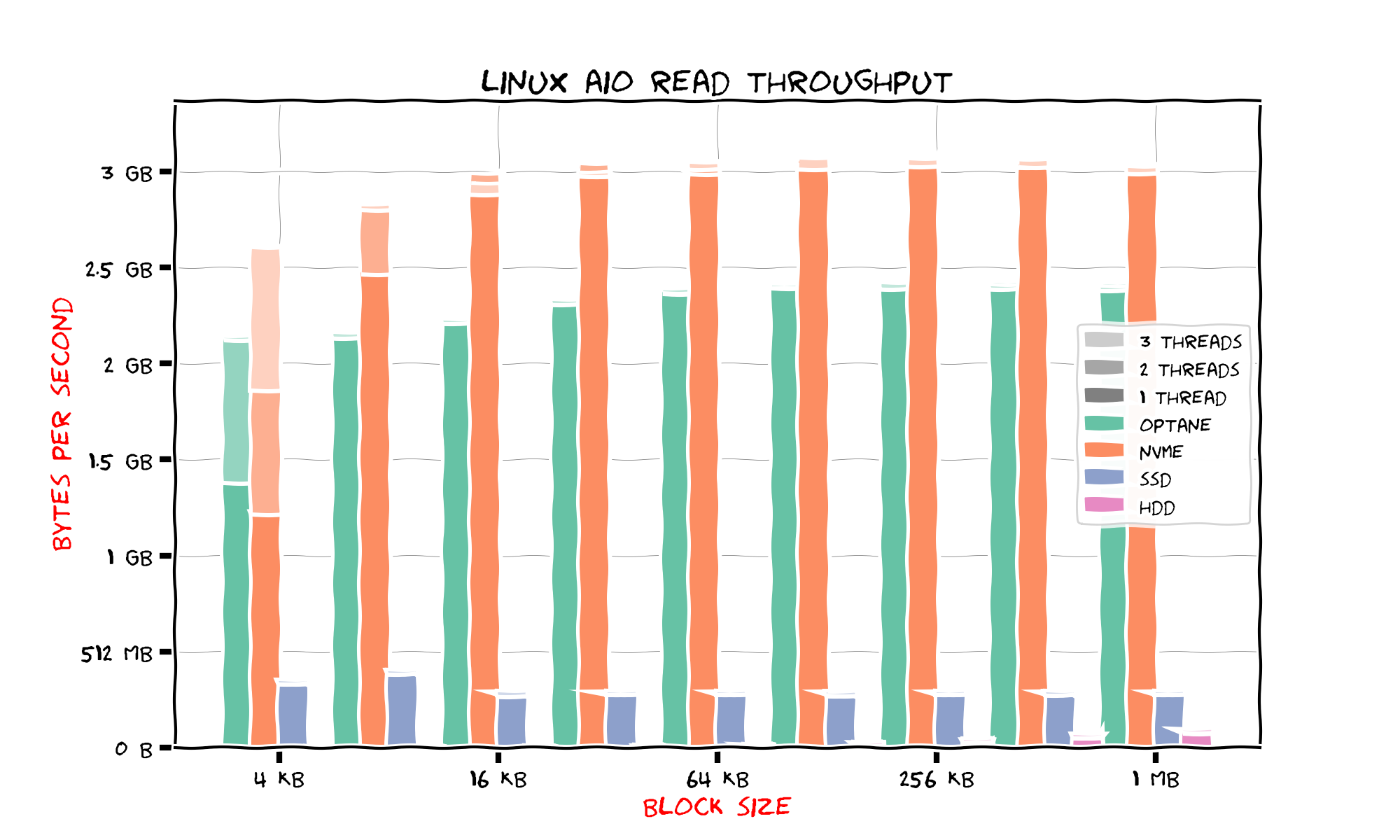}
    \caption{Asynchronous reading throughput for different block sizes.}
    \label{fig:throughputaio:all}
\end{figure}
\begin{figure}
    \includegraphics[width=\linewidth]{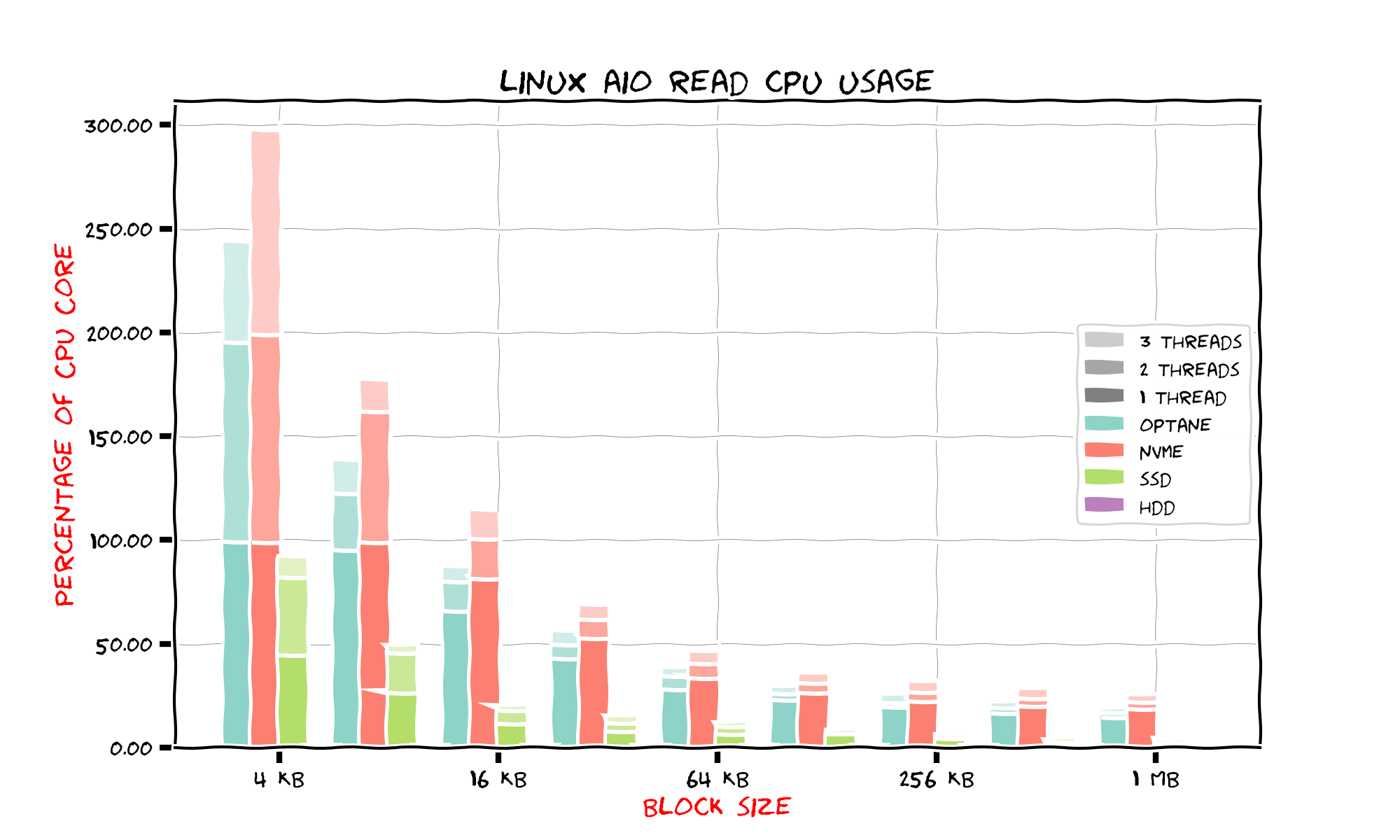}
    \caption{Asynchronous reading CPU usage for different block sizes.}
    \label{fig:cpuaio:all}
\end{figure}

Figure~\ref{fig:throughputaio:all} shows throughput when reading using Linux aio.
As we have discussed we need three threads to saturate NVMe SSD and Optane bandwidth.
Since it is also interesting what throughput is achieved with one and two threads the results are also shown on the same figure.
Each bar is split into three sections and they are shown with different luminosity.
Sections that correspond to two and three threads are less colorful.

What catches the eye first is that for 32 kilobytes block size the bar is above 3 gigabytes per second.
It is also very interesting that we are able to achieve rather high throughput on NMVe without an enormous number of threads.
The bar is slightly above 2.5 gigabytes per second for just three threads.
Recall that with the synchronous interface we required 256 threads for a similar result.
As for Optane not much new here.
The results look like what we achieved with synchronous interface and multiple threads.

Results for SSD are strictly worse.
With synchronous interface we were able to achieve throughput of more than 500 megabytes per second when using 1 megabyte block.
However with asynchronous interface the result looks like 300 megabytes per second for all block sizes.
An interesting exception is 8 kilobytes where for some reason the bar looks like 400 megabytes per second.
For 4 kilobytes and 8 kilobytes block size the achieved throughput is relatively same for both synchronous and asynchronous interfaces.

Next we look at the CPU usage. Figure~\ref{fig:cpuaio:all} shows the results.
Similar to throughput chart bar sections color intensity decrease depicts increase in the number of threads. 
The figure clearly shows that CPU usage is high for small blocks in general.
For NVMe SSD asynchronous interface is strictly better than synchronous not only by the number of threads but also by the CPU utilization.
It requires three full cores of CPU time instead of five to read 2.5 gigabytes per second using 4 kilobytes blocks.
As for Optane the CPU utilization is also lower.
We need slightly smaller than two and a half cores to read 2.2 gigabytes per second with 4 kilobytes blocks while it required more than three full cores of CPU time to achieve the same throughput with synchronous interface.

\begin{figure}
    \includegraphics[width=\linewidth]{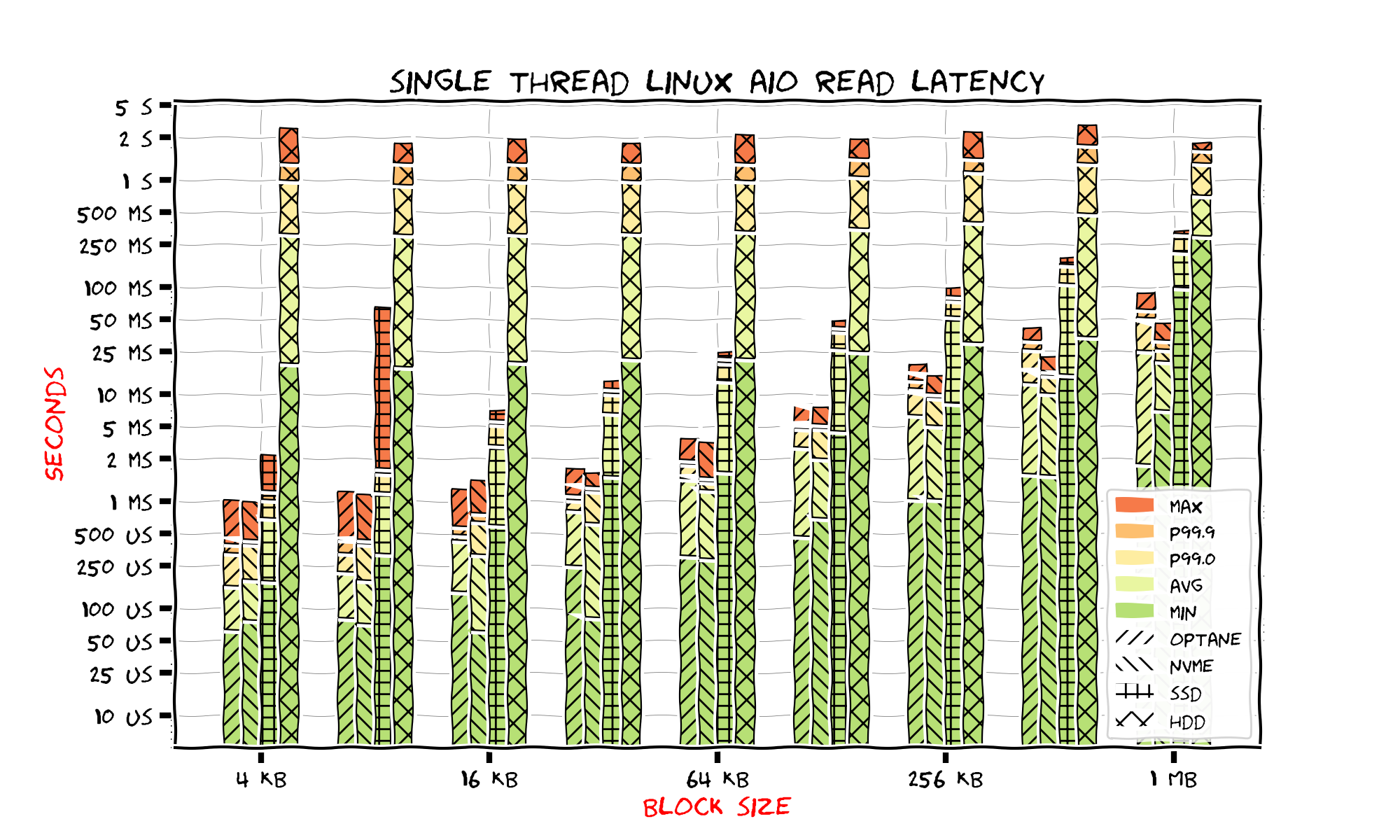}
    \caption{Single-threaded asynchronous reading latency for different block sizes.}
    \label{fig:latencysingle:all}
\end{figure}

\begin{figure}
    \includegraphics[width=\linewidth]{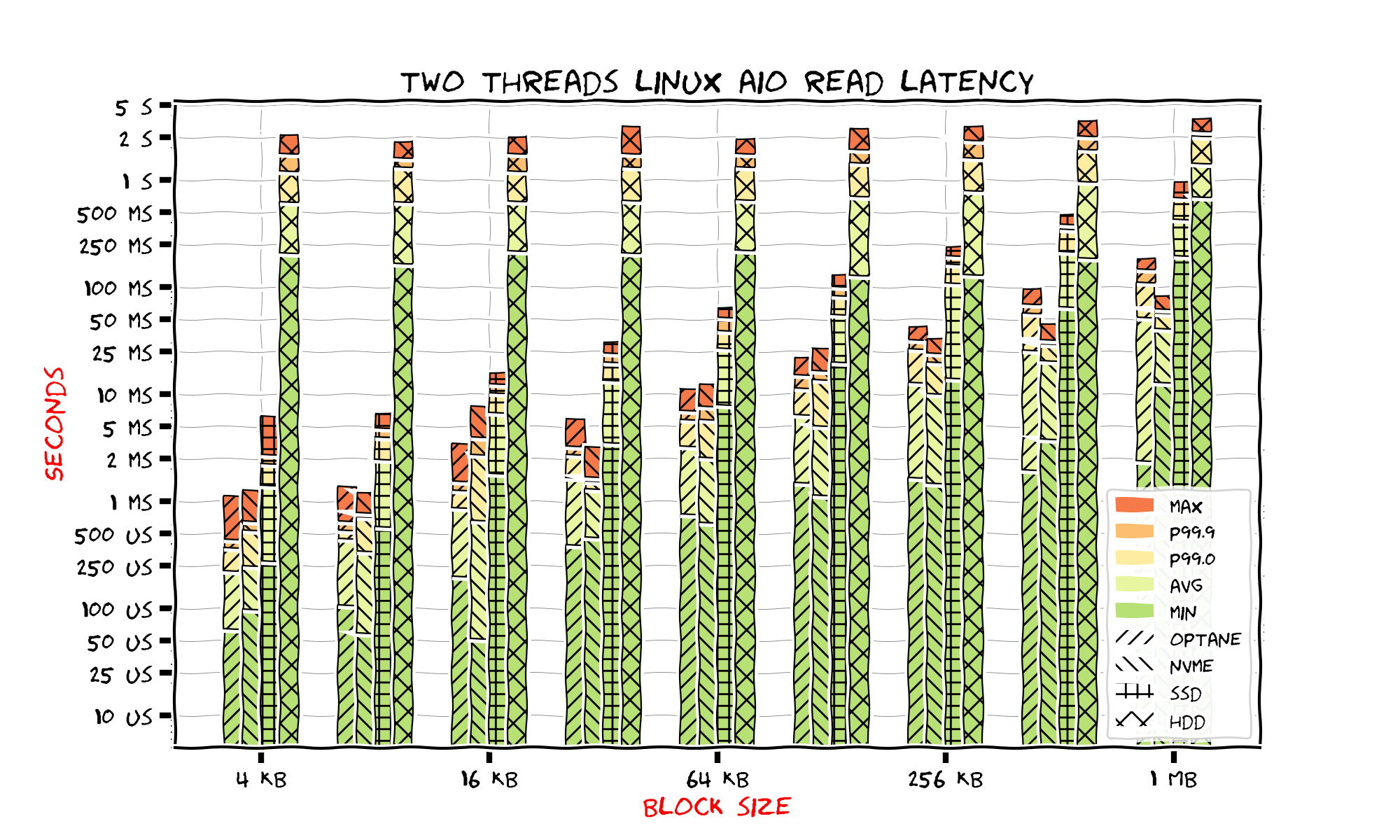}
    \caption{Two-threaded asynchronous reading latency for different block sizes.}
    \label{fig:latencytwo:all}
\end{figure}

\begin{figure}
    \includegraphics[width=\linewidth]{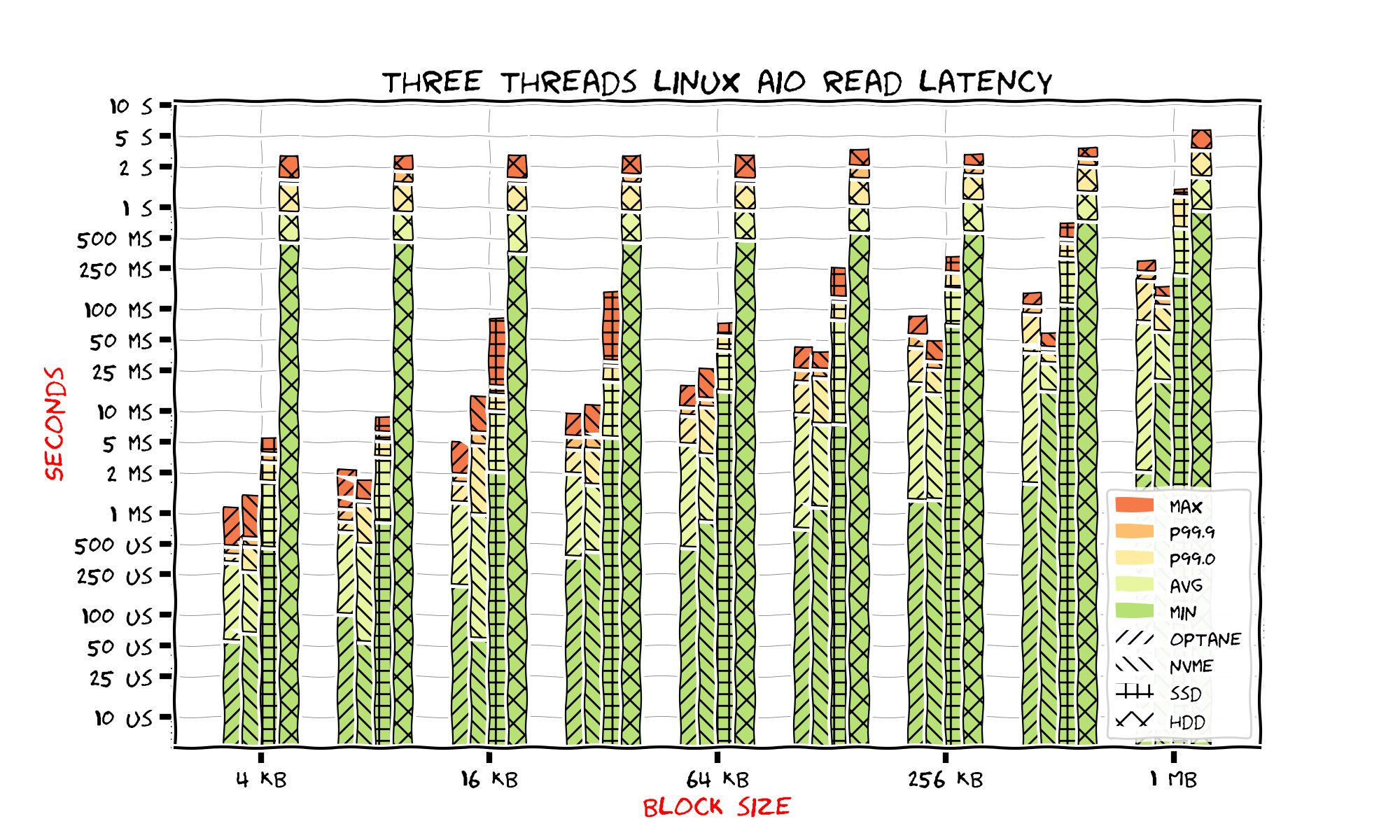}
    \caption{Three-threaded asynchronous reading latency for different block sizes.}
    \label{fig:latencythree:all}
\end{figure}

Now we move to latency. Unfortunately it is intractable to depict results on a single figure so we use separate charts for different number of threads.
Figures~\ref{fig:latencysingle:all},~\ref{fig:latencytwo:all}, and~\ref{fig:latencythree:all} show the results.

If we look at 99.9 percentile we'll see that for 4 kilobytes blocks latency is about 300-400 microseconds for Optane and about 500 microseconds for NVMe SSD (recall that we need two threads for Optane and three threads for NMVe).
For Optane this is strictly worse than the latency achieved by reading from multiple threads with synchronous interface.
That result was slightly above 50 microseconds which is nearly ten times lower.
On the other hand latency for reading with the same throughput from NVMe SSD via synchronous interface is above millisecond.
It turns out that for NVMe SSD it is better to use Linux aio by all means.

Surprisingly results for SATA SSD behave similar to those for Optane.
Usage of Linux aio increases latency for small block sizes.
If we take 64 kilobytes block the latency is lower for Linux aio and not slightly lower but several times lower.
Recall that in this case the throughput is also reduced significantly.

For HDD the latency is measured by seconds which is enormously high.
Everything suggests that the best way to read from HDD is via synchronous interface and from a single thread.


\subsection{Single reading of multiple blocks}

With asynchronous interface we can perform an operation which is impossible with a synchronous one: read several random blocks at once in a single request.
Even though system calls \texttt{preadv} and \texttt{preadv2} accept a vector of buffers they access blocks on a device sequentially.
With asynchronous interface we can insert into the queue a number of absolutely arbitrary requests and wait for all of them to complete.
Evidently the request duration depend on the number of blocks.
However this dependency is sub-linear.
Thus a single asynchronous call would be faster than several synchronous in a row.

We show latencies for reading 4 kilobytes blocks.
As with the single reading experiment in section~\ref{sec:single} we execute some requests and show minimum, maximum, average and percentiles.
Since in terms of data the request complexity grows linear with the number of blocks increased we use linear scale for latency.
This is different from all other sections where latency is shown at the logarithmic scale.
It is easy to see that although the growth is linear the derivative is less than one.

\begin{figure}
    \begin{minipage}[c]{.47\linewidth}
        \includegraphics[width=\linewidth,trim=25 5 70 15, clip]{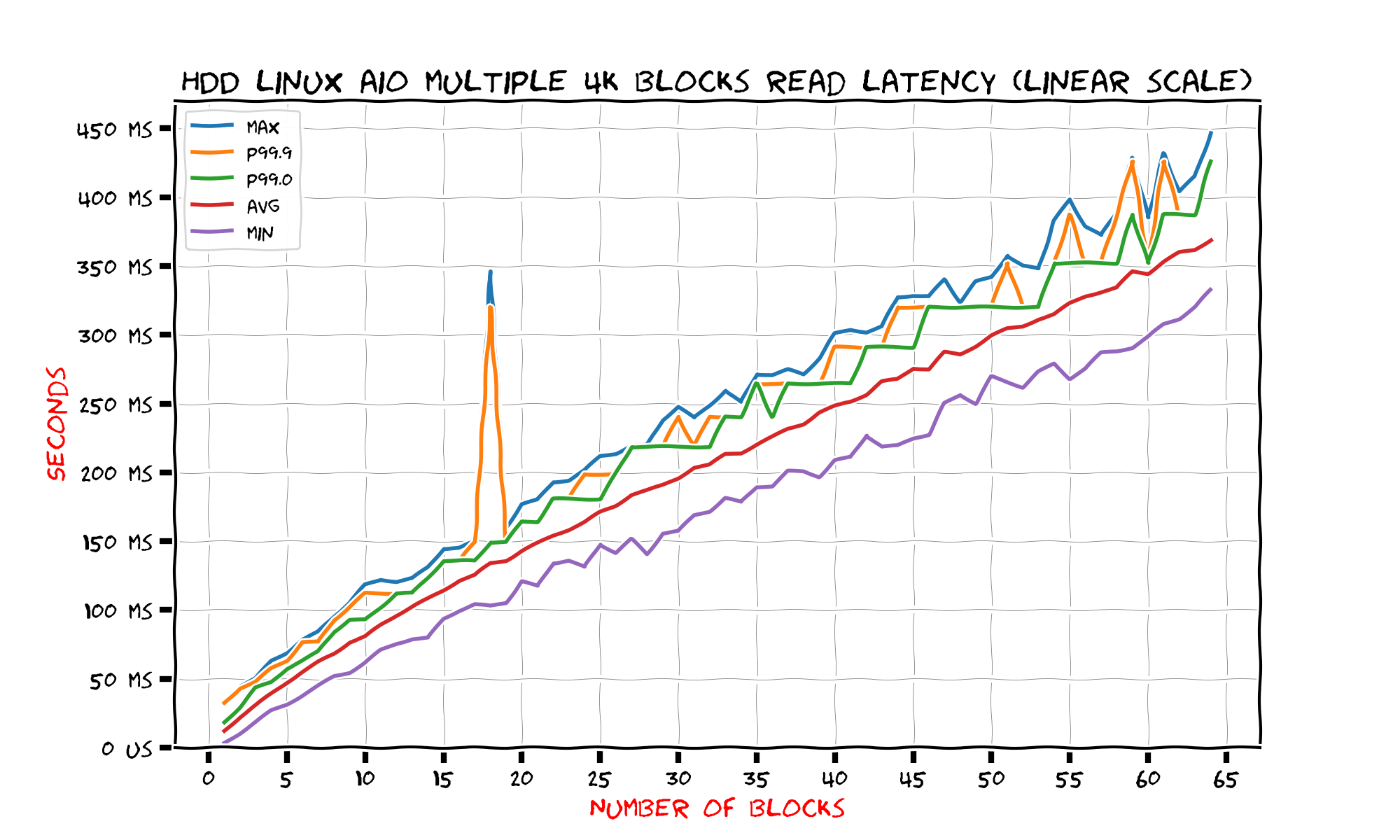}
        \caption{HDD multiple blocks read.}
        \label{fig:multblockaio:hdd}
    \end{minipage}
    \hfill
    \begin{minipage}[c]{.47\linewidth}
        \includegraphics[width=\linewidth,trim=25 5 70 15, clip]{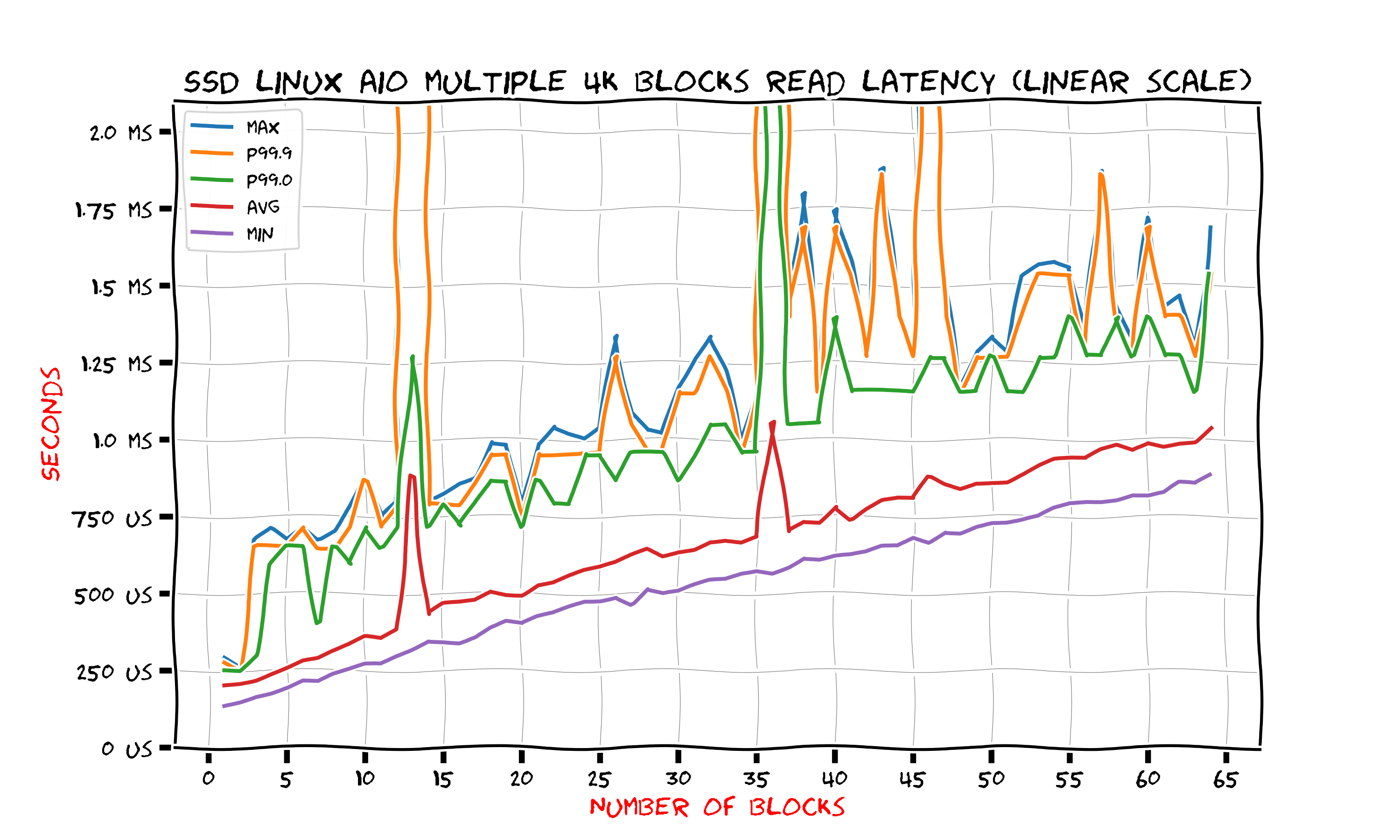}
        \caption{SSD multiple blocks read.}
        \label{fig:multblockaio:ssd}
    \end{minipage}
\end{figure}

Figure~\ref{fig:multblockaio:hdd} shows the results of HDD.
It is easy to calculate that reading multiple blocks at once with asynchronous interface is faster than multiple synchronous readings.
Recall that a single random read from HDD takes 12 milliseconds on average.
Here we see that five blocks are read in 50 milliseconds on average which is slightly faster.
If we take 20 blocks we get 150 milliseconds.
With synchronous interface the result would be 60 and 240 milliseconds respectively.
It is easy to notice that the response time grows faster for the synchronous interface.

\begin{figure}
    \begin{minipage}[c]{.47\linewidth}
        \includegraphics[width=\linewidth,trim=25 5 70 15, clip]{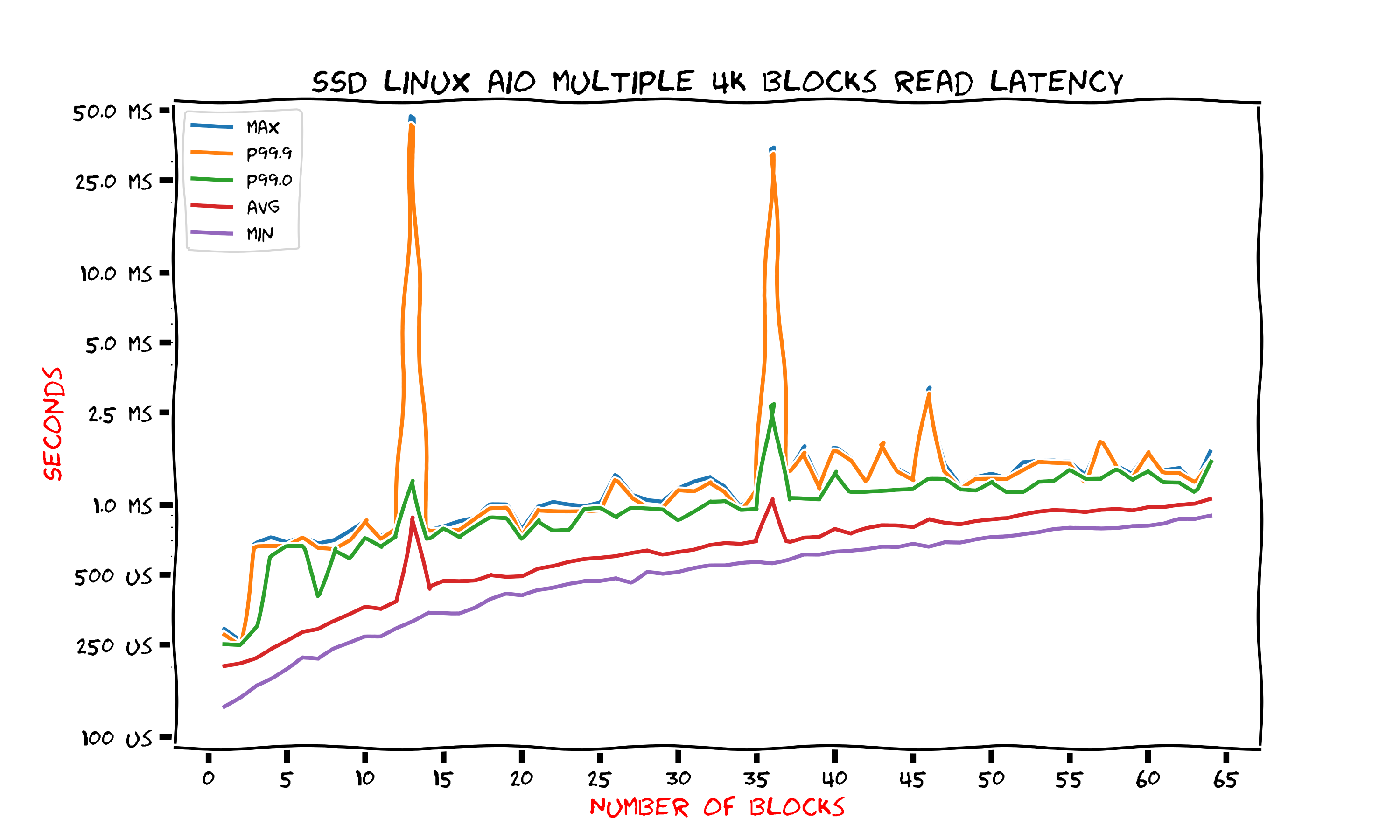}
        \caption{SSD multiple blocks read (logarithm scale).}
        \label{fig:multblockaiolog:ssd}
    \end{minipage}
    \hfill
    \begin{minipage}[c]{.47\linewidth}
        \includegraphics[width=\linewidth,trim=25 5 70 15, clip]{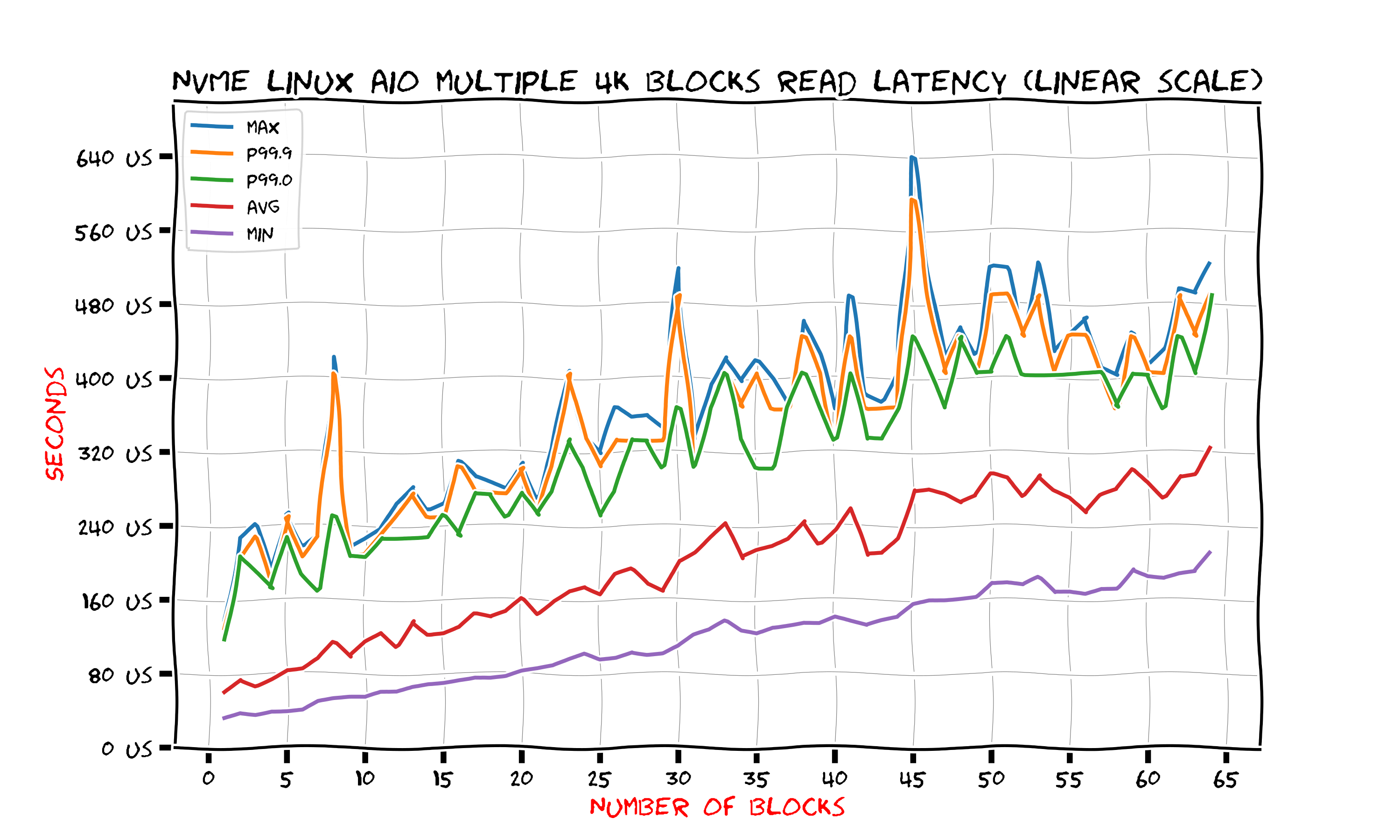}
        \caption{NVMe SSD multiple blocks read.}
        \label{fig:multblockaio:nvme}
    \end{minipage}
\end{figure}

Now let's look at SSD.
Figure~\ref{fig:multblockaio:ssd} shows the results.
Here we see that latency growth is really low.
It raises high at the beginning when we step from two blocks to three and then it only doubles while the number of blocks is increased by order of magnitude.
There are three spikes on this figure which go up to almost 50 milliseconds.
Recall that we already observed them in previous experiments.
For the sake of completeness the results are also shown on a figure~\ref{fig:multblockaiolog:ssd} with logarithmic scale.
One could also ascertain why we don't use logarithmic scale in this section.
It is really hard to grasp the correlation between the number of blocks and latency on such a chart.

Results for NVMe SSD are shown on figure~\ref{fig:multblockaio:nvme}.
They appear to behave similarly to SSD.
The latency doubles at the beginning when the number of thread is increased from one to two.
After that the latency smoothly increases from 25 microseconds to 400 microseconds while the number of blocks raises from three up to sixty.
In the end the line is almost horizontal.

\begin{figure}
    \begin{minipage}[c]{.47\linewidth}
        \includegraphics[width=\linewidth,trim=25 5 70 15, clip]{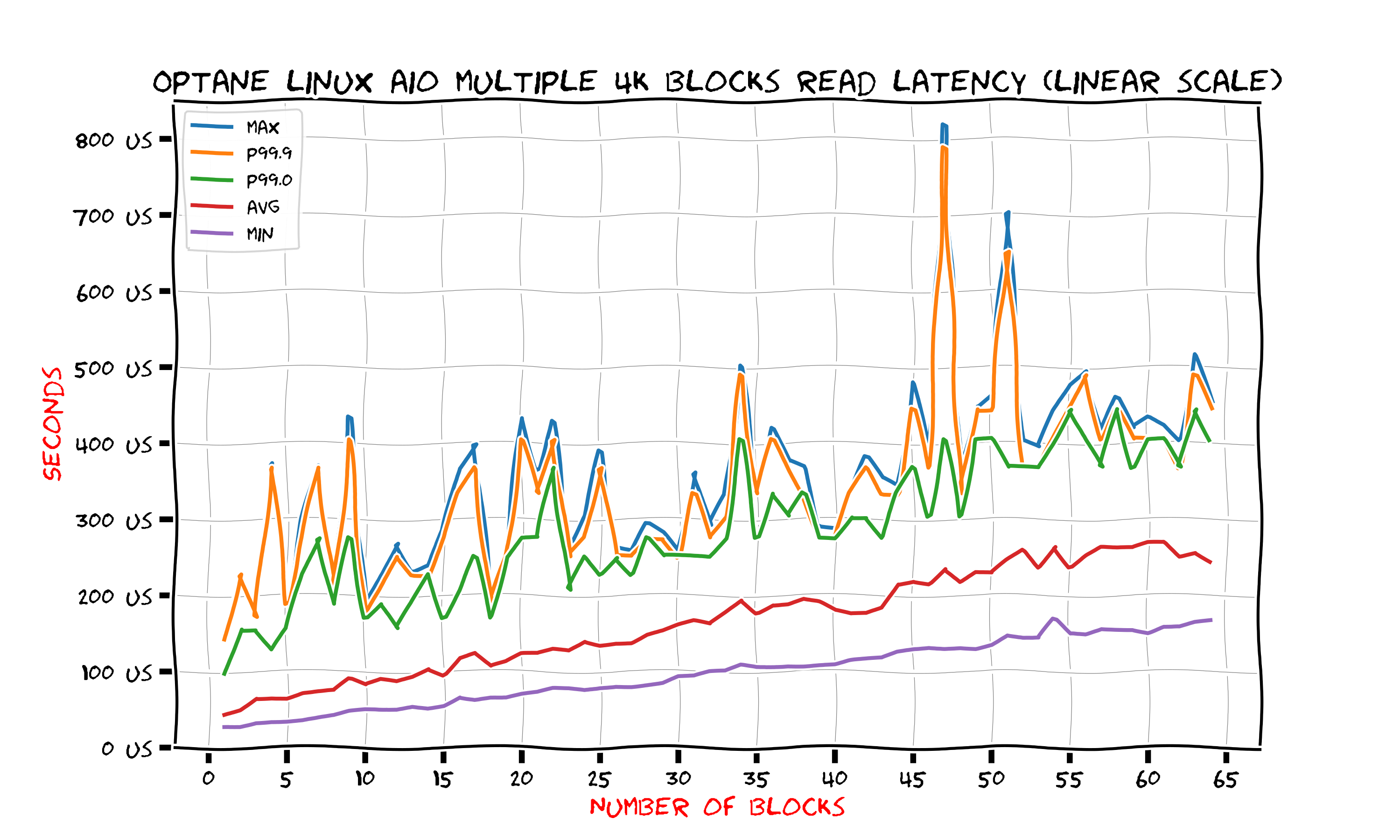}
        \caption{Optane multiple blocks read.}
        \label{fig:multblockaio:optane}
    \end{minipage}
    \hfill
    \begin{minipage}[c]{.47\linewidth}
        \includegraphics[width=\linewidth,trim=25 5 70 15, clip]{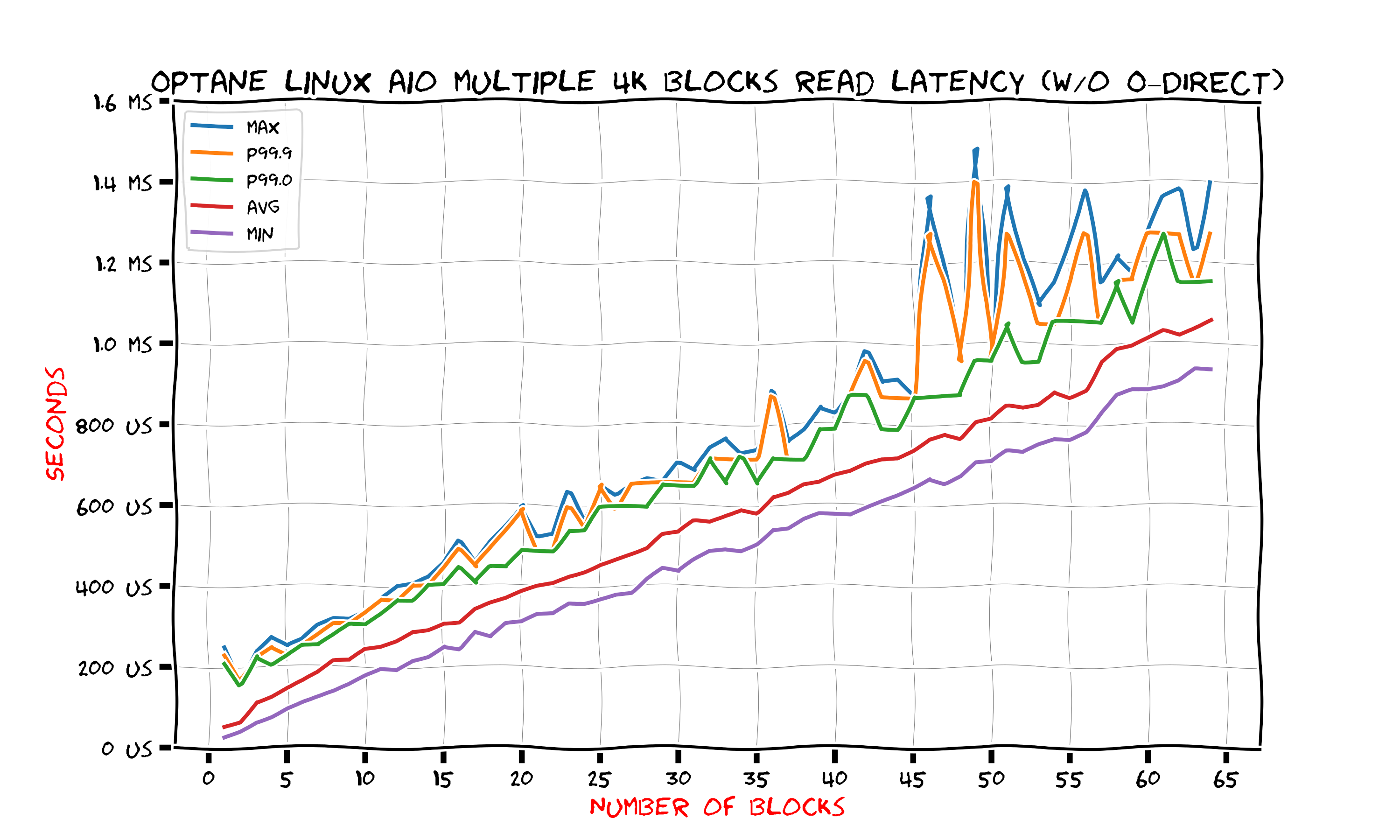}
        \caption{Optane multiple blocks read without \texttt{O\_DIRECT}.}
        \label{fig:multblockaioundirect:optane}
    \end{minipage}
\end{figure}

Optane performs better than NVMe SSD here.
As shown on figure~\ref{fig:multblockaio:optane} there is no annoying doubling of latency in the beginning.
We can read almost 50 blocks and keep ourselves in the limit of 250 microseconds.
However the overall performance is not enormously better than NVMe SSD.

Recall that we disable the kernel cache in our experiments.
With asynchronous interface the benefit is the most noticeable.
For example figure~\ref{fig:multblockaioundirect:optane} shows results for Optane when the file we read from is opened without \texttt{O\_DIRECT} flag.
It is easy to see that latencies are several times worse.

\subsection{Summary}

We looked at Linux aio interface.
In contrast to synchronous interface we were able to read from NVMe SSD up to 2.5 gigabytes per second with 4 kilobytes blocks using only three threads.
The CPU usage is also three full cores.
Surprisingly further increase in the number of threads doesn't help.

We finished our overview of Linux 4.19 interfaces. Starting from the next section we move to the kernel version 5.4.
At first we and compare Linux aio performance in 5.4 and 4.19 kernels.
So old kernel user can see the benefit of updating the kernel.


\section{Asynchronous Input-Output in Linux 5.4}
\label{sec:aio54}

Rather surprisingly Linux aio has not been adopted widely.
In Linux version 5.4 kernel developers make another attempt to create asynchronous interface.
This new interface is called~\texttt{uring}
We will study~\texttt{uring} in the next section.
To deliver an extensive comparison between Linux aio and \texttt{uring} interfaces we need to achieve the best performance for both of them (or show various trade-offs).
But first we run our tests with the same parameters as in the previous section to compare different kernel versions.

\subsection{Comparing with Linux 4.19}

In this subsection we run tests with precisely the same parameters as in section~\ref{sec:aio}.
We do this to compare different kernel versions with each other and show that it is worth updating the Linux kernel on production systems.

\begin{figure}
    \includegraphics[width=\textwidth]{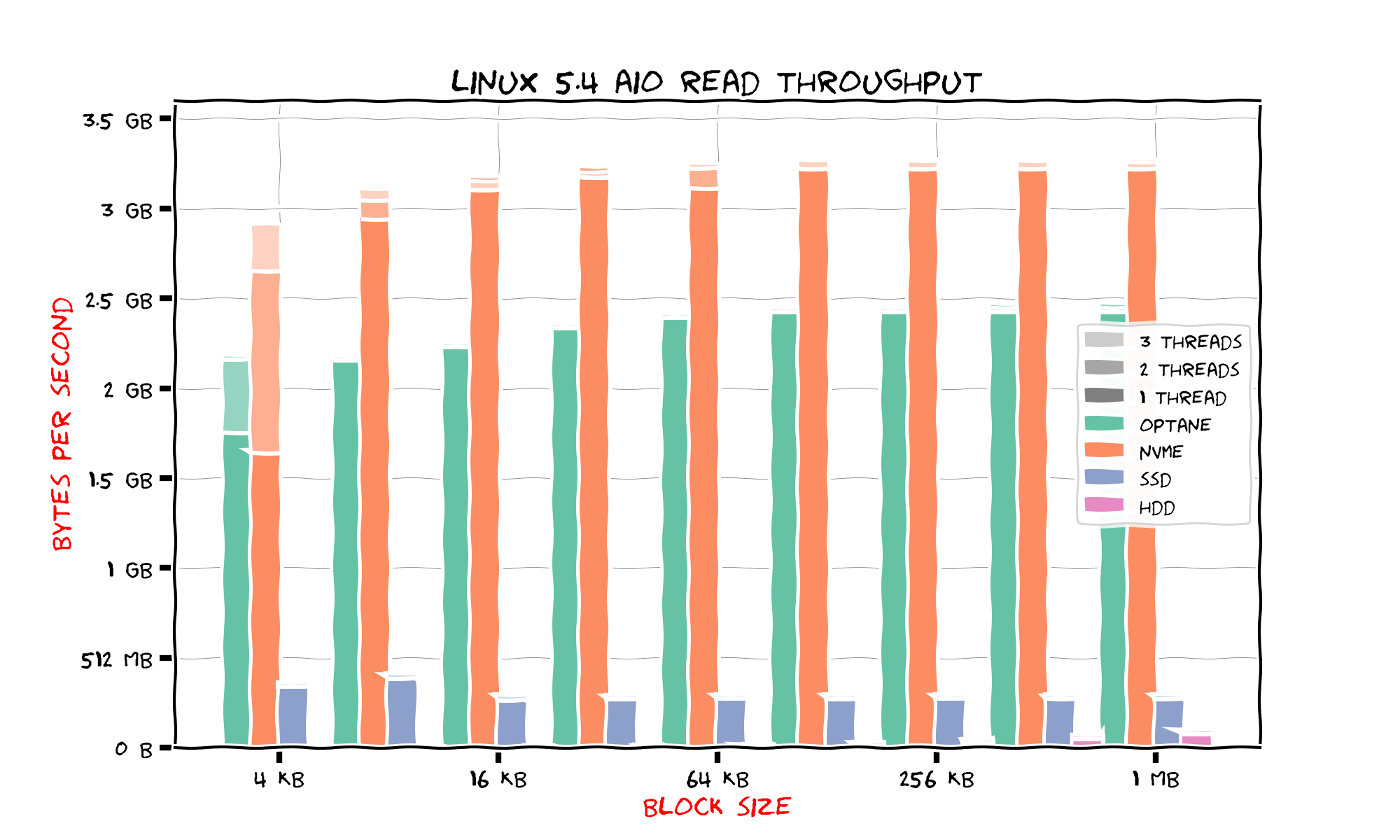}
    \caption{Asynchronous reading throughput on Linux 5.4.}
    \label{fig:allaio54:throughput}
\end{figure}

Figure~\ref{fig:allaio54:throughput} shows throughput for all four storage devices.
Note that most of the bars are way above 3 gigabytes per second.
In all our previous results 3 gigabytes per second appeared as an unreachable goal and we succeeded to reach it only with a low margin using asynchronous interface.
With kernel version 5.4 we see 3.2 gigabytes per second for rather small block size of 32 kilobytes.
As for 4 kilobytes blocks we almost reach 2.9 gigabytes per second.
When executed with kernel 4.19 the result was noticeably worse.
So it happens that sole kernel upgrade increases NVMe SSD throughput by 7\%.
Anyone who uses Linux aio interface and saturates NVMe SSD bandwidth should consider updating their kernel.

\begin{figure}
    \includegraphics[width=\textwidth]{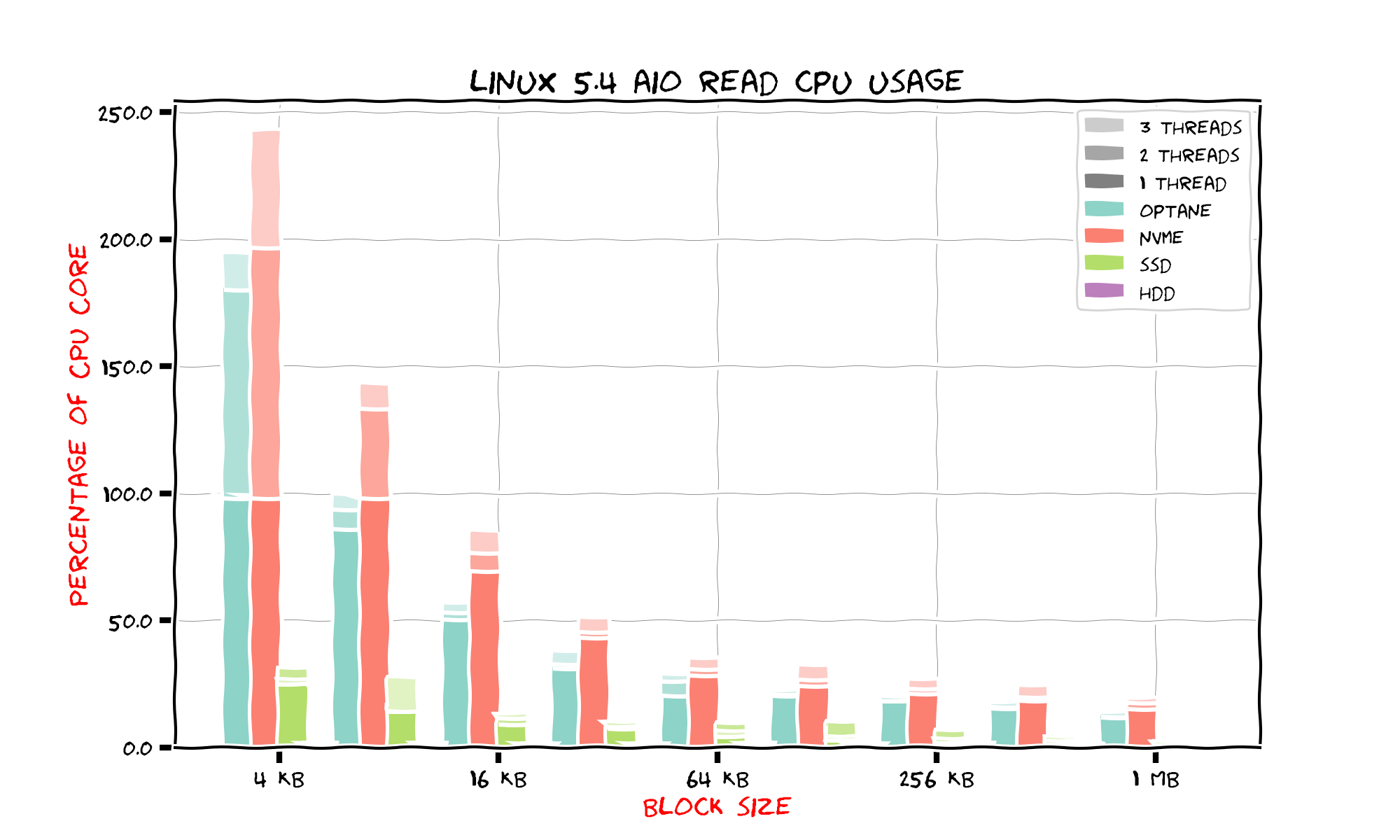}
    \caption{Asynchronous reading CPU usage on Linux 5.4.}
    \label{fig:allaio54:cpu}
\end{figure}

As figure~\ref{fig:allaio54:cpu} shows, CPU usage also drops.
Reading with 4 kilobytes blocks requires half a core of CPU time less.
Again, if a production system saturates NVMe SSD bandwidth with 4 kilobytes blocks the kernel upgrade results in 17\% CPU usage reduction.

\begin{figure}
    \includegraphics[width=\textwidth]{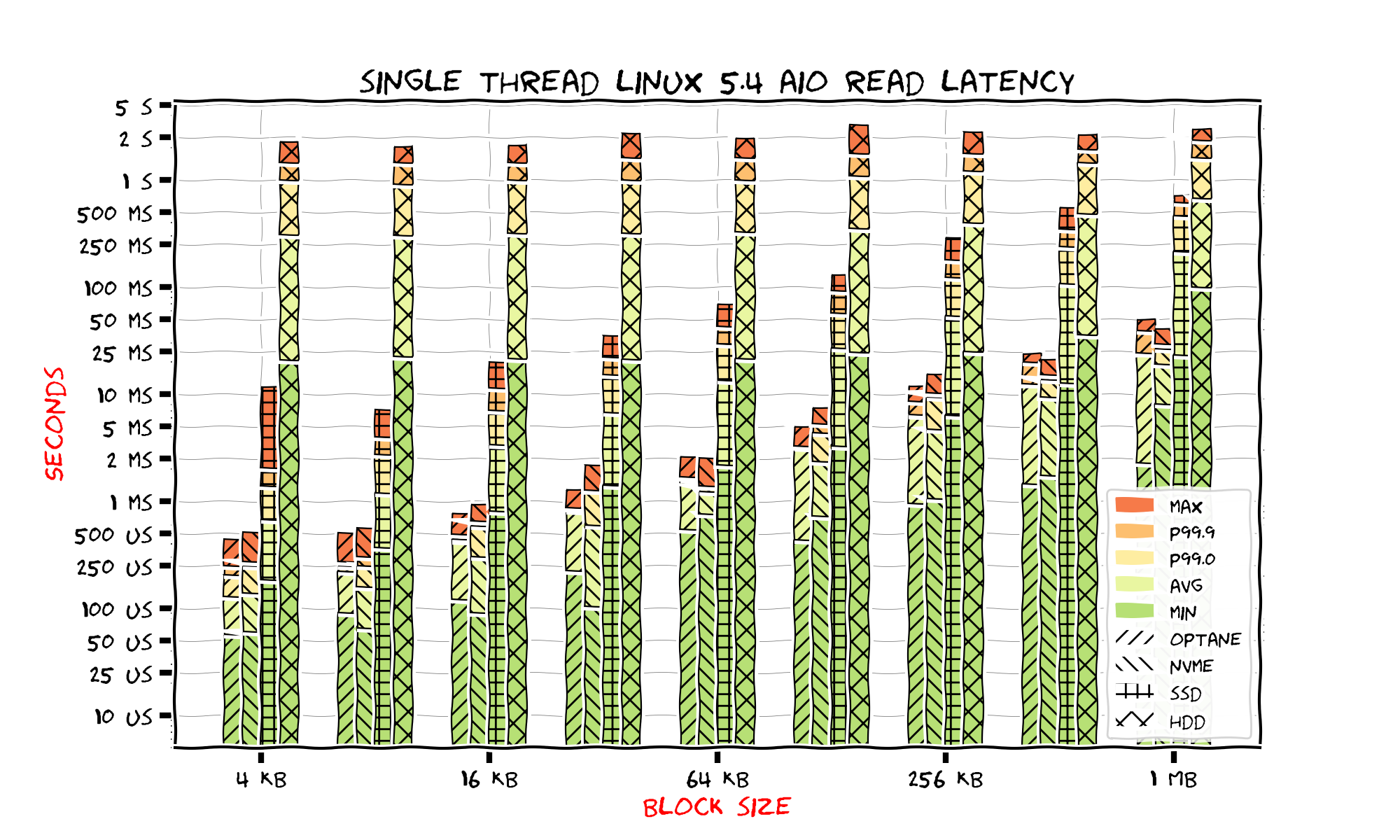}
    \caption{Single-threaded asynchronous reading latency on Linux 5.4.}
    \label{fig:allaio54thread1:latency}
\end{figure}

\begin{figure}
    \includegraphics[width=\textwidth]{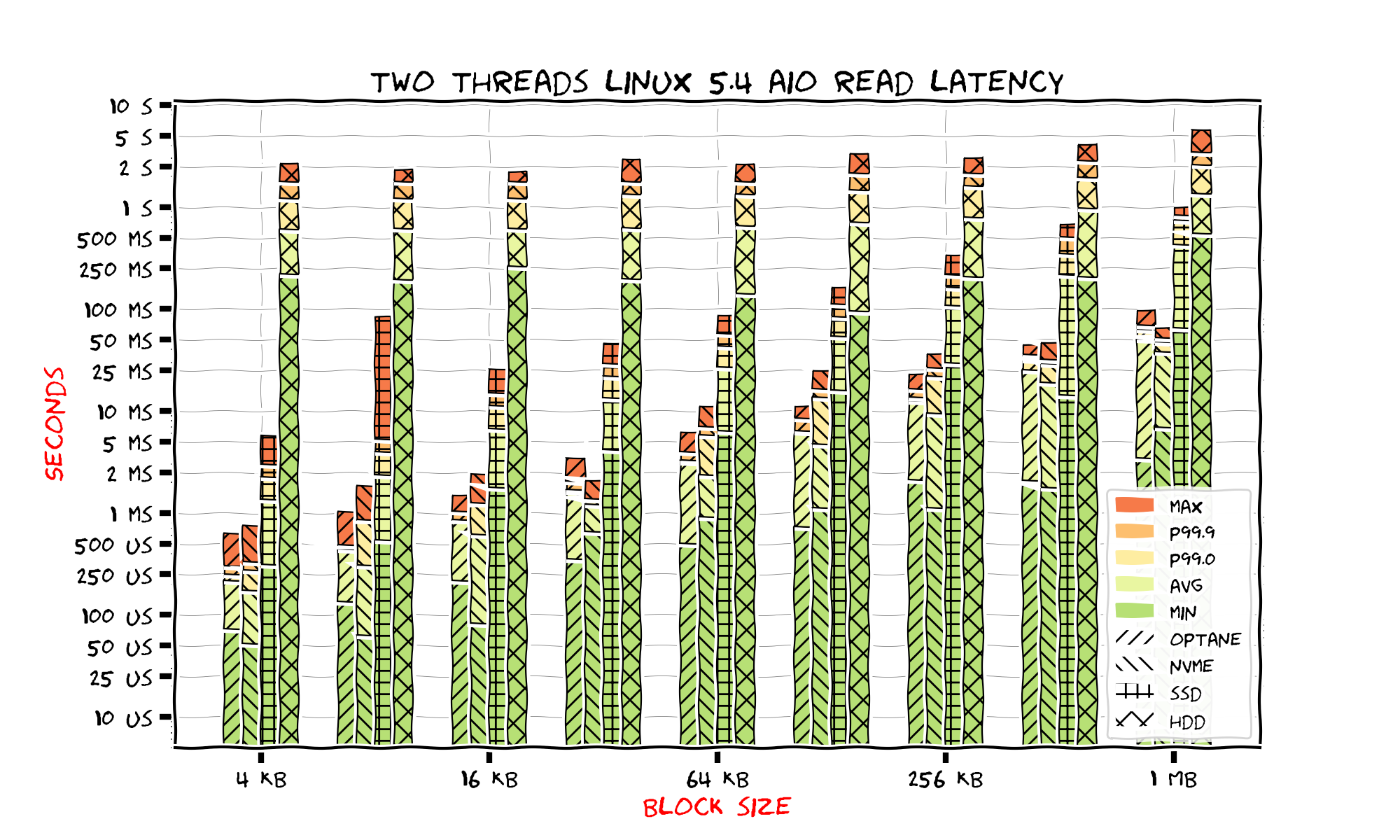}
    \caption{Two-threaded asynchronous reading latency on Linux 5.4.}
    \label{fig:allaio54thread2:latency}
\end{figure}

\begin{figure}
    \includegraphics[width=\textwidth]{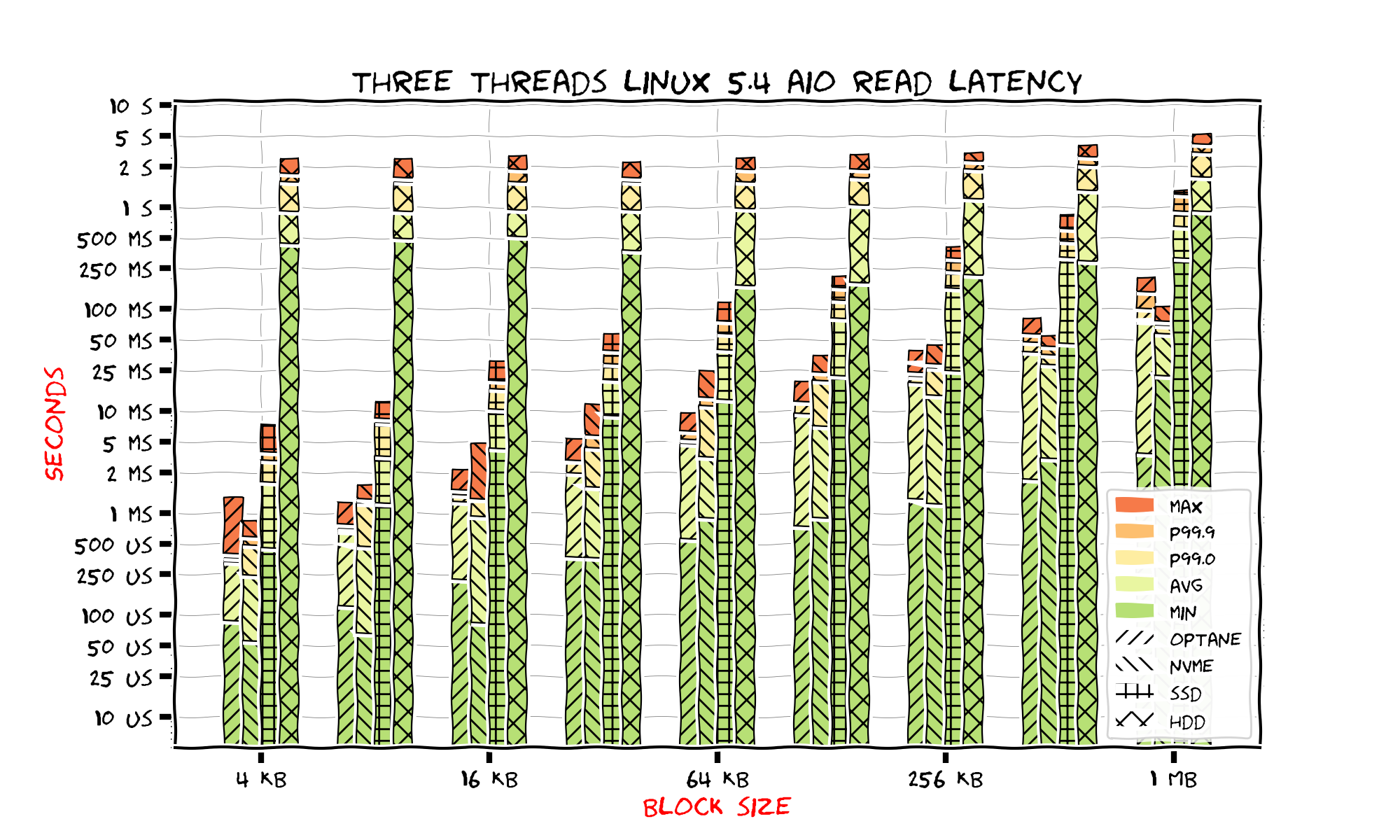}
    \caption{Three-threaded asynchronous reading latency on Linux 5.4.}
    \label{fig:allaio54thread3:latency}
\end{figure}

Finally, let's look at latency.
Figures~\ref{fig:allaio54thread1:latency},~\ref{fig:allaio54thread2:latency}, and~\ref{fig:allaio54thread3:latency} show latencies for one, two, and three threads respectively.
Compared to previous results we see that 99.9 percentile is better for middle sized blocks, but for blocks of size 4 and 8 kilobytes it looks relatively the same for all three solid devices.
The results for HDD are the same, however it should be obvious that HDD is not of our concern here.
Furthermore the block sizes represented on the figure are inadequately small for modern HDDs.
One concerned about HDD throughput should pick block size 16 megabytes and larger.


\subsection{Selecting optimal parameters}

When creating asynchronous input-output queue one has to specify its length.
Usually programmers pick some number which seems to be adequate or good enough.
Our test were no exception here.
Strictly speaking the queue length affects throughput and latency.
It seems obvious in the corner cases: a queue of length one should behave like synchronous interface while a very large queue should demonstrate large latency.

\begin{figure}
    \includegraphics[width=\textwidth]{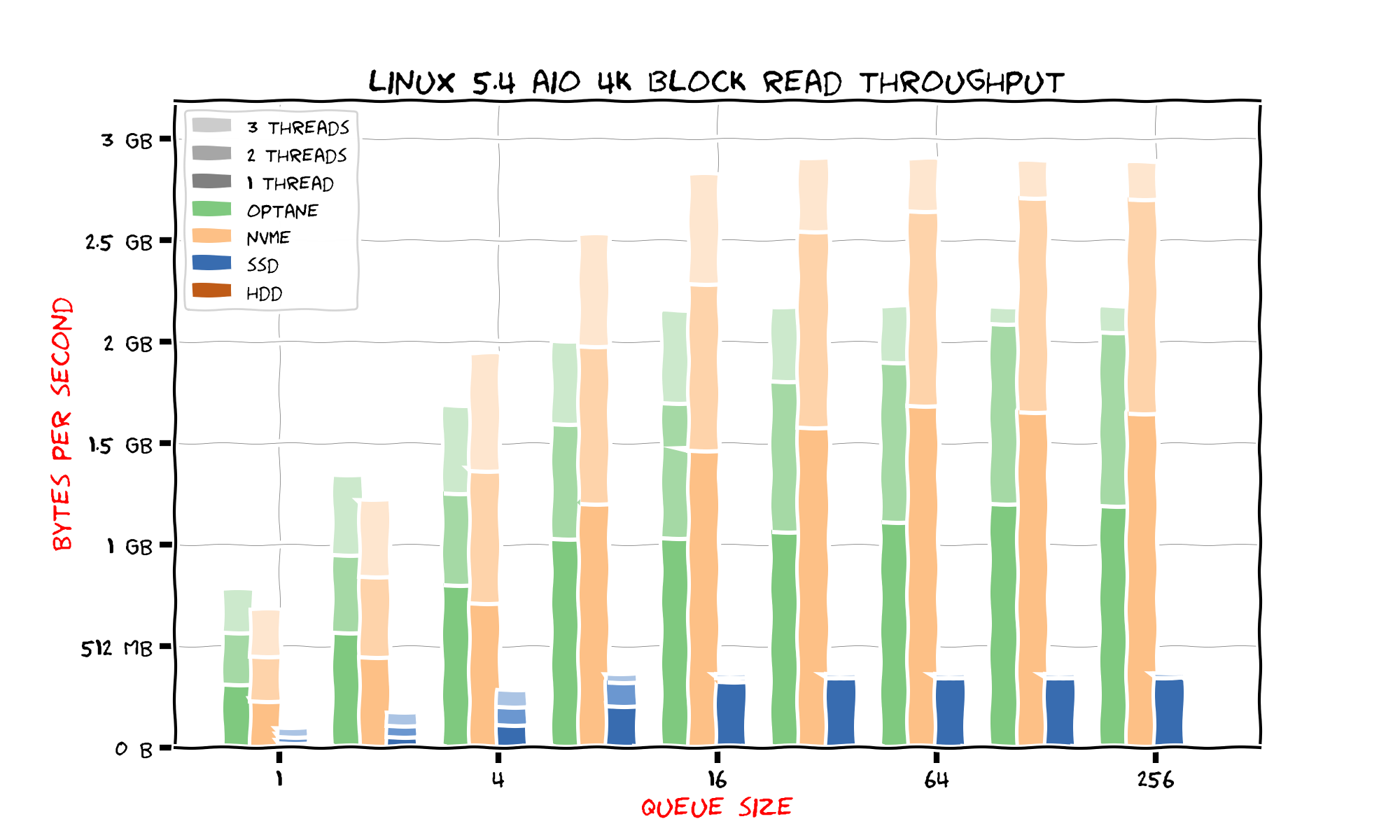}
    \caption{Asynchronous reading throughput on Linux 5.4.}
    \label{fig:allaio54queue:throughput}
\end{figure}

Let's find the optimum queue size.
We read with block size of 4 kilobytes and vary the queue size from 1 up to 256 elements.
Figure~\ref{fig:allaio54queue:throughput} shows throughput observed in this experiment.
The figure looks familiar but the bars correspond to different queue sizes instead of block sizes.

Looking at the figure we could say that any queue size is distinguishable.
In particular queues with 16 and 32 elements stand out the most since they achieve the best throughput for Optane and NMVe respectively.
Also queue of 128 elements looks rather interesting since it allows to use only two threads to read from Optane and achieve almost maximum throughput.

\begin{figure}
    \includegraphics[width=\textwidth]{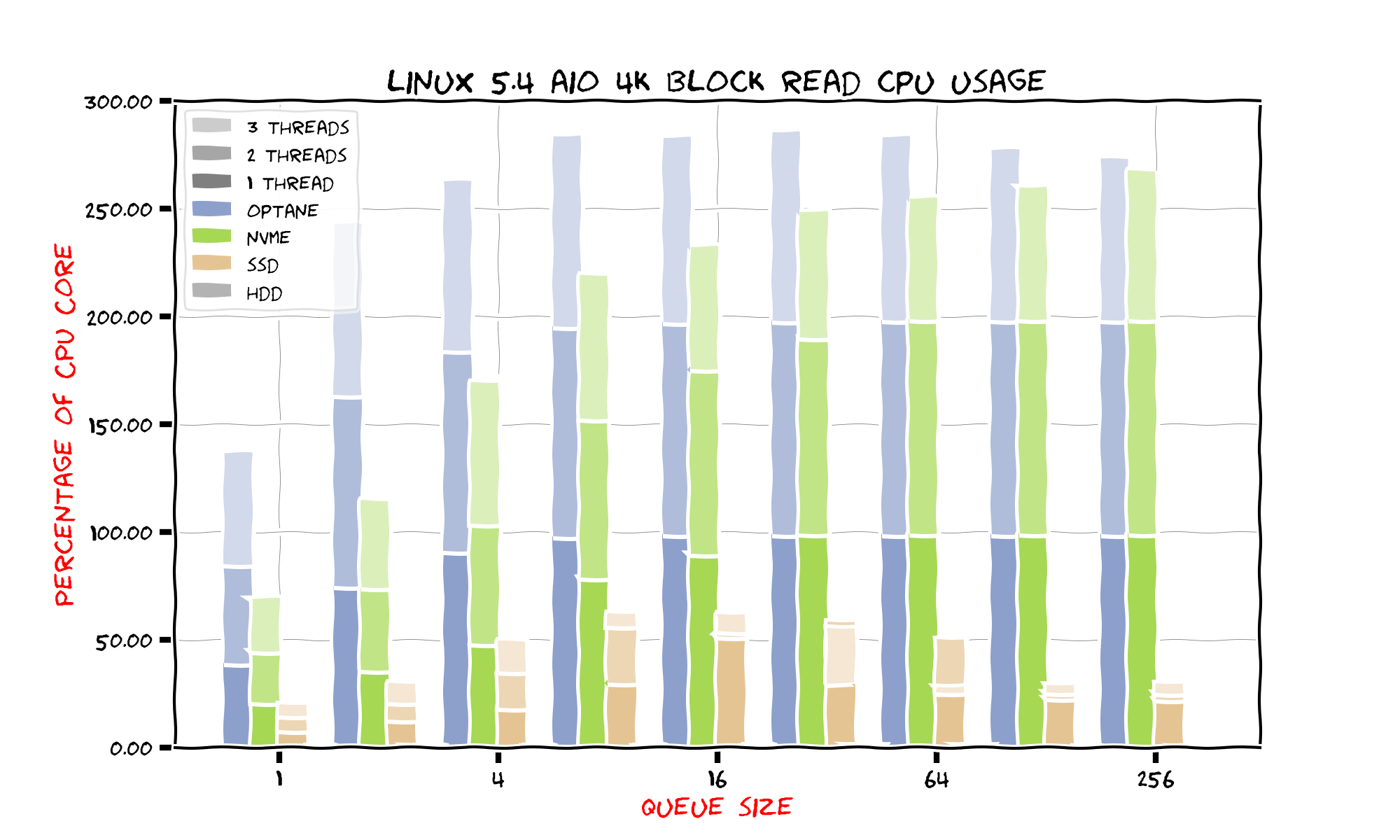}
    \caption{Asynchronous reading CPU usage on Linux 5.4.}
    \label{fig:allaio54queue:cpu}
\end{figure}

Figure~\ref{fig:allaio54queue:cpu} shows observed CPU usage.
It doesn't seem like it would affect our choice of the best queue size.
The CPU usage grows alongside throughput growth and this should be expected.
We won't dive in 10\% oscillations of the CPU usage.
However the reader may find this important for some particular case.

\begin{figure}
    \includegraphics[width=\textwidth]{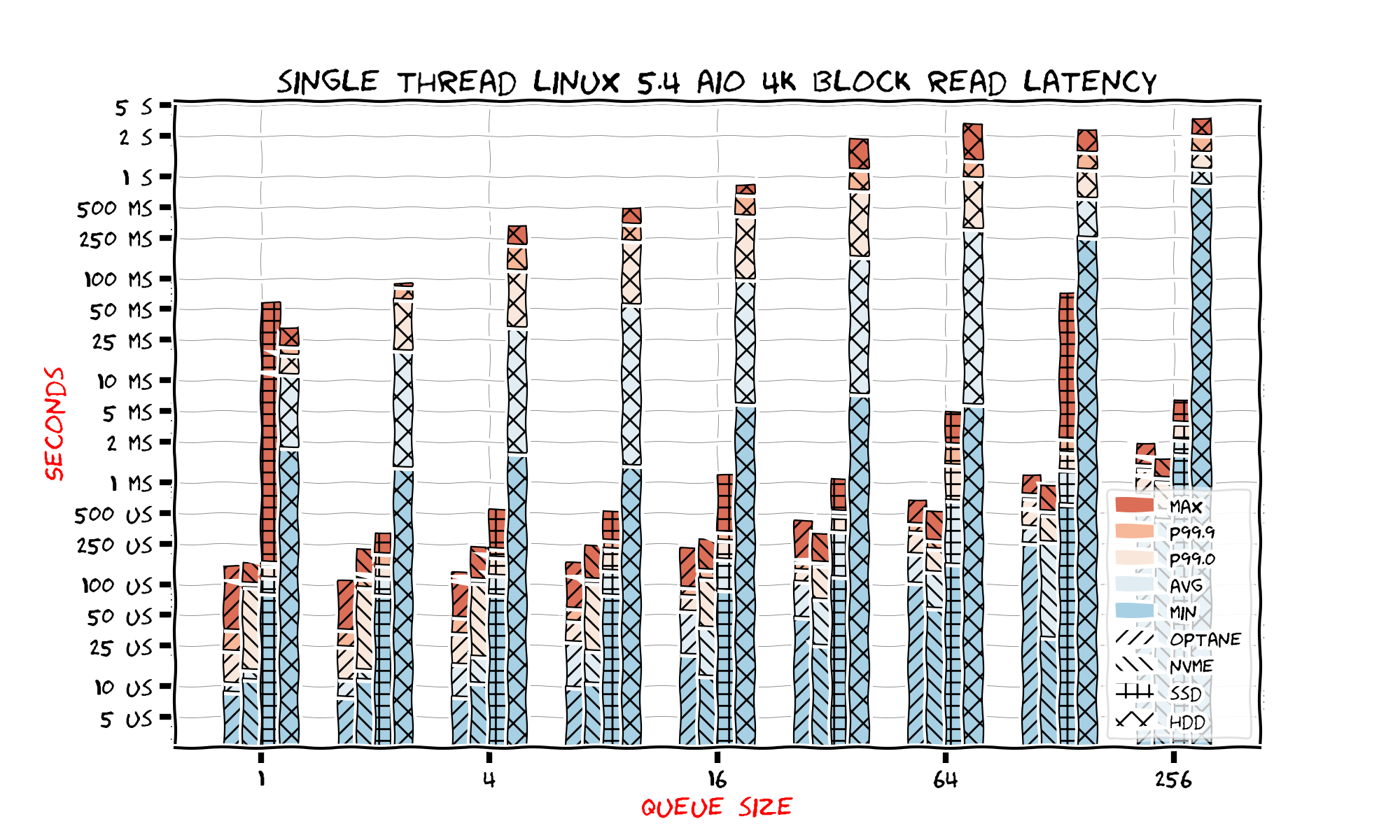}
    \caption{Single-threaded asynchronous reading latency on Linux 5.4.}
    \label{fig:allaio54thread1queue:latency}
\end{figure}

\begin{figure}
    \includegraphics[width=\textwidth]{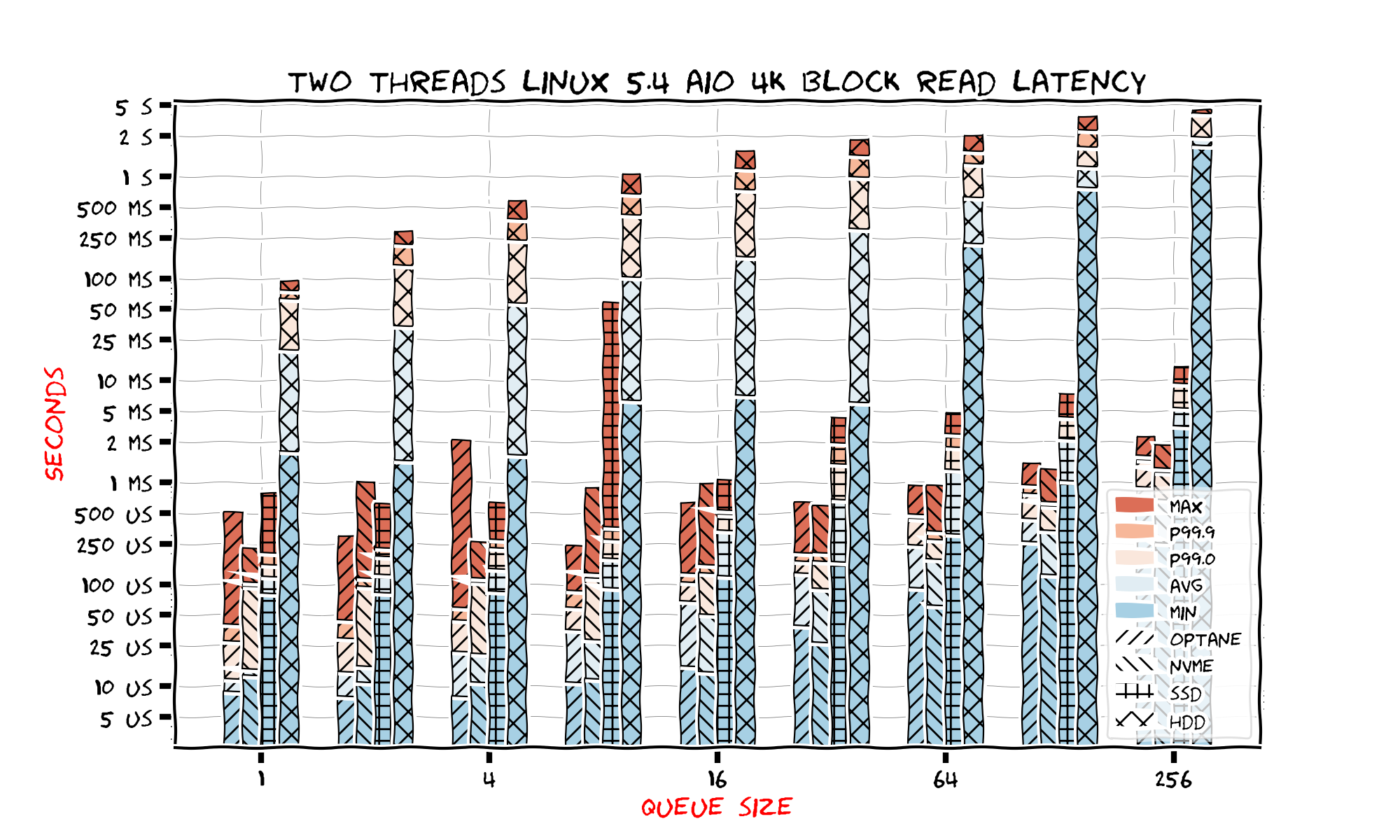}
    \caption{Two-threaded asynchronous reading latency on Linux 5.4.}
    \label{fig:allaio54thread2queue:latency}
\end{figure}

\begin{figure}
    \includegraphics[width=\textwidth]{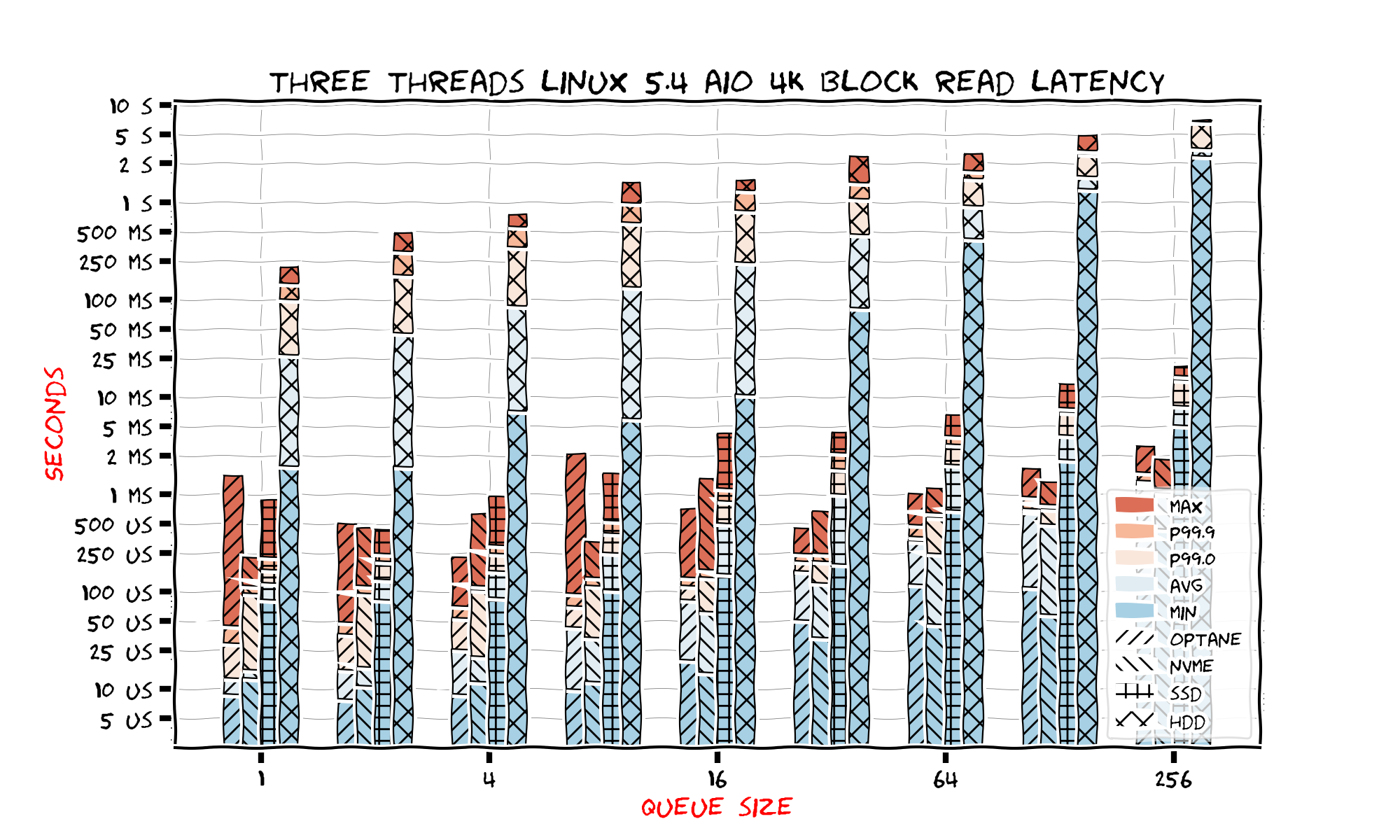}
    \caption{Three-threaded asynchronous reading latency on Linux 5.4.}
    \label{fig:allaio54thread3queue:latency}
\end{figure}

Latency is more important to us.
Figures~\ref{fig:allaio54thread1queue:latency},~\ref{fig:allaio54thread2queue:latency}, and~\ref{fig:allaio54thread3queue:latency} show the results for one, two, and three threads respectively.
It is straightforward to see that latency increase when the queue size is increased.
In particular for the 128 elements queue Optane latency is quite bad.
It is 1 millisecond in 99.9 percentile.
If we pick a smaller queue of 16 elements the latency would be 100 microseconds for the same 99.9 percentile.
For queue size 32 the latency for both NVMe SSD and Optane is 250 microsecond.
Recall that we should look at the third figure since for this queue size we need three threads to saturate the bandwidth.


It seems that we need a queue of 16 elements and three threads when we read from either Optane or NMVe.
For SSD we need the same queue size and a single thread.
However it is too early to finish our configuration at this point.

When \texttt{io\_getevents} is called an application specifies how many request should be awaited for completion.
Clearly one could wait for a single request to complete or for all requests in the queue.
In some sense this parameter determines how often we add new requests into the queue.
If we use this in a real application that won't be our choice since this usually depends on user behavior.
However in synthetic experiments we both can and should tweak this parameter since this could show us what performance we can hope for in real life.

It is interesting to look how latency and throughput depend on the number of requests we insert into the queue at a single step (a batch).
We select two best queue sizes of 16 and 32 elements and try to vary size of the batch.
Earlier when we tried to find the best queue size the batch size was equal to one.
Next we look at the throughput, CPU usage and latency observed in these experiments and finally pick up the best parameters.

It can be hard to read new charts since legend is extended.
We add one new dimension for the queue size which is either 16 or 32.
Each bar is now split in two and each half-bar represents the different queue size.
Thus results for a specific storage device and a number of threads but with different queue sizes are shown by adjacent bars.

\begin{figure}
    \includegraphics[width=\textwidth]{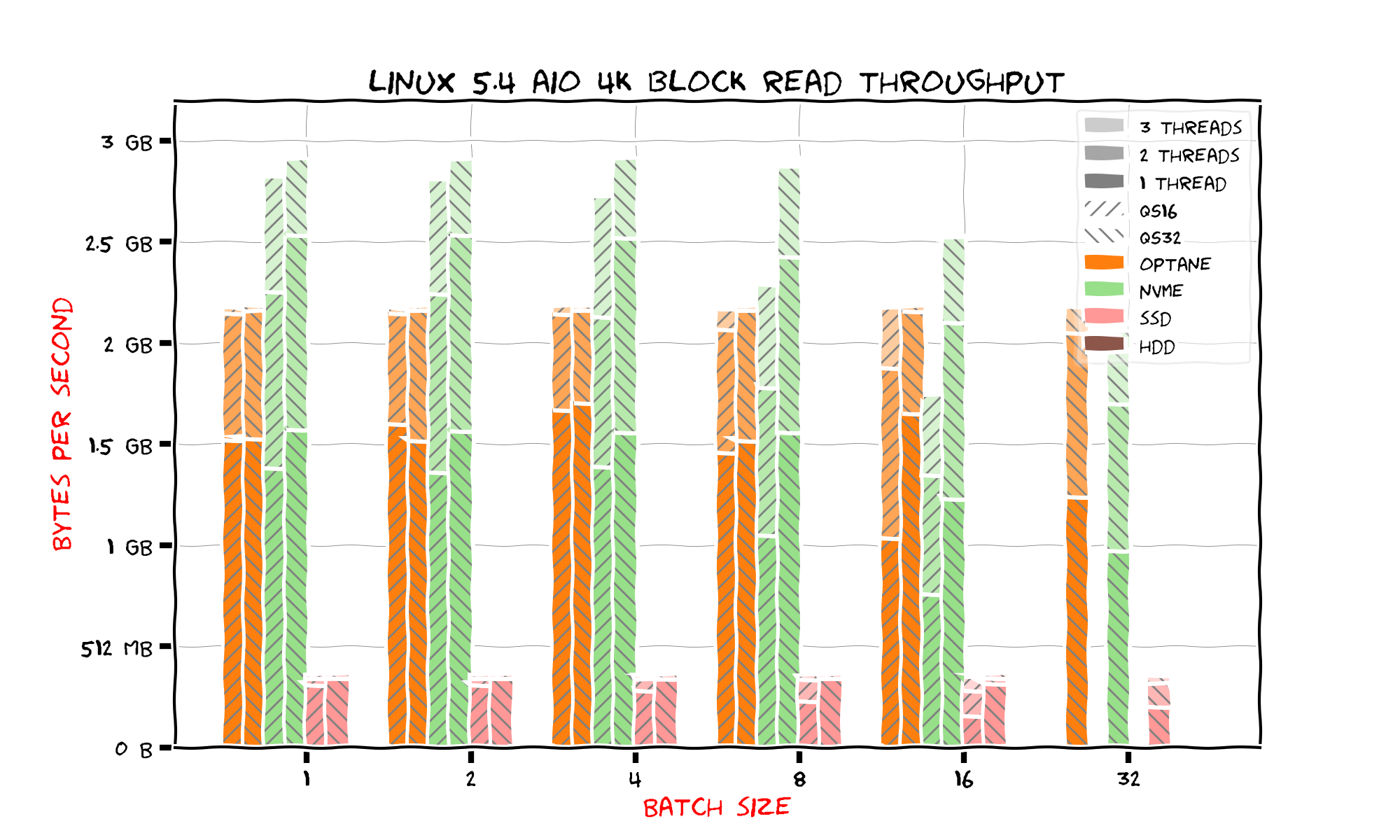}
    \caption{Asynchronous reading throughput on Linux 5.4.}
    \label{fig:allaio54batch:throughput}
\end{figure}

\begin{figure}
    \includegraphics[width=\textwidth]{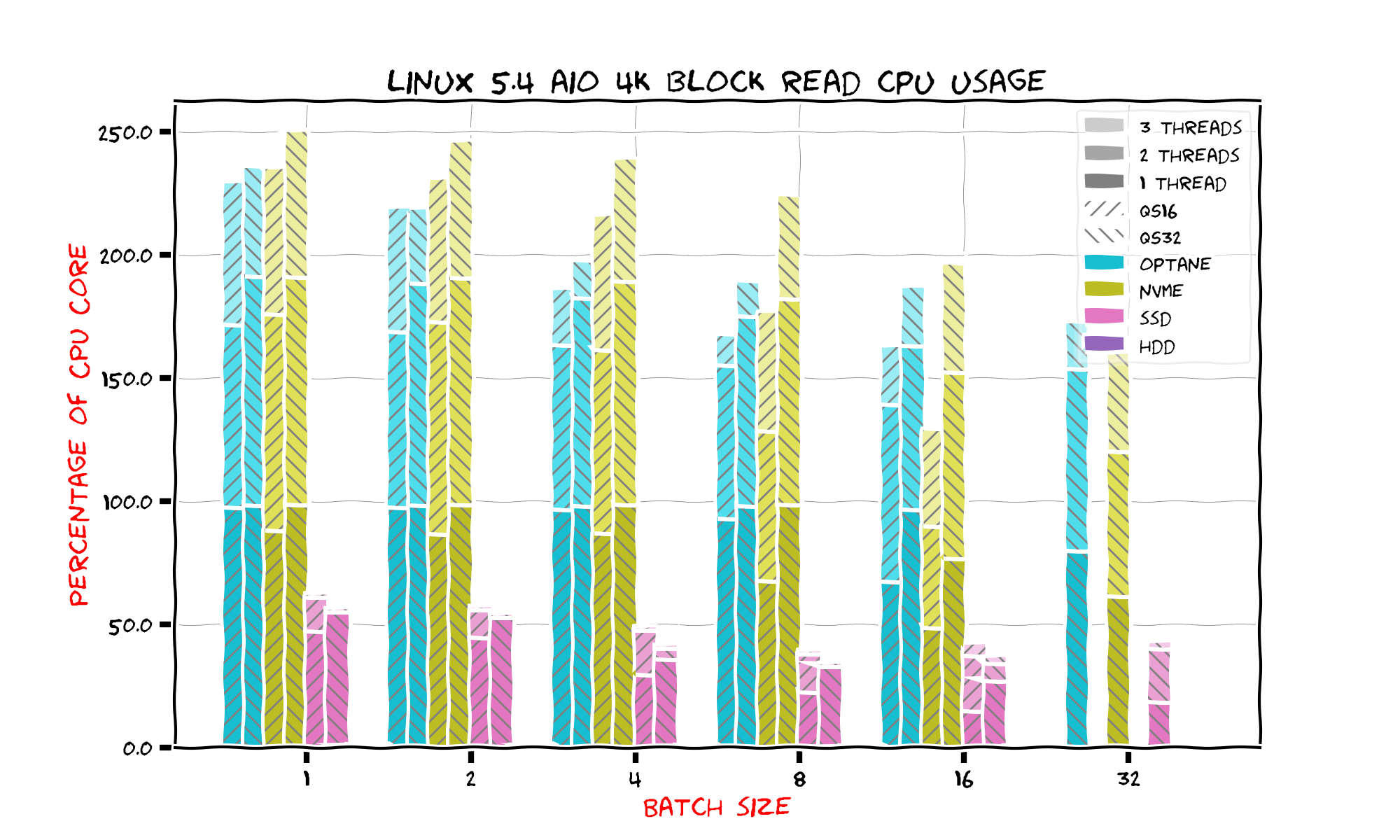}
    \caption{Asynchronous reading CPU usage on Linux 5.4.}
    \label{fig:allaio54batch:cpu}
\end{figure}

Figures~\ref{fig:allaio54batch:throughput} and~\ref{fig:allaio54batch:cpu} show throughput and CPU usage respectively.
Different queue sizes can be distinguished by hatches.
In the legend queue sizes of 16 and 32 are labeled as QS16 and QS32 respectively.
Different groups of bars differ by batch size thus the horizontal axis represent the batch size.

\begin{figure}
    \includegraphics[width=\textwidth]{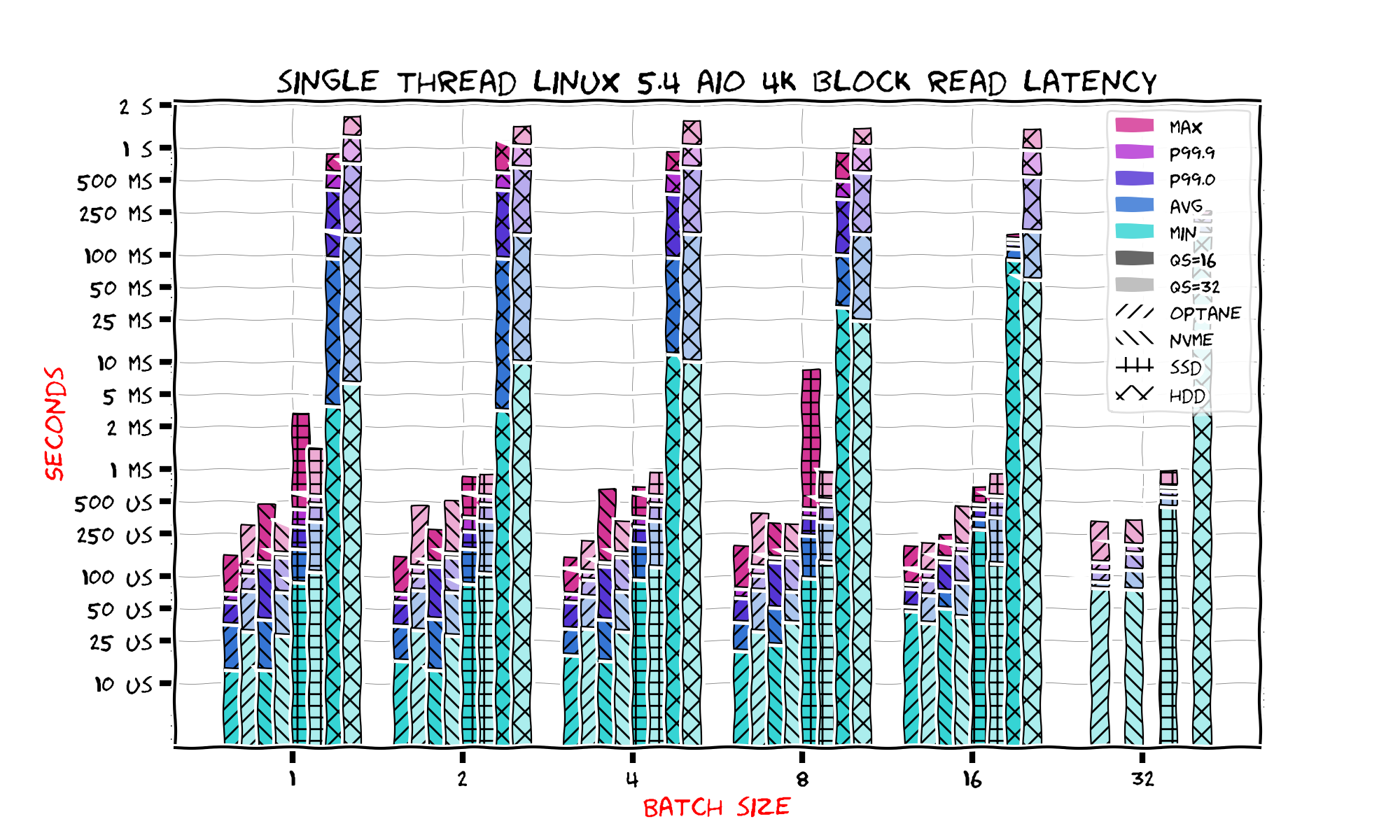}
    \caption{Single-threaded asynchronous reading latency on Linux 5.4.}
    \label{fig:allaio54thread1batch:latency}
\end{figure}

\begin{figure}
    \includegraphics[width=\textwidth]{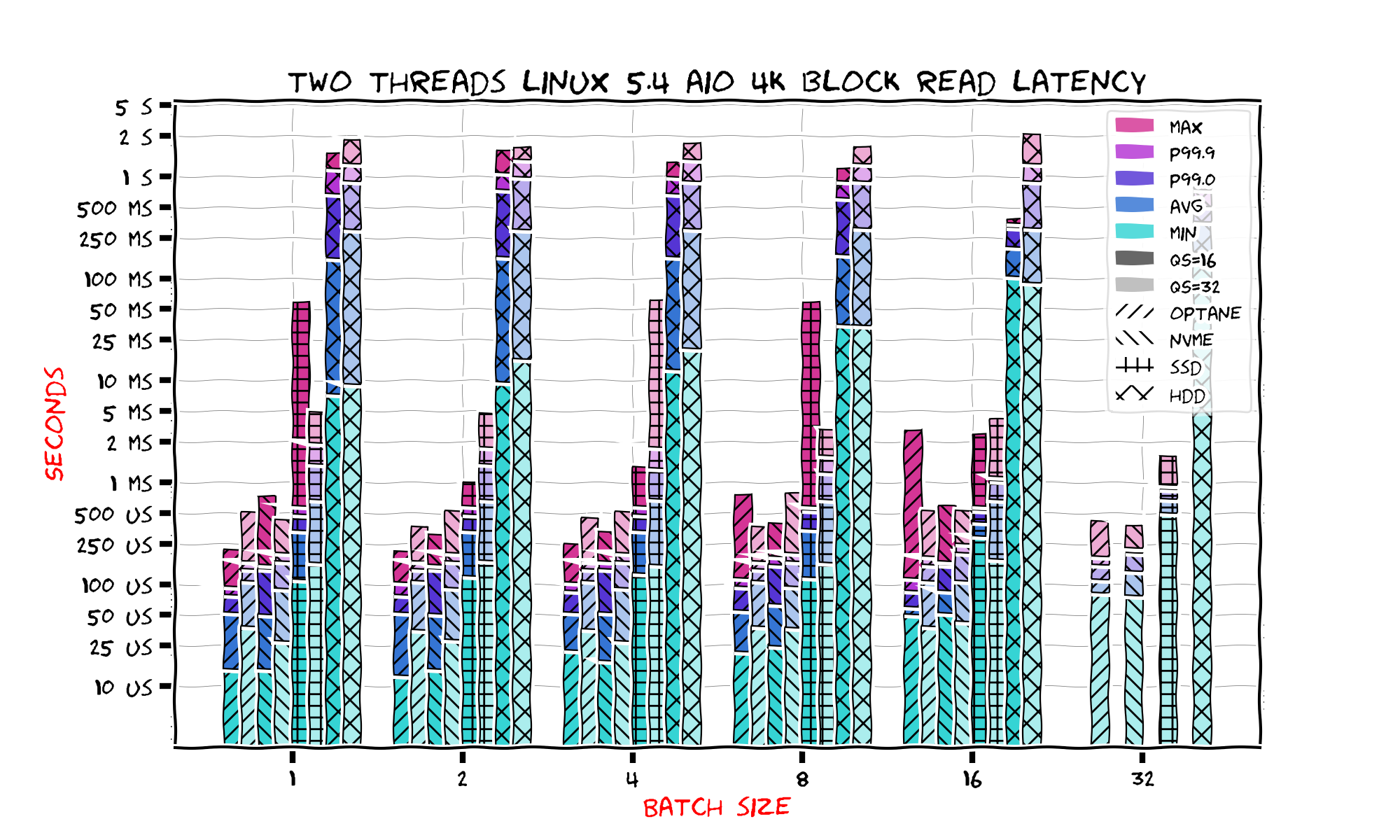}
    \caption{Two-threaded asynchronous reading latency on Linux 5.4.}
    \label{fig:allaio54thread2batch:latency}
\end{figure}

\begin{figure}
    \includegraphics[width=\textwidth]{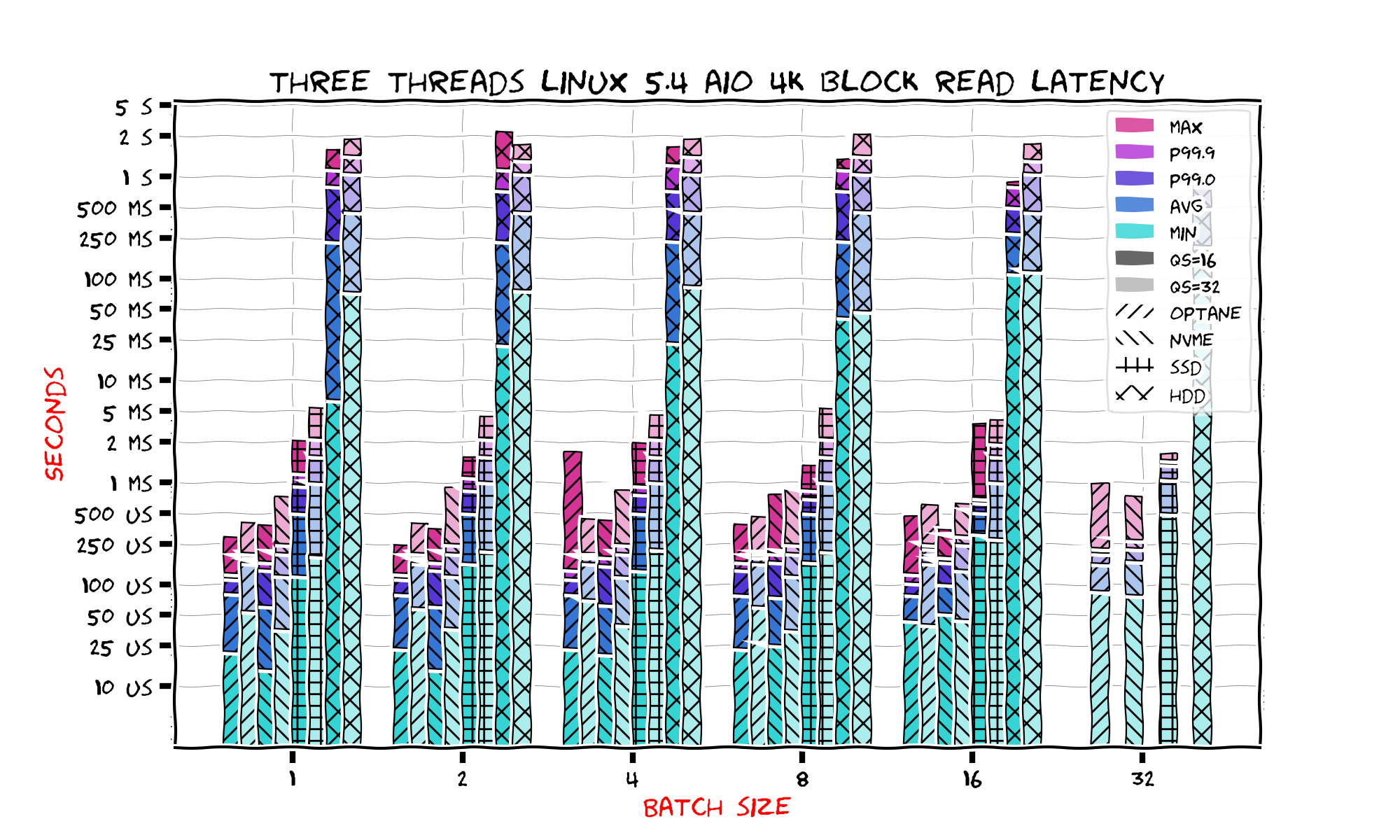}
    \caption{Three-threaded asynchronous reading latency on Linux 5.4.}
    \label{fig:allaio54thread3batch:latency}
\end{figure}

Figures~\ref{fig:allaio54thread1batch:latency},~\ref{fig:allaio54thread2batch:latency}, and~\ref{fig:allaio54thread3batch:latency} show latencies for one, two, and three threads respectively.
Unfortunately hatches have already been in use on latency charts and represent the storage type.
Different queue sizes are represented by color saturation.
More colorful bars correspond to the queue of 16 elements while less colorful correspond to the queue of 32 elements.


Let's deduce the best batch size.
First we look at latency.
It is straightforward to notice that in almost all cases the queue of 32 elements leads to larger latencies than the queue of 16 elements.
The only advantage of the 32 elements queue is larger throughput when reading from NVMe SSD.
However difference of 50 or even 100 megabytes per second doesn't justify 10\% and higher increase in latency.
Therefore we will look at 16 elements queue.

Let's look at throughput on figure~\ref{fig:allaio54batch:throughput}.
Notice that for Optane if the batch size is 4 or lower it suffices to use only two threads to saturate the bandwidth.
Almost all these variants also have the same latency.
We can pick batch size 4 here to save some CPU time.

As for NVMe SSD the throughput chart shows that we need to choose between batch sizes 1, 2, and 4 with preference for the smaller size.
It happens again that all three bars look similar therefore for NVMe SSD we pick batch size 1.
This leads to the highest throughput.
However the throughput is 2.8 gigabytes per second.
If we want to reach 2.9 gigabytes per second we need to choose 32 elements queue and batch size 4.
But this will lead to nearly 50\% latency increase in 99.9 percentile.

The charts clearly show that it is better to read from SSD using a single thread since this leads to 300 microsecond latency in 99.9 percentile.
If we use two threads the latency would be 500 microseconds for the same percentile. For three threads the latency is around 1 millisecond, however it drops to around 600 microseconds when the bath size is 16.
The lowest latency for a single thread is with batch size 2 or 4.
If we take throughput into account the batch size 2 is strictly better.
The throughput is rather close to what we can achieve.
If one would like to sacrifice latency and get the maximum throughput then they should use three threads, queue size 16, and batch size 8.

HDD is almost unnoticeable on throughput and CPU usage figures.
This is not surprising since we read using 4 kilobyte blocks and for HDD all the time is spent on seeking.
Therefore we just pick the lowest latency: single thread, 16 elements queue and batch size also 16.
The resulting latency is slightly above 100 milliseconds.

This concludes selecting the parameters. We summarize our choices in the table below.

\begin{center}
\begin{tabular}{|l|l|l|l|}
\hline
Storage & Threads & Queue Size & Batch size \\ \hline
Pptane  & 2       & 16         & 4 \\ \hline
NVMe    & 3       & 16         & 1 \\ \hline
SSD     & 1       & 16         & 2 \\ \hline
HDD     & 1       & 16         & 16 \\ \hline
\end{tabular}
\end{center}

\subsection{Executing with optimal parameters}

In the last subsection we execute our usual experiments with different block sizes but with the queue and batch sizes picked in previous subsection.

\begin{figure}
    \includegraphics[width=\textwidth]{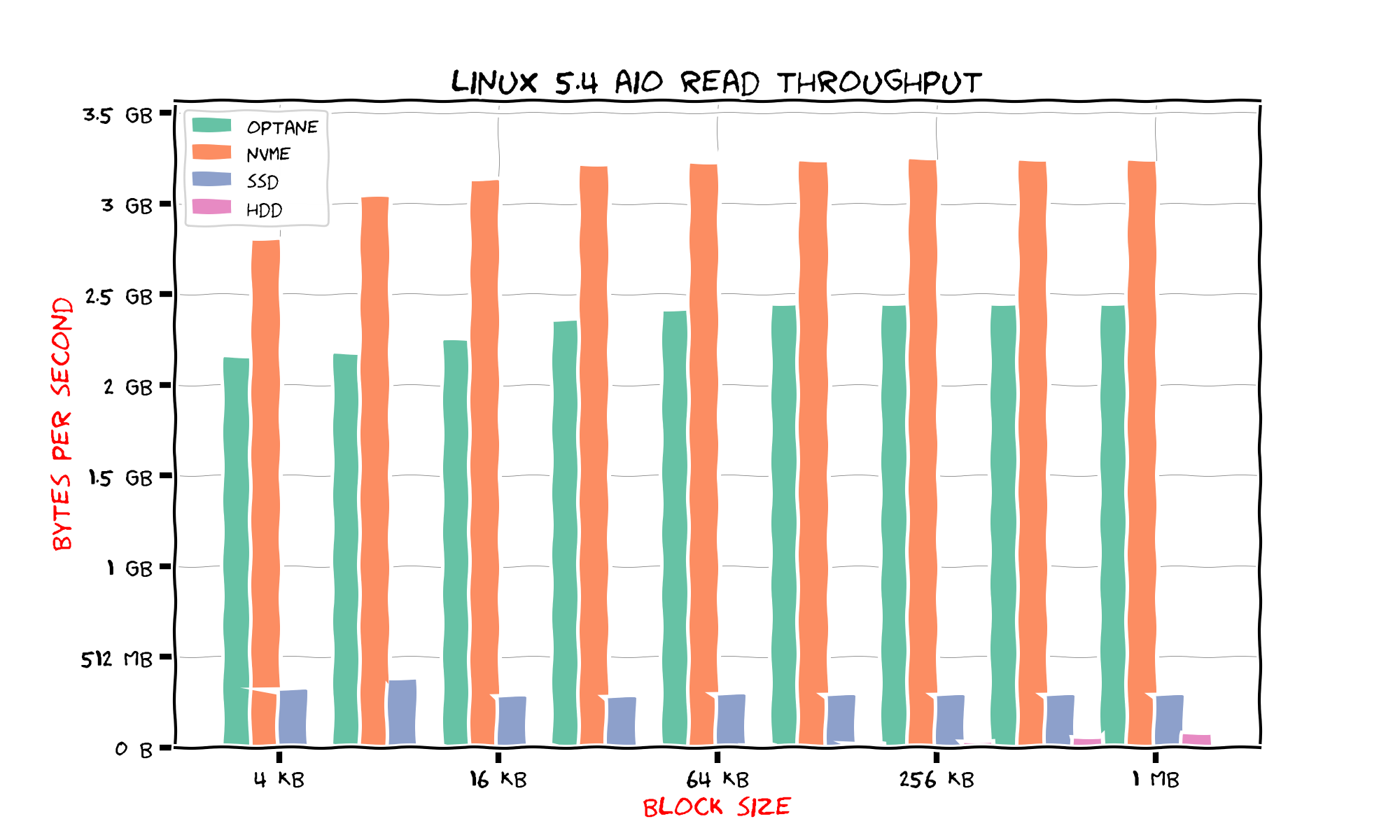}
    \caption{Best asynchronous reading throughput on Linux 5.4.}
    \label{fig:allaio54best:throughput}
\end{figure}

Figure~\ref{fig:allaio54best:throughput} shows the resulting throughput.
It seems like NVMe SSD performs slightly worse for block sizes 4 and 8.
For block sizes 16 and larger the results look similar to observed previously.
As for the other storage types the results look the same.

\begin{figure}
    \includegraphics[width=\textwidth]{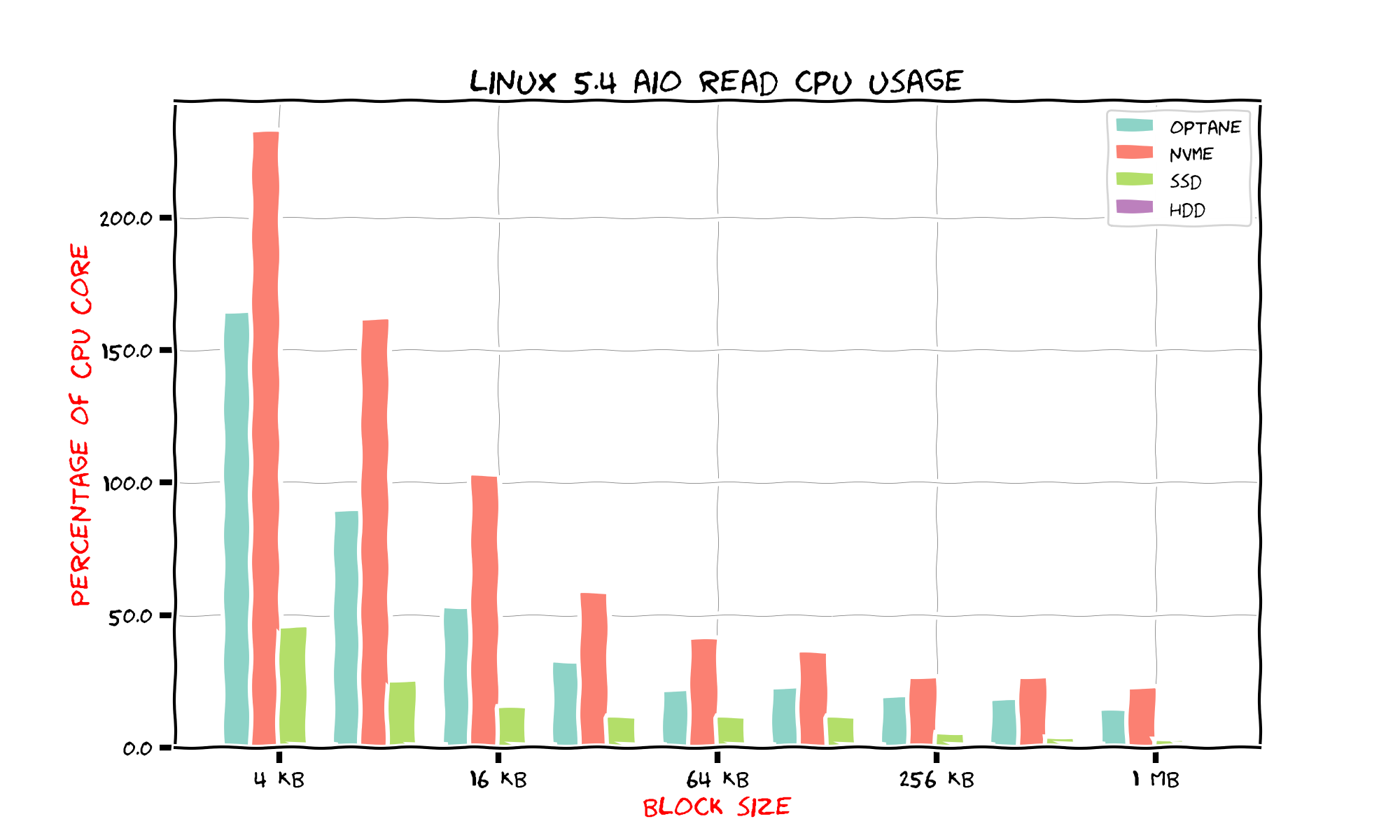}
    \caption{Best asynchronous reading CPU usage on Linux 5.4.}
    \label{fig:allaio54best:cpu}
\end{figure}

The CPU usage is shown on figure~\ref{fig:allaio54best:cpu}.
It appears to be slightly better for 4 kilobytes block size and slightly worse for 8 kilobytes block size but the overall results are also nearly the same.

\begin{figure}
    \includegraphics[width=\textwidth]{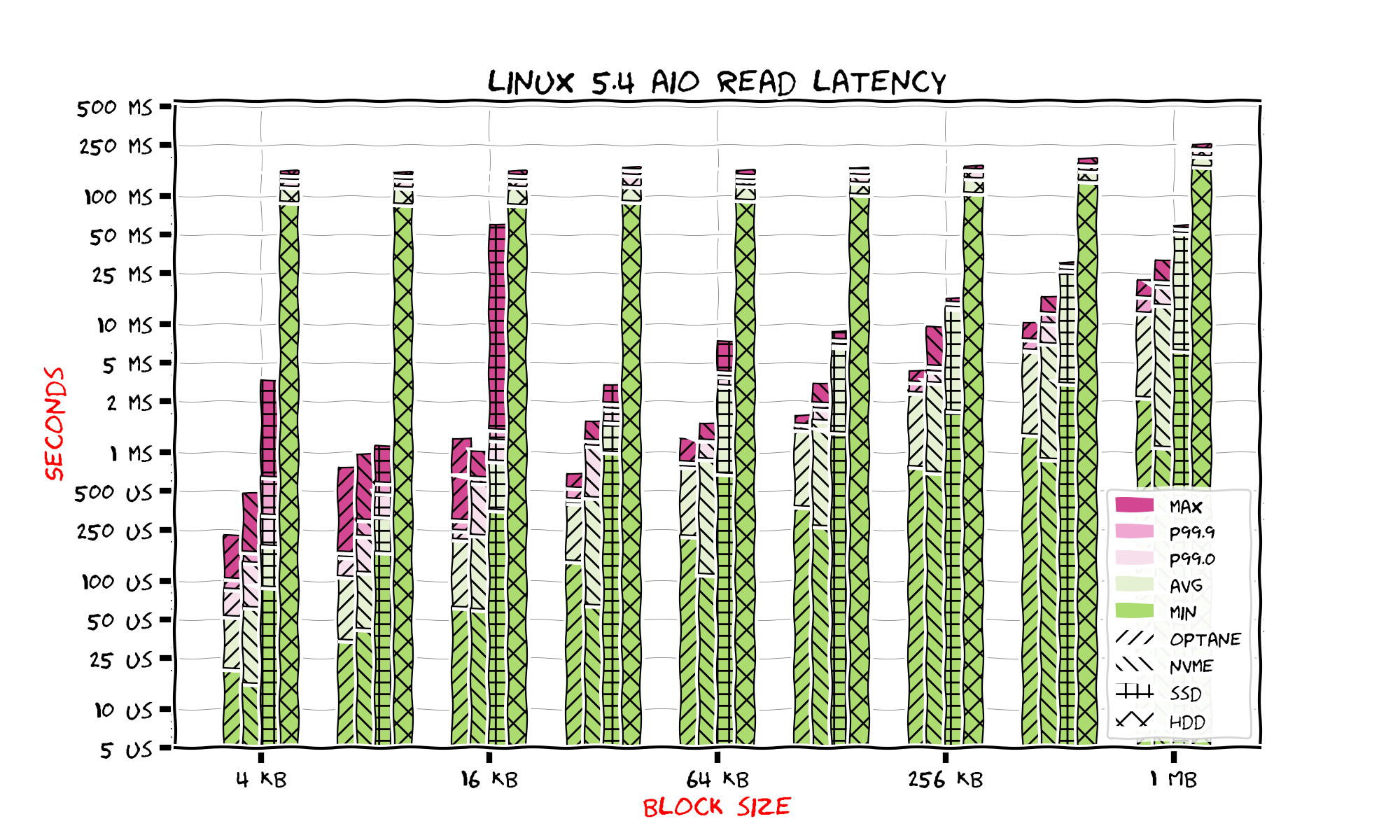}
    \caption{Best asynchronous reading latency on Linux 5.4.}
    \label{fig:allaio54best:latency}
\end{figure}

On the other hand latency difference is enormous.
The results are shown on figure~\ref{fig:allaio54best:latency}.
Latency is several times better for all storage types and for all block sizes.
For example reading from Optane with block size 4 kilobytes demonstrates latency of 100 microseconds in 99.9 percentile.
In the beginning of this section it was 250 microseconds.
For NVMe SSD the improvement is even more significant.
We started with 600 microseconds in 99.9 percentile and now we see about 150 microseconds.

\subsection{Summary}

We looked at Linux aio interface with kernel version 5.4, compared it to version 4.19 and selected optimal parameters.
We were able to significantly improve latency.
For small blocks it begins to look like single block reading.
For NVMe SSD and Optane this difference is within the factor of two.
For SSD it is slightly worse, more like a factor of 2.5.

Recall that our ultimate goal is to compare Linux aio with \texttt{uring}. In the next section we finally start reading via \texttt{uring}. 

\section{Linux 5.4 \texttt{uring} interface}
\label{sec:uring}

As we stated several times already Linux kernel version 5.4 introduced new asynchronous input-output interface called \texttt{uring}.
The interface itself slightly resembles NVMe SSD (refer to~\ref{subsec:nvme}).
There are two queues: submission queue (SQ) and completion queue (CQ).
Both queues reside in a process address space and could be accessed without issuing a system call.
Nevertheless it is required to notify the kernel that there are new elements in SQ by using \texttt{uring\_enter} system call.
One could also wait for completion with the same call.
On the other hand an application could look at CQ and wait for new elements to appear without making any system call.
It is quite tricky to use these system calls and developers purpose \texttt{liburing} library which slightly simplifies \texttt{uring} usage.
We use \texttt{liburing} in our experiments.
For a detailed overview of \texttt{uring} refer to~\cite{Uring-Intro} and~\texttt{liburing} source code.

\subsection{Choosing queue parameters}

\begin{figure}
    \includegraphics[width=\textwidth]{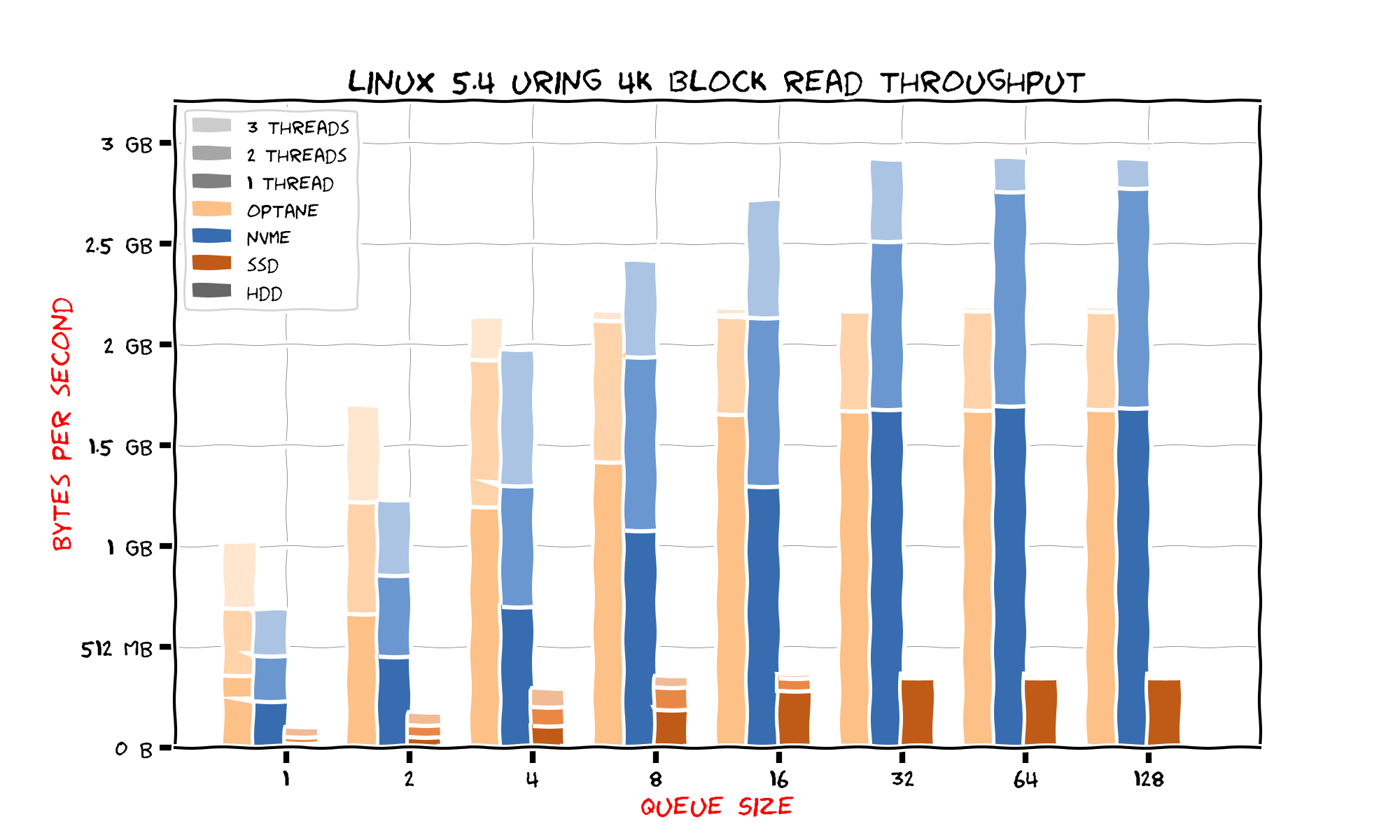}
    \caption{Uring reading throughput.}
    \label{fig:alluringqueue:throughput}
\end{figure}

\begin{figure}
    \includegraphics[width=\textwidth]{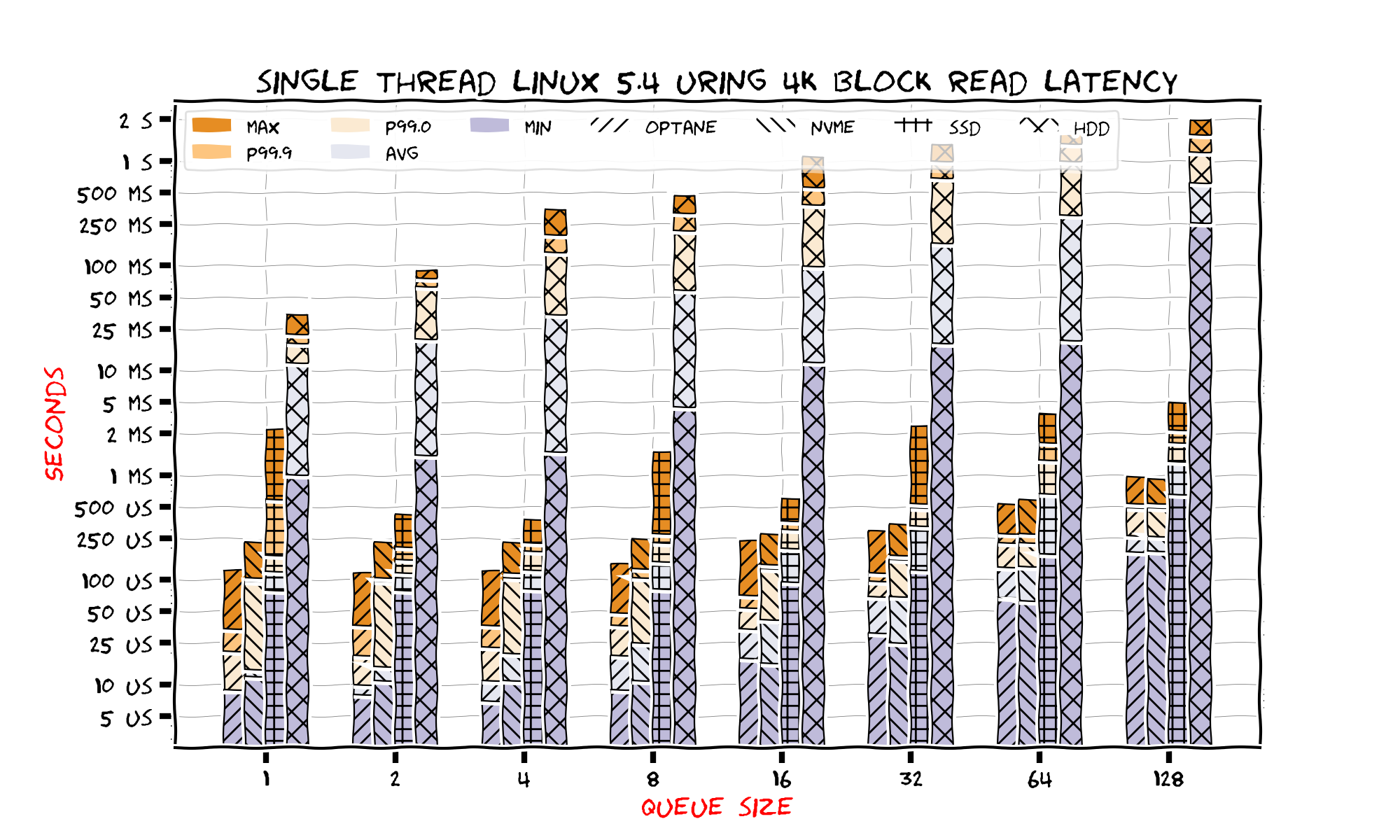}
    \caption{Single-threaded uring reading latency.}
    \label{fig:alluringthread1queue:latency}
\end{figure}

\begin{figure}
    \includegraphics[width=\textwidth]{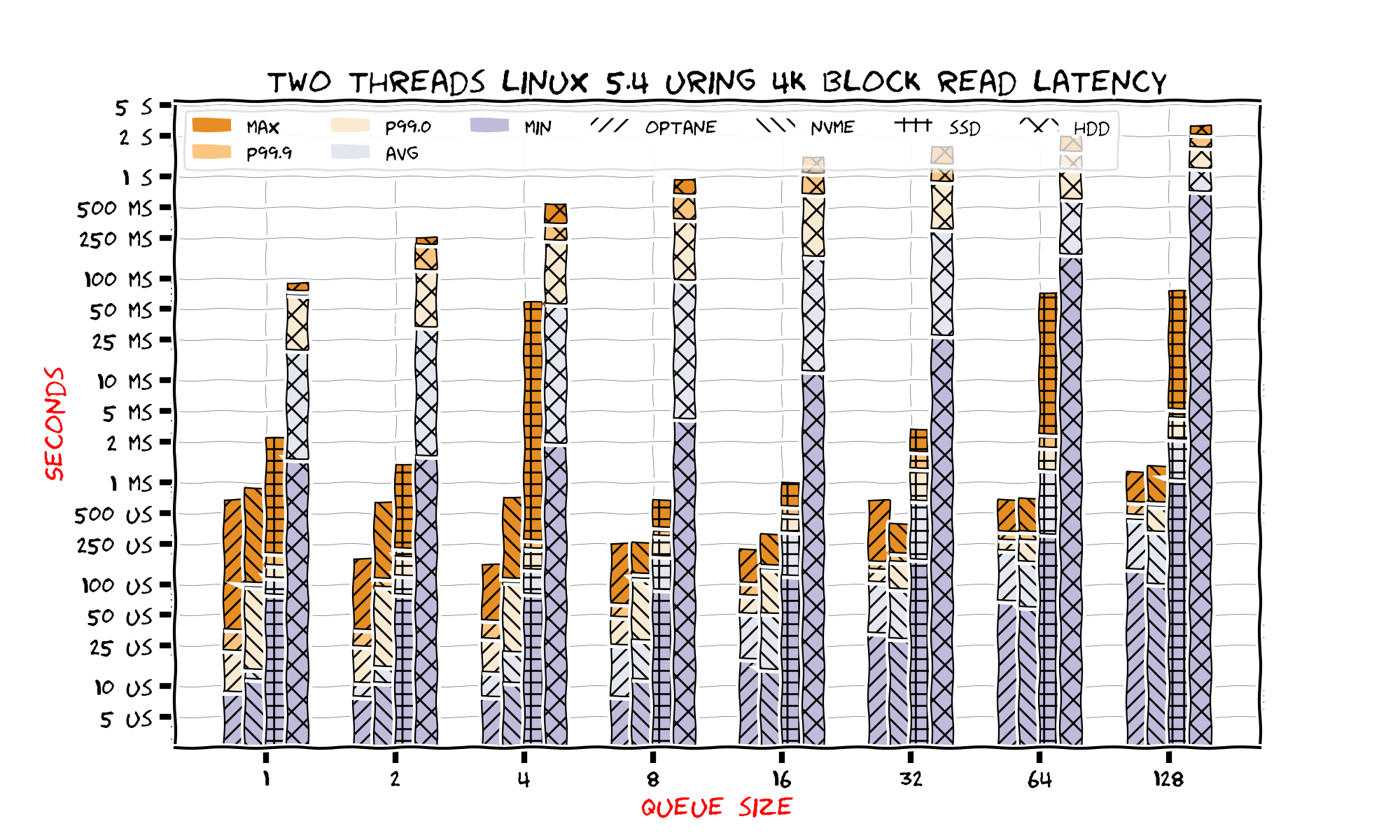}
    \caption{Two-threaded uring reading latency.}
    \label{fig:alluringthread2queue:latency}
\end{figure}

\begin{figure}
    \includegraphics[width=\textwidth]{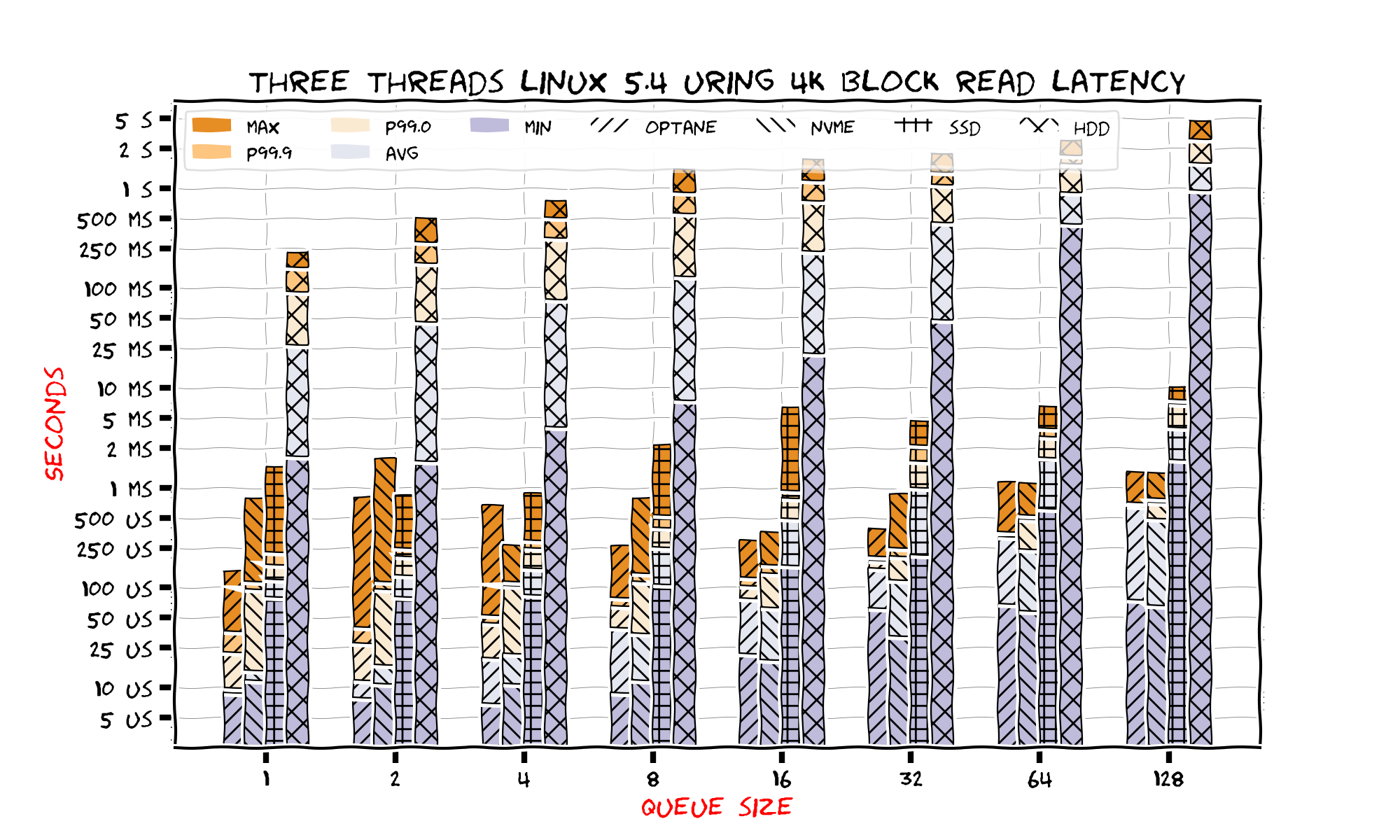}
    \caption{Three-threaded uring reading latency.}
    \label{fig:alluringthread3queue:latency}
\end{figure}

Now we start testing \texttt{uring} interface.
Similar to what we did in previous section for Linux aio we begin with choosing the optimum queue size.
Figures~\ref{fig:alluringqueue:throughput},~\ref{fig:alluringqueue:throughput},~\ref{fig:alluringthread1queue:latency},~\ref{fig:alluringthread2queue:latency}, and~\ref{fig:alluringthread3queue:latency} show observed throughput, CPU usage, and latencies respectively.
Each group of bars correspond to the same queue size and the number of elements in the queue is increased from left to right.

\begin{figure}
    \includegraphics[width=\textwidth]{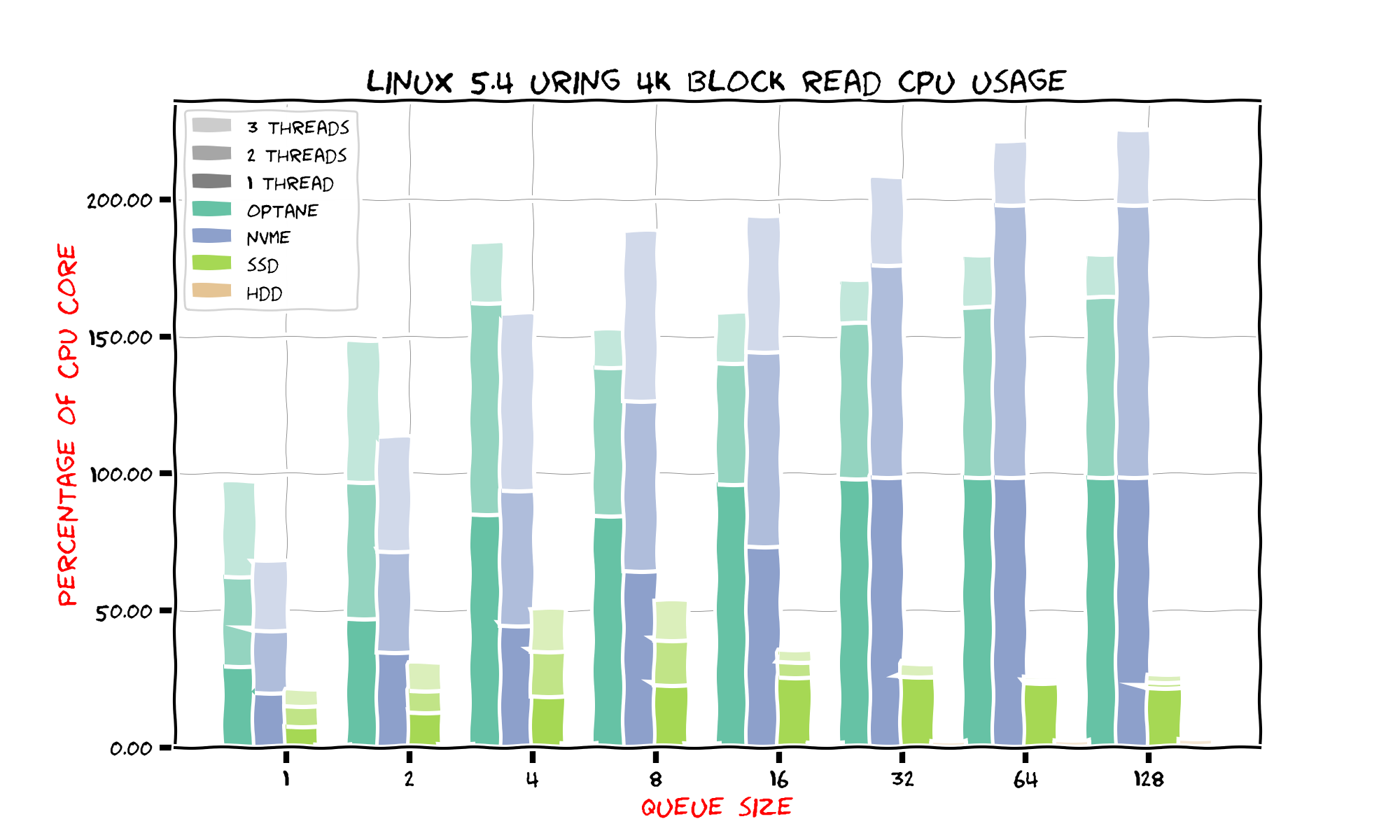}
    \caption{Uring reading CPU usage.}
    \label{fig:alluringqueue:cpu}
\end{figure}


As we can see for NVMe SSD and SSD the choice is between queue sizes 16 and 32 while for Optane best variants are 4 and 8.
As for HDD the throughput is too small to make a difference in these experiments.
Let's pick a single element queue for HDD and read from a single thread.
This should lead to lower latency.

Next we need to choose the batch size (that is the number of requests we insert into the queue each time).
This time we are interested in four queue sizes therefore there are too many bars on the figures.
To save some space we leave out HDD.

\begin{figure}
    \includegraphics[width=\textwidth]{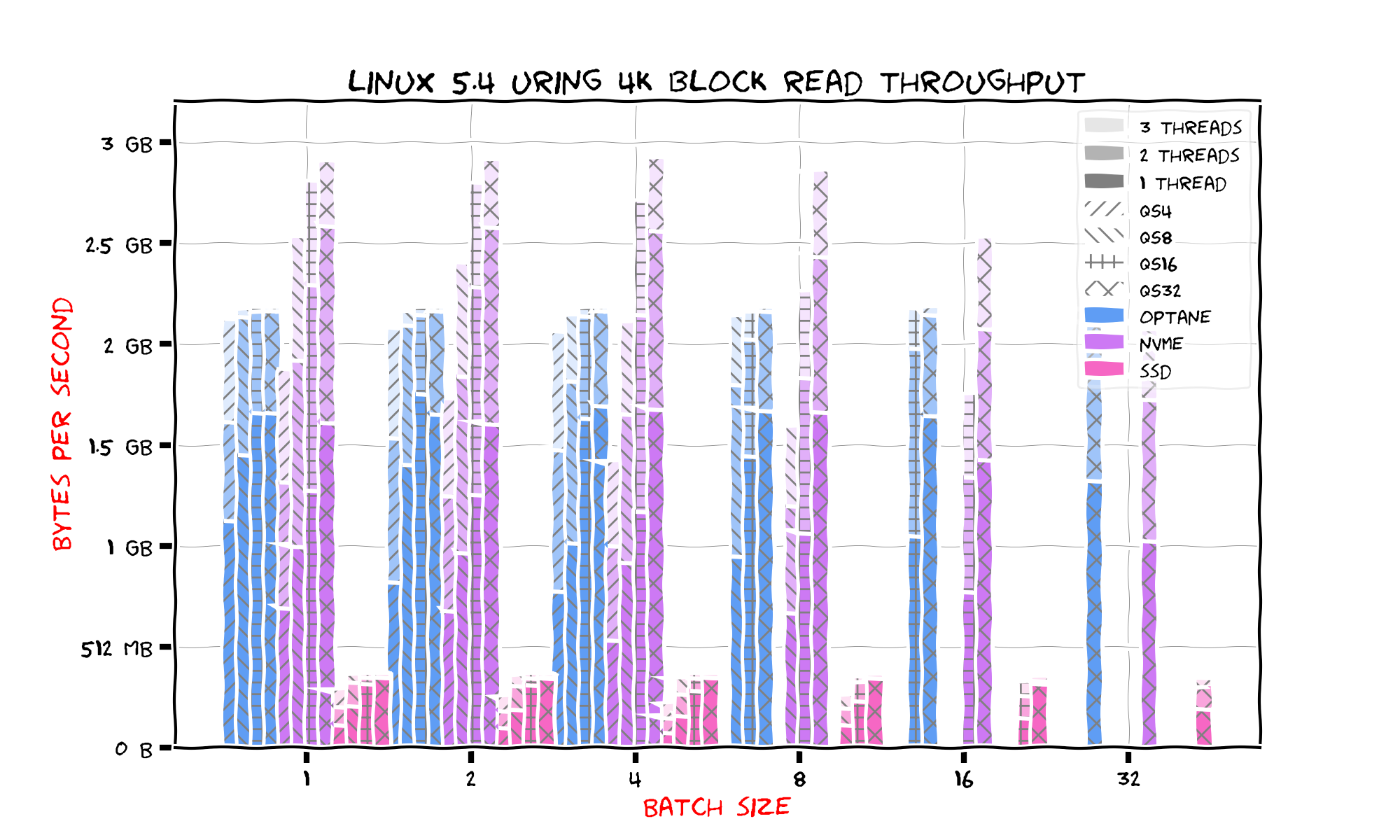}
    \caption{Uring reading throughput.}
    \label{fig:alluringbatch:throughput}
\end{figure}

\begin{figure}
    \includegraphics[width=\textwidth]{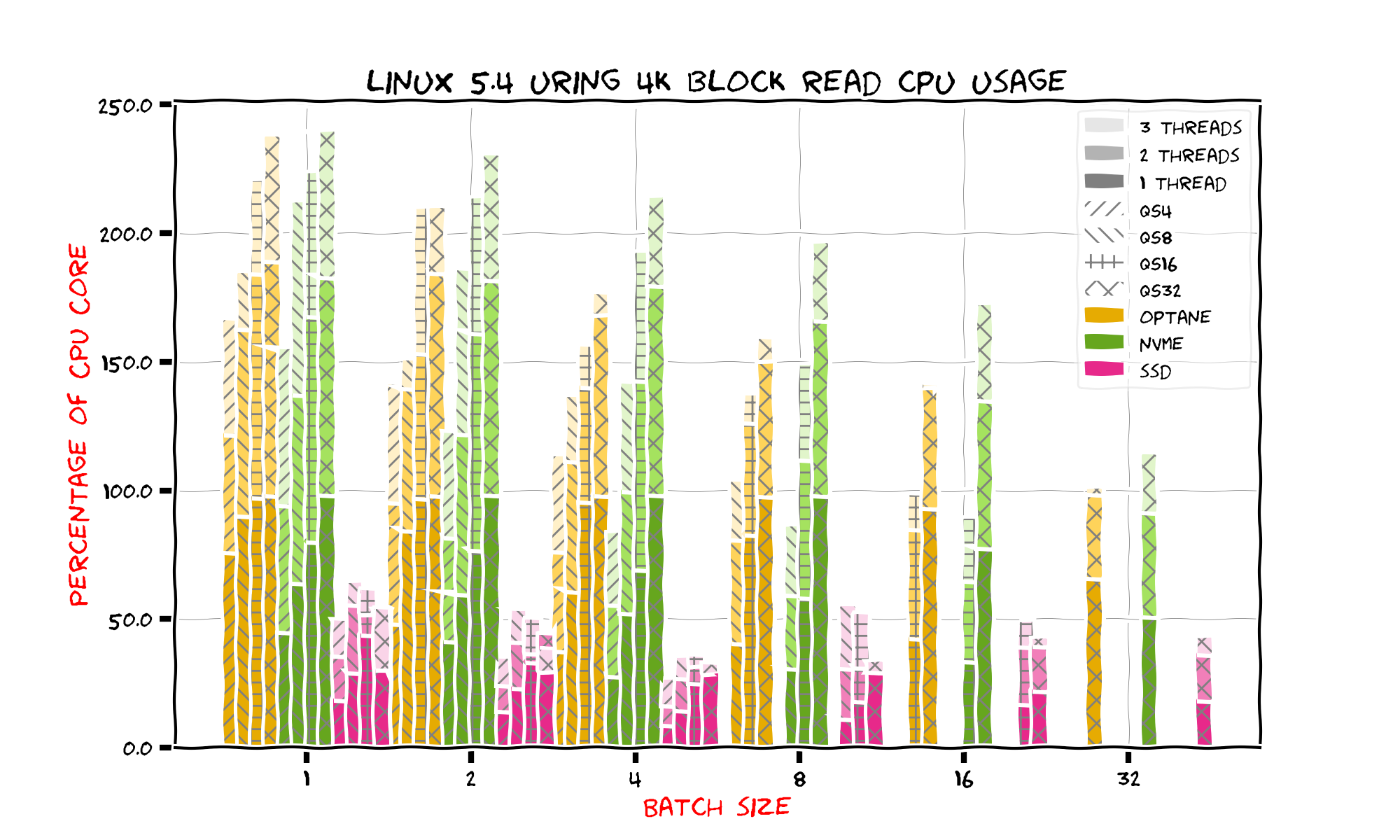}
    \caption{Uring reading CPU usage.}
    \label{fig:alluringbatch:cpu}
\end{figure}

\begin{figure}
    \includegraphics[width=\textwidth]{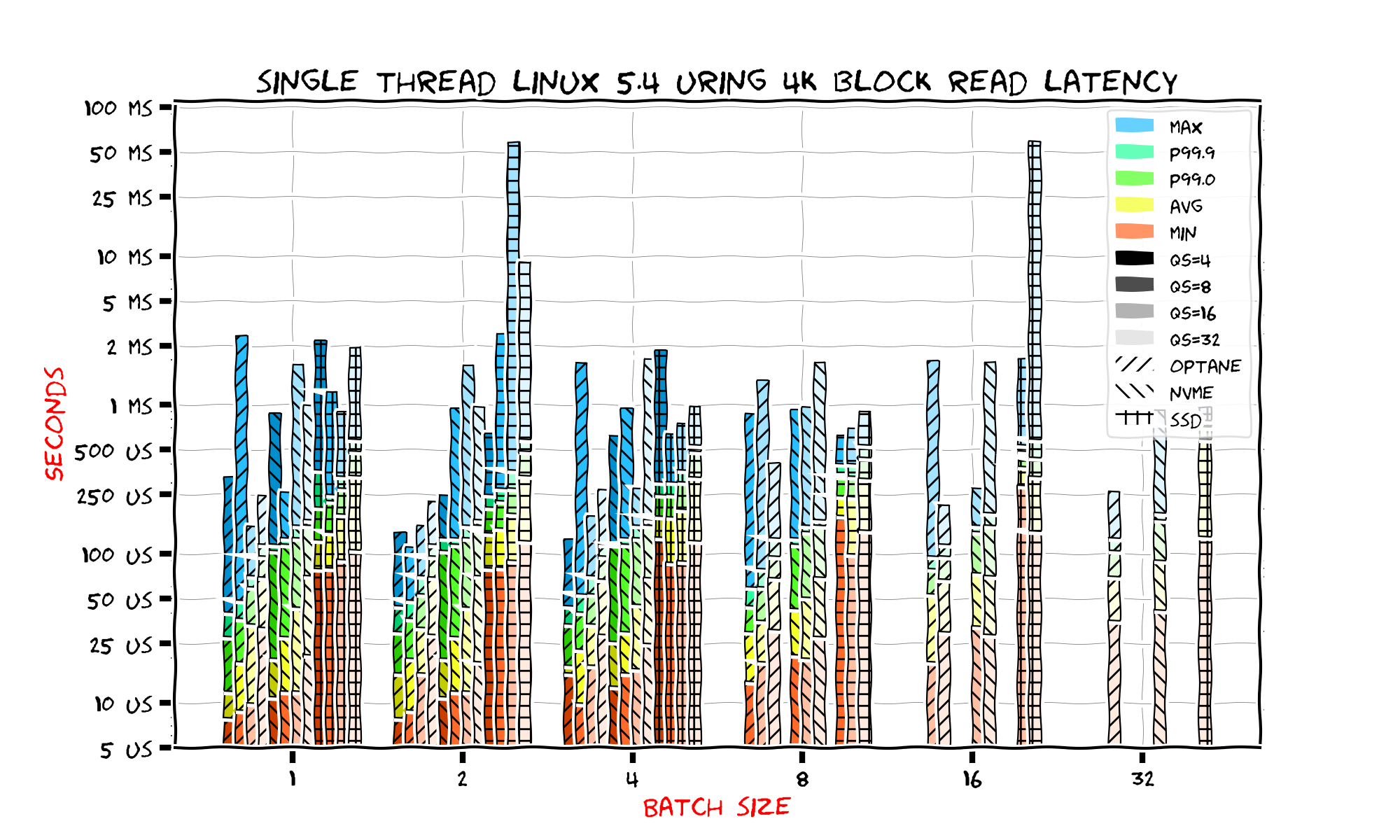}
    \caption{Single-threaded uring reading latency.}
    \label{fig:alluringthread1batch:latency}
\end{figure}

\begin{figure}
    \includegraphics[width=\textwidth]{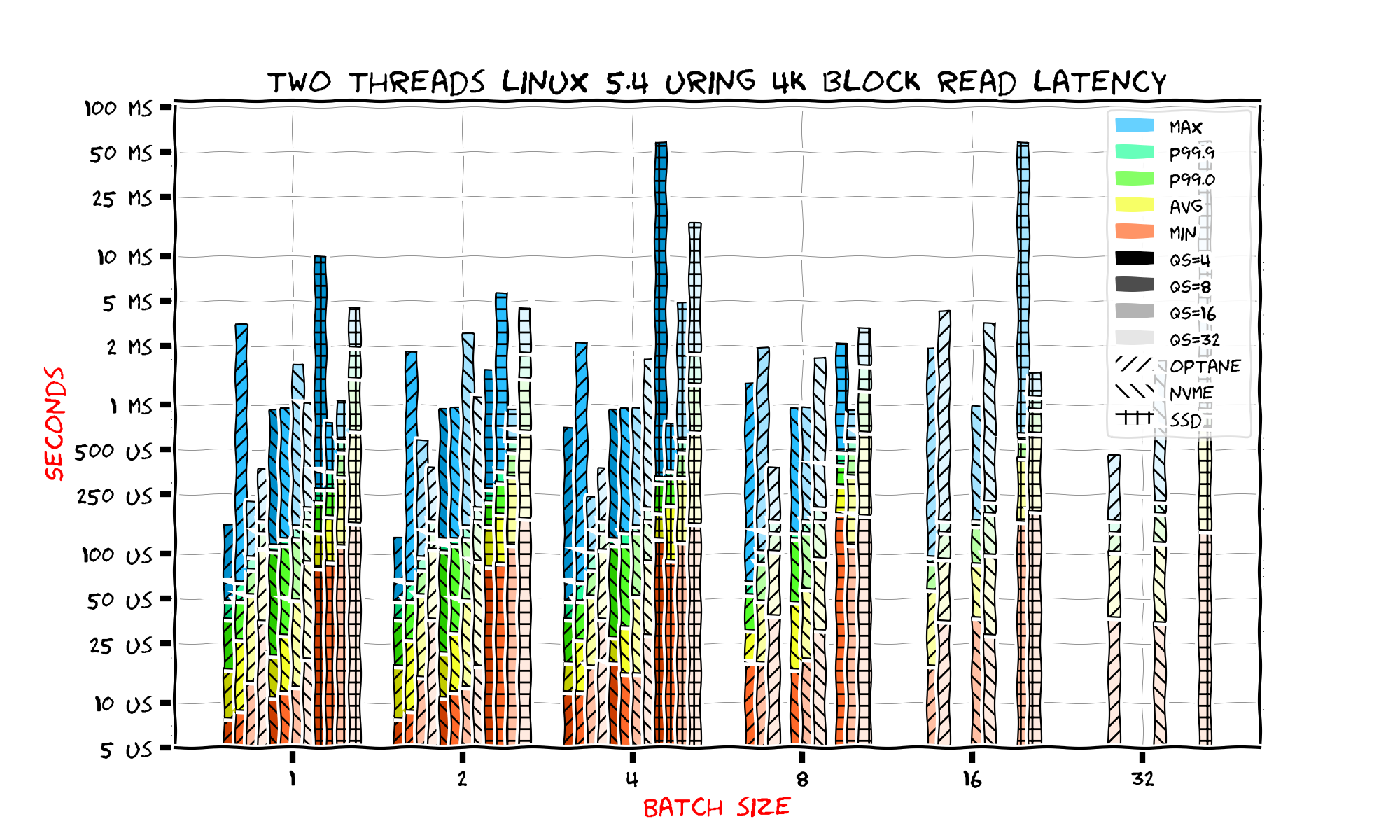}
    \caption{Two-threaded uring reading latency.}
    \label{fig:alluringthread2batch:latency}
\end{figure}

\begin{figure}
    \includegraphics[width=\textwidth]{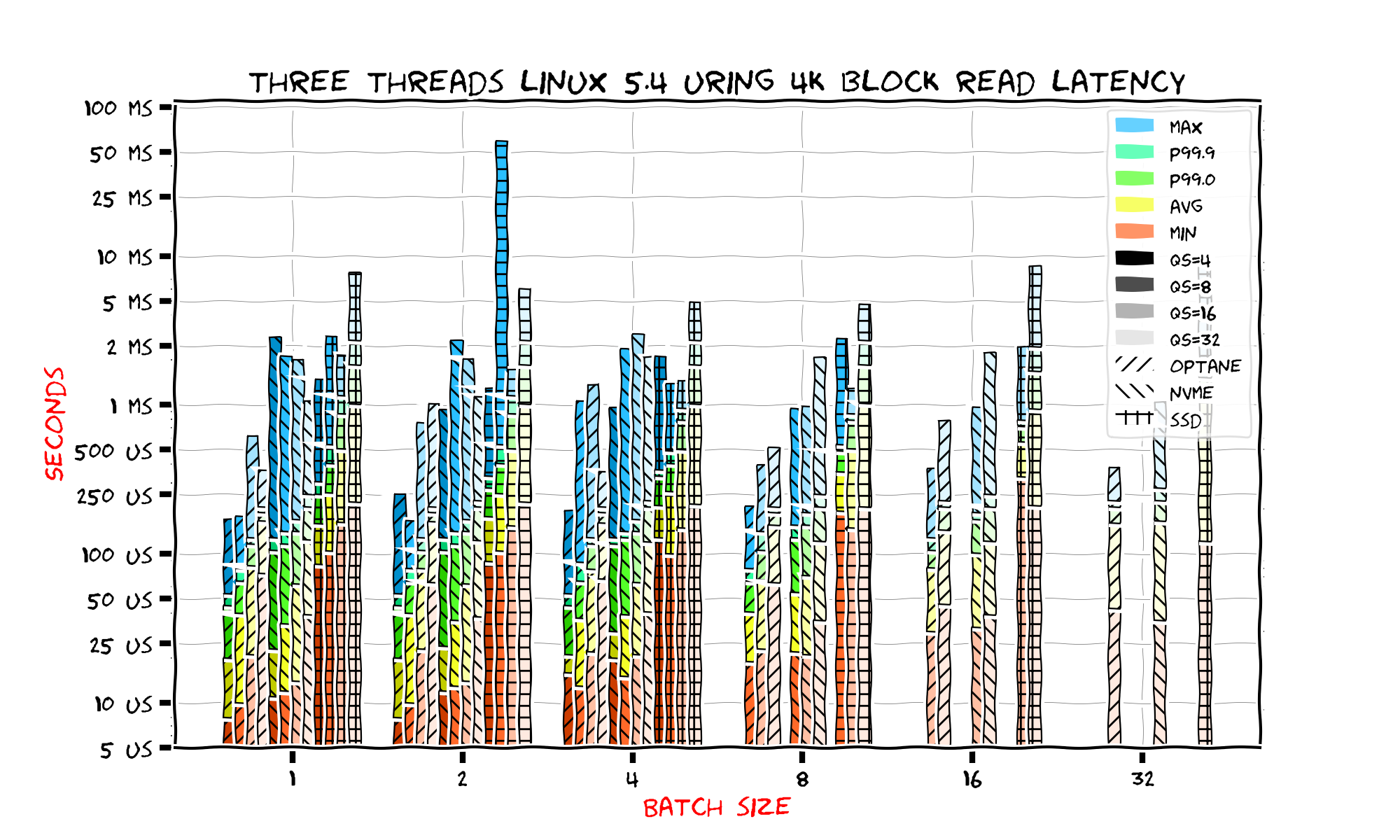}
    \caption{Three-threaded uring reading latency.}
    \label{fig:alluringthread3batch:latency}
\end{figure}

Figures~\ref{fig:alluringbatch:throughput},~\ref{fig:alluringbatch:cpu},~\ref{fig:alluringthread1batch:latency},~\ref{fig:alluringthread2batch:latency}, and~\ref{fig:alluringthread3batch:latency} show the results.


For Optane we should look at queue sizes 4 and 8.
Queue of 4 elements demonstrates better latency: 50 microseconds even when reading from three threads.
Throughput is just a few tens of megabytes per second lower than the optimum.
It is rather hard to pick the batch size.
We choose batch size 2 to save CPU usage by 25\% of a single core.

When reading from NVMe SSD we choose between queues of 16 and 32 elements.
We need to use three threads to achieve the desired throughput therefore let's look at figure~\ref{fig:alluringthread3batch:latency}.
For both queue sizes the 99.9 percentile is between 100 and 200 microseconds but for 16 elements queue the latency is lower.
Although we have to sacrifice some throughput.
As for the batch size we pick 2 also by looking at the CPU usage even though the difference is slightly noticeable.
If one would like to maximize throughput they should pick queue size 32 and batch size 4.
This results in 2.9 gigabytes per second but the price is nearly doubled latency.

As for the SSD it seems that we need to cope with the fact that all variants which lead to maximum throughput exhibit latency of 500 microseconds and higher.
If we look at the CPU usage the winner is the queue size 32 and batch size 8.

As stated earlier, for HDD we pick queue size 1 just because this makes sense.
Our selected parameters are presented in the table below.

\begin{center}
\begin{tabular}{|l|l|l|l|}
\hline
Storage & Threads & Queue Size & Batch size \\ \hline
Optane  & 3       & 4          & 2 \\ \hline
NVMe    & 3       & 16         & 2 \\ \hline
SSD     & 1       & 32         & 8 \\ \hline
HDD     & 1       & 1          & 1 \\ \hline
\end{tabular}
\end{center}

\subsection{Executing with selected parameters}

%%%%

\begin{figure}
    \includegraphics[width=\textwidth]{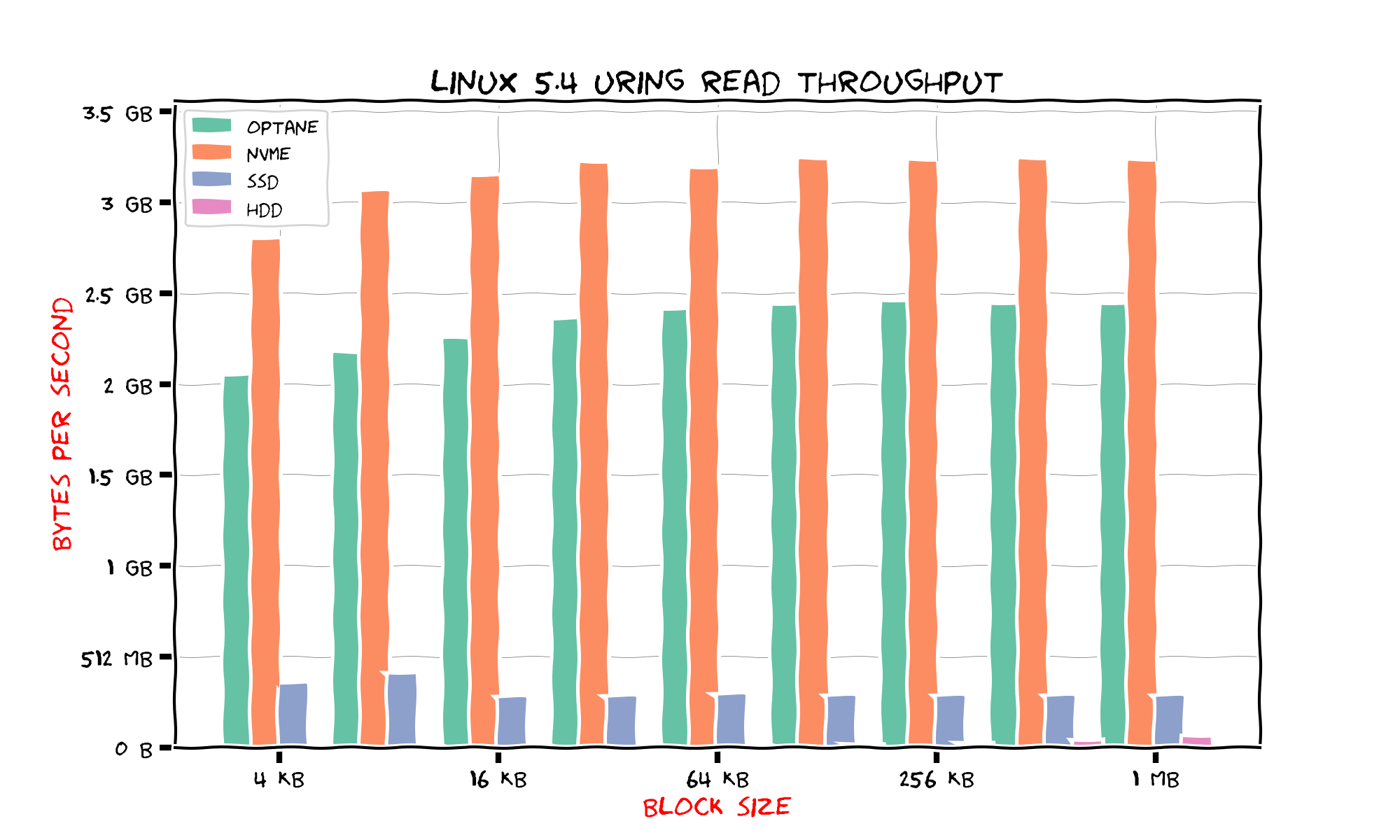}
    \caption{Best uring reading throughput.}
    \label{fig:alluringbest:throughput}
\end{figure}

\begin{figure}
    \includegraphics[width=\textwidth]{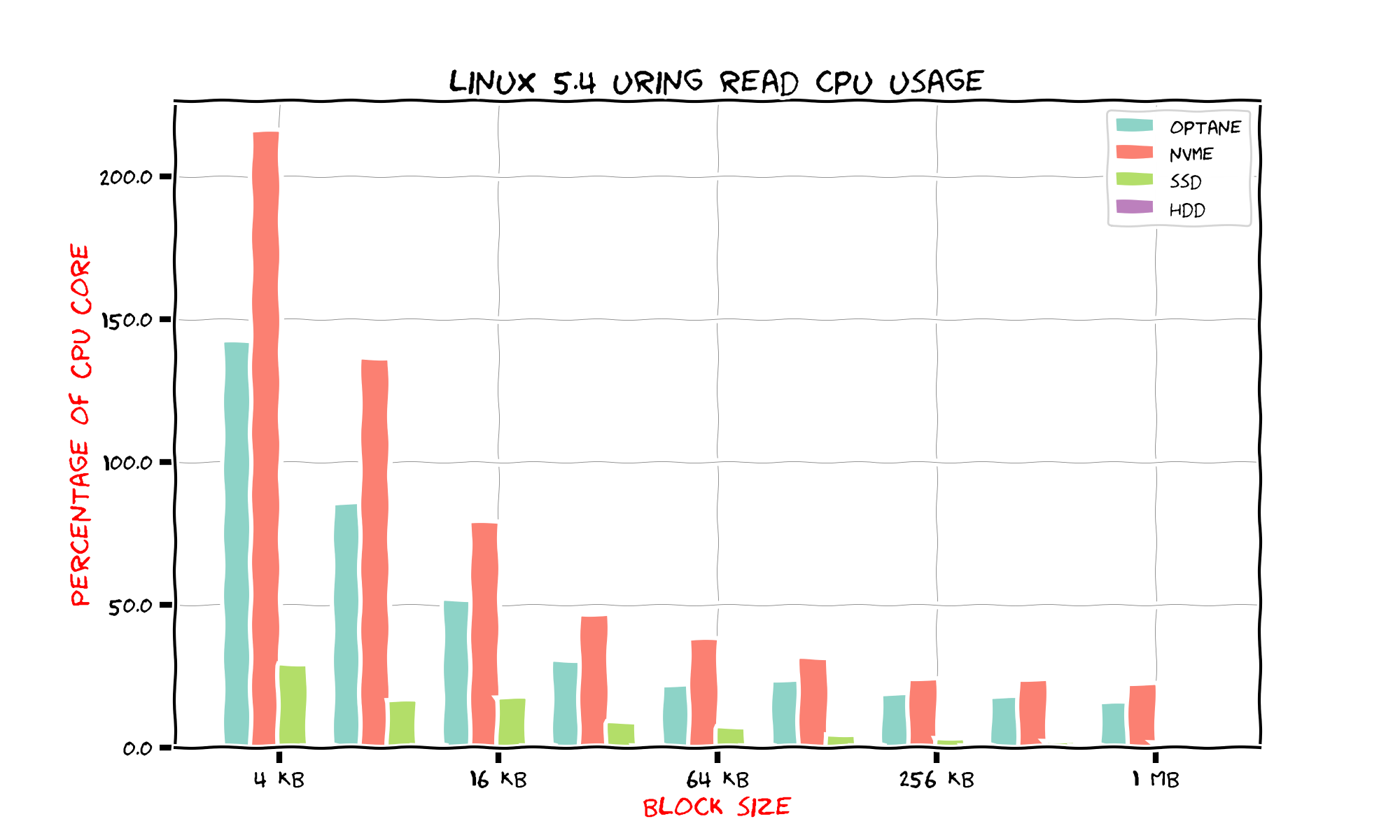}
    \caption{Best uring reading CPU usage.}
    \label{fig:alluringbest:cpu}
\end{figure}

\begin{figure}
    \includegraphics[width=\textwidth]{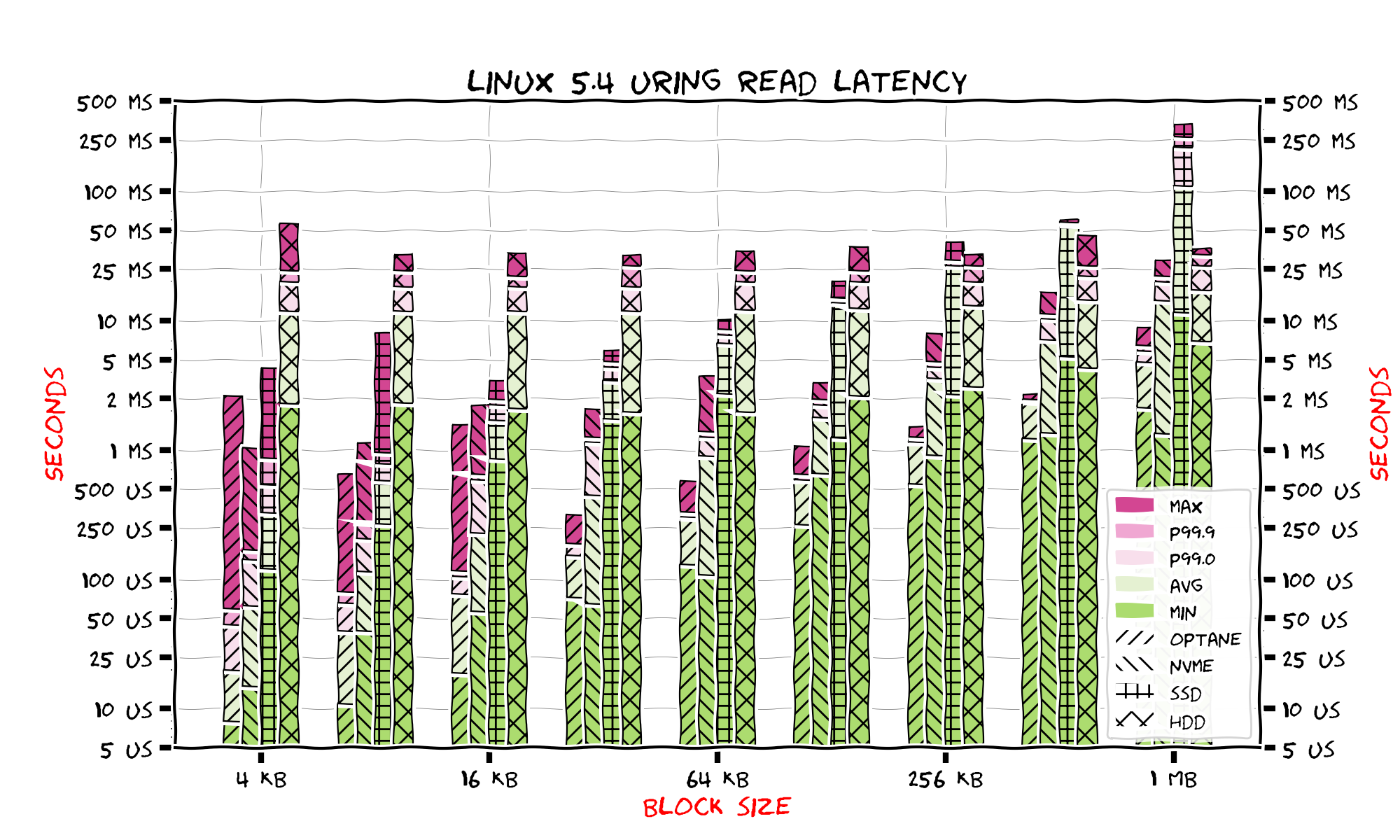}
    \caption{Best uring reading latency.}
    \label{fig:alluringbest:latency}
\end{figure}

Here we present results of our experiments with parameters selected in the previous subsection.
Figures~\ref{fig:alluringbest:throughput},~\ref{fig:alluringbest:cpu}, and~\ref{fig:alluringbest:latency} show throughput, CPU usage and latency respectively.

It seems that in terms of throughput there is almost no difference.
A notable exception is Optane for which the result is worse for 4 kilobytes blocks.
However this drop in throughput is compensated by latency reduction.
It could be that HDD performance is also worse but recall that we selected rather unusual setting on purpose.
With a single element queue \texttt{uring} works as synchronous interface.

For small block sizes CPU usage is lower by 10\% compared to Linux aio.
This is observed for all storage types.

Also for small blocks we observe latency improvement for all solid state devices.
Optane latency finally reaches 50 microseconds in 99.9 percentile for 4 kilobytes blocks.
Recall that the last time we observed such latency was when we used synchronous interface from multiple threads.
However that time it required CPU power of more than three full cores.
Now we need even lower than one and a half of a core.
So the improvement is by a factor of two.

Reading from NVMe SSD latency is now closer to 100 microseconds than with Linux aio interface but in general the difference is not really noticeable on our logarithmic scale charts.
As for SSD it seems that there is no change at all.
Both \texttt{uring} and Linux aio the latency is slightly above 500 microseconds in 99.9 percentile when reading with small block sizes.
We won't talk about HDD latency here however it's worth to note that since the latency is not so embarrassingly high latencies for other drives are represented more clearly on the chart.

\subsection{Summary}

We looked at \texttt{uring} asynchronous interface and compared it to Linux aio.
At last we were able to repeat success of synchronous interface: 50 microsecond latency in 99.9 percentile when reading from Optane with 4 kilobytes blocks.
NVMe SSD latency is also better now.
NVMe SSD throughput is the same while Optane throughput is slightly worse for \texttt{uring}.
We made this sacrifice to gain some improvement in latency.

There are many parameters in \texttt{uring} interface.
One could fix files and buffers and even utilize a special kernel thread to process requests.
We didn't try these features here and leave this for the next section.

\section{Tuning \texttt{uring}}
\label{sec:uringtune}

In previous section we found optimal queue parameters for default \texttt{uring} mode.
There are some new features in \texttt{uring} itself namely fixed buffers and kernel file objects and even special kernel thread for processing requests. 
These tweaks have potential to improve performance.
In particular one could expect that using a kernel thread would liberate the process from issuing system calls and thus improve performance.
However in our experiments we weren't able to see any justification for this.
In this section we try to tune \texttt{uring} and show the results to purify false expectations.

\subsection{Fixed files}

If you read from the same files over and over you should attach them to the queue.
As stated in \texttt{uring} documentation when a file descriptor is passed into the kernel the corresponding file object has to be locked before performing the operation.
After the request is finished the file object is unlocked.
To avoid unnecessary locking and unlocking the file object can be in some sense attached to the queue.
The documentation calls this feature \emph{fixed files}.

\begin{figure}
    \includegraphics[width=\textwidth]{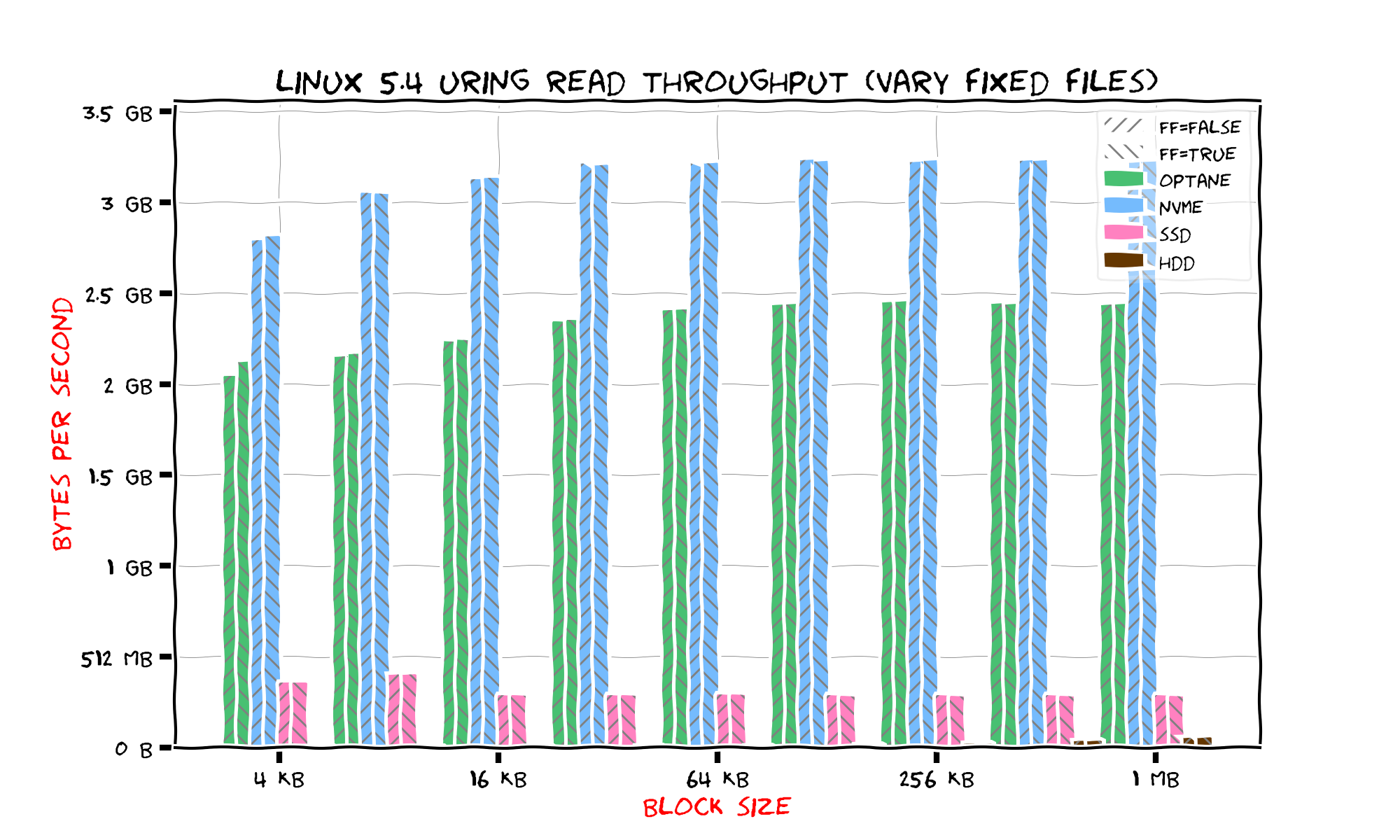}
    \caption{Uring reading with fixed files throughput.}
    \label{fig:alluringfixedfiles:throughput}
\end{figure}

\begin{figure}
    \includegraphics[width=\textwidth]{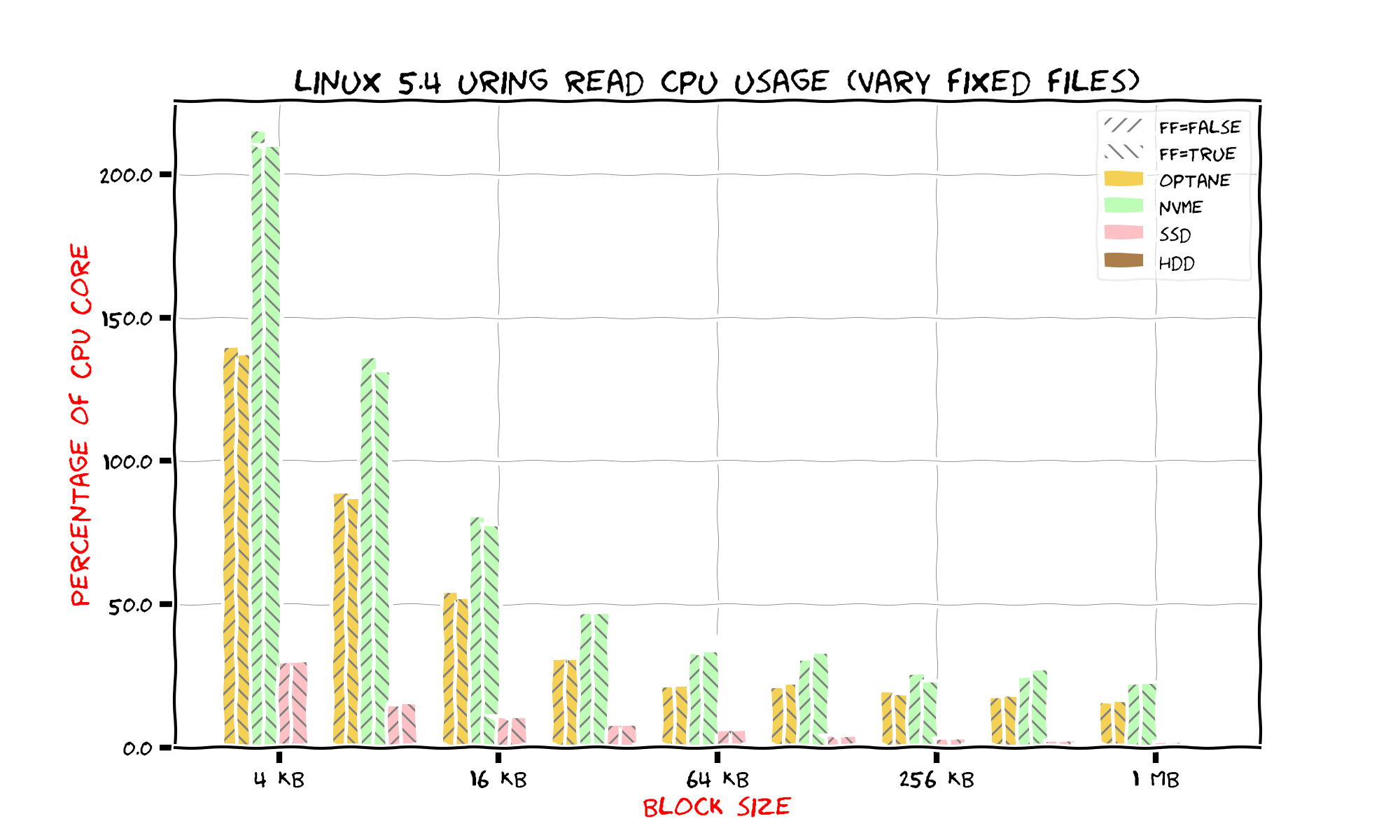}
    \caption{Uring reading with fixed files CPU usage.}
    \label{fig:alluringfixedfiles:cpu}
\end{figure}

\begin{figure}
    \includegraphics[width=\textwidth]{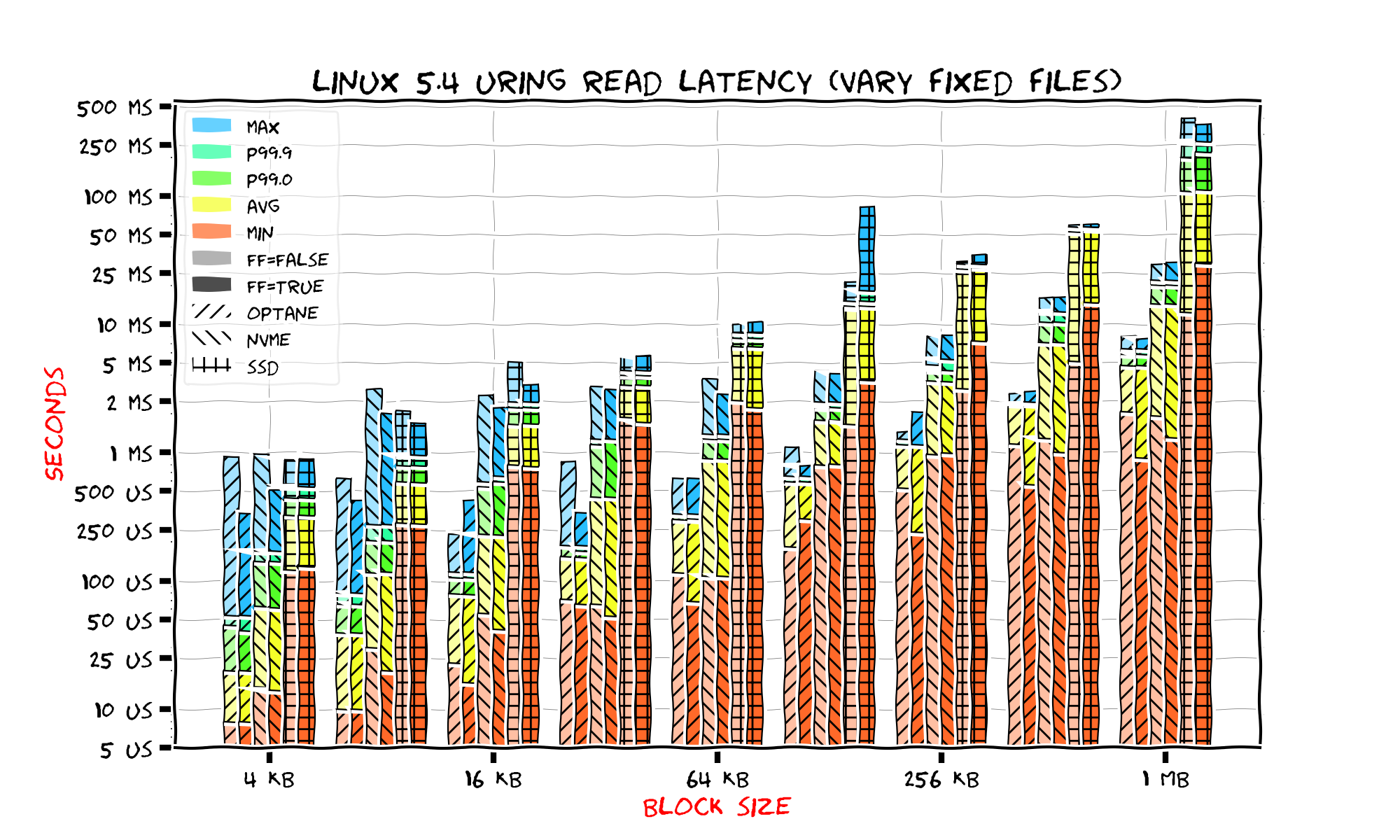}
    \caption{Uring reading with fixed files latency.}
    \label{fig:alluringfixedfiles:latency}
\end{figure}

To measure the effect we execute previously obtained optimal configuration with and without fixing.
Figures~\ref{fig:alluringfixedfiles:throughput},~\ref{fig:alluringfixedfiles:cpu}, and~\ref{fig:alluringfixedfiles:latency} show throughput, CPU usage and latency comparison respectively.
As we can see this results in slightly less CPU usage when reading from NVMe SSD and Optane with small blocks.
Also there is a huge reduction in maximum latency for Optane and NVMe SSD and 4 kilobyte blocks.
In case of Optane the drop is from nearly a millisecond to 300 microseconds.
We didn't take maximum latency into account previously but it is pleasant to see improvement anyway.
Especially when it could be achieved with such a small effort.
All other values look pretty much the same.

\subsection{Fixed buffers}

Another thing which originates from internal kernel organization is buffer registration.
As \texttt{uring} documentation says when \texttt{O\_DIRECT} is used the kernel has to map the data buffer inside the kernel address space.
After data is transferred into the memory the buffer is unmapped from the kernel address space.
This could take even more time than locking a file object.
Thus uring has a special option to fix a buffer once and reuse it afterwards without the necessity of extra memory management for each request.

\begin{figure}
    \includegraphics[width=\textwidth]{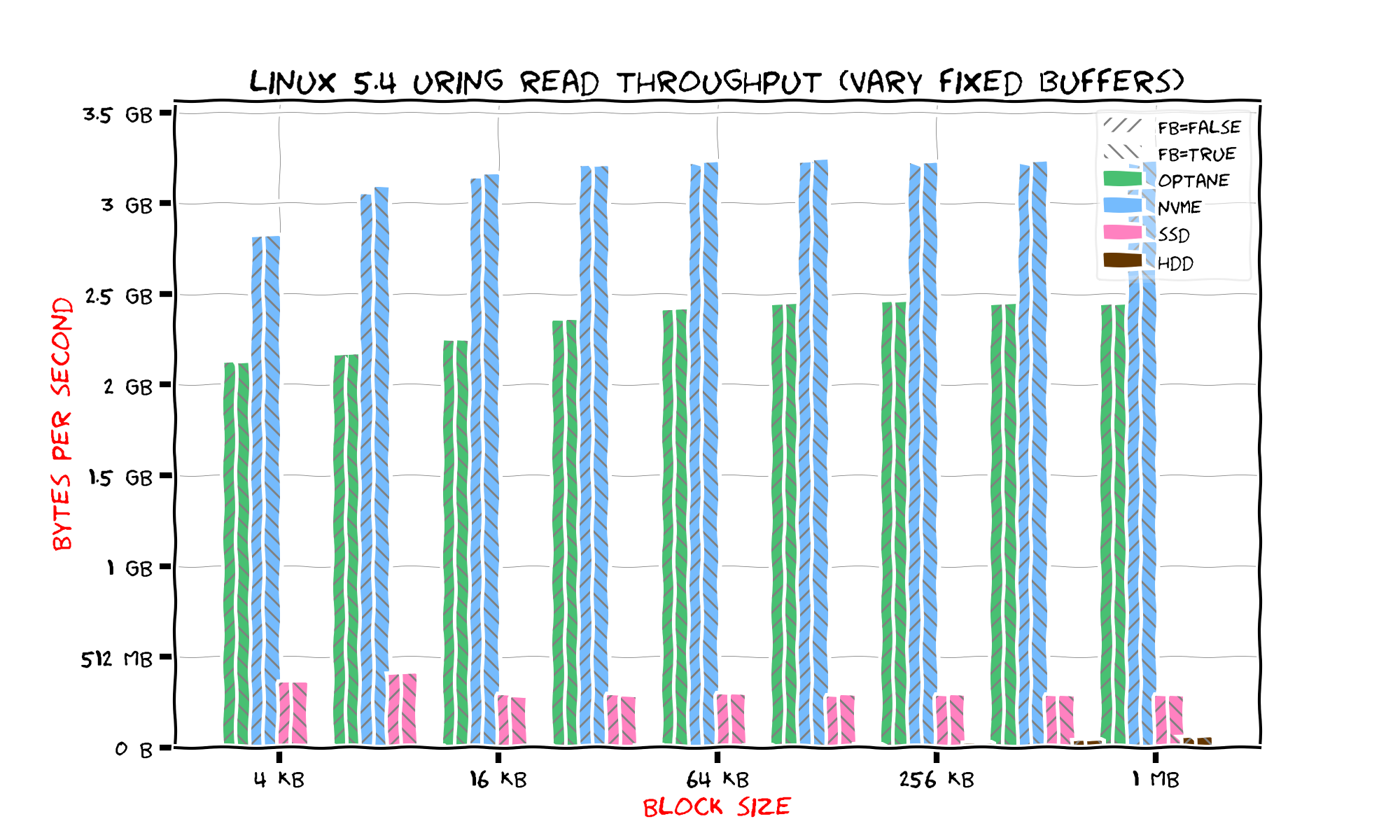}
    \caption{Uring reading with fixed buffers throughput.}
    \label{fig:alluringfixedbuffers:throughput}
\end{figure}

\begin{figure}
    \includegraphics[width=\textwidth]{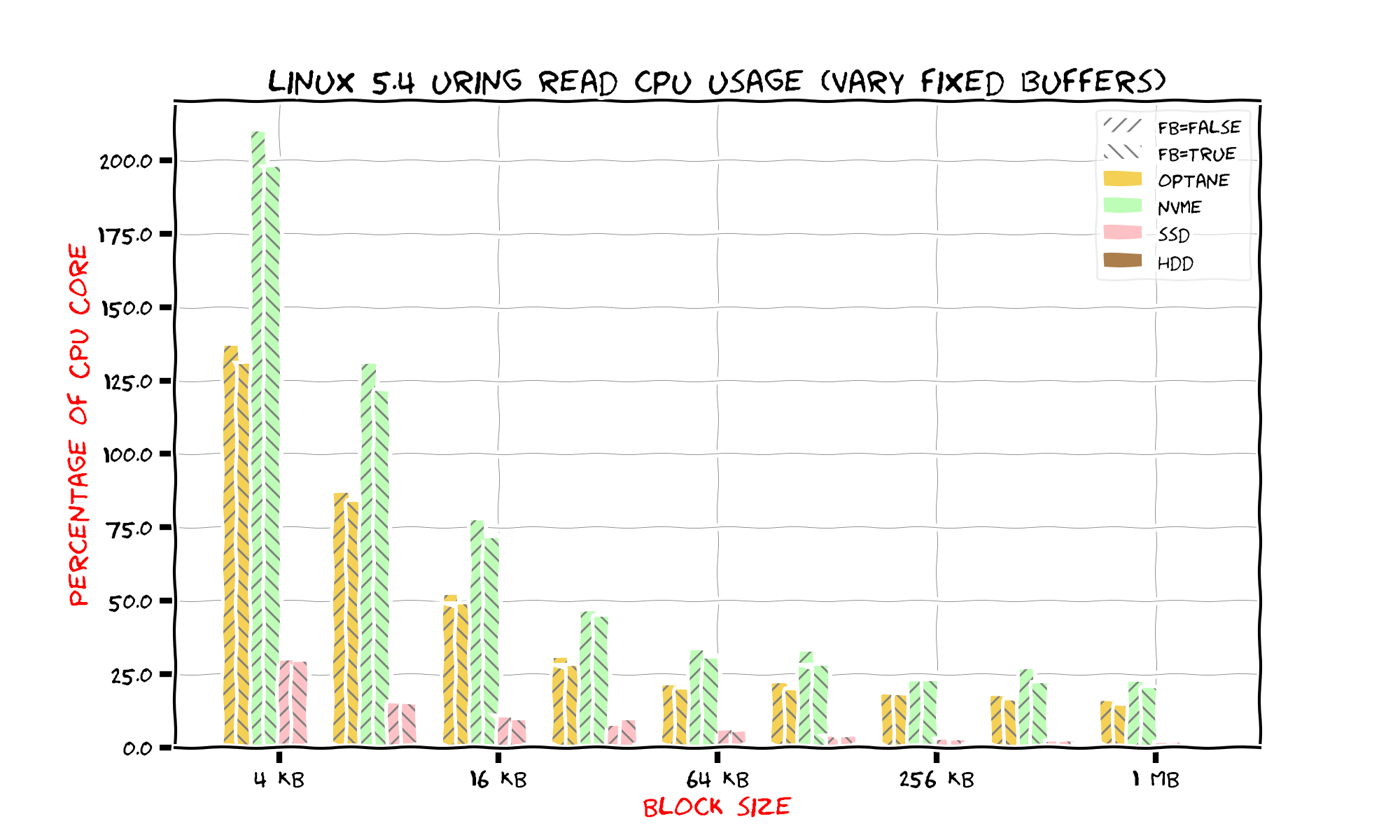}
    \caption{Uring reading with fixed buffers CPU usage.}
    \label{fig:alluringfixedbuffers:cpu}
\end{figure}

\begin{figure}
    \includegraphics[width=\textwidth]{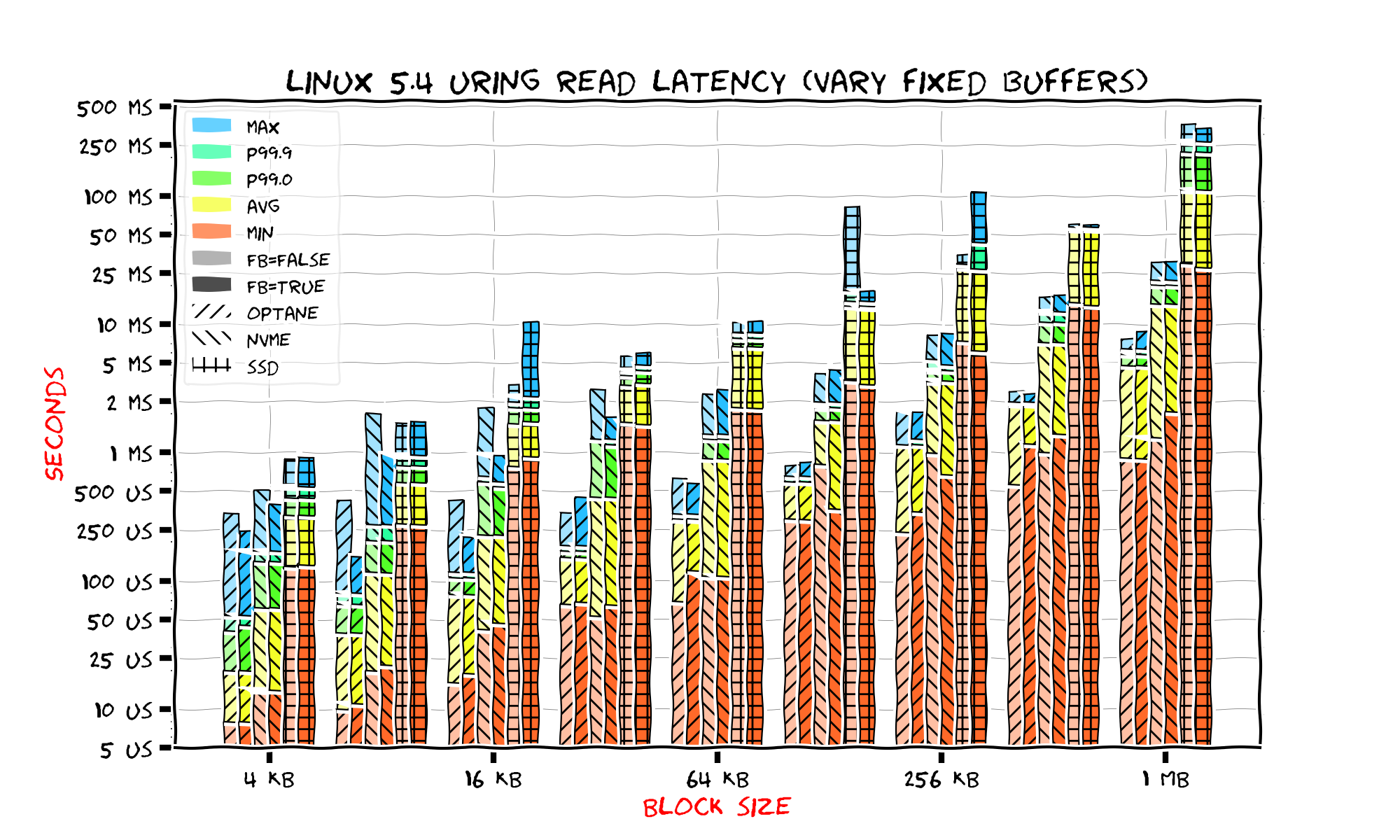}
    \caption{Uring reading with fixed buffers latency.}
    \label{fig:alluringfixedbuffers:latency}
\end{figure}

Figures~\ref{fig:alluringfixedbuffers:throughput},~\ref{fig:alluringfixedbuffers:cpu}, and~\ref{fig:alluringfixedbuffers:latency} show the throughput, CPU usage and latency comparison when buffers are fixed and when they are not.
Throughput is not changed at all for 4 kilobytes blocks and is slightly improved for 8 and 16 kilobytes blocks.
CPU usage is better for all block sizes.
Finally for small block sizes we again see maximum latency decrease.

\subsection{SQ poll thread}

Finally, probably the most interesting feature is the ability to create a special kernel thread which would read requests from the submission queue, execute them and write notification into the completion queue.
Thus a userspace program doesn't need to make any system call at all and making context switch between user and kernel modes to submit a request could be avoided.
All the application has to do is to look at completion queue in its own memory.
One could hope that this will lead to better latency.

Unfortunately the results are disappointing.
In general they are similar to simple \texttt{uring} and sometimes even worse.
Maybe there is something wrong with our experiments.
It would be nice to look at a report for more successful attempt.

Anyway we try our best and present the results.
The CPU usage measurement should be taken carefully.
Obviously kernel poll thread consumes CPU.
However since it is a special thread inside the kernel which doesn't relate to a process in common sense its resource consumption is not shown in standard usage statistics.
The kernel poll thread is neither part of a process nor its child.
In our experiments we carefully find this thread in \texttt{\/proc} using its name and gathered CPU consumption from there.
We hope that in future kernel versions this will be more straightforward.

If we execute the experiments with the same queue parameters as selected in previous section the results are remarkably bad.
In particular the latency is too large while the throughput stays the same.
It should come with no surprise: after all we changed the execution significantly and the workload is distributed differently across contexts and probably even CPU cores.
Therefore we once again find optimal queue parameters using the same method as in previous sections.
We hope that this approach makes the comparison fair enough.

%TODO: subscription block size but should be queue size, color saturation.

\begin{figure}
    \includegraphics[width=\textwidth]{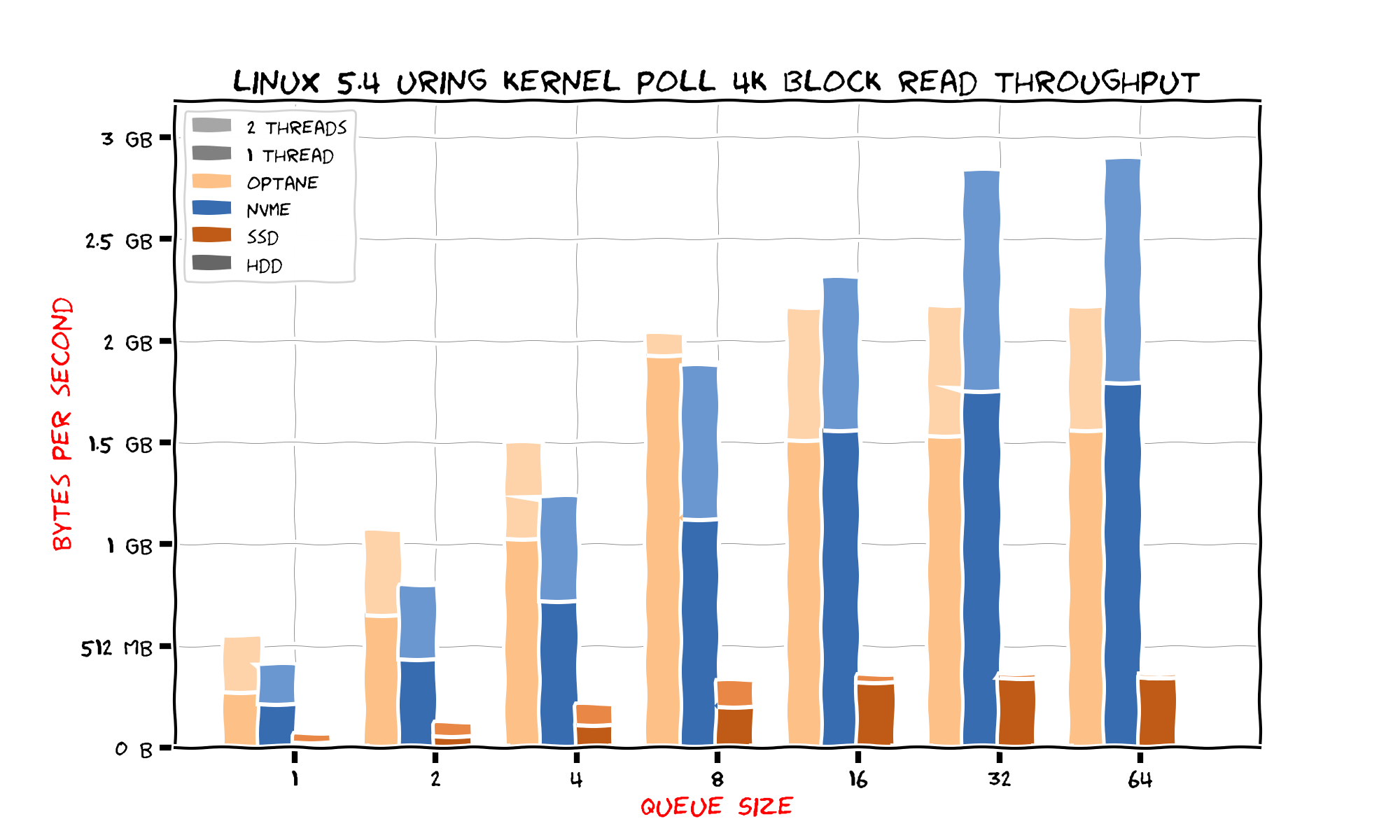}
    \caption{Uring reading with kernel poll thread throughput.}
    \label{fig:alluringpollqueue:throughput}
\end{figure}

\begin{figure}
    \includegraphics[width=\textwidth]{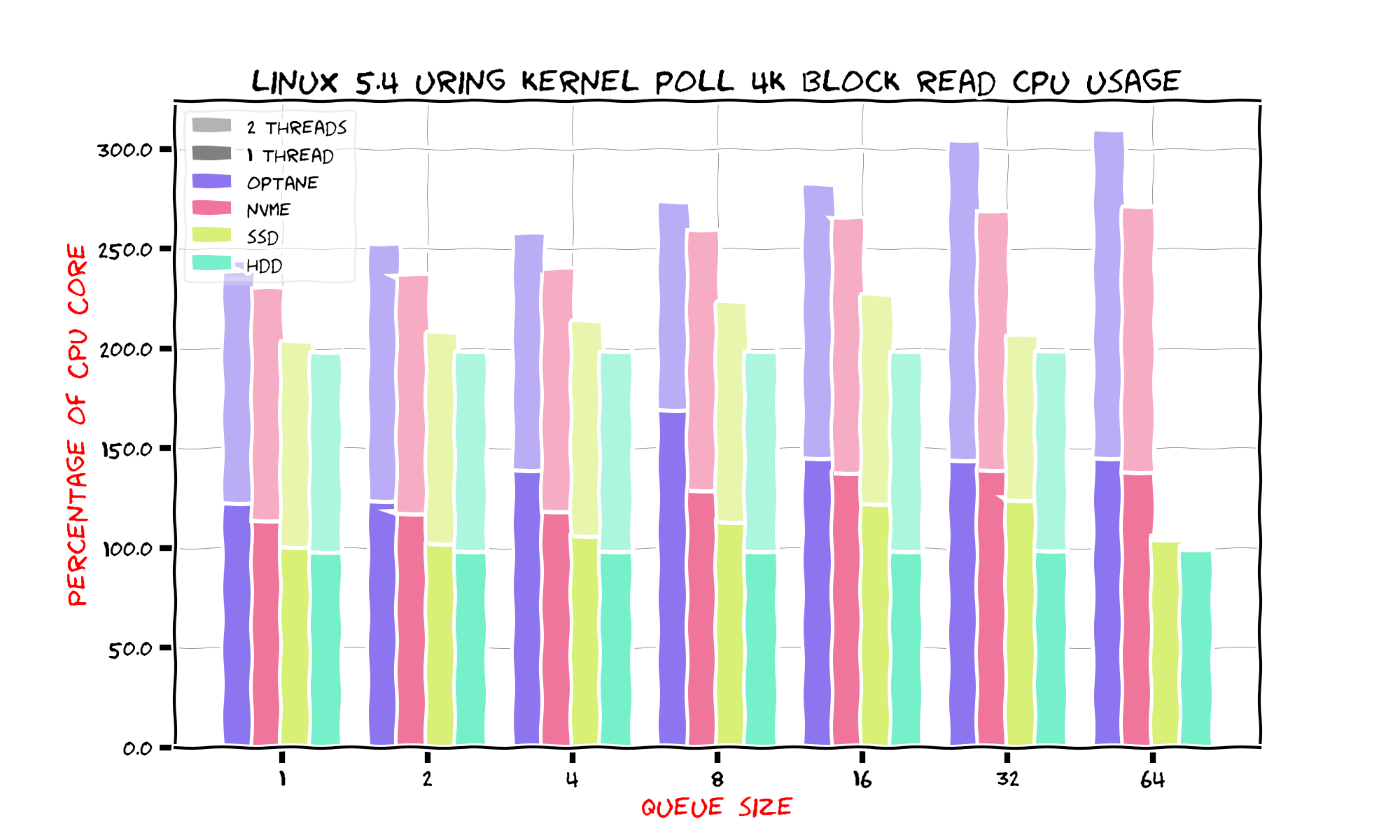}
    \caption{Uring reading with kernel poll thread CPU usage.}
    \label{fig:alluringpollqueue:cpu}
\end{figure}

\begin{figure}
    \includegraphics[width=\textwidth]{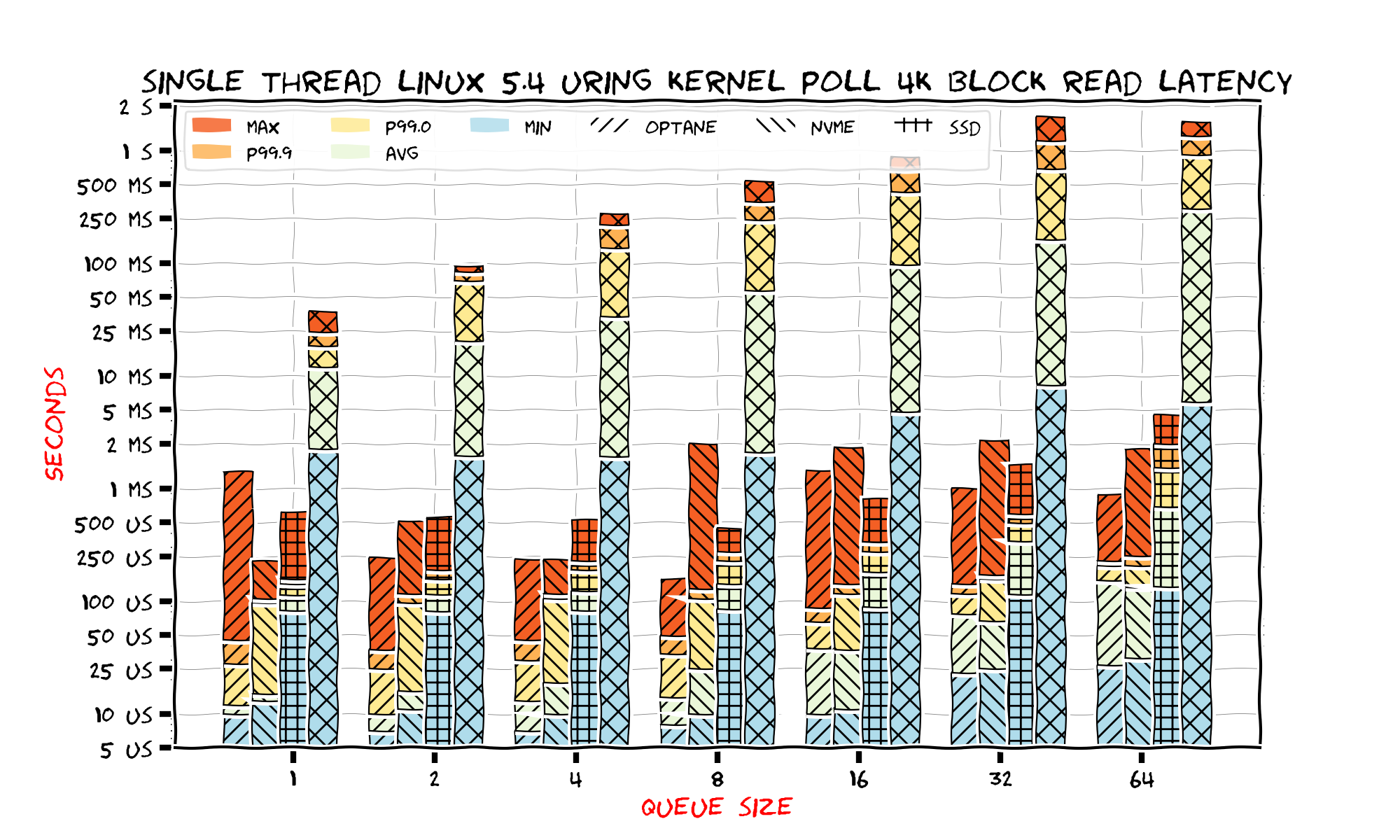}
    \caption{Single-threaded uring reading with kernel poll thread latency.}
    \label{fig:alluringpollthread1queue:latency}
\end{figure}
\begin{figure}
    \includegraphics[width=\textwidth]{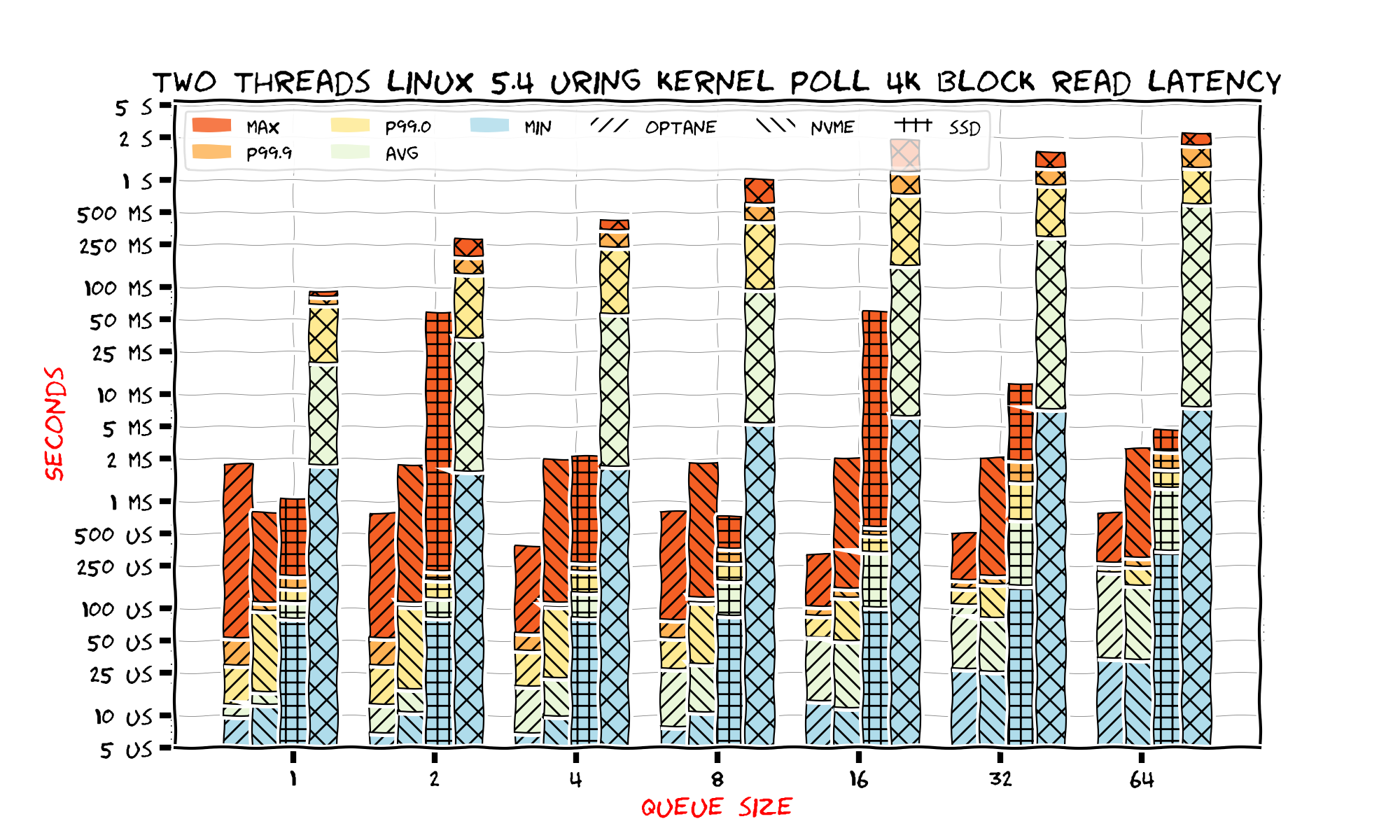}
    \caption{Two-threaded uring reading with kernel poll thread latency.}
    \label{fig:alluringpollthread2queue:latency}
\end{figure}

%%%%%%

\begin{figure}
    \includegraphics[width=\textwidth]{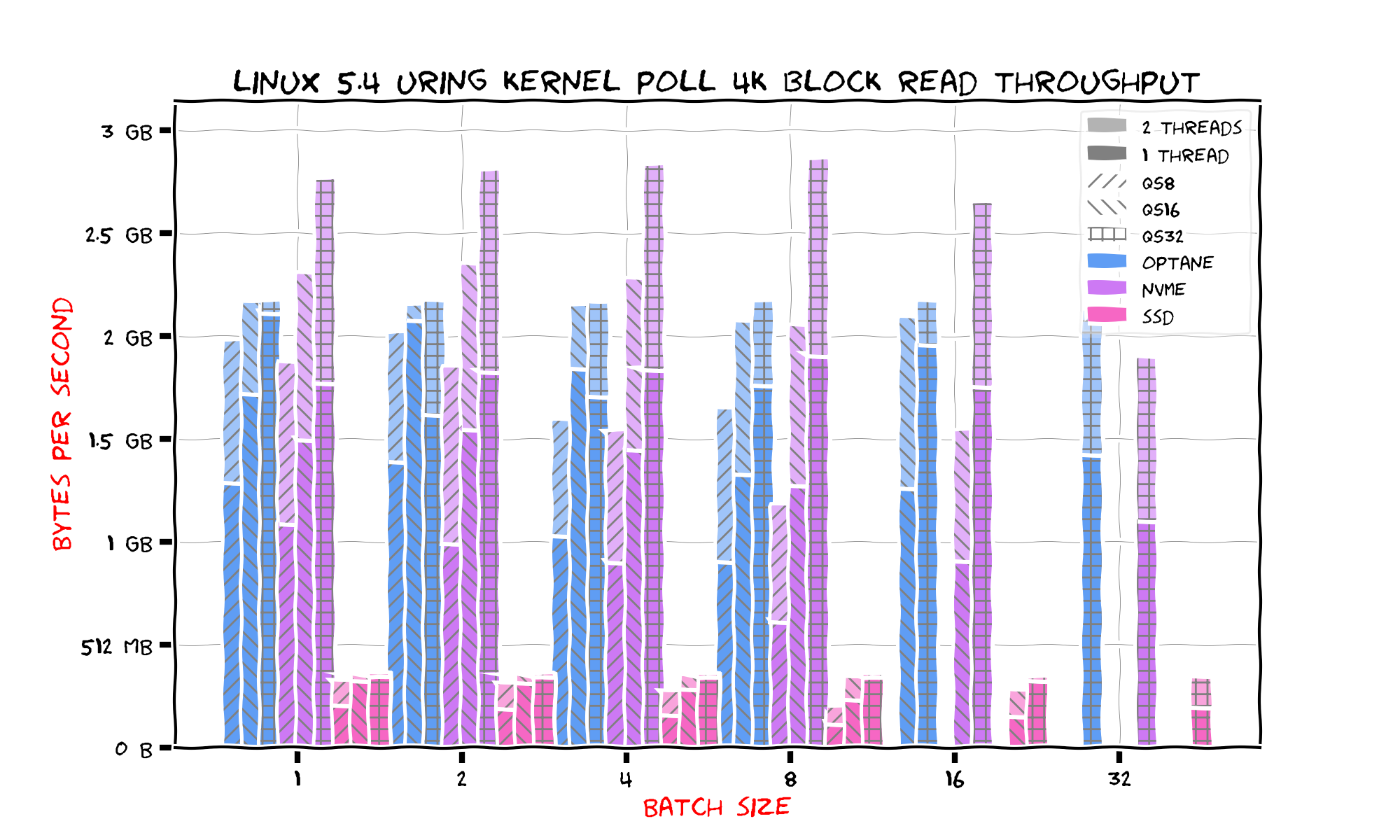}
    \caption{Uring reading with kernel poll thread throughput.}
    \label{fig:alluringpollbatch:throughput}
\end{figure}

\begin{figure}
    \includegraphics[width=\textwidth]{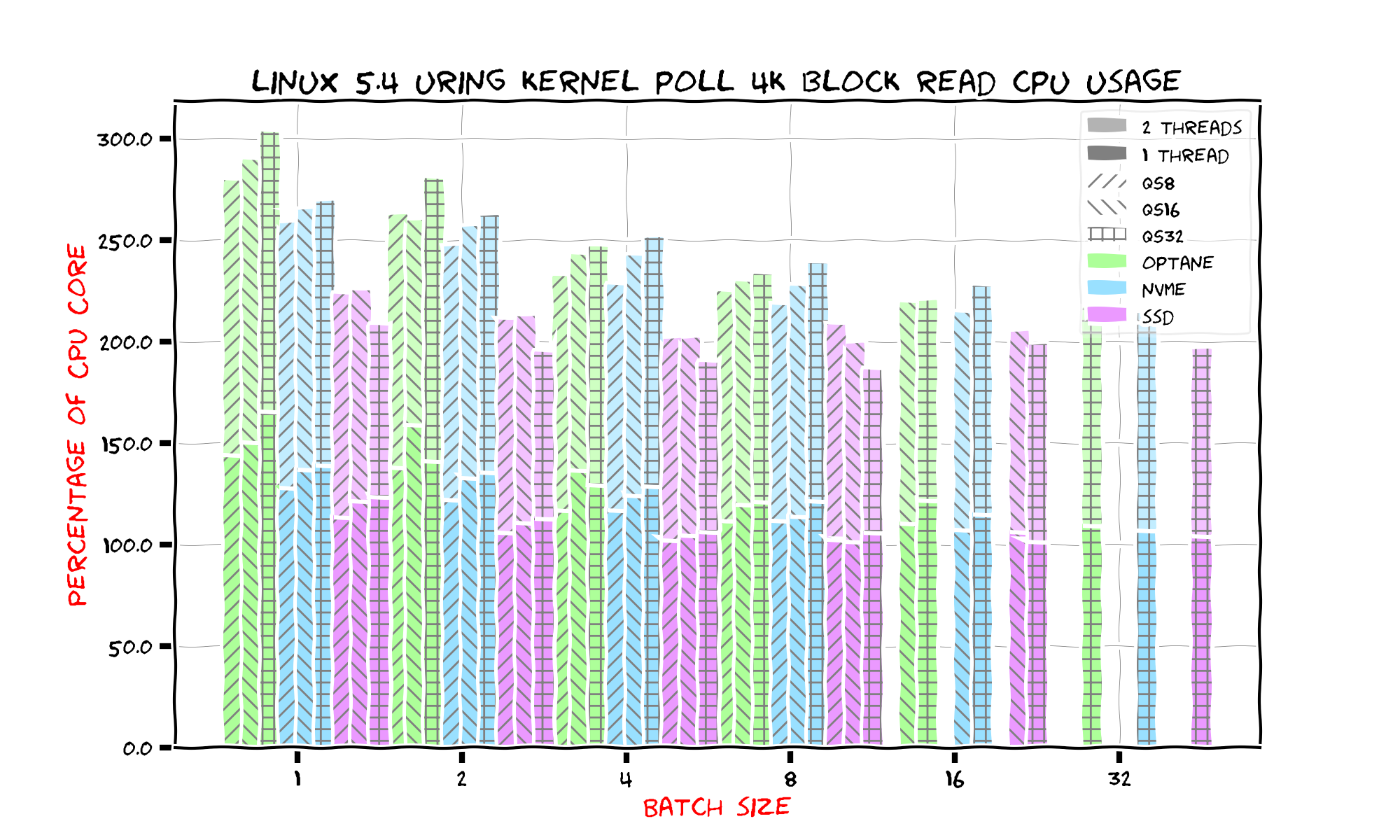}
    \caption{Uring reading with kernel poll thread CPU usage.}
    \label{fig:alluringpollbatch:cpu}
\end{figure}

\begin{figure}
    \includegraphics[width=\textwidth]{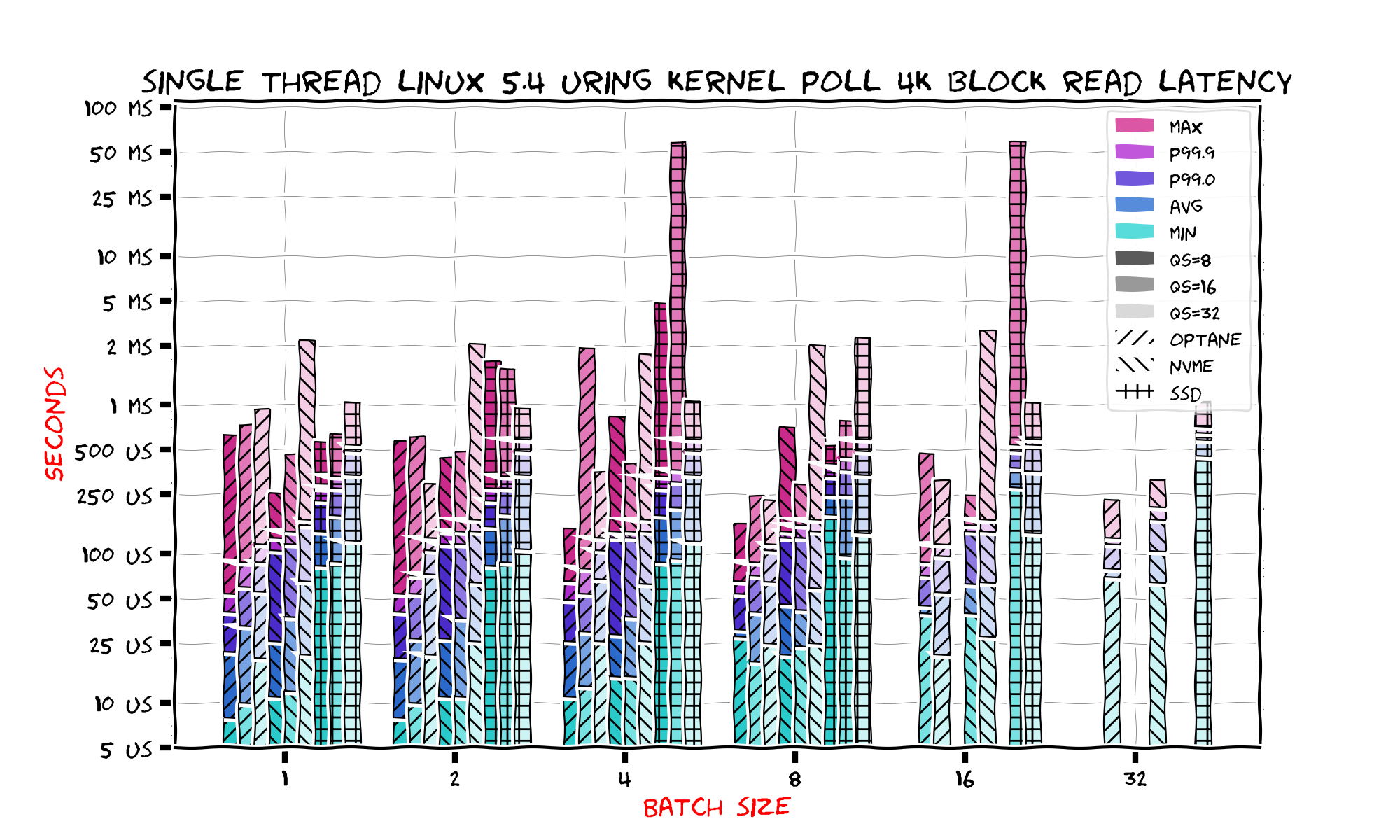}
    \caption{Single-threaded uring reading with kernel poll thread latency.}
    \label{fig:alluringpollthread1batch:latency}
\end{figure}

\begin{figure}
    \includegraphics[width=\textwidth]{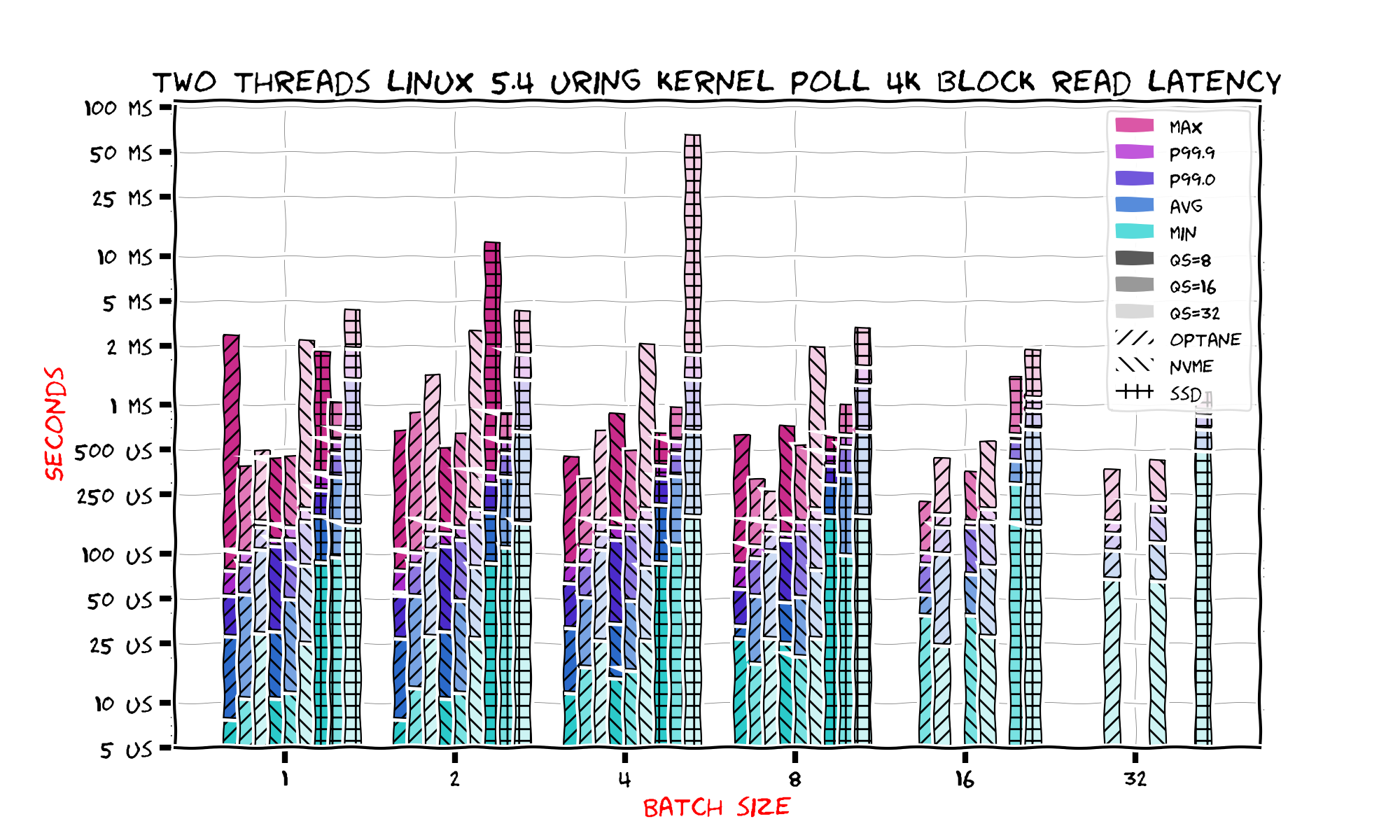}
    \caption{Two-thread uring readingwith kernel poll thread latency.}
    \label{fig:alluringpollthread2batch:latency}
\end{figure}

At first we try to vary the queue size.
Figures~\ref{fig:alluringpollqueue:throughput} and~\ref{fig:alluringpollqueue:cpu} show throughput and CPU usage respectively while figures~\ref{fig:alluringpollthread1queue:latency} and~\ref{fig:alluringpollthread2queue:latency} shows latencies for single and two threads.
We pick queue sizes 8, 16, and 32 as the most interesting.
Let's look how batch size affects the performance.
Figures~\ref{fig:alluringpollbatch:throughput},~\ref{fig:alluringpollbatch:cpu},~\ref{fig:alluringpollthread1batch:latency}, and~\ref{fig:alluringpollthread2batch:latency} show the results.


There are a few options for Optane with throughput more than 2 gigabytes per second.
However all of them lead to latency more than 50 microseconds in 99.9 percentile.
We choose queue size 16, batch size 2 and two threads.
Queue size 8 would give slightly better latency but throughput will be worse.
The reader should consider this option if the latency is more important.
The CPU usage is slightly above 150\% of a single core.

For NVMe SSD throughput chart shows that we need to choose 32 elements queue and two threads.
All other variants fall behind.
Now we need to select batch size from 1 to 8.
Latencies for all these four variants look alike and are around 200 microseconds.
So we pick batch size 8 by the lowest CPU usage which is slightly above 250\% of a single core.

For SSD we pick 16 elements queue, single thread, and batch size 4.
This leads to maximum throughput, latency is about 300 microseconds and two full cores of CPU time.

For HDD we again take single element queue and single thread.
The selected parameters are summarized in the table below.

\begin{center}
\begin{tabular}{|l|l|l|l|}
\hline
Storage & Threads & Queue Size & Batch size \\ \hline
Optane  & 2       & 16         & 2 \\ \hline
NVMe    & 2       & 32         & 8 \\ \hline
SSD     & 1       & 16         & 4 \\ \hline
HDD     & 1       & 1          & 1 \\ \hline
\end{tabular}
\end{center}

% Single thread for Optane: all-upg1m-throughput.png

\subsection{Executing with selected parameters}

\begin{figure}
    \includegraphics[width=\textwidth]{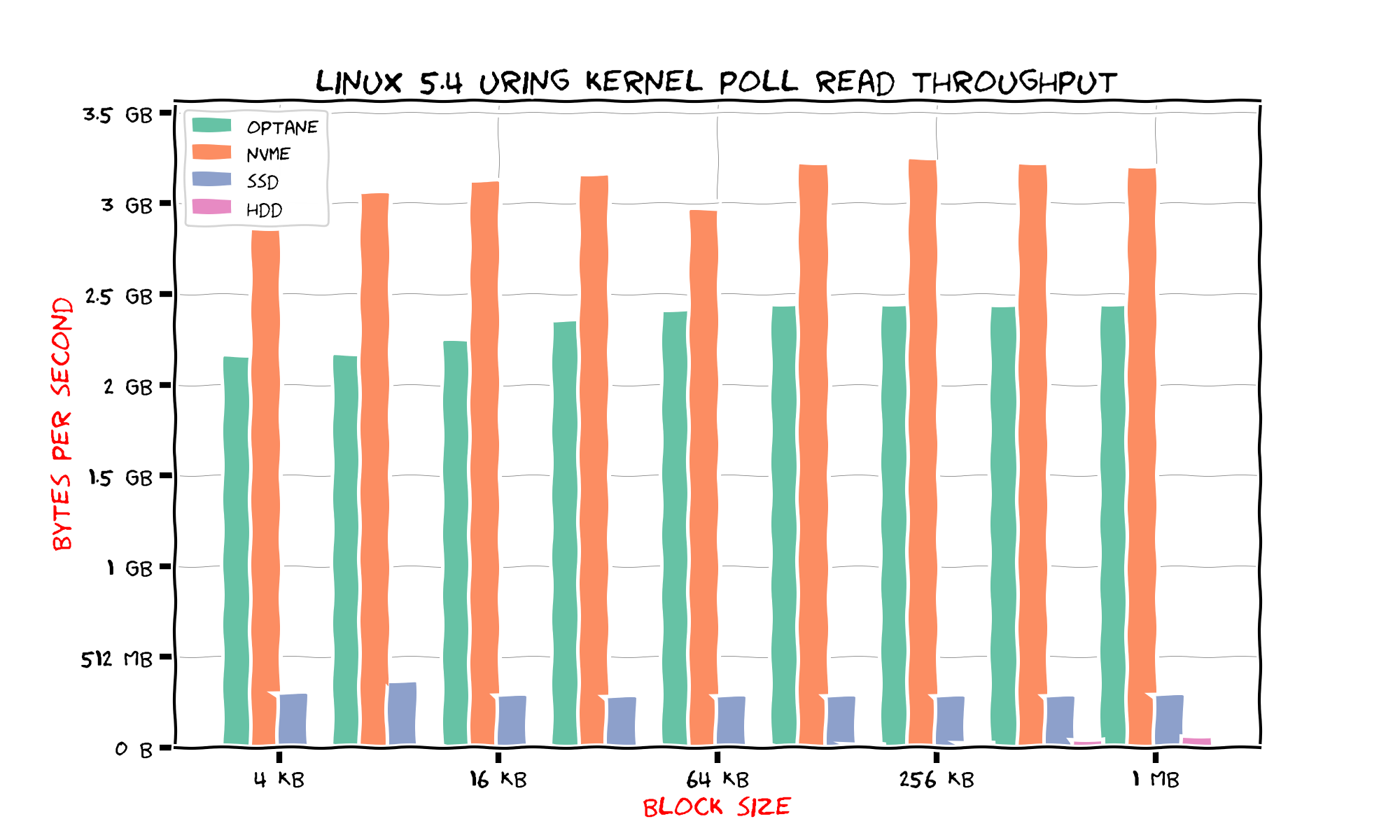}
    \caption{Best uring reading with kernell poll thread throughput.}
    \label{fig:alluringpollbest:throughput}
\end{figure}

\begin{figure}
    \includegraphics[width=\textwidth]{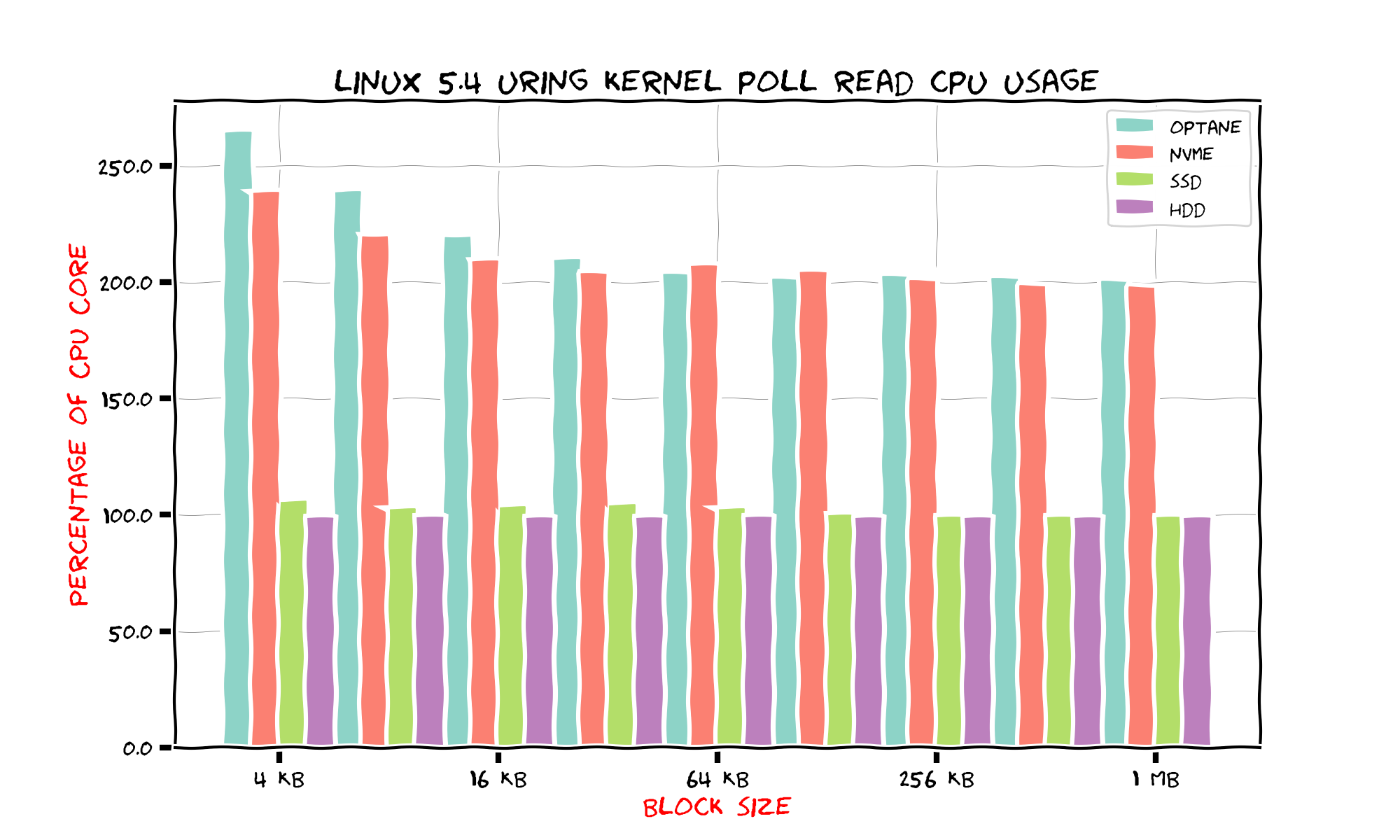}
    \caption{Best uring reading with kernel poll thread CPU usage.}
    \label{fig:alluringpollbest:cpu}
\end{figure}

\begin{figure}
    \includegraphics[width=\textwidth]{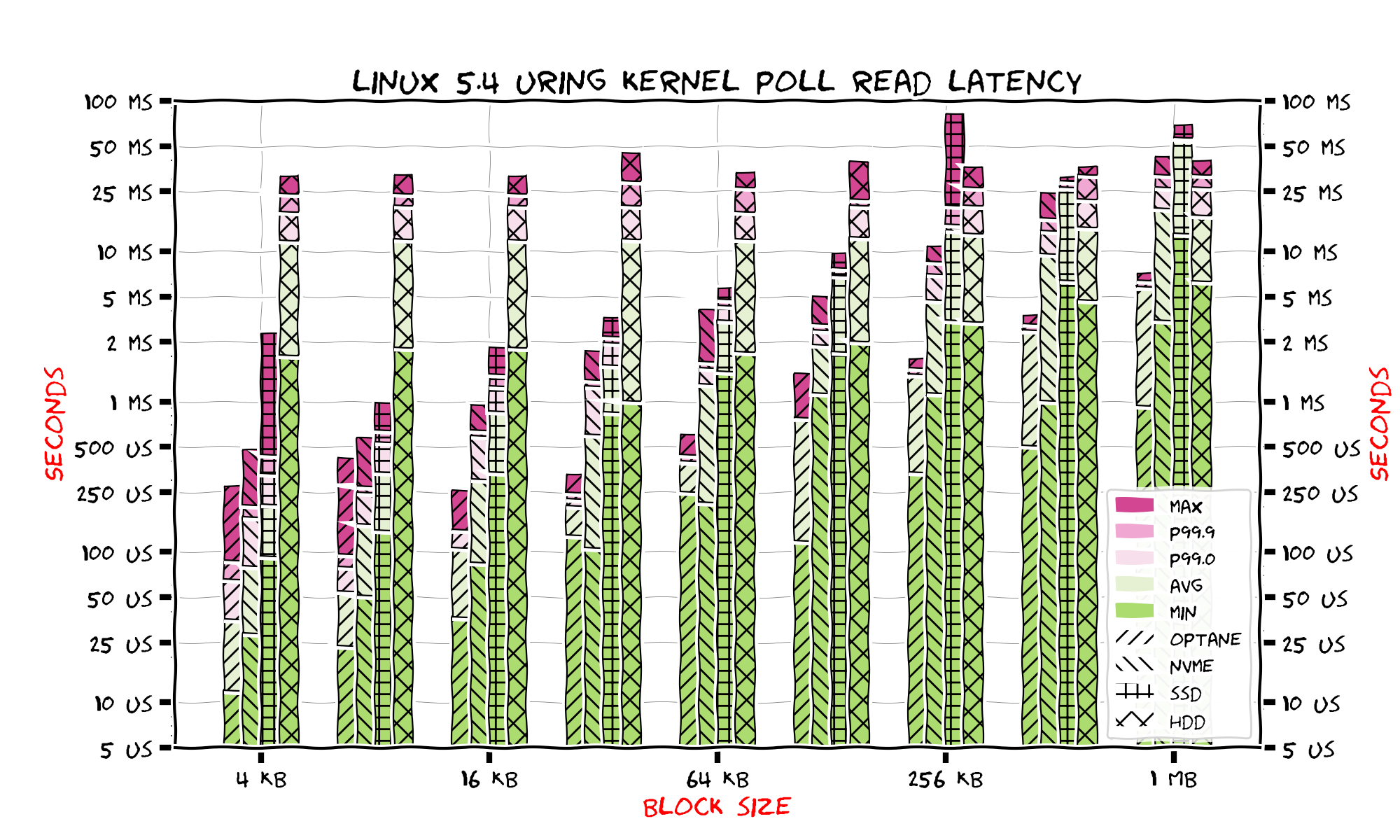}
    \caption{Best uring reading with kernel poll thread latency.}
    \label{fig:alluringpollbest:latency}
\end{figure}

In this subsection we present the results of running our experiments with the selected parameters.
Figures~\ref{fig:alluringpollbest:throughput},~\ref{fig:alluringpollbest:cpu}, and~\ref{fig:alluringpollbest:latency} show throughput, CPU usage and latency respectively.

At first let's look at throughput.
There is an annoying drop for 64 kilobytes blocks.
Our parameters have been chosen for 4 kilobytes blocks and may be not so good for other block sizes.
For other block sizes the results are similar to those without the kernel poll thread.

As for the CPU usage everything is bad now.
Kernel poll threads consume 100\% of a core and where the usage was close to zero previously we now see 100\% or even 200\%.
Thus using kernel poll thread leads to more CPU consumption.

Finally, let's look at latency.
For Optane latency is worse for 4 kilobytes blocks.
We were able to reach 50 microseconds in 99.9 percentile without using kernel poll thread.
Now this latency is about 100 microseconds.
Results for NVMe SSD look similar.
There is an improvement for SSD: 30 microseconds against 500.
However the throughput is noticeably worse now.
Just in case recall that for SSD it is better to read with synchronous interface from several threads.
It is possible to reach 500 megabytes per second for large block sizes.
However we were unable to repeat this result with asynchronous interfaces.

\subsection{Summary}

We investigated additional \texttt{uring} capabilities and observed that they don't help much.
Nevertheless it may be worth to use fixed files and buffers to reduce maximum latency.
On the other hand enabling kernel poll thread could lead to a fiasco.

Now we've finished our overview of Linux input-output interfaces.
In the next section we give a brief summary of all our experiments.

\section{All experiments combined}
\label{sec:summary}

We looked at both modern external memory hardware and Linux application interfaces to read from them.
Intel Optane shows the lowest latency of 12 microsecond.
NVMe SSD shows the highest throughput of 3.2 gigabytes per second.
It appears that under load Optane and NVMe SSD demonstrate latencies within the factor of two from each other (for a single read the difference is by the order of magnitude).
So we could say that they perform as devices from the same class.

In previous sections we studied extensively program interfaces to access the storage.
It turns out that it is not easy to pick an appropriate interface and to select adequate parameters.
Our experiments result in lots of data only fraction of which was selected for the presentation.
Now we try to show all our data on a few plots.

We are interested in three parameters: throughput, latency and CPU usage.
We usually concern with 99.9 percentile and the CPU usage is a second class parameter compared to throughput and latency.
Therefore we show our data on a plot where axes are throughput and latency 99.9 percentile.
Instead of points we use figures which shape represent block size and color represent CPU usage.
For each block size we picked a few best choices and made their figures larger.

To find out the specific setting we use labels that encode the interface, number of threads and queue parameters.
First letter represents the interface: ``P'' for \texttt{pread}, ``A'' for Linux aio, and ``U'' for \texttt{uring}.
For asynchronous interfaces next is queue size followed by ``B'' and batch size.
For \texttt{uring} there could be additional ``F'' and ``M'' which stand for fixed files and buffers respectively.
Finally any label can end with ``T'' followed by the number of threads if executed from more than one thread.

\begin{figure}
    \includegraphics[width=\textwidth]{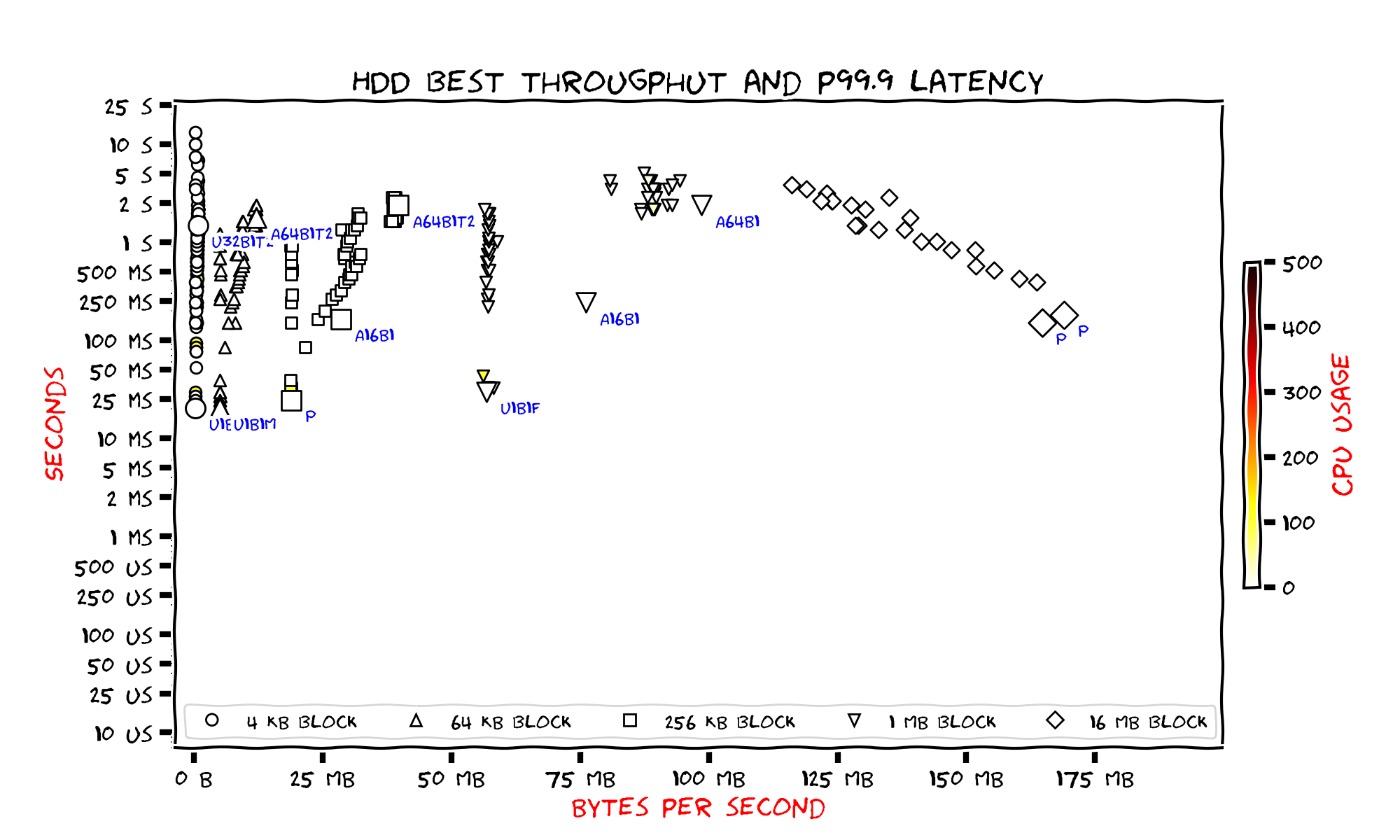}
    \caption{All experiments for HDD.}
    \label{fig:everything:hdd}
\end{figure}

Figure~\ref{fig:everything:hdd} shows results for HDD.
Now it is really straightforward to realize that for HDD it is better to read with synchronous interface from a single thread.
It should not be surprising that the same setting could result in slightly different points on the plot.
Especially for HDD where everything depends on luck and data placement.

Another interesting thing to notice is that larger queue size leads to more throughput for asynchronous interfaces.
This should not be surprising since larger queue allows more possibility for reordering.
On the other hand the latency could be as large as a few seconds.
The best example here is 1 megabyte block and settings U1B1F, A16B1, and A64B1.
One could also notice the same behavior for smaller blocks. 

It is not an error that almost all figures are white.
In our experiments reading from HDD we never utilizes more than 5\% of a single core.
The notable exception is the case when \texttt{uring} kernel poll thread is enabled which results in 100\% utilization of a single core.
Careful eye should be able to find a few yellow figures.

\begin{figure}
    \includegraphics[width=\textwidth]{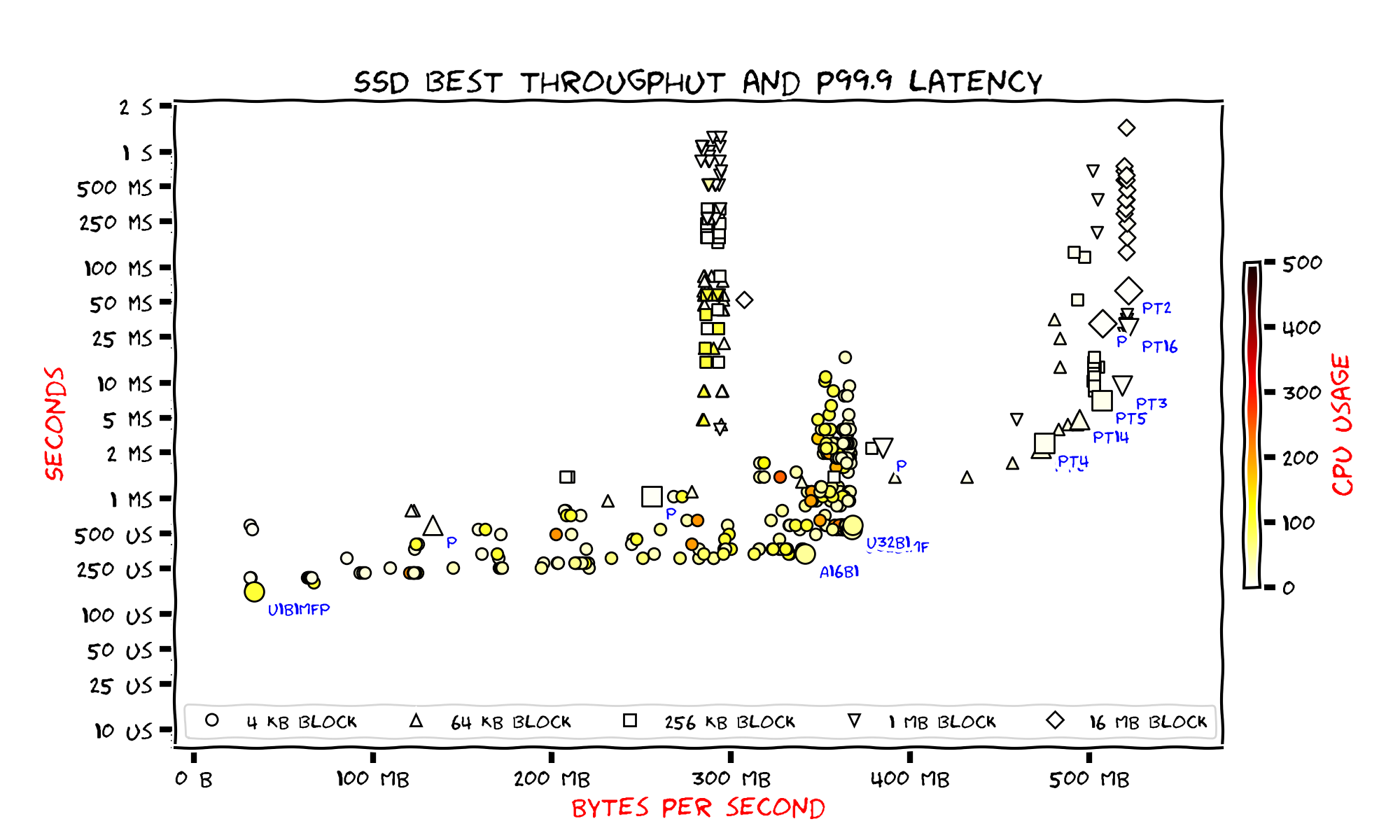}
    \caption{All experiments for SSD.}
    \label{fig:everything:ssd}
\end{figure}

Figure~\ref{fig:everything:ssd} shows results for SSD.
There are three noticeable clusters on the plot: one with the largest throughput about 500 megabytes per second, another one which holds indistinguishably best combination of throughput and latency for 4 kilobytes blocks, and a strange vertical cluster with throughput 300 megabytes per second and large latency.
It is straightforward to notice that one can reach 500 megabytes per seconds only when reading with block size larger than 4 kilobytes.
It's better to use synchronous interface from multiple threads in this case.
For 4 kilobytes blocks the best choice is \texttt{uring} with queue size 32 and batch size 1 (label U32B1).
An interesting observation is that 32 is exactly the number of requests in NCQ, the hardware request queue.
The latency is about 500 microseconds.
Neighbor label A16B1 shows better latency which is 250 microseconds but this comes with the sacrifice of some throughput.

\begin{figure}
    \includegraphics[width=\textwidth]{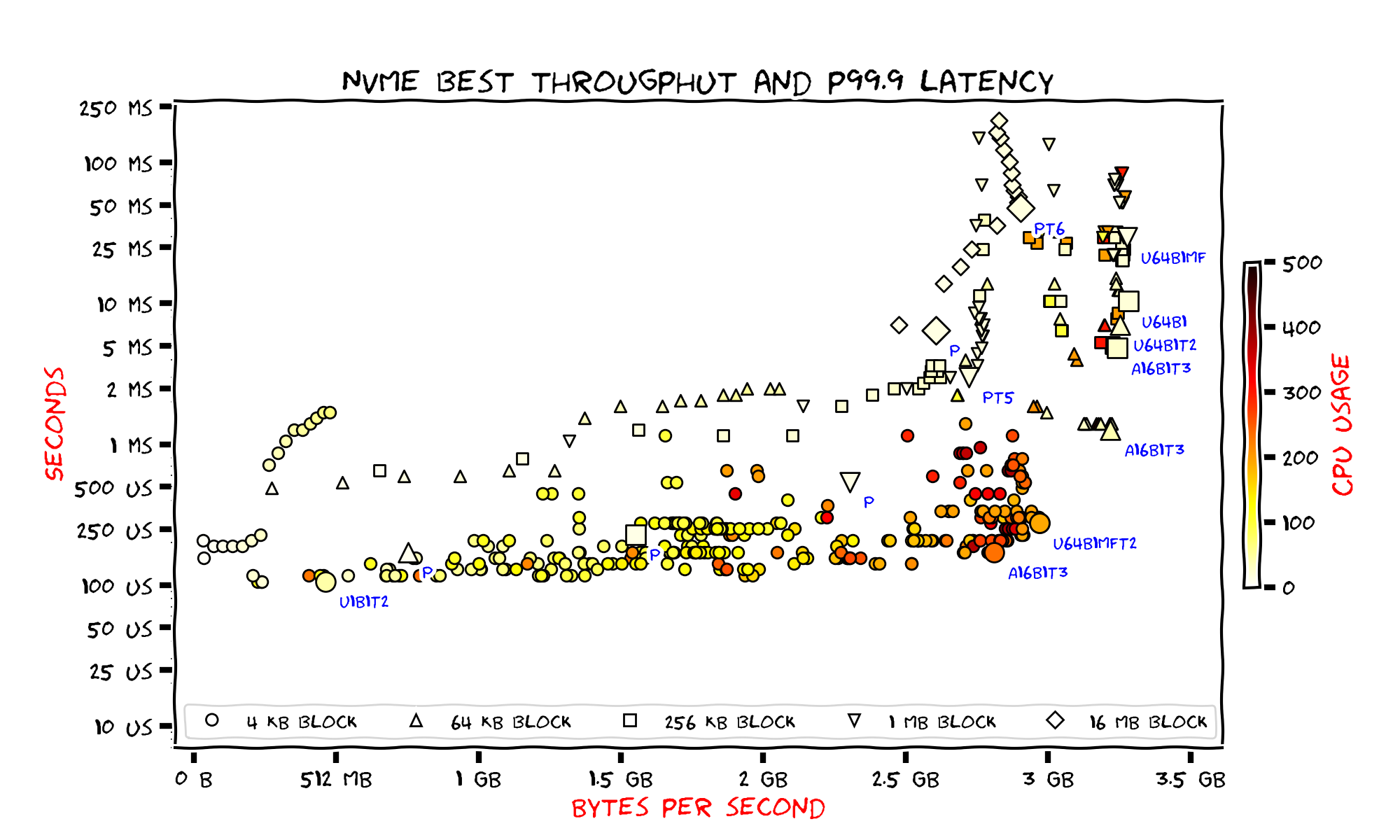}
    \caption{All experiments for NVMe SSD.}
    \label{fig:everything:nvme}
\end{figure}

Figure~\ref{fig:everything:nvme} shows all our experiments for NVMe SSD.
One could also note that the peak throughput of 3.2 gigabytes per second is possible only for large enough blocks.
This looks similar to SSD however this time we need to use asynchronous interface from multiple threads.

For 4 kilobytes blocks there are two interesting points labeled by A16B1T3 and U64B1MFT2.
One can choose between them according to their desired trade-off between throughput and latency.
If CPU usage is also of concern it should be noted that one of the circles is darker than the other which represents more CPU consumption.

\begin{figure}
    \includegraphics[width=\textwidth]{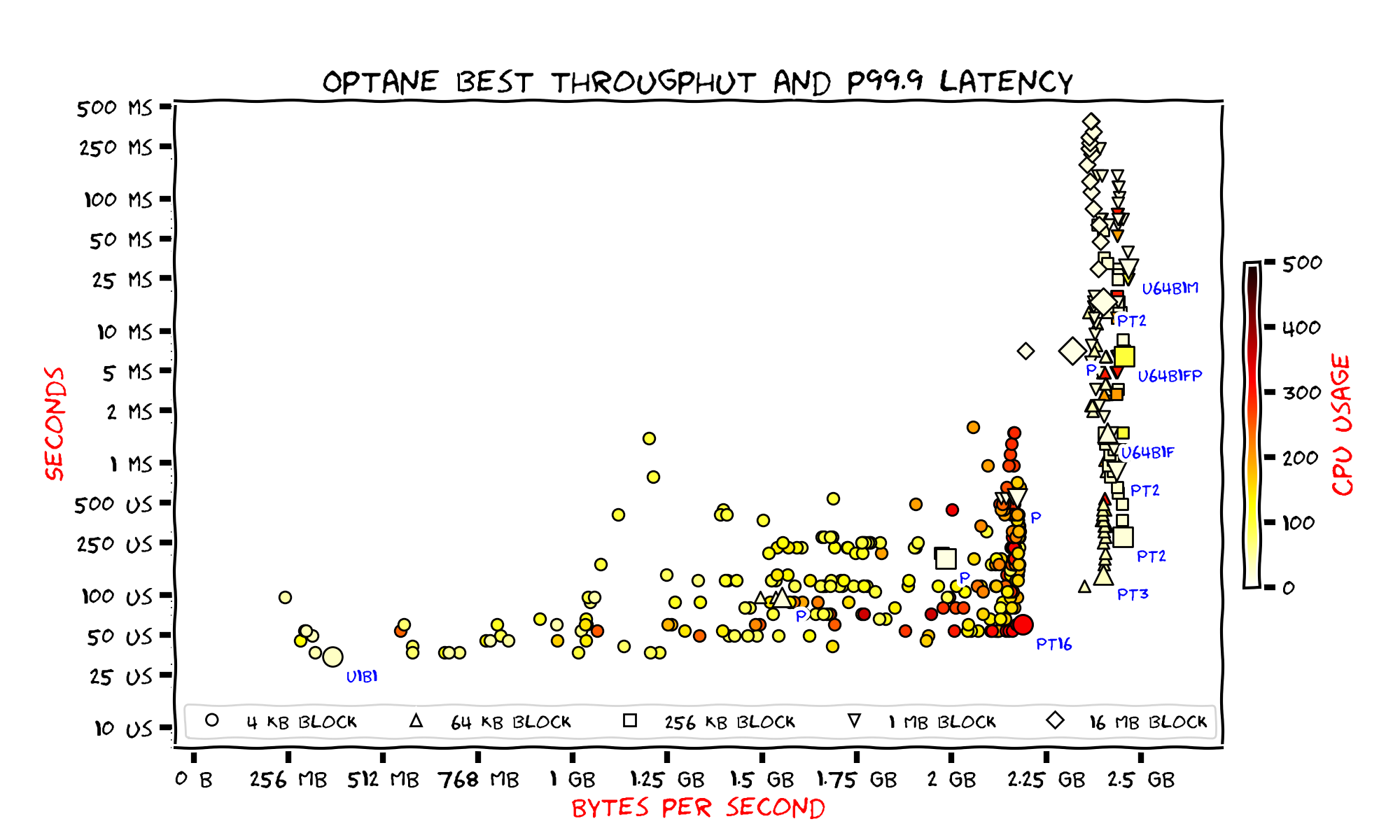}
    \caption{All experiments for Optane.}
    \label{fig:everything:optane}
\end{figure}

Figure~\ref{fig:everything:optane} shows results for Optane.
Once again we see that the maximum throughput overall is by some step further than the maximum throughput for 4 kilobytes blocks.
Here the absolute winner is multi-threaded synchronous interface.
For blocks larger than 4 kilobytes a few threads are enough.
We need three threads for 64 kilobytes blocks, single thread for 16 megabytes blocks and two threads for all other block sizes in the middle.
For 4 kilobytes blocks we need 16 threads.
The CPU consumption is equal to three full CPU cores but with respect to trade-off between throughput and latency this point looks as the unrivaled best option.

This concludes our presentation. In the next section we cover related work.

%VReading from external memory turned out to be hard.
%We think our main contribution is to bring some evidence that the task of reading from external memory is unexpectedly hard.
%There are various storage types and various interfaces.
%There is no exact winner and it may be difficult to make an appropriate choice for a specific task.
%Probably the best an application writer could do is to implement several methods and choose between them. % either automatically or manually.

\section{Related Work}
\label{sec:relatedwork}

In this section we give a brief overview of recent storage performance reports and related materials.
Benchmarking is quite a broad topic~\cite{10.1145/1367829.1367831}.
Many technical websites publish measurements of new devices when they appear.
Here we focus mostly on the academic papers.

% devices

% hdd

HDD internal operation is described in~\cite{Ruemmler94Modeling} and~\cite{Anderson03SCSIvsATA} with some (outdated) performance characteristics.
Preference for sequential workload has been stated in \cite{Schlosser04mems} in the form of \emph{unwritten contract} for filesystems and applications.
Internal geometry reconstruction through microbenchmarking has received attention ever since explicit cylinder-head-sector interface has been replaced by logical block
addressing:~\cite{Schindler1999AutomatedDD}, \cite{Talagala1999MicrobenchmarkbasedEO}, \cite{10.1145/1807060.1807063}, and recently \cite{HDDmicro}.
HDD performance heavily depends on scheduling~\cite{Worthington1994SchedulingAF}. Mentioned Linux BFQ scheduler is presented in~\cite{BFQ2010}.
Shingled disks, which appeared in the last decade as a response to increased capacity demands, are studied in~\cite{Shingle2015}.

% ssd

Mass market SSDs are made of NAND flash memory cells~\cite{SSDwork}.
NAND flash doesn't exactly match the block-device interface and a complex software called \emph{flash translation layer} is running as a part of a device firmware.
Different algorithms for flash translation layer are surveyed in~\cite{FTLSurvey2014}.
Although writing is not covered in our report at all, the major concern when working with SSD is write performance~\cite{Stoica2013FlashWrite} and endurance~\cite{Boboila2010FlashEndurance}.

In~\cite{He2017TheUC} authors describe SSD architecture and present some patterns that developers should follow.
They further investigate them using filesystems and databases as examples.
Since SSD behave differently from HDD a new benchmarking methodology has been purposed as a set of representative I/O patterns~\cite{Bouganim09uflip}.
Raw NAND memory cell has been benchmarked in~\cite{Desnoyers10Empirical}.

% nvme

First performance comparison which simultaneously studies HDD, SSD, and NVMe SSD appears in~\cite{Xu2015PerformanceAO}.
They state that NMVe SSD would be useful for databases.
They also conclude that the kernel stack (at the time of their writing) should be rewritten to comply with NVMe SSD.

There is a demand to move flash translation layer to a host. \emph{Open-channel} SSDs with exposed internals became available.
Since there are more options to coordinate with an application such a device can demonstrate better performance in particular for a mixed workload~\cite{Bjorling2017LightNVM}.

% Optane

Intel Optane SSD is a notable example of \emph{ultra-low latency SSD}.
Its performance is investigated in~\cite{Wu2017EarlyEwO}, however is only compared to HDD.
In~\cite{Wu2019TowardsAU} authors study peculiarities of Optane SSD and try to deduce some rules for developers.
There is an interesting use-case where Optane SSD is used as a replacement for a DRAM cache~\cite{Eisenman2018DRAMFootprint}.
While the problem itself is more involved detailed storage benchmarks are also provided.

% Optane PMM
Intel Optane SSD and Optane PMM are both based on the same technology~\cite{FoongOptane2016,HadyOptane2017}.
DIMM-located Optane PMM is studied in~\cite{Basic-Optane-PMM} and~\cite{OSTI-Optane-PMM}.
The former provides reader with detailed latencies and throughputs with comparison to both volatile memory and SSD.
It further proceeds with benchmarking of various filesystems and databases.
The latter investigates the appliance of such a storage for memory-consuming graph algorithms.

% interfaces

Linux block I/O has been studied for a long time~\cite{LinuxBlockIO2004}.
Asynchronous I/O was presented in~\cite{Bhattacharya2003AsynchronousIS} and analyzed in~\cite{Bhattacharya2004LinuxAP}.
Databases such as ScyllaDB and MySQL successfully adopted asynchronous I/O.
ScyllaDB developers described investigation for the best I/O method in their blog~\cite{ScyllaIO}.

Uring motivation and description is presented in~\cite{Uring-Intro}.
Although it's new it has been already started gaining polularity~\cite{ScyllaUring}.
To the best of our knowledge our report is the first detailed comparison between uring and other methods.
Since uring is still in active development we expect more benchmarks to appear in the future.

Justification for polled I/O can be found on Linux kernel mail list~\cite{PollIO}.
Prototype study appeared earlier in the literature~\cite{Yang12Poll}.
Although polling reduce average latency it can severely damage high percentile~\cite{Koh2018UltraLow}.
Delayed polling which saves CPU cycles has also been considered~\cite{Eisenman2018DRAMFootprint}.

New kernel queue for block I/O is described in~\cite{Bjrling2013LinuxBI}.
Recently, schedulers were added to this queue as well~\cite{Corbet2017}.
There is an attempt to reduce latency even further by overlapping block I/O path with itself~\cite{Gyusun2019AsyncSSD}.
It should be noted however that other layers of the sotfware stack can affect latency and break the expectations~\cite{EnlighteningIOKim}.

Another approach is to use completely userspace framework such as SPDK~\cite{SPDK2017} or NVMeDirect~\cite{Kim2016NVMeDirectAU}.
Howerver this would require using an SSD as a raw device or pick a user-level filesystem such as BlobFS~\cite{BlobFS}.
It is yet an open challenge to create a high-performance system-wide userspace filesystem~\cite{LiuFsProc2019}.

\bibliography{bibl} 

\end{document}